\RequirePackage[l2tabu, orthodox]{nag} 

\documentclass[11pt, a4paper, oneside]{book}

\usepackage[utf8]{inputenc}
\usepackage[main = english]{babel} 			
\usepackage[T1]{fontenc}							

\usepackage[ 
		layoutoffset=0pt,						
		bottom=4cm, 
		top=2.5cm,	
		left=4cm, 
		right=2.5cm,			
		bindingoffset=0pt,						
		headheight=26pt				
		]{geometry}	
					
\usepackage{microtype, fancyhdr}	 
\usepackage{setspace}					 
\usepackage{pdflscape} 					 

\usepackage{amsfonts,amsmath,amsthm,amssymb, mathrsfs, amscd}	 
\numberwithin{equation}{chapter}	
\usepackage{faktor}
\allowdisplaybreaks

\usepackage{url, enumitem} 			
\usepackage[noadjust]{cite}				
\usepackage{epigraph}						
\usepackage[nottoc]{tocbibind}			
\usepackage{blindtext}						
\usepackage{imakeidx}					
\makeindex

\usepackage{graphicx} 				
\graphicspath{{Pictures/}}				
\usepackage{xcolor}
\usepackage{longtable} 				
\usepackage{lscape} 					
\usepackage{booktabs} 				
\usepackage{multicol} 					
\usepackage{multirow}		
\usepackage{wrapfig} 					

\usepackage{rotating}
\usepackage[percent]{overpic}

\let\headruleORIG\headrule
\renewcommand{\headrule}{\color{black} \headruleORIG}

\usepackage{colortbl}
\arrayrulecolor{black}

\fancypagestyle{plain}{
  \fancyhead{}
  \fancyfoot{}
  
}

\fancypagestyle{otherplain}{
  \fancyhead{}
  \fancyfoot[C]{\thepage}
  
}

\makeatletter
\def\cleardoublepage{\clearpage\if@twoside \ifodd\c@page\else%
    \hbox{}%
    \thispagestyle{empty}
    \newpage%
    \if@twocolumn\hbox{}\newpage\fi\fi\fi}
\makeatother

\usepackage{etoolbox}
\patchcmd{\chapter}{plain}{otherplain}{}{}
\makeatletter
  \let\ps@plain\ps@otherplain
\makeatother

\usepackage[
hidelinks,bookmarksnumbered
]{hyperref} 
\usepackage{breqn} 
\usepackage{pdfpages} 
\usepackage{float}
\usepackage{parskip}

\theoremstyle{plain}

\theoremstyle{remark}

\theoremstyle{definition}

\def\blankpage{%
      \clearpage%
      \thispagestyle{empty}%
      \addtocounter{page}{-1}%
      \null%
      \clearpage}

\begin{document}
\pagenumbering{roman}


\begin{titlepage}

\newgeometry{left=40mm, right=40mm, top=40mm, bottom=40mm}

\begin{center}

\begin{spacing}{2.5}
{\Huge \textsc{\textbf{Strong gravity beyond General Relativity}}}
\end{spacing}

\vspace{1.0 cm}

\begin{center}
\end{center}
\vspace{3cm}

{\LARGE Llibert Aresté Saló} 

\vspace{2cm}

{\large \sffamily
Submitted in partial fulfillment of the
requirements \\ of the Degree of Doctor of Philosophy\\
\vspace{5cm}
School of Mathematical Sciences\\
Queen Mary University of London\\
\vspace{0.5cm}
March 2024 
}

\end{center}

\end{titlepage}

\restoregeometry  
\doublespacing		

\addtocounter{page}{2}%
\blankpage

\renewcommand{\baselinestretch}{1.0}\normalsize
\chapter*{Declaration}
\label{C:Statement}
\addcontentsline{toc}{chapter}{\nameref{C:Statement}}

I, Llibert Aresté Saló, confirm that the research included
within this thesis is my own work or that where it has been carried out in
collaboration with, or supported by others, that this is duly acknowledged
below and my contribution indicated. Previously published material is also
acknowledged below.

I attest that I have exercised reasonable care to ensure that the work is
original, and does not to the best of my knowledge break any UK law, infringe
any third party’s copyright or other Intellectual Property Right, or contain any
confidential material.

I accept that the College has the right to use plagiarism detection software to
check the electronic version of the thesis.

I confirm that this thesis has not been previously submitted for the award of a
degree by this or any other university.

The copyright of this thesis rests with the author and no quotation from it or
information derived from it may be published without the prior written consent
of the author.

Signature: 

\

\

Date: $28^{\text{th}}$ March 2024




\chapter*{List of publications}
\label{C:Publications}
\addcontentsline{toc}{chapter}{\nameref{C:Publications}}

\vspace{-0.53cm}

Chapters \ref{C:Wellposedness} and \ref{C:BBHin4dST} of this thesis are based on the following manuscripts:
\begin{itemize}
    \item L.~Arest\'e Sal\'o, K.~Clough and P.~Figueras,
\textit{Well-posedness of the Four-Derivative Scalar-Tensor Theory of Gravity in Singularity Avoiding Coordinates}, 
Phys. Rev. Lett. \textbf{129}, 261104 (2022) [arXiv:2208.14470 [gr-qc]].
    \item L.~Arest\'e Sal\'o, K.~Clough and P.~Figueras,
\textit{Puncture gauge formulation for Einstein-Gauss-Bonnet gravity and four-derivative scalar-tensor theories in $d+1$ spacetime dimensions}, 
Phys. Rev. D \textbf{108}, 084018 (2023) [arXiv:2306.14966 [gr-qc]].
\item L.~Arest\'e Sal\'o, K.~Clough, D.~D.~Doneva, P.~Figueras and S.~S.~Yazadjiev,
\textit{Unequal-mass binaries in the Four-Derivative Scalar-Tensor theory of gravity}, to appear.
\end{itemize}
Chapter \ref{C:Hyp} is based on
\begin{itemize}
    \item D.~D.~Doneva, L.~Arest\'e Sal\'o, K.~Clough, P.~Figueras and S.~S.~Yazadjiev,
\textit{Testing the limits of scalar-Gauss-Bonnet gravity through nonlinear evolutions of spin-induced scalarization}, 
Phys. Rev. D \textbf{108}, 084017 (2023) [arXiv:2307.06474 [gr-qc]].
\end{itemize}

On the other hand, all the numerical results obtained in this thesis and in the papers mentioned above have been obtained through our code \texttt{GRFolres}, of which I have been the core developer, which is reviewed in Chapter \ref{C:GRFolres}. It has recently been made open-source and all its capabilities are described in the following publication,
\begin{itemize}    
    \item L.~Arest\'e Sal\'o, S.~E.~Brady, K.~Clough et al
    , \textit{GRFolres: A code for modified gravity simulations in strong gravity}, Journal of Open Source Software \textbf{9}, 6369 (2024) [arXiv:2309.06225 [gr-qc]].
\end{itemize}
\texttt{GRFolres} is an extension of the \texttt{GRChombo} code. I am as well a member and developer of the GRTL Collaboration, in charge of maintaining this code, as for the following paper, 
\begin{itemize}
\item T.~Andrade, L.~Arest\'e Sal\'o, J.C.~Aurrekoetxea et al, 
\textit{GRChombo: An adaptable numerical relativity code
for fundamental physics}, Journal of Open Source Software \textbf{6}, 3703 (2021) [arXiv:2201.03458 [gr-qc]].
\end{itemize}

Finally, I participated in several other projects related to the topic of this thesis (but not included in it) as a co-author, namely
\begin{itemize}
    \item S.~E.~Brady, L.~Arest\'e Sal\'o, K.~Clough, P.~Figueras and A.~P.~S, \textit{Solving the initial conditions problem for modified gravity theories},
Phys. Rev. D \textbf{108}, 104022 (2023) [arXiv:2308.16791 [gr-qc]].
    \item D.~D.~Doneva, L.~Arest\'e Sal\'o and S.~S.~Yazadjiev,
\textit{$3+1$ nonlinear evolution of Ricci-coupled scalar-Gauss-Bonnet gravity}, Phys. Rev. D \textbf{110}, 024040 (2024) [arXiv:2404.15526 [gr-qc]]. 
\end{itemize}

\chapter*{Abstract}
\label{C:Abstract}
\addcontentsline{toc}{chapter}{\nameref{C:Abstract}}


This thesis focuses on the application of numerical relativity methods to the solutions of problems in strong gravity. Our goal is the study of mergers of compact objects in the strong field regime where non-linear dynamics manifest and deviations from General Relativity could be found.

We develop a new formulation of the Einstein equations in $d+1$ spacetime dimensions in the moving punctures approach, which leads to a well-posed set of equations for the Einstein-Gauss-Bonnet gravity (EGB), as well as for the most general parity-invariant scalar-tensor theory of gravity up to four derivatives ($4\partial$ST).

Using this formulation, we have implemented the equations of the $4\partial$ST theory in \texttt{GRFolres}, an open-source extension of our numerical relativity code \texttt{GRChombo}. This has enabled us to evolve equal and unequal-mass Binary Black Hole mergers in this effective field theory of gravity, as well as to study the loss of hyperbolicity  and its relation to the weak coupling condition, among other topics of interest left for further study.


 
\chapter*{Acknowledgements}
\label{C:Acknowledgements}
\addcontentsline{toc}{chapter}{\nameref{C:Acknowledgements}}

First I want to start especially thanking Pau Figueras and Katy Clough for all their guidance and unconditional support during all the stages of my Ph.D. I have learned from them not only all the physics and how to always challenge myself but also in general how to be a successful academic. 
I am really grateful for having been introduced to their extensive network, which has enabled me to start really fruitful collaborations.

I am really proud of having been part of the GAnG centre in Queen Mary. I have enjoyed all our social activities, meetings, coffee breaks and the relaxed and friendly atmosphere within it. I owe a huge thanks to my academic older siblings -Tiago, Lorenzo and Chenxia, who have always been available to offer their help, as well as the younger ones -Sam, Shunhui and Areef-, with whom I have had very useful discussions. I have also benefited very much from discussion with other members of the group: \'Aron, Juan, Mahdi... And I cannot forget to thank all the rest of colleagues from maths who contributed to the friendly atmosphere of the Ph.D. office.

Throughout the last stage of my Ph.D. I have had the pleasure to work closely with external collaborators, among them Daniela Doneva, Dina Traykova, Miguel Bezares and Thomas Sotiriou, who I also want to acknowledge. I owe as well a special thanks to all the \texttt{GRChombo} collaboration, which has been really welcoming from the very beginning and which has always provided me a great support. 

On the non-academic side, it has been an honour to form part during all those years of \textit{Castellers of London}. I have really enjoyed all our activities and I have found there my family in London. Finally, I am as well very grateful to all the rest of my friends that have kept the contact with me throughout this time, with whom I have shared many unforgettable moments. Last but not least, I want to express my gratitude to my parents, brother and the rest of my family, who have always supported me in this path.

The simulations presented used the ARCHER2 UK National Supercomputing Service\footnote{\href{https://www.archer2.ac.uk}{\texttt{https://www.archer2.ac.uk}}} under the EPSRC HPC project no. E775. 
I acknowledge the use of Athena at HPC Midlands+, which was funded by the EPSRC on grant EP/P020232/1, in this research, as part of the HPC Midlands+ consortium, 
as well as the Gauss Centre for Supercomputing e.V.\footnote{\href{www.gauss-centre.eu}{\texttt{www.gauss-centre.eu}}} for providing computing time on the GCS Supercomputer SuperMUC-NG at Leibniz Supercomputing Centre.\footnote{\href{www.lrz.de}{\texttt{www.lrz.de}}}
This work used the DiRAC Data Intensive service (DIaL2 / DIaL3) at the University of Leicester, managed by the University of Leicester Research Computing Service on behalf of the STFC DiRAC HPC Facility.\footnote{\href{www.dirac.ac.uk}{\texttt{www.dirac.ac.uk}}} The DiRAC service at Leicester was funded by BEIS, UKRI and STFC capital funding and STFC operations grants. DiRAC is part of the UKRI Digital Research Infrastructure.
Calculations were performed using the Sulis Tier 2 HPC platform hosted by the Scientific Computing Research Technology Platform at the University of Warwick. Sulis is funded by EPSRC Grant EP/T022108/1 and the HPC Midlands+ consortium. 
This research also utilised Queen Mary’s Apocrita HPC facility, supported by QMUL Research-IT \cite{apocrita}. 
For some computations we have also used the Young Tier 2 HPC cluster at UCL; we are grateful to the UK Materials and Molecular Modelling Hub for computational resources, which is partially funded by EPSRC (EP/P020194/1 and EP/T022213/1).


\renewcommand{\contentsname}{Table of contents}
\tableofcontents
\listoffigures
\listoftables

\addtocounter{page}{1}%
\blankpage

\pagenumbering{arabic}

\part{Background material}

\addtocounter{page}{1}%
\blankpage


\chapter{Introduction}
\label{C:Intro}

\section{Motivation}

More than one hundred years after it was proposed, Einstein's gravitational theory of General Relativity (GR) remains the key pillar of modern cosmology, the physics of black holes and neutron stars and all other gravity-related fields. 

All tests carried out so far have found GR to be an accurate description of gravitational phenomena. However, these cover mainly the weak field regime and only very recently have we started to probe the strong field regime where non-linear dynamics manifest, in particular from the observation of the mergers of compact objects, after the first detections of gravitational waves (GWs) \cite{LIGOScientific:2016aoc,LIGOScientific:2016lio,LIGOScientific:2016dsl,LIGOScientific:2017vwq,LIGOScientific:2018dkp} released by the LIGO-Virgo Collaboration (LVC), which is now the LIGO-Virgo-KAGRA (LVK) Collaboration. 

In order to interpret these waveforms, Numerical Relativity (NR) has become an essential tool, allowing us to predict them from the theory itself using computational power. In comparison to perturbative approaches (post-Newtonian theory, general relativistic perturbation theory...), NR allows us to test the regime where the highest curvatures manifest, as long as we employ a well-posed formulation. This guarantees that, given some suitable initial data, the solution to the equations of motion exists, is unique and depends continuously on the initial data. 

Finding a well-posed formulation has been a challenge for NR since its beginnings a few decades ago. Multiple different formulations with several gauge choices, constraint damping methods and initial data solvers have been required before being able to evolve the first binary black hole evolution in 2005 \cite{Pretorius:2005gq}. From then on, other well-posed formulations have been found to be successful in NR for not only evolving black hole spacetimes and other compact objects in GR, but also for being applied to cosmology and modified theories of gravity.

Current observations indicate that the deviations from GR in the strong field regime are small \cite{LIGOScientific:2021sio}. Therefore, it makes sense to consider theories arising as small modifications of GR and effective field theory (EFT) provides an organising principle.  In EFT, one adds all possible terms allowed by symmetry to the leading order GR Lagrangian. These terms are organised in a derivative expansion and appear multiplied by dimensionful coupling constants that encode the effects of the underlying (unknown) microscopic theory. 

Motivated by an EFT approach, GR can be modified by adding new field content or invariants among other possibilities and, in this sense, some particular classes of higher derivative theories of gravity that have received attention in recent years are the Lovelock theories \cite{Lovelock:1971yv} in the case of pure gravity, and the Horndeski theories \cite{Horndeski:1974wa} in the case of scalar-tensor theories. Lovelock theories are especially relevant in high energy physics, since they arise in the low energy limit of string theory, and Horndeski theories have been widely studied in cosmology given that scalar fields are already used in models of inflation, dark matter and dark energy. Observations both from electromagnetic and GW detections have recently enabled to constrain the values of the parameters in the Horndeski Lagrangian, in particular by comparing the speed of propagation of the GWs with the one predicted from the theory.

Progress in numerically studying these theories has been made for Horndeski theories of gravity using order-reduced methods that  evolve the scalar equation of motion on a fixed GR background \cite{Richards:2023xsr,R:2022tqa,Okounkova:2022grv,Elley:2022ept,Doneva:2022byd,Okounkova:2020rqw,Silva:2020omi,Okounkova:2019zjf,Okounkova:2019dfo,Witek:2018dmd,Evstafyeva:2022rve}. These do not have any well-posedness issues, as long as a certain regularity in the background metric is satisfied
. Such simulations can provide an estimate of the scalar dynamics and associated energy losses, but may miss information about the fully non-linear impact on the metric, and potentially suffer from the accumulation of secular errors over long inspirals. Despite their limitations, these studies put initial constraints on the coupling from the merger signal, and have identified many interesting effects such as dynamical descalarisation \cite{Silva:2020omi}, in which initially scalarised black holes form an unscalarised remnant, and so-called ``stealth dynamical scalarisation'' \cite{Elley:2022ept}, in which the late merger of initially unscalarised black holes can cause scalar
hair to grow due to the spin of the remnant. \footnote{These works typically neglect the four-derivative scalar term, which we see from our work is justified since it is always subdominant to the effect of the Gauss-Bonnet term.}

Both Lovelock and Horndeski theories have second order equations of motion, and hence there was hope that suitable well-posed formulations could be found. There have been attempts to cure the loss of hyperbolicity by ‘fixing’ the system of evolution equations \cite{Franchini:2022ukz,Cayuso:2023aht,Lara:2024rwa} or by modifying the theory, e.g., by adding an extra coupling of the scalar field to the Ricci scalar \cite{Liu:2022fxy,Thaalba:2023fmq}. In parallel, Kov\'acs and Reall have shown that these theories are indeed well-posed in the weak coupling regime in a modified version of the harmonic gauge \cite{Kovacs:2020pns,Kovacs:2020ywu}. Subsequently, work has begun to numerically study some specific scalar-tensor theories within these classes in their highly dynamical and fully non-linear regimes \cite{East:2020hgw,East:2021bqk,East:2022rqi,Corman:2022xqg}.

These studies rely on the generalised harmonic coordinates, which are appealing because of the manifest wave-like structure of the equations, but their practical implementation in numerical simulations necessitates excision of the interior domain of the black hole. The latter, whilst conceptually straightforward, can be difficult to implement in practise. As a consequence, many groups in the numerical relativity community have opted to use singularity avoiding coordinates such as the BSSN \cite{Nakamura:1987zz,Shibata:1995we,Baumgarte:1998te}, Z4C \cite{Bona:2003fj,Bernuzzi:2009ex} or CCZ4 \cite{Alic:2011gg} formulations in the puncture gauge \cite{Campanelli:2005dd,Baker:2005vv}, which do not require the explicit excision of the interior of black holes from the computational domain. Instead, excision is achieved by the suitable use of dynamical coordinates. This strongly motivates the extension of the results of \cite{Kovacs:2020pns,Kovacs:2020ywu} to singularity avoiding coordinates, to allow such groups to generate waveforms in these models.

In this thesis we develop and implement a new formulation of NR to singularity avoiding coordinates that allows us to study the effect of modified gravity in the strong gravity regime, as previously presented in \cite{AresteSalo:2022hua,AresteSalo:2023mmd,Doneva:2023oww,AresteSalo:2023hcp}. This contribution will play a vital role in the application of NR tools to scenarios beyond GR and the Standard Model and will be crucial in the characterisation of signals coming from black hole mergers (and other compact objects) in such scenarios. 

We follow the conventions in Wald’s book \cite{Wald:1984rg}. Greek letters $\mu$, $\nu$... denote spacetime indices and they run from 0 to $d$ (being $d$ the number of spatial dimensions); Latin letters i, j,... denote indices on the spatial hypersurfaces and they run from 1 to $d$. We set $G=c=1$.

\section{Outline of the thesis}

We start with two introductory chapters, Chapters \ref{C:MG} and \ref{C:NR}, which comprise an overview of the main concepts of Modified Gravity, with a special emphasis in Lovelock and Horndeski theories of gravity, and Numerical Relativity, where we define the main NR formulations, briefly review gauge conditions and initial data and introduce the concepts of strong hyperbolicity and well-posedness.

The original research work of this thesis starts in Chapter \ref{C:Wellposedness}, whee we discuss extensively about well-posedness, introduce our formulation and prove that it is well-posed in the weak coupling regime of Einstein-Gauss-Bonnet gravity (EGB) and the Four-Derivative Scalar-Tensor theory of gravity ($4\partial$ST). In Chapter \ref{C:BBHin4dST} we present the main results regarding Binary Black Hole mergers in the $4\partial$ST theory and Chapter \ref{C:Hyp} deals with the hyperbolicity loss of this theory when going beyond the weak coupling regime. Finally we include in Chapter \ref{C:GRFolres} a review of the capabilities of \texttt{GRFolres}, our NR code in modified gravity.

We summarise our conclusions and discuss about further directions in Chapter \ref{C:Conclusions}. Finally we include in the Appendices long computations and equations that complement Chapter \ref{C:Wellposedness} of our manuscript, as well as a convergence test for one of our simulations presented in Chapter \ref{C:BBHin4dST}.

\addtocounter{page}{1}%
\blankpage

\chapter{Modified Gravity}
\label{C:MG}

Despite its successful history, General Relativity is a classical theory which is not yet known to admit a quantised description. Therefore, we expect it to break down at higher energy scales than those that have been tested so far, particularly in those for which an extension or modification would be needed in order to properly understand the behaviour of gravity, such as singularities.

\section{Classification of modified gravity theories}

\begin{figure}[hbt]
\centering
\includegraphics[scale=0.92]{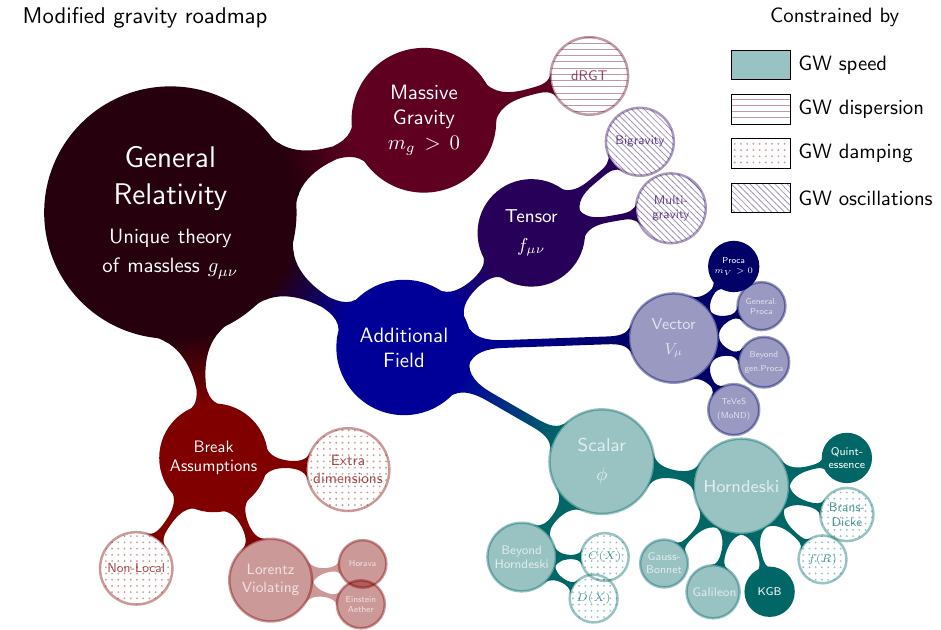}
\caption[Modified gravity road-map.]{Modified gravity road-map showing a classification of the various alternative branches and theories of gravity coming from possible extensions of GR. Figure taken from \cite{Ezquiaga:2018btd}.}
\label{F:modgrav}
\end{figure}

In recent decades, numerous different extensions or modifications to General Relativity, driven by various theoretical motivations, have been suggested. 
Lovelock’s theorem states that GR is the unique four-dimensional, local, second derivative theory for a massless spin-2 field \cite{Lanczos:1938sf,Lovelock:1971yv,Lovelock:1972vz}.
Therefore, modifications to GR require one of these ``pillars'' to be broken. 
In Figure \ref{F:modgrav} we display a schematic categorisation of some of those theories, which can be classified into the following types:
\begin{itemize}
    \item Adding new field content: It is often considered the easiest and most natural approach and the new fields can be either:
\begin{itemize}
    \item Scalar fields, which can be both real and complex. Some examples are Horndeski gravity or Cherns-Simons gravity.
\item Vector fields, such as Einstein-\ae ther gravity or generalised Proca gravity.
\item Tensor fields, such as massive gravity, bigravity and bimetric MOND (Modified Newtonian Dynamics) gravity.
\end{itemize}
\item Adding invariants, which includes adding extra
dimensions, such as Lovelock theories, or adding
higher order terms to the action, such as $f(R)$, quadratic gravity or Ho\v{r}ava-Lifshitz gravity, some of which break locality.
\item Emergent approaches of gravity, which are based on the idea that gravity is not fundamental but instead can be derived from some underlying structure. Some examples are causal dynamical triangulation (CDT) and loop quantum gravity (LQG).
\end{itemize}

\section{Effective field theories}

Among all these possibilities, we are especially interested in those that can be seen as effective field theories with GR as their low energy limit. From a minimal perspective, one might consider the addition of higher derivatives of the metric to be the most well-motivated.

In order to understand this, let's look at a complex scalar field $\Phi$ with a spontaneously-broken $U(1)$ symmetry with the following action \cite{Burgess:2014lwa,Solomon:2017nlh,Allwright:2018rut}
\begin{equation}
    S=\int d^4x[-\partial_{\mu}\Phi^{\dagger}\partial^{\mu}\Phi-V(\Phi^{\dagger}\Phi)]\,,
\end{equation}
with the potential given by
\begin{equation}
V(\Phi)=\frac{\lambda}{2}(\Phi^{\dagger}\Phi-v^2)\,.
\end{equation}
which is symmetric under $\Phi\to e^{i\omega}\Phi$, but for which the symmetry is spontaneously broken in the vacuum $\Phi^{\dagger}\Phi=v^2/2$. The spectrum of fluctuations about this solution contains a massive
mode with $M^2 = \lambda v^2$ and a massless Goldstone boson, which can be made explicit by defining two real fields $\rho$ and $\theta$ such that 
\begin{equation}
    \Phi=\frac{v}{\sqrt{2}}(1+\rho)e^{i\theta}\,.
\end{equation}
Then we can integrate out the heavy field $\rho$ by solving perturbatively in $M$ its equation of motion, namely
\begin{equation}
    \Box\rho = (1+\rho)(\partial_{\mu}\theta)(\partial^{\mu}\theta)+V'(\rho)\,,
\end{equation}
and inserting this solution back into the action to obtain a theory
for $\theta$ alone, leaving us with the action
\begin{align}
    \frac{S}{v^2}=&~\int d^4x\Big[-\frac{1}{2}(\partial_{\nu}\theta)(\partial_{\nu}\theta)+\frac{1}{2M^2}[(\partial_{\nu}\theta)(\partial_{\nu}\theta)]^2\nonumber\\
    &\hspace{1.4cm}-\frac{2}{M^4}(\partial_{\mu}\partial_{\nu}\theta)(\partial^{\mu}\partial^{\sigma}\theta)(\partial^{\nu}\theta)(\partial_{\sigma}\theta) +{\mathcal O}\Big(\frac{1}{M^6}\Big)\Big]\,,
\end{align}
where we have truncated the action at some finite order in $1/M$, which is valid for energies well below the mass of $\rho$, $E\ll M$. This is the key point of the effective field theory, namely that the $\Box\rho$ part of the equation of motion becomes subdominant for those energies. When one probes higher energies, we exit the regime of validity of the EFT and one has to add more terms in the expansion. The resulting equation of motion that comes from this EFT yields
\begin{align}\label{highorder}
    \Box\theta=&~ \frac{2}{M^2}\big[(\partial_{\nu}\theta)(\partial^{\nu}\theta)\Box\theta + 2(\partial_{\mu}\partial_{\nu}\theta)(\partial^{\nu}\theta)(\partial^{\mu}\theta) \big]\nonumber\\
    &+\frac{4}{M^4}\big[(\partial^{\nu}\partial^{\sigma}\Box\theta)(\partial_{\nu}\theta)(\partial_{\sigma}\theta)+(\partial^{\sigma}\Box\theta)(\Box\theta)(\partial_{\sigma}\theta)+(\partial^{\sigma}\Box\theta)(\partial^{\nu}\theta)(\partial_{\nu}\partial_{\sigma}\theta)\nonumber\\
    &\hspace{1.2cm}+(\partial^{\mu}\partial^{\sigma}\theta)(\Box\theta)(\partial_{\mu}\partial_{\sigma}\theta)+2(\partial^{\nu}\partial^{\mu}\partial^{\sigma}\theta)(\partial_{\mu}\partial_{\nu}\theta)(\partial_{\sigma}\theta)\big]+{\mathcal O}\Big(\frac{1}{M^6}\Big)\,.
\end{align}

Hence, we see that the term ${\mathcal O}(M^{-4})$ contains third- and fourth-order derivatives
of $\theta$ to the equation of motion, which can lead to runaway instability. However, we cannot produce the ghost associated to this instability while we remain within the regime of validity of the EFT, meaning that the ghost can only be excited (thus inducing a runaway instability) when probing energies high enough, which would require us to consider more terms in the expansion.

Note that when evolving the equation of motion for long timescales, one has to take into account the scale of the time in order to assure that we are within the regime of validity of the EFT \cite{Reall:2021ebq}. When considering a time interval $[0,T)$ with $T$ growing with $M$ (for example $T\propto M^{\lambda}$ for some $\lambda>0$) it could be that the accuracy of the EFT approximation is lessened (see Section 2.7 of \cite{Reall:2021ebq} for further details).

This explicitly shows that an EFT with high derivatives is not unphysical, given that it can be the result of the truncation of a healthy UV physical theory. In some cases it is possible to address, to some extent, certain potential sources of ill-posedness arising from EFTs. Indeed, through field redefinitions higher-order derivatives can be removed, or pushed to higher orders. In this case, one can use a field redefinition to push the higher-order time derivatives in Eq. \eqref{highorder} to the next order in the EFT expansion, namely by considering \cite{Allwright:2018rut}
\begin{equation}
    \theta\to\theta+\frac{2}{M^4}(\partial_{\mu}\partial_{\nu}\theta)(\partial^{\mu}\theta)(\partial^{\nu}\theta)\,.
\end{equation}

After having discussed this EFT toy model, let's return to the theories of gravity. The example just discussed shows us that it makes sense to express any EFT as an expansion of terms with increasing mass dimension (or increasing number of derivatives). Therefore, we consider the addition of all possible terms allowed by symmetry to the leading order GR Lagrangian and organise these terms in a derivative expansion which appear multiplied by dimensionful coupling constants that encode the effects of the underlying (unknown) microscopic theory. 

When we add the restriction of second order equations of motion, so that we avoid the appearance of  Ostrogadskii instabilities \cite{Woodard:2015zca,Motohashi:2014opa,Donoghue:2021eto}, we are left with the broad families of Lovelock and Horndeski theories of gravity, which will be defined in the next sections. In particular, if we cut the EFT expansion to four derivatives of the metric, we are led (up to field redefinitions) to their respective subclasses Einstein-Gauss-Bonnet gravity (EGB) and the Four-Derivative Scalar-Tensor theory of gravity ($4\partial$ST), which will be the main focus of this thesis.

\section{Lovelock theories of gravity}

Lovelock theories of gravity \cite{Lovelock:1971yv} are the most general theories in which the gravitational field is described by a single metric tensor satisfying a diffeomorphism invariant second order equation of motion. They have lately attracted attention due to their possible relevance to high energy physics \cite{Camanho:2014apa} and cosmology \cite{Bueno:2024fzg}. 

Lovelock theories of gravity describe string theory-inspired ultraviolet corrections to the Einstein-Hilbert action. In particular, the five-dimensional Lovelock theory arises in the low energy limit of M-theory (and, consequently, of string theory) when the theory is compactified from 11D to 5D \cite{Garraffo:2008hu}.

The black hole solutions of Lovelock theory are also especially relevant. They can be regarded as generalisations of the Boulware-Deser solution of Einstein-Gauss-Bonnet gravity \cite{Boulware:1985wk} and possess many features that are not present in GR.

The recent results studying the causality and hyperbolicity of Lovelock gravity \cite{Reall:2014pwa,Reall:2021voz,Kovacs:2020ywu,AresteSalo:2023mmd} will enable to further study its implications and test its relevance.

\subsection{Definition}

In vacuum, Lovelock's equation of motion is \cite{Kovacs:2020ywu}
\begin{align}
G^{\mu}_{~\nu}+\Lambda\delta^{\mu}_{\nu}+\sum\limits_{p\geq2}k_p\,\delta^{\mu\rho_1...\rho_{2p}}_{\nu\sigma_1...\sigma_{2p}}R_{\rho_1\rho_2}^{~~~~\sigma_1\sigma_2}...R_{\rho_{2p-1}\rho_{2p}}^{~~~~~~~~~\rho_{2p-1}\rho_{2p}}=0\,,
\end{align}
where $g_{\mu\nu}$ is the metric tensor\footnote{All spacetime indices are raised and lowered with the metric $g_{\mu\nu}$.}, $G_{\mu\nu}=R_{\mu\nu}-\frac{1}{2}g_{\mu\nu}R$ is the Einstein tensor\footnote{The Riemann tensor, Ricci tensor and Ricci scalar are respectively
\begin{subequations}
\begin{align}
    R^{\rho}_{~\sigma\mu\nu}&=\partial_{\mu}\Gamma^{\rho}_{~\nu\sigma}-\partial_{\nu}\Gamma^{\rho}_{~\mu\sigma}+\Gamma^{\rho}_{~\mu\lambda}\Gamma^{\lambda}_{~\nu\sigma}-\Gamma^{\rho}_{~\nu\lambda}\Gamma^{\lambda}_{~\mu\sigma}\,,\\
    R_{\mu\nu}&=R^{\rho}_{~\mu\rho\nu}\,, \\
    R&=g^{\mu\nu}R_{\mu\nu}\,,
\end{align}
\end{subequations}
with $\Gamma^{\rho}_{\mu\nu}=\tfrac{1}{2}g^{\rho\sigma}(\partial_{\mu}g_{\nu\sigma}+\partial_{\nu}g_{\sigma\mu}-\partial_{\sigma}g_{\mu\nu})$ being the Christoffel symbols associated to the metric $g_{\mu\nu}$.}, $\Lambda$ is the cosmological constant, $k_p$ are dimensionful coupling constants and the generalised Kronecker delta is $\delta^{\rho_1...\rho_q}_{\sigma_1...\sigma_q}=q!\delta^{\rho_1}_{[\sigma_1}\delta^{\rho_2}_{\sigma_2}...\delta^{\rho_q}_{\sigma_q]}$. 

Note that, because of the antisymmetrisation, those equations are only different to GR when the number of spacetime dimensions is higher than four.

\subsection{Einstein-Gauss-Bonnet gravity}

Einstein-Gauss-Bonnet gravity (EGB) is the particular case of Lovelock theories of gravity when setting $k_p=0$  $\forall p>2$, which comes from the action
\begin{align}\label{eq:action_Lovelock}
    S=\tfrac{1}{2\kappa}\int d^{d+1}x\sqrt{-g}(R+\lambda^{\text{GB}}\,{\mathcal L}^{\text{GB}})\,,
\end{align}
where $\kappa=8\pi G$, ${\mathcal L}^{\text{GB}}=\frac{1}{4}\delta^{\mu_1\mu_2\mu_3\mu_4}_{\nu_1\nu_2\nu_3\nu_4}R_{\mu_1\mu_2}^{~~~~~\nu_1\nu_2}R_{\mu_3\mu_4}^{~~~~~\nu_3\nu_4}=R_{\mu\nu\rho\sigma}R^{\mu\nu\rho\sigma}-4R_{\mu\nu}R^{\mu\nu}+R^2$ is the Gauss-Bonnet curvature, $d$ is the number of spatial dimensions,  and $\lambda^{\text{GB}}$ is an arbitrary coupling with dimensions of $[\text{length}]^2$. 

This theory is especially relevant from our viewpoint given that it can be motivated as an effective field theory in the sense that it includes all the four-derivative terms (up to field redefinitions) \cite{Burgess:2003jk}. The Gauss-Bonnet term is also predicted by some string theories \cite{Camanho:2014apa}.

The equations of motion that follow from varying \eqref{eq:action_Lovelock} are given by 
\begin{align}
    G^{\mu\nu}=\lambda^{\text{GB}}{\mathcal H}^{\mu\nu}\,,
\end{align}
where
\begin{align} \label{eq:Hmunu_EGB}
    {\mathcal H}_{\mu\nu}=-2\big(R\,R_{\mu\nu}-2\,R_{\mu\alpha}\,R^{\alpha}_{~\nu}-2\,R^{\alpha\beta}\,R_{\mu\alpha\nu\beta}+R_{\mu}^{~\alpha\beta\gamma}\,R_{\nu\alpha\beta\gamma}\big)+\tfrac{1}{2}g_{\mu\nu}{\mathcal L}^{\text{GB}}\,.
\end{align}

\section{Horndeski theories of gravity}

Horndeski theories of gravity \cite{Horndeski:1974wa} are the most general theories of a metric tensor coupled to a scalar field $\phi$,
with second order equations of motion, arising from a diffeomorphism-invariant action in four spacetime dimensions. They have been widely studied both from  cosmology \cite{Baker:2019gxo,Baker:2020apq} and strong gravity \cite{Richards:2023xsr,R:2022tqa,Okounkova:2022grv,Elley:2022ept,Doneva:2022byd,Okounkova:2020rqw,Silva:2020omi,Okounkova:2019zjf,Okounkova:2019dfo,Witek:2018dmd,Evstafyeva:2022rve,East:2020hgw,East:2021bqk,East:2022rqi,Corman:2022xqg,AresteSalo:2022hua,AresteSalo:2023mmd,Doneva:2023oww} perspectives. 

The addition of a scalar field degree of freedom that couples to gravity is especially relevant in a cosmological set-up given that scalar fields are already used in models of inflation, dark matter and dark energy. Furthermore, the Horndeski Lagrangian provides a concrete, yet flexible, system in which to connect electromagnetic (EM) and gravitational wave (GW) constraints on a common set of parameters. Its
parameters have been widely tested using electromagnetic observables \cite{Bellini:2015xja,Kreisch:2017uet,Noller:2018wyv,SpurioMancini:2019rxy}. 

Some of the additional terms in the Horndeski action imply a speed of propagation of the GWs different to the velocity of light, which has been constrained by the recent GW observations.  
In particular, the LIGO-Virgo's Binary Neutron Star Merger GW170817 \cite{LIGOScientific:2017vwq,LIGOScientific:2017ync} bounded the fractional relative difference in propagation
speeds of GWs and light to be less than $10^{-15}$ \cite{LIGOScientific:2017ync,Baker:2017hug,Ezquiaga:2017ekz,Creminelli:2017sry,Sakstein:2017xjx,Boran:2017rdn,Mastrogiovanni:2020gua} at redshift zero. And this has led to the conclusion that those terms in the Horndeski action are no longer viable \cite{Baker:2019gxo}.

However, according to de Rham and Melville \cite{deRham:2018red}, when a dark energy EFT has the typical parameter values needed to give it interesting dynamics on cosmological scales, the energy scale of GW170817
potentially lies within its strongly coupled regime. Moreover, if a low-energy effective theory is to admit a Lorentz-invariant completion, one would actually expect the action of
operators above the strong coupling scale to return the speed of tensor modes to the velocity of light. This raises the possibility that the constraints reported above were obtained in a regime where
modified gravity effects are already suppressed, and may differ at other scales such as those probed by LISA.

In the context of black holes solutions, Horndeski theories of gravity are also particularly interesting given that they violate general relativity's no-hair theorem \footnote{The no-hair theorem states that all stationary black hole solutions of the Einstein–Maxwell equations of gravitation and electromagnetism in general relativity can be completely characterised by only three independent externally observable classical parameters: mass, electric charge, and angular momentum.}, giving rise to black hole solutions with a non-trivial configuration of the scalar field.

\subsection{Definition}

The explicit form of Horndeski's action yields
\begin{align}
    {\mathcal S}=\tfrac{1}{2\kappa}\int d^4x\sqrt{-g}\left({\mathcal L}_1+{\mathcal L}_2+{\mathcal L}_3+{\mathcal L}_4+{\mathcal L}_5\right)\,,
\end{align}
where \footnote{The d'Alembertian operator is defined as $\Box=g^{\mu\nu}\nabla_{\mu}\nabla_{\nu}$, where $\nabla_{\mu}$ is the covariant derivative that acts on an arbitrary tensor as
\begin{align}    \nabla_{\rho}T^{\mu_1...\mu_m}_{\qquad~~~\nu_1..\nu_n}=\;&\partial_{\rho}T^{\mu_1...\mu_m}_{\qquad~~~\nu_1..\nu_n}+\Gamma^{\mu_1}_{\rho\sigma}T^{\sigma...\mu_m}_{\qquad~~\nu_1...\nu_n}+...+\Gamma^{\mu_m}_{\rho\sigma}T^{\mu_1...\sigma}_{\qquad~\nu_1...\nu_n}\nonumber\\&-\Gamma^{\sigma}_{\rho\nu_1}T^{\mu_1...\mu_m}_{\qquad~~~\sigma...\nu_n}-...-\Gamma^{\sigma}_{\rho\nu_m}T^{\mu_1...\mu_m}_{\qquad~~~\nu_1...\sigma}\,.
\end{align}}
\begin{subequations}
\begin{align}
{\mathcal L}_1&=R+X-V(\phi)\,, \\
{\mathcal L}_2&=G_2(\phi,X)\,,\label{eqg2}\\
{\mathcal L}_3&=G_3(\phi,X)\Box\phi\,,\\
{\mathcal L}_4&=G_4(\phi,X)R+\partial_XG_4(\phi,X)\left[(\Box\phi)^2-(\nabla_{\mu}\nabla_{\nu}\phi)(\nabla^{\mu}\nabla^{\nu}\phi)\right]\,,\\
{\mathcal L}_5&=G_5(\phi,X)G_{\mu\nu}\nabla^{\mu}\nabla^{\nu}\phi-\frac{1}{6}\partial_XG_5(\phi,X)\left[(\Box\phi)^3-3\Box\phi(\nabla_{\mu}\nabla_{\nu}\phi)(\nabla^{\mu}\nabla^{\nu}\phi)\right.\nonumber\,\\
&\hspace{6.5cm}\left.+2(\nabla_{\mu}\nabla_{\nu}\phi)(\nabla^{\nu}\nabla^{\rho}\phi)(\nabla_{\rho}\nabla^{\mu}\phi)\right]\,,
\end{align}    
\end{subequations}
where $\{G_i(\phi,X)\}_{i=2}^5$ are arbitrary functions, with $X=-\tfrac{1}{2}(\partial\phi)^2\equiv -\frac{1}{2}g^{\mu\nu}\partial_{\mu}\phi\,\partial_{\nu}\phi$ and $V(\phi)$ being the potential of the scalar field. 

\subsection{Four-Derivative Scalar-Tensor theory of gravity}

For some particular choice of the functions $\{G_i(\phi,X)\}_{i=2}^5$ \cite{Kobayashi:2011nu}, namely
\begin{subequations}
    \begin{align}
        G_2(\phi,X)=&~g_2(\phi)X^2+8\lambda^{(4)}X^2(3-\ln X)\,,\\
        G_3(\phi,X)=&-4\lambda^{(3)}X(7-3\ln X)\,,\\
        G_4(\phi,X)=&~4\lambda^{(2)}X(2-\ln X)\,,\\
        G_5(\phi,X)=&-4\lambda^{(1)}\ln X\,,
    \end{align}
\end{subequations}
where $\lambda^{(n)}:=\partial^n\lambda/\partial\phi^n$, one gets the action of the Four-Derivative Scalar-Tensor theory of gravity ($4\partial$ST) \cite{Kovacs:2020ywu}, which yields
\begin{align}\label{eq:4dST}
    S=\tfrac{1}{2\kappa}\int d^4x\sqrt{-g}\left(R+X-V(\phi)+g_2(\phi)X^2+\lambda(\phi){\mathcal L}^{GB}\right)\,,
\end{align}
where $\lambda(\phi)$ and $g_2(\phi)$ \footnote{The notation $g_2(\phi)$ has been used since this term comes from the Horndeski term \eqref{eqg2} with $G_2(\phi,X)=g_2(\phi)X^2$.  For $g_2(\phi)=V(\phi)=0$ we are left with the well-known Einstein-scalar-Gauss-Bonnet (EsGB) theory of gravity.}  are arbitrary functions of $\phi$.

The Four-Derivative Scalar-Tensor theory of gravity is an effective field theory given that it contains all possible parity-invariant scalars (up to field redefinitions) in four dimensions built from the metric $g_{\mu\nu}$ and a scalar field $\phi$ minimally coupled to gravity up to four derivatives \cite{Weinberg:2008hq}. It was first introduced when studying generic theories of inflation with a single inflaton field \cite{Weinberg:2008hq}, in which the terms quadratic in the curvature arise from the leading corrections to the correlation function for tensor modes. For an inflationary theory with an inflaton $\phi$ with potential $V(\phi)$ minimally coupled to the Einstein-Hilbert action, the leading correction to the Lagrangian will consist of a sum of all generally covariant terms with four spacetime derivatives and coefficients of order unity, which yields by getting rid of total derivatives the following expression,
\begin{align} \label{correction4dst}
    \Delta{\mathcal L}&=\sqrt{-g}\Big[f_1(\phi)\,X^2+f_2(\phi)\,\Box\phi +f_3(\phi)\,(\Box\phi)^2+f_4(\phi)\,R^{\mu\nu}\,\nabla_{\mu}\phi\,\nabla_{\nu}\phi+f_5(\phi)\,R\,X\nonumber\\ 
    &+ f_6(\phi)\,R\,\Box\phi+f_7(\phi)\,R^2+f_8(\phi)\,R^{\mu\nu}R_{\mu\nu}+f_9(\phi)\,{\mathcal L}^{\text{GB}}
    +f_{10}(\phi)\,\epsilon^{\mu\nu\rho\sigma}R_{\mu\nu}^{~~\kappa\lambda}R_{\rho\sigma\kappa\lambda}\Big]\,,
\end{align}
where $\{f_i(\phi)\}_{i=1}^{10}$ are arbitrary functions of $\phi$. One can see that with some redefinitions we can eliminate all of the terms above except the first one and the last
two. Specifically, the second term in Eq. \eqref{correction4dst} just provides a field-dependent correction to the kinematic term, which can be eliminated by a redefinition of the inflaton field; the third term just provides a correction
to the potential, which can be absorbed into a redefinition of $V(\phi)$; the fourth and fifth terms supply corrections to both $f_1(\phi)$ and the kinematic term; the sixth term provides corrections to the
kinematic term and the potential; and the seventh and eighth terms
provide corrections to the kinematic term and potential and to $f_1(\phi)$. Furthermore, the last term vanishes if one assumes parity-invariance. Hence, we obtain the action in Eq. \eqref{eq:4dST}.

Its equations of motion are
    \begin{subequations}\label{eq:4dst_eq}
    \begin{align}
& G^{\mu\nu}= \tfrac{1}{2}T^{\phi\,\mu\nu} +{\mathcal H}^{\mu\nu} 
+T^{X\,\mu\nu} -\tfrac{1}{2}V(\phi)g^{\mu\nu}\,, \label{eq:eom_metric} \\
&[1+2g_2(\phi)X]\,\Box \phi -V'(\phi)-3X^2g_2'(\phi) 
- 2g_2(\phi)(\nabla^{\mu}\phi)(\nabla^{\nu}\phi)\nabla_{\mu}\nabla_{\nu}\phi = -\lambda'(\phi){\mathcal L}^{\text{GB}} \,, \label{eq:eom_sf}
\end{align}
    \end{subequations}
where \footnote{We have considered the coupling of gravity to the scalar field as in the canonical Horndeski Lagrangian, which differs from the normalisation used in \cite{East:2020hgw}.}
\begin{subequations}
\begin{align}
    T^{\phi}_{\mu\nu}=&~(\nabla_{\mu}\phi)(\nabla_{\nu}\phi) +g_{\mu\nu}\,X\,,
    \label{eq:scalar_st}\\
    T^X_{\mu\nu}=&~ g_2(\phi)X(\nabla_{\mu}\phi)(\nabla_{\nu}\phi)+\tfrac{1}{2}g_2(\phi)X^2g_{\mu\nu}\,, \label{eq:X_tensor}\\
    {\mathcal H}_{\mu\nu}= & -4\,\big[2R^{\rho}_ {~(\mu}{\mathcal C}_{\nu)\rho}-{\mathcal C}(R_{\mu\nu}-\frac{1}{2}R\,g_{\mu\nu})-\frac{1}{2}R\,{\mathcal C}_{\mu\nu}+{\mathcal C}^{\alpha\beta}\left(R_{\mu\alpha\nu\beta}-g_{\mu\nu}R_{\alpha\beta}\right)\big]\,,\label{eq:H_tensor}
\end{align}
\end{subequations}
with 
\begin{align}\label{Cmunu}
    {\mathcal C}_{\mu\nu}\equiv\lambda'(\phi)\nabla_{\mu}\nabla_{\nu}\phi+\lambda''(\phi)(\nabla_{\mu}\phi)(\nabla_{\nu}\phi)\,,
\end{align}
and ${\mathcal C}\equiv g^{\mu\nu}\mathcal{C}_{\mu\nu}$.

Recent interest of this theory has resulted in multiple studies concerning the current observational and theoretical constraints on its couplings. In Table I of \cite{Evstafyeva:2022rve} (see also references therein) there is a summary of the current constraints on a shift-symmetric Einstein-scalar-Gauss-Bonnet theory, i.e. with $g_2(\phi)=0$ and $\lambda(\phi)=\beta_0\,\phi$ \footnote{This is called shift-symmetric because constant shifts in the scalar field preserve the action, since the Gauss-Bonnet curvature is a total derivative in four dimensions.}. The strongest constraint is the one derived in \cite{Lyu:2022gdr} from several neutron star black hole binaries and binary black hole events, namely
\begin{equation}
    \sqrt{\beta_0}\approx 0.66 \text{ km} \left(\frac{\sqrt{\beta_0}}{M_{\odot} } \right) \lesssim 1.18\text{ km}\,,
\end{equation}
where $M_{\odot}$ is the mass of the Sun. We caveat that these constraints differ when taking into account the effect of a quadratic coupling or a mass term.

\subsection{Other Horndeski theories}

Apart from the $4\partial$ST theory there are other sub-classes of Horndeski that have been studied in the literature, which we briefly summarise here:
\begin{itemize}
    \item Quintessence, which refers to a scalar field minimally coupled to gravity with $G_2=G_3=G_4=G_5=0$. 
    \item k-essence, by setting $G_3=G_4=G_5=0$, with the usual choice of $G_2(\phi,X)=f(\phi)g(X)$. 
This has recently been studied in the context of kinetic screening \cite{Bezares:2020wkn,Lara:2022gof,Boskovic:2023dqk}, which suppresses the
scalar force in the vicinity of a massive body, allowing the theory to pass local experimental tests (at the scale of the solar
system) whilst potentially accounting for dark energy \cite{Will:1993hxu,Will:2014kxa}.
    \item Galileon theories \cite{Nicolis:2008in}, with $G_3\neq0$ and $G_4=G_5=0$, which involve an internal ``Galilean” invariance, under which the gradient of the scalar field shifts by a constant.
    \item Brans-Dicke theory, where $G_3=G_5=0$, $G_2(\phi,X)=\big(\frac{\omega}{\phi} -1\big)X$ (with $\omega$ being a dimensionless constant bounded by observational constraints)  and $G_4(\phi,X)=\phi$, which includes both Mach’s principle and Dirac’s large number hypothesis \cite{Brans:1961sx}.
    \item Damour-Esposito-Far\`ese (DEF) gravity \cite{Damour:1992kf,Damour:1993hw}, which is the first model leading to spontaneous scalarisation \footnote{Spontaneous scalarisation is the mechanism by which constant scalar configurations are rendered tachyonically unstable when the curvature becomes significant enough.}, is a generalisation of the Brans-Dicke theory with $G_3=G_5=0$, $G_2(\phi,X)=X\frac{\omega(\phi)}{\phi}$ (with $\omega(\phi)$ an arbitrary function of $\phi$) and $G_4=\phi$. It can be written in the Einstein's frame with a conformal transformation of the metric and a field redefinition (see \cite{Franchini:2019npi,Doneva:2022ewd}).
    \item Regularised 4D Einstein-Gauss-Bonnet theory of gravity \cite{Glavan:2019inb}, which re-scales the coupling constant in EGB gravity and introduces an extra scalar gravitational degree of freedom, thus becoming a Horndeski theory which is non-trivial in four spacetime dimensions. This theory has also been recently studied in the context of cosmology \cite{Fernandes:2021ysi,Fernandes:2022zrq}.
\end{itemize}

\section{Other modified gravity theories}

In this section we are going to give a brief overview of some of the other modified gravity theories mentioned in the classification at the beginning of this chapter.

\begin{itemize}
    \item Einstein-\ae ther theory \cite{Eling:2004dk,Jacobson:2007veq} is a generally covariant theory of gravity coupled to a dynamical, unit timelike vector field that breaks local Lorentz symmetry. Even though numerous observations severely limit the possibility of Lorentz violating
physics among the standard model fields \cite{Mattingly:2005re}, the constraints on Lorentz violation in the gravitational sector are generally weaker, which motivates the relevance of this theory. 

Its action is derived as the most general diffeomorphism-invariant derivative expansion (up to second order in derivatives) for the metric $g_{\mu\nu}$ and \ae ther $u^{\mu}$, which yields
\begin{equation}
    S=\frac{1}{2\kappa}\int d^4x\sqrt{-g}(R+K_{~~\rho\sigma}^{\mu\nu}\nabla_{\mu}u^{\rho}\nabla_{\nu}u^{\sigma}+\lambda(u^{\mu}u_{\mu}-1))\,,
\end{equation}
where $K_{~~\rho\sigma}^{\mu\nu}=c_1g^{\mu\nu}g_{\rho\sigma}+c_2\delta^{\mu}_{\rho}\delta^{\nu}_{\sigma}+c_3\delta^{\mu}_{\sigma}\delta^{\nu}_{\rho}+c_4u^{\mu}u^{\nu}g_{\rho\sigma}$, with the coefficients $\{c_i\}_{i=1}^4$ being dimensionless constants and $\lambda$ a Lagrange multiplier that enforces the unit constraint. Its equations of motion are
\begin{subequations}
    \begin{align}
        &\nabla_{\nu}J^{\nu}_{~\mu}-c_4\dot{u}_{\nu}\nabla_{\mu}u^{\nu}=\lambda\,u_{\nu}\,,\\
        &G_{\mu\nu}=\nabla_{\rho}(J_{(\mu}^{~~\rho}u_{\nu)}-J^{\rho}_{~(\mu}u_{\nu)}-J_{(\mu\nu)}u^{\rho})+c_1[(\nabla_{\rho}u_{\mu})(\nabla^{\rho}u_{\nu})-(\nabla_{\mu}u_{\rho})(\nabla_{\nu}u^{\rho})]\nonumber\\
        &\hspace{1cm}+c_4\dot{u}_{\mu}\dot{u}_{\nu}+[u_{\sigma}(\nabla_{\rho}J^{\rho\sigma})-c_4\dot{u}^2]u_{\mu}u_{\nu}+\frac{1}{2}g_{\mu\nu}K_{~~\rho\sigma}^{\alpha\beta}\nabla_{\alpha}u^{\rho}\nabla_{\beta}u^{\sigma}\,,
    \end{align}
\end{subequations}
where $J^{\mu}_{~\nu}=K^{\mu\rho}_{~~\nu\sigma}\nabla_{\rho}u^{\sigma}$ and $\dot{u}^{\mu}=u^{\nu}\nabla_{\nu}u^{\mu}$. The coupling parameters of Einstein-\ae ther gravity have been constrained in a recent work \cite{Gupta:2021vdj} for isolated/binary pulsars.

\item Ho\v{r}ava-Lifshitz gravity \cite{Horava:2008ih,Horava:2009uw} is another theory of gravity that breaks as well Lorentz symmetry, which is equivalent to Einstein-\ae ther theory in the infrared limit \cite{Sotiriou:2010wn}.

Ho\v{r}ava-Lifshitz gravity came out as as attempt to add higher order spatial derivatives without adding higher order time derivatives. This could presumably lead to a theory with improved ultraviolet (UV) behaviour without having to face the consequences of having higher order time derivatives. This obviously clashes with Lorentz invariance, but one can make sure that Lorentz violations in the IR could stay below current experimental constraints.

Its action is written down in terms of the $3+1$ variables from the ADM decomposition defined in Chapter \ref{C:NR}, where $\alpha$ is the lapse function, $\gamma_{ij}$ is the induced metric on hypersurfaces with $t=$ const. and $K_{ij}$ is the extrinsic curvature,
\begin{equation}
    S=\frac{1}{2\kappa}\int d^3x\,dt\,\alpha\sqrt{\gamma}[K^{ij}K_{ij}-\lambda\,K^2-V(\gamma_{ij},\alpha)]\,.
\end{equation}
Here $\lambda$ is a dimensionless running coupling and $V(\gamma_{ij},\alpha)$ can generically
depend on $\gamma_{ij}$ and $\alpha$ and their spatial derivatives, but does not contain time derivatives or the shift $\beta^i$.
Because of power counting renormalisability $V$ is required to contain terms which are at least sixth order in spatial derivatives. There are other assumptions that one can consider in order to reduce the number of invariants contained in $V$, some of which are detailed in \cite{Sotiriou:2010wn}.

\item Ghost-free massive gravity, known as dRGT (de Rham, Gabadaadze and Tolley) massive gravity \cite{deRham:2010kj,deRham:2011qq}, is a
four-dimensional covariant non-linear theory of massive gravity which is ghost-free in the decoupling limit to all orders. The theory re-sums explicitly all the nonlinear terms
of an effective field theory of massive gravity and has recently been studied numerically in \cite{deRham:2023ngf}.

Its action is
\begin{equation}
    S=\frac{1}{2\kappa}\int d^4x\sqrt{-g}\Big(R+\frac{m^2}{2}\sum\limits_{n=0}^4\alpha_n\,{\mathcal L}_n({\mathcal K})\Big)\,,
\end{equation}
where $\{\alpha_n\}_{n=0}^4$ are arbitrary coefficients, $m$ is the graviton mass and
\begin{equation}
    {\mathcal L}_n({\mathcal K})=\epsilon^{\mu_1...\mu_4}\epsilon_{\nu_1...\nu_4}{\mathcal K}^{\nu_1}_{~~\mu_1}...{\mathcal K}^{\nu_n}_{~~\mu_n}\delta^{\nu_{n+1}}_{~~~~\mu_{n+1}}...\delta^{\nu_4}_{~~\mu_4}\,,
\end{equation}
with ${\mathcal K}^{\mu}_{~\nu}=\delta^{\mu}_{\nu}-E^{\mu}_{~\nu}$, with $E^{\mu}_{~\nu}$ being a viel-bein such that $g_{\mu\nu}=\eta^{\rho\sigma}E_{\rho\mu}E_{\sigma\nu}$, where $\eta^{\mu\nu}$ is the Minkowski metric.

\item Quadratic gravity \cite{Stelle:1976gc} is an EFT built from the four-dimensional operators $R^2$, $R_{\mu\nu}R^{\mu\nu}$ and $R_{\mu\nu\rho\sigma}R^{\mu\nu\rho\sigma}$, which are the leading order curvature operators when matter is present. Such quadratic-curvature corrections are widely expected to arise from quantum fluctuations. The action takes the following form,
\begin{equation}
    S=\int d^4x\sqrt{-g}\Big(\frac{R}{2\kappa}+\alpha\,R_{\mu\nu}R^{\mu\nu}-\beta\,R^2+{\mathcal L}_{\text{matter}}\Big),
\end{equation}
with $\alpha$ and $\beta$ arbitrary coupling constants, where we have used that in four dimensions the Gauss-Bonnet curvature is a total derivative for getting rid of the $R_{\mu\nu\rho\sigma}R^{\mu\nu\rho\sigma}$ term. There has been recent work performing fully nonlinear numerical simulations in \cite{Held:2021pht,Held:2023aap,Cayuso:2023dei}. 
In this context, neutron star (NS) binaries become one of the most relevant scenarios.

The theory of quadratic gravity propagates the usual graviton, which is a massless spin-2 mode; a massive spin-0 mode with mass $m_0^2=-\frac{1}{2\kappa(3\beta-\alpha)}$; and a massive spin-2 mode with mass $m_2^2=-\frac{1}{2\kappa\alpha}$ which is an Ostrogadskii ghost \cite{Held:2023aap}.

If we neglect matter fields, then the next leading correction to the Einstein-Hilbert action keeping diffeomorphism invariance leads to \cite{Endlich:2017tqa}
\begin{equation}
    S=\frac{1}{2\kappa}\int d^4x\sqrt{-g}\Big(R-c_3\frac{R_{\mu\nu\rho\sigma}R^{\mu\nu}_{~~~\alpha\beta}R^{\alpha\beta\rho\sigma}}{\Lambda^4}-\tilde{c}_3\frac{\tilde{R}_{\mu\nu\rho\sigma}R^{\mu\nu}_{~~~\alpha\beta}R^{\alpha\beta\rho\sigma}}{\tilde{\Lambda}^4}\Big)\,,
\end{equation}
where $\tilde{R}_{\mu\nu\rho\sigma}=\epsilon_{\alpha\beta\rho\sigma}R_{\mu\nu}^{~~~\alpha\beta}$, $\Lambda$ is the energy scale of the EFT and $c_3$ and $\tilde{c}_3$ are ${\mathcal O}(1)$ coefficients, which gives rise to the following field equation,
\begin{align}
    &G_{\mu\nu}=\frac{6c_3}{\Lambda^4}(\nabla_{\alpha}R_{\mu\beta\rho\sigma})(\nabla^{\beta}R_{\nu}^{~\alpha\rho\sigma})+\frac{2\tilde{c}_3}{\Lambda^4}\big\{\epsilon_{\mu\delta\rho\sigma}R_{\nu}^{~\alpha\beta\gamma}(\nabla^{\sigma}\nabla_{\alpha}R_{\beta\gamma}^{~~\delta\rho}) \nonumber\\
    &\hspace{1cm}+\epsilon_{\nu\delta\rho\sigma}R_{\mu}^{~\alpha\beta\gamma}(\nabla^{\sigma}\nabla_{\alpha}R_{\beta\gamma}^{~~\delta\rho})+\epsilon_{\mu\delta\rho\sigma}(\nabla_{\alpha}R_{\beta\gamma}^{~~\rho\sigma})(\nabla^{\delta}R_{\nu}^{~\alpha\beta\gamma}))\nonumber\\
    &\hspace{1cm}+\epsilon_{\beta\gamma\rho\sigma}(\nabla_{\alpha}R_{\mu\delta}^{~~\rho\sigma})(\nabla^{\delta}R_{\nu}^{~\alpha\beta\gamma})+\epsilon_{\nu\delta\rho\sigma}(\nabla_{\alpha}R_{\beta\gamma}^{~~\rho\sigma})(\nabla^{\delta}R_{\mu}^{~\alpha\beta\gamma})\big\}\,.
\end{align}
However, according to \cite{Camanho:2014apa}, non-trivial values of $c_3$ and $\tilde{c}_3$, under certain assumptions about the UV completion, would require an infinite tower of higher spin particles coupled to standard model fields
with gravitational strength. Given that the mass of the lightest of those particles has to be of order $\Lambda$ and the couplings must allow mediation of long range forces of gravitational strength
between any matter fields, the $c_3$ and $\tilde{c}_3$ terms must be suppressed by a much higher scale, since we have not observed any additional long range forces on sub-kilometer distances. 

Therefore, on this assumption the first non-trivial correction in a vacuum high derivative EFT would correspond to the contributions to eight derivatives, namely \cite{Endlich:2017tqa}
\begin{equation}
S=\frac{1}{2\kappa}\int d^4x\sqrt{-g}\Big(R-\frac{1}{\Lambda^6}{\mathcal C}^2-\frac{1}{\tilde{\Lambda}^6}{\tilde{\mathcal C}}^2-\frac{1}{\Lambda_-^6}{\mathcal C}\tilde{\mathcal C}\Big)\,,    
\end{equation}
where ${\mathcal C}=R_{\mu\nu\rho\sigma}R^{\mu\nu\rho\sigma}$ and $\tilde{\mathcal C}=\tilde{R}_{\mu\nu\rho\sigma}R^{\mu\nu\rho\sigma}$, and $\Lambda$, $\tilde{\Lambda}$ and $\Lambda_-$ are arbitrary coupling scales with units of inverse mass, which leads to
\begin{align}
    &G^{\mu\nu}=\frac{1}{\Lambda^6}\Big(8\,R^{\mu\rho\nu\sigma}\nabla_{\rho}\nabla_{\sigma}{\mathcal C}+\frac{1}{2}g^{\mu\nu}{\mathcal C}^2 \Big)+\frac{1}{\tilde{\Lambda}^6}\Big(8\,\tilde{R}^{\mu\rho\nu\sigma}\nabla_{\rho}\nabla_{\sigma}\tilde{\mathcal C}+\frac{1}{2}g^{\mu\nu}\tilde{\mathcal C}^2\Big)\nonumber\\
    &\hspace{2cm}+\frac{1}{\Lambda_-^6}\Big(4\,\tilde{R}^{\mu\rho\nu\sigma}\nabla_{\rho}\nabla_{\sigma}{\mathcal C}+4\,R^{\mu\rho\nu\sigma}\nabla_{\rho}\nabla_{\sigma}\tilde{\mathcal C}+\frac{1}{2}g^{\mu\nu}\tilde{\mathcal C}{\mathcal C}\Big)\,.
\end{align}
These theories have recently been studied in \cite{Cayuso:2023aht} using the ``fixing the equations'' mechanism. The main idea of the ``fixing the equations'' technique \cite{Cayuso:2017iqc}, which was inspired by dissipative relativistic
hydrodynamics \cite{Israel:1976efz,Israel:1976tn,Muller:1967zza}, is to modify the higher-order contributions to the equations of motion by replacing them with some auxiliary
fields, and let the latter relax towards their correct value through a driver equation.

\end{itemize}

\addtocounter{page}{1}%
\blankpage

\chapter{Numerical Relativity}
\label{C:NR}

Einstein's equations can only be solved analytically for highly symmetric cases. In particular, a Binary Black Hole (BBH) merger cannot be described analytically in GR, much less in Modified Gravity. Therefore, one needs to find a well-posed initial value problem in order to solve the equations numerically. 

\section{ADM decomposition}

\begin{figure}[hbt]
\centering
\includegraphics[scale=.5]{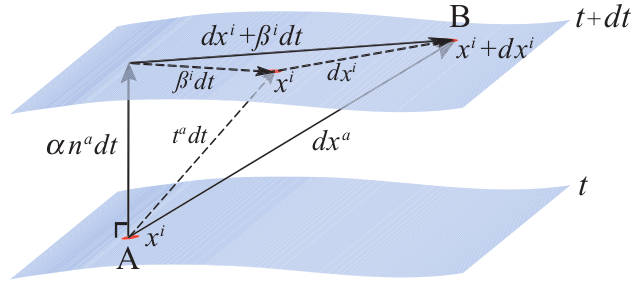}
\caption[Foliation of the spacetime.]{Foliation of the spacetime between two neighbouring slices. Figure taken from \cite{Baumgarte:2010ndz}.}
\label{F:slicing}
\end{figure}

In order to evolve our equations we first need to define a time coordinate by performing a $d+1$ decomposition of the spacetime metric (where $d$ is the number of spatial dimensions) in the following form,
\begin{equation}
    ds^2=-\alpha^2 dt^2+\gamma_{ij}(dx^i+\beta^i dt)(dx^j+\beta^j dt)\,,
    \label{eq:adm_metric}
\end{equation}
where $\alpha$ and $\beta^i$ are the lapse function and shift vector respectively (see Figure \ref{F:slicing}) and $\gamma_{ij}$ is the $d$-dimensional metric (which is purely spatial) induced on the $S_t\equiv\{t\equiv x^0=\text{const.}\}$ hypersurfaces, which are orthogonal to the unit timelike vector given by $n^{\mu}=\frac{1}{\alpha}(\delta_t^{\mu}-\beta^i\delta_i^{\mu})$.\footnote{All the spatial indices are raised and lowered with the physical spatial metric $\gamma_{ij}$.} 

This is known as the ADM decomposition, after Arnowitt, Deser and Misner \cite{Arnowitt:1959ah}.

In order to understand the implications of the Einstein equations on the $S_t$ hypersurfaces, one needs to decompose the spacetime Riemann tensor $R_{\mu\nu\rho\sigma}$ into spatial and timelike parts. This decomposition naturally involves the spatial tensor $R_{ijkl}$ (induced by the spatial metric $\gamma^{ij}$), which measures the intrinsic curvature of the hypersurface $S_t$. However, this tensor provides no information about how $S_t$ fits in the spacetime manifold. The missing information is contained in the so-called extrinsic curvature $K_{ij}$, which is the projection of the spacetime covariant derivative of the normal timelike vector $n^{\mu}$ into the hypersurfaces $S_t$, defined as
\begin{equation}
    \partial_{\perp}\gamma_{ij}=-2\alpha\,K_{ij}+2\gamma_{k(i}\partial_{j)}\beta^k\,,
\end{equation}
where we use the shorthand notation $\partial_{\perp}\equiv \partial_t-\beta^i\partial_i$. The tensors that we have just introduced are related to each other through the following well-known identities,
\begin{subequations}\label{eq:codazzi}
    \begin{align}
        &R_{ijkl}+K_{ik}K_{jl}-K_{il}K_{jk}=\gamma_i^{\mu}\gamma_j^{~\nu}\gamma_k^{~\rho}\gamma_l^{~\tau}R_{\mu\nu\rho\tau} \,,\label{eq:gauss}\\
        &D_jK_{ik}-D_iK_{ij}=\gamma_i^{~\mu}\gamma_j^{~\nu}\gamma_k^{~\rho}n^{\tau}R_{\mu\nu\rho\tau} \,,
    \end{align}
\end{subequations}
where $D_i$ is the covariant derivative respect to the spatial metric $\gamma_{ij}$ and $\gamma_i^{~\mu}$ projects spacetime covectors into spatial covectors. These equations are respectively known as the Gauss-Codazzi and Codazzi-Mainardi equations.

\section{Derivation of the ADM equations}

Once the foliation has been defined as a function of the spatial metric, the lapse and the shift, we can proceed to derive the evolution equations coming from the Einstein field equations.

Contracting some indices in the equations \eqref{eq:codazzi} and using the Einstein field equations to replace the r.h.s. terms with an arbitrary stress-energy tensor, we are left with the Hamiltonian and momentum constraints, namely
\begin{subequations}
    \begin{align}
        &R+K^2-K_{ij}K^{ij}=2\,\kappa\,\rho\,,\\
        &D^j(K_{ij}-\gamma_{ij}K)=\kappa\,J_i\,,
    \end{align}
\end{subequations}
where $\rho=n^{\mu}n^{\nu}T_{\mu\nu}$ and $J_i=-\gamma_i^{\mu}n^{\nu}T_{\mu\nu}$. Then, using again the equation \eqref{eq:gauss} together with the field equations $G_{\mu\nu}=\kappa\,T_{\mu\nu}$ and the following identity \cite{Alcubierre:2008},
\begin{equation}
    \gamma_i^{~\mu}\gamma_j^{~\nu}n^{\rho}n^{\tau}R_{\mu\nu\rho\tau}={\mathcal L}_nK_{ij}+K_{ik}K_{~j}^k+\frac{1}{\alpha}D_iD_j\alpha\,,
\end{equation}
where ${\mathcal L}_n$ denotes the Lie derivative along $n^{\mu}$, we obtain the ADM equations, which provide the first formulation of Numerical Relativity,
\begin{align}
    \partial_{\perp}K_{ij}=&~\alpha\,\big[R_{ij}-2K_{ik}K^k_{~j}+K\,K_{ij}]-\kappa\Big(S_{ij}-{\frac{\gamma_{ij}}{d-1}}(S-\rho)\Big)\nonumber\\&-D_iD_j\alpha+2\,K_{k(i}\partial_{j)}\beta^k\,,
\end{align}
where $K$ is the trace of the extrinsic curvature of the spatial slices, $S_{ij}=\gamma_i^{~\mu}\gamma_j^{~\nu}T_{\mu\nu}$ and $S=\gamma^{ij}S_{ij}$ its trace.

\section{Formulations of Numerical Relativity}

As was discovered in early work on Numerical Relativity, the ADM evolution equations are not suitable for numerical evolution. The reason is that these equations are only weakly hyperbolic but not strongly hyperbolic and, thus, not well-posed (see Section \ref{sec:wellposedness} for further details). In order to overcome this problem several other formulations have been proposed.

\subsection{BSSN formalism}

 Baumgarte, Shapiro, Shibata and Nakamura developed the BSSN formalism \cite{Nakamura:1987zz,Shibata:1995we,Baumgarte:1998te}, which is a conformal reformulation of ADM with strong hyperbolicity properties that enables NR simulations. The following conformal decomposition of the evolution variables is applied,
\begin{subequations}
\begin{align}
\chi&=\det(\gamma_{ij})^{-\frac{1}{d}}\,, \\
\tilde{\gamma}_{ij}&=\chi\,\gamma_{ij}\,, \\ \tilde{A}_{ij}&=\chi\Big(K_{ij}-{\frac{1}{d}}\,\gamma_{ij}K\Big)\,.
\end{align}
\end{subequations}

Then the equations of motion yield as follows,
\begin{subequations}
\begin{align}
\partial_{\perp}\tilde{\gamma}_{ij} =& -2\alpha\tilde{A}_{ij}+2\tilde\gamma_{k(i}\partial_{j)}\beta^k-{\frac{2}{d}}\tilde{\gamma}_{ij}\partial_k\beta^k\,, \\
\partial_{\perp}\chi =&~ {\frac{2}{d}}\,\chi\,\big(\alpha K - \partial_k\beta^k\big)\,, \\
\partial_{\perp}K=&-D^iD_i\alpha +\alpha\left[R +K^2\right]+{\frac{\kappa\,\alpha}{d-1}}\big[S-d\,\rho\big]\,,\\
\partial_{\perp}\tilde{A}_{ij}=&~\alpha\,[\tilde{A}_{ij}(K-2\Theta)-2\,\tilde{A}_{ik}\tilde{A}^k_{~j}]+2\tilde{A}_{k(i}\partial_{j)}\beta^k- {\frac{2}{d}}(\partial_k\beta^k)\tilde{A}_{ij}\nonumber\\&+\chi\left[\alpha\left(R_{ij}-\kappa\,S_{ij}\right)-D_iD_j\alpha\right]^{\text{TF}} \,,\\
\partial_{\perp}\tilde{\Gamma}^i
=&~2\,\alpha\,\Big(\tilde\Gamma^i_{\phantom i kl}\tilde A^{kl}-{\frac{d-1}{d}}\tilde\gamma^{ij}\partial_jK-{\frac{d}{2\,\chi}}\,\tilde A^{ij}\partial_j\chi\Big) 
-2\,\tilde A^{ij}\partial_j\alpha-\tilde{\Gamma}^j\partial_j\beta^i + {\frac{2}{d}}\,\tilde{\Gamma}^i\partial_j\beta^j \nonumber\\&+ {\frac{d-2}{d}}\,\tilde\gamma^{ik}\partial_k\partial_j\beta^j + \tilde\gamma^{jk}\partial_j\partial_k\beta^i 
-2\,\kappa\,\alpha\,\tilde\gamma^{ij}J_j\,.
\end{align}
\end{subequations}
where the superscript TF refers to the trace-free part and $\tilde{\Gamma}^i\equiv\tilde\gamma^{kl}\tilde \Gamma^i_{kl}$, with $\tilde \Gamma^i_{kl}$ being the Christoffel symbols associated to the conformal spatial metric $\tilde{\gamma}_{ij}$.

The key points that make this formulation strongly hyperbolic in comparison to the ADM equations are the introduction of the auxiliary variable $\tilde{\Gamma}^i$ and the addition of the momentum constraints to the equation for $\tilde{A}_{ij}$.

\subsection{The Z systems}

In early $2000$s, a new well-posed formalism known as Z4 was proposed in \cite{Bona:2003fj} that introduced a constraint vector $Z_{\mu}$. In \cite{Gundlach:2005eh}, it was shown that these constraints could be damped in Einstein's vacuum field equations by adding some suitable lower order terms, namely
\begin{align}
    R_{\mu\nu}+2\nabla_{(\mu}Z_{\nu)}-\kappa_1\,\Big[2n_{(\mu}Z_{\nu)}-\tfrac{2}{d-1}(1+\kappa_2)g_{\mu\nu}\,n_{\sigma}Z^{\sigma}\Big]=\kappa\Big(T_{\mu\nu}-\frac{1}{2}g_{\mu\nu}T\Big)\,,
\end{align}
where $\kappa_1$ and $\kappa_2$ are two damping coefficients with the stability bounds $\kappa_1>0$ and $\kappa_2>-1$, which will be revisited in Chapter \ref{C:Wellposedness}, and $T=g^{\mu\nu}T_{\mu\nu}$. Then, the equations of motion in $d+1$ form yield
\begin{subequations}
\begin{align}
\partial_{\perp}\gamma_{ij} =& -2\alpha K_{ij}+2\gamma_{k(i}\partial_{j)}\beta^k, \\
\partial_{\perp}K_{ij}=&-D_iD_j\alpha +\alpha\left[R_{ij}+2\,D_{(i}Z_{j)} +K_{ij}(K-2\Theta) -2\,K_{ik}K^k_{~j}\right] \nonumber\\
&-\kappa\Big(S_{ij}-\frac{\gamma_{ij}}{d-1}(S-\rho)\Big)-\gamma_{ij}\kappa_1(1+\kappa_2)\,\alpha\,\Theta+2K_{k(i}\partial_{j)}\beta^k\,,\\
\partial_{\perp}\Theta =&~{\frac{\alpha}{2}}\big[R-K_{ij}\,K^{ij}+\,K(K-2\Theta) 
+2\,D^iZ_i]-Z_i\,D^i\alpha\nonumber\\&-\kappa\,\alpha\,\rho-{\frac{\kappa_1}{2}}\big(d+1+(d-1)\kappa_2\big)\,\alpha\,\Theta \,,\\
\partial_{\perp}Z_i
=&~\alpha\big(D_j(K_{i}^{~j}-\delta_i^jK)+D_i\Theta-2K_i^{~j}Z_j\big) -\Theta\,D_i\alpha-\kappa\,J_i\,,
\end{align}
\end{subequations}
where $\Theta\equiv-n_{\mu}Z^{\mu}=\alpha\,Z^0$.

Some variants to the Z4 system have been suggested, such as the Z3 system \cite{Bona:2003qn}, obtained by a symmetry breaking mechanism in which
\begin{equation}
    \tilde{K}_{ij}\equiv K_{ij}+\frac{n}{2}\,\Theta\,\gamma_{ij}\,,
\end{equation}
where $n$ is a free parameter, which has led to successful numerical simulations in \cite{Alic:2008pw}.

The most popular version of the Z system is the CCZ4 formulation \cite{Alic:2011gg,Alic:2013xsa}, which is a conformal and covariant extension of Z4, which provides additional stability  properties. Its variables are the conformal ones from the BSSN approach ($\chi$, $\tilde{\gamma}_{ij}$, $\tilde{A}_{ij}$ and $K$), as well as $\Theta$ and $\hat{\Gamma}^i\equiv\tilde{\Gamma}^i+2\,\tilde{\gamma}^{ij}Z_j$ from the Z4 decomposition. Its equations of motion in the $d+1$ form will be given in Chapter \ref{C:Wellposedness}.

\subsection{Generalised Harmonic Coordinates}

Parallelly, another well-posed $3+1$ formulation of the Einstein equations was presented in \cite{Pretorius:2004jg} and it led to the first Binary Black Hole evolution in \cite{Pretorius:2005gq}. 

This approach made use of Generalised Harmonic Coordinates, which consist in evolving a harmonic constraint function $C^{\mu}\equiv H^{\mu}-\Box x^{\mu}$, where $H^{\mu}$ are source functions that can be independently chosen. It is related to the Z4 formalism by setting $C^{\mu}=-2Z^{\mu}$. The trace-reversed version of the Einstein field equations in this gauge yields a system of quasi-linear wave equations,
\begin{align}
    R_{\mu\nu}-\partial_{(\mu}H_{\nu)}=-\frac{1}{2}g^{\alpha\beta}\partial_{\alpha}\partial_{\beta}g_{\mu\nu}+{\mathcal N}_{\mu\nu}(g,\partial g)\,,
\end{align}
where ${\mathcal N}_{\mu\nu}(g,\partial g)$ contains all lower order terms in the equations. As for the Z4 formalism, the damped harmonic gauge introduced in \cite{Szilagyi:2009qz} imposes that the constraints are damped in numerical simulations.

The variables that are usually employed in the harmonic gauge are the components of the spacetime metric $g^{\mu\nu}$. However, there exists a formulation in terms of the $3+1$ variables in a similar way to the BSSN and Z4 approaches \cite{Brown:2011qg}, which yields the following evolution equations:
\begin{subequations}
\begin{align}
\partial_{\perp}\gamma_{ij} =& -2\alpha K_{ij}+2\gamma_{k(i}\partial_{j)}\beta^k, \\
\partial_{\perp}K_{ij}=&-D_iD_j\alpha +\alpha[R_{ij}-\,D_{(i}{\mathcal C}_{j)}-2\,K_{ik}K^k_{~j} -\pi\,K_{ij}]+2K_{k(i}\partial_{j)}\beta^k \nonumber\\
&-\kappa\,\alpha\Big(S_{ij}-\frac{\gamma_{ij}}{d-1}\,\big(S-\rho\big)\Big)-\gamma_{ij}\kappa_1(1+\kappa_2)\,\alpha\,{\mathcal C}_{\perp}\,,\\
\partial_{\perp}\pi =&-\alpha\,K_{ij}\,K^{ij}+D_iD^i\alpha+{\mathcal C}^iD_i\alpha -{\frac{\kappa}{d-1}}\,\alpha\,(S+(d-2)\rho)
\nonumber\\&-\kappa_1\big(1-\kappa_2\big)\,\alpha\,{\mathcal C}_{\perp} \,,\\
\partial_{\perp}\rho^i
=&~g^{kl}\partial_k\partial_l\beta^i+\alpha\,D^i\pi-\pi D^i\alpha-2K_i^{~j}D_j\alpha+2\,\alpha\,K^{jk}\,\Gamma^i_{jk}
\nonumber\\& -\kappa\,\alpha\, J_i +\kappa_1\,\alpha\,{\mathcal C}^i   \,,
\end{align}
\end{subequations}
where the variables $\pi$ and $\rho^i$ come from the decomposition of the constraints ${\mathcal C}^{\mu}$,
\begin{subequations}
\begin{align}
    &{\mathcal C}_{\perp}\equiv n_{\mu}{\mathcal C}^{\mu}=\pi+K\,,\\
    &{\mathcal C}^i\equiv\gamma^i_{~\mu}{\mathcal C}^{\mu}=-\rho^i+\Gamma^i\,,
\end{align}    
\end{subequations}
where $\Gamma^i\equiv\gamma^{kl}\Gamma^i_{kl}$, with $\Gamma^i_{kl}$ being the Christoffel symbols associated to the spatial metric $\gamma_{ij}$. These expressions, together with appropriate choices of $H^{\mu}$, define the equations for the gauge variables, as we will thoroughly explain in the next chapter with a CCZ4 formulation.

\section{Gauge conditions}

The choice of the coordinate system is given in terms of the gauge variables, the lapse function $\alpha$ and the shift vector $\beta^i$. The simplest choice, which is known as \emph{geodesic slicing}, corresponds to $\alpha=1$ and $\beta^i=0$. However, this is not suitable when dealing with singularities or in a cosmological set-up where overdensities are present. Thus, finding a good gauge has been and continues being one of the most difficult problems in NR.

Throughout the end of the $20^{\text{th}}$ century, there were many attempts to find hyperbolic reformulations of the $3+1$ Einstein evolution equations by using evolution type slicing conditions. This resulted in the \emph{Bona-Mass\'o} family of slicing conditions for the lapse \cite{Bona:1994dr}, namely
\begin{align}
    \partial_{\perp}\alpha=-\alpha^2\,f(\alpha)\,K\,,
\end{align}
with $f(\alpha)$ being an arbitrary positive function of $\alpha$. It is a generalisation of the harmonic slicing (recovered when $f(\alpha)=1$), which is derived by imposing $\Box x^{\mu}=0$. 

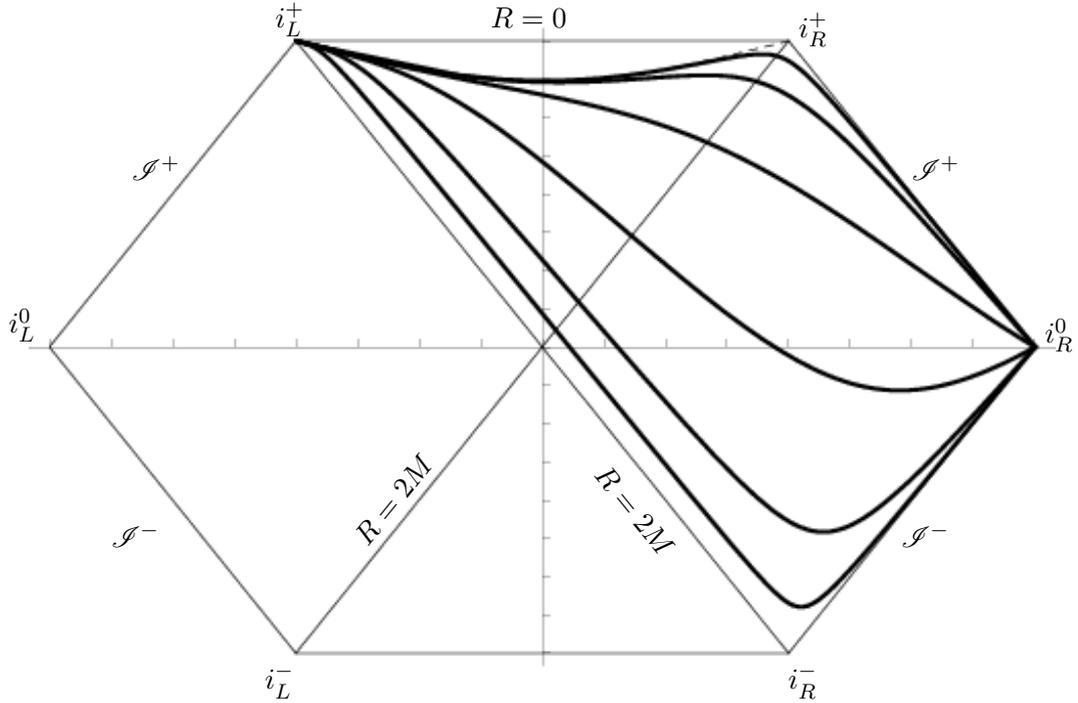
\begin{figure}[hbt]
\vspace{0.5cm}
\centering
\begin{overpic}[width=135mm]{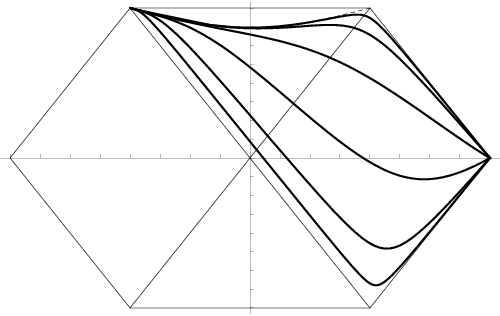} 
\put(75,62){$i_R^+$}
\put(24,63){$i_L^+$}
\put(86,48){$\mathscr{I}^+$}
\put(10,48){$\mathscr{I}^+$}
\put(8,12){$\mathscr{I}^-$}
\put(85,12){$\mathscr{I}^-$}
\put(74,-2){$i^-_R$}
\put(23,-2){$i^-_L$}
\put(-2,33){$i_L^0$}
\put(99,32){$i_R^0$}
\put(33,12){\begin{rotate}{51}
  $R = 2M$
\end{rotate}}
\put(55,19){\begin{rotate}{-51}
  $R = 2M$
\end{rotate}}
\put(45,63){$R = 0$}
\end{overpic}
\vspace{0.5cm}
\caption[Penrose diagram of the 1+log condition.]{Penrose diagram of the slices defined by the stationary
  solution of a black hole of mass $M$ with the $1+\log$ condition. 
  Every slice approaches future timelike infinity $i_L^+$ along the curve $R = R_0 \approx 1.31 M$ and spatial
  infinity $i_R^0$ along a curve of constant time. 
  Figure taken from \cite{Hannam:2008sg}.} 
\label{F:Penanalyt}
\end{figure}

The particular case of $f(\alpha)=2/\alpha$ was shown to be very robust in practise and to have singularity avoiding properties (see Figure \ref{F:Penanalyt}). It is known as the $1+\log$ slicing, given that for zero shift the solution for the lapse is $\alpha=h(x^i)+\ln(\det(\gamma_{ij}))$, with $h(x^i)$ being an arbitrary positive time-independent function.

Regarding the shift condition, the harmonic slicing $\Box x^{\mu}=0$ leads to 
\begin{align}
    \partial_{\perp}\beta^i=-\alpha D^i\alpha+\alpha^2\Gamma^i\,,
\end{align}
where $\Gamma_i=\gamma^{jk}\Gamma^i_{jk}$, being $\Gamma^i_{jk}$ the Christoffel symbols of the physical metric $\gamma_{ij}$. Another robust slicing condition for the shift is the Gamma-driver, which forms the puncture gauge equations together with the $1+\log$ slicing for the lapse. In the CCZ4 formalism, they have the following form, \footnote{Here the Gamma-driver equation \eqref{eq:gamma_driver} has been written in the integrated form, which contains an integration constant that we did not include. If the initial data is not conformally flat, one has to take this constant into account to obtain smooth coordinates throughout the evolution. See for instance \cite{Figueras:2015hkb,Figueras:2017zwa} for examples where this is important.}
\begin{subequations}\label{gauge}
\begin{align}
\partial_{\perp}\alpha  &= - 2\alpha(K-2\Theta), \\
\partial_{\perp}\beta^i &=\tfrac{d}{2(d-1)}\hat{\Gamma}^i -\eta\beta^i\,,\label{eq:gamma_driver}
\end{align}
\end{subequations}
where the parameter $\eta>0$ is used to damp oscillations in the shift and the factor $\frac{d}{2(d-1)}$  in \eqref{eq:gamma_driver} comes from imposing that the shift propagates at the speed of light in the asymptotic region \cite{Alcubierre:2002kk} \footnote{This factor is a gauge choice and in some instances, e.g., higher dimensions, other choices may be more convenient for numerical stability.}. The moving puncture gauge permits evolution of black hole spacetimes in the BSSN and CCZ4 formulations and, given that it is singularity-avoiding, there is no need for explicit excision of the inner part of the black hole, unlike in the harmonic gauge.

\section{Initial data}

A key problem in NR, as essential as successfully evolving the system, is being able to construct initial data that satisfies the Hamiltonian and momentum constraints, which have an elliptic nature.

The elements of the spatial metric together with their time derivatives, the lapse and shift constitute $(d+1)^2$ components that must be specified initially, but we have only $d+1$ constraint equations. Only $d+1$ of the remaining $d(d+1)$ free components relate to physical degrees of freedom and the remainder are gauge. Therefore, there are an infinite number of ways to set the free data, some of which can lead to uniqueness and existence problems.

Many different approaches have been suggested in the literature (see \cite{Alcubierre:2008,Baumgarte:2010ndz,Shibata:2015} for a review). One of them is the conformal transverse-traceless decomposition (CTT), which decomposes the extrinsic curvature into its trace $K$ and a traceless tensor $A_{ij}$. Using a conformal factor $\psi$ such that $\gamma_{ij}=\psi^4\bar{\gamma}_{ij}$ with $\det(\bar{\gamma}_{ij})=1$, the conformal $\bar{A}_{ij}=\psi^{-2}A_{ij}$ is further decomposed into a transverse-traceless part $\bar{A}^{\text{TT}}_{ij}$ and a vector potential $W_i$ such that
\begin{equation}
    \bar{A}_{ij}=\bar{A}^{\text{TT}}_{ij}+\bar{D}_iW_j+\bar{D}_jW_i-\frac{2}{d}\bar{\gamma}_{ij}\bar{D}_kW^k\,,
\end{equation}
where $\bar{D}_i$ is the covariant derivative with respect to the conformal metric $\bar{\gamma}_{ij}$. This method is successful in vacuum GR but problematic when the sources are fundamental fields (such as scalar fields), since reconstructing the fields' configuration from the rescaled quantities can induce a loss of uniqueness of solutions \cite{Baumgarte:2006ug}. However, this is for example not a problem at all when dealing with fluids, since its energy density can be specified as a fundamental quantity independently of the choice of the initial conformal factor (see \cite{Aurrekoetxea:2022mpw} for a further discussion on this).

This was recently solved by the CTTK approach presented in \cite{Aurrekoetxea:2022mpw}, in which instead of solving the
Hamiltonian constraint as a second order elliptic equation for a choice of the mean curvature $K$, the authors solve an algebraic equation for $K$ for a choice of the conformal factor. We have also recently applied it to the $4\partial$ST theory \cite{Brady:2023dgu}.

\section{Hyperbolicity and well-posedness}\label{sec:wellposedness}

As we have already mentioned in the previous sections, well-posedness is an essential prerequisite to be able to evolve the equations of motion numerically. The property of well-posedness guarantees that, given some suitable initial data, the solution to the equations of motion exists, is unique and depends continuously on the initial data. 

Let's consider a first order equation on $\mathbb{R}^{d+1}$ of the form \cite{Kovacs:2020ywu} \footnote{Those results also hold for second order equations, see Appendix A of \cite{Kovacs:2020ywu} for further details.}
\begin{align} \label{ode}
    \partial_0U=B(x,U,\partial_iU)\,,
\end{align}
where $U$ is an $N$-component vector and $x^{\mu}=(x^0,x^i)$ are the spacetime coordinates. Its principal part is defined as 
\begin{equation}
    {\mathcal M}(x,U,\xi_i)=(\partial_{\partial_jU}B)(x,U,\partial_iU)\xi_j\,,
\end{equation}
where $\xi_i$ is an arbitrary vector of unit norm with respect to a smooth positive definite (inverse) metric $\gamma^{ij}$ on surfaces of constant $x^0$. 

Then \eqref{ode} is said to be weakly hyperbolic if all eigenvalues of ${\mathcal M}(x,U,\xi_i)$ are real for
any such $\xi_i$. It will be further strongly hyperbolic if there exists an $N\times N$ Hermitian matrix valued function ${\mathcal K}(x,U,\xi_i)$ (called the symmetriser) that is positive definite with smooth dependence on its arguments, and a positive constant $\Lambda$ satisfying the conditions
\begin{align}\label{symmetriser}
    {\mathcal K}(x,U,\xi_i){\mathcal M}(x,U,\xi_i)={\mathcal M}^{\dagger}(x,U,\xi_i){\mathcal K}(x,U,\xi_i)
\end{align}
and
\begin{align}
    \Lambda^{-1}I\leq{\mathcal K}(x,U,\xi_i)\leq\Lambda I\,,
\end{align}
with $I$ being the identity matrix of size $N\times N$. 

In particular, if ${\mathcal M}(x,U,\xi_i)$ is diagonalisable with real eigenvalues and a complete set of linearly independent and bounded eigenvectors that depend
smoothly on the variables $(x,U,\xi_i)$ for any $\xi_i$, then one can use those eigenvectors to construct a suitable symmetriser: if $S$ denotes the matrix whose columns are the eigenvectors of ${\mathcal M}(x,U,\xi_i)$, then it holds that ${\mathcal K}=(S^{-1})^{\dagger}S^{-1}$ is a positive definite, smooth and bounded symmetriser that satisfies the conditions above.

Strongly hyperbolic first order systems of the form \eqref{ode} with initial data $U(0,x^i)=f(x^i)$ are well-posed in Sobolev spaces $H^s$ with $s>s_0$ for some constant $s_0$ \footnote{The Sobolev space $H^k(\Omega)$ for an arbitrary open set $\Omega\subset{\mathbb R}^d$ are defined as 
\begin{equation}
    H^k(\Omega)=\{u\in L^2(\Omega), \ \partial^{\alpha}u\in L^2(\Omega) \ \forall\alpha\in{\mathbb N}^d,\ |\alpha|\leq k \}
\end{equation} 
and are endowed with the norm $||u||_{H^k}=\sqrt{\sum\limits_{|\alpha|\leq k}\int_{\Omega}|\partial^{\alpha}u|^2 }$.}, namely there exists a unique local solution $U\in C([0,T], H^s({\mathbb R}^{d}))$ with $T>0$ depending on the $H^s$-norm of the initial data, meaning that
\begin{equation}
    ||U||_{H^s}(x^0)\leq \Lambda||f(x^i)||_{H^s}\,.
\end{equation}

Note that well-posedness would also hold in the case of constant coefficient linear equations (but not for non-linear equations) when we have the following bound on the solution of the system \cite{Kreiss2004InitialBoundaryVP},
\begin{equation}
    ||U||_{H^s}(x^0)\leq \Lambda e^{\lambda x^0}||f(x^i)||_{H^s}\,,
\end{equation}
for a real constant $\lambda$, which follows from assuming the weaker condition on the symmetriser (replacing Eq. \eqref{symmetriser})
\begin{equation}
    {\mathcal K}(\xi_i)\,i\,{\mathcal M}(\xi_i)-{\mathcal M}^{\dagger}(\xi_i)\,i\,{\mathcal K}(\xi_i)\leq 2\,\lambda\,{\mathcal K}(\xi_i)\,,
\end{equation}
where both ${\mathcal K}$ and ${\mathcal M}$ now only depend on $\xi_i$ given that this is only valid for constant coefficient linear equations.

\section{GRChombo}

All the numerical results in this thesis have been obtained in the framework of \texttt{GRChombo} \cite{Andrade:2021rbd,Radia:2021smk}, an AMR\footnote{By Adaptive Mesh Refinement (AMR), we mean that the spatial domain in our simulations is covered by a uniform coarse grid, on top of which a hierarchy of Cartesian grids of
increasing resolution is stacked in areas in which more refinement is required.} based open-source code for performing numerical relativity simulations which uses hybrid MPI/OpenMP parallelism in order to achieve good performance in HPC (High Performance Computing) clusters. This code is developed and maintained by the GRTL (former GRChombo) collaboration (for further details see \href{https://github.com/GRTLCollaboration}{\texttt{https://github.com/GRTLCollaboration}}).

\texttt{GRChombo} implements both the BSSN and CCZ4 formalisms with puncture gauge. It is able to evolve Binary Black Hole mergers, but it is particularly suited for applications in fundamental physics, and has also been extensively used for Boson Stars \cite{Croft:2022bxq,Evstafyeva:2022bpr}, cosmological spacetimes \cite{Aurrekoetxea:2019fhr,Aurrekoetxea:2023jwd}, higher dimensional black strings and black rings \cite{Figueras:2015hkb,Figueras:2017zwa,Bantilan:2019bvf,Andrade:2020dgc}, Primordial Black Holes \cite{deJong:2021bbo,deJong:2023gsx} and cosmic strings \cite{Helfer:2018qgv,Drew:2022iqz,Drew:2023ptp} among other topics of interest in NR. An apparent horizon finder
is as well provided in order to locate black holes and deduce their masses and angular momenta. It is also possible to extract gravitational waves and compute other relevant diagnostics.

As one of the main achievements of this thesis, we have developed an extension of \texttt{GRChombo}, called \texttt{GRFolres} \footnote{Folres (pronounced fol-res) is a word meaning covers or linings in the Catalan language. It has a specific application in the tradition
of \textit{Castells} (Human Towers), denoting the second layers of reinforcement above the base \textit{pinya}. We use it here in analogy to our
understanding of effective field theories (EFTs) of gravity as an infinite sum of terms organised as a derivative expansion, in which the
\textit{pinya} corresponds to GR (with up to two derivatives), and the \textit{folre} to modified theories up to four derivatives, which are those that
we are able to simulate with \texttt{GRFolres}.}, which has recently been made publicly available \cite{AresteSalo:2023hcp} (see \href{https://github.com/GRTLCollaboration/GRFolres}{\texttt{https://github.com/GRTLCollaboration/GRFolres}}), which we will review in Chapter \ref{C:GRFolres}.
It is a tool that implements the equations of motion of the Four-Derivative Scalar-Tensor theory of gravity and some other modified gravity theories in a well-posed formulation that we develop in next chapter.

\addtocounter{page}{1}%
\blankpage

\part{Research work}

\addtocounter{page}{1}%
\blankpage

\chapter{Well-posedness in Modified Gravity}
\label{C:Wellposedness}

This chapter is based on the work on well-posedness in Einstein-Gauss-Bonnet gravity and the Four-Derivative Scalar-Tensor theory of gravity presented in \cite{AresteSalo:2022hua,AresteSalo:2023mmd} using an adaptation of the modified harmonic gauge to singularity-avoiding coordinates.

As we discussed in Chapter \ref{C:NR}, having a well-posed set of equations is essential for Numerical Relativity, especially when dealing with strong gravity. Many real life physical theories have non well-posed equations, such as inverse problems. Therefore, not having a well-posed formulation does not necessary mean that a theory is invalid and in some cases (such as GR) non well-posed equations can be re-expressed in a suitable well-posed formulation.

As shown in \cite{Papallo:2017qvl} the formulations that are well-posed in GR lose this property in some modified theories of gravity, in particular the family of Horndeski and Lovelock theories that were presented in Chapter \ref{C:MG}.

\section{Modified harmonic gauge}

\begin{figure}[hbt]
\centering
\includegraphics[scale=1]{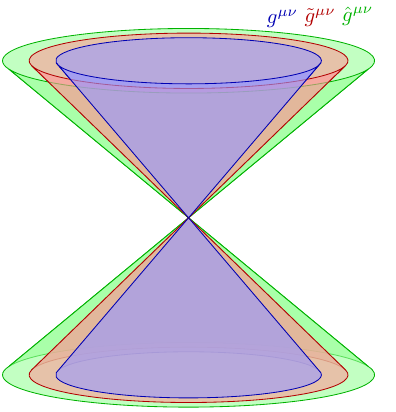}
\caption[Null cones of the physical and auxiliary metrics in the modified harmonic gauge.]{Cotangent space showing the null cones of $g^{\mu\nu}$, $\tilde{g}^{\mu\nu}$ and $\hat{g}^{\mu\nu}$ in the modified harmonic gauge. Figure taken from \cite{Kovacs:2020ywu}.}
\label{F:cone}
\end{figure}

Recently a new formalism was proposed by Kov\'acs and Reall in \cite{Kovacs:2020pns,Kovacs:2020ywu}, known as the modified harmonic gauge (MHG), which was shown to be well-posed in all Horndeski and Lovelock theories in the weak coupling regime. The vacuum Einstein field equations in this formulation yield 
\begin{align}\label{mhg}
G^{\mu\nu}+\hat{P}_{\alpha}^{~\beta\mu\nu}\nabla_{\beta}H^{\alpha}=0\,, \qquad H^{\mu}\equiv \tilde{g}^{\rho\sigma}\Gamma^{\mu}_{\rho\sigma}=0\,,
\end{align}
where $\hat{P}_{\alpha}^{~\beta\mu\nu}=\delta_{\alpha}^{(\mu}\hat{g}_{\phantom{\mu}}^{\nu)\beta}-\frac{1}{2}\delta_{\alpha}^{\beta}\hat{g}_{\phantom{a}}^{\mu\nu}$, $H^{\mu}$ are the gauge conditions and $\hat{g}^{\mu\nu}$ and $\tilde{g}^{\mu\nu}$ \footnote{The auxiliary spacetime metric $\tilde g_{\mu\nu}$ is unrelated to the conformal spatial metric $\tilde \gamma_{ij}$ defined in the BSSN formalism.} are two auxiliary Lorentzian metrics whose null cones do not intersect with each other and lie outside the light cone of the physical metric $g^{\mu\nu}$ as displayed in Figure \ref{F:cone} \footnote{The introduction of these auxiliary metrics changes the propagation speed of the unphysical modes, which is essential for avoiding a degeneracy that triggers the ill-posedness.}.

The simplest way to define these auxiliary metrics (which is especially useful when deriving the $d+1$ equations) is by setting \cite{Kovacs:2020ywu}
\begin{subequations}\label{abfunc}
\begin{align}
\tilde{g}^{\mu\nu}&=g^{\mu\nu}-a(x)\,n^{\mu}n^{\nu}\,,\\ \quad \hat{g}^{\mu\nu}&=g^{\mu\nu}-b(x)\,n^{\mu}n^{\nu}\,,
\end{align}
\end{subequations}
where $n^{\mu}$ is the unit timelike vector orthogonal to $t=$const. hypersurfaces, with $a(x)$ and $b(x)$ being two functions dependent on the spacetime coordinates $x^{\mu}$ that satisfy $0<a(x)<b(x)$, $0<b(x)<a(x)$ or $-1<a(x)<0<b(x)$, but are otherwise arbitrary\footnote{Note that some additional constraint is necessary when adapting this modified gauge to the moving punctures approach, as we will discuss later on in this chapter.}.

This formulation has been implemented in Generalised Harmonic Coordinates (GHC) in Einstein-scalar-Gauss-Bonnet gravity \cite{East:2020hgw,Corman:2022xqg} and has shown to work in practise for equal-mass Binary Black Hole mergers.

\section{Modified puncture gauge}

The main contribution of this thesis, on which all the following results are based, is the adaptation of the modified harmonic gauge introduced above to singularity avoiding coordinates, namely to the puncture gauge in a CCZ4 formalism \cite{AresteSalo:2022hua,AresteSalo:2023mmd}. This formulation ensures well-posedness of Einstein-Gauss-Bonnet gravity and the Four-Derivative Scalar-Tensor theory of gravity in the weak coupling regime, as we will show later in this chapter.

In the CCZ4 formalism discussed in Chapter \ref{C:NR}, Einstein's field equations with an arbitrary stress-energy tensor $T^{\mu\nu}$, in the modified harmonic gauge introduced by \cite{Kovacs:2020ywu, Kovacs:2020pns} and supplemented by constraint damping terms \footnote{We will analyse this further in Section \ref{sec:damping}, where we will derive the stability bounds of the $\kappa_1$ and $\kappa_2$ constants. Note that the $b(x)$ explicitly appearing in equation \eqref{ESFE} has been manually inserted so that these stability bounds do not depend on the dimension $d$.}, yield
\begin{align} \label{ESFE}
&G^{\mu\nu}+2\hat{P}_{\alpha}^{~\beta\mu\nu}\nabla_{\beta}Z^{\alpha}
-\kappa_1\Big[2n^{(\mu}Z^{\nu)}+\Big(\tfrac{d-3}{2+b(x)}+\tfrac{d-1}{2}\kappa_2\Big)\,n^{\alpha}Z_{\alpha}\,g^{\mu\nu}\Big]
= \kappa\,T^{\mu\nu} \,,
\end{align}
where, as in equation \eqref{mhg}, $\hat{g}^{\mu\nu}$ and $\tilde{g}^{\mu\nu}$ are the two auxiliary Lorentzian metrics defined in \eqref{abfunc} and $Z^{\mu}$ is the vector of constraints,
\begin{equation}\label{eq:Zmu}
    Z^{\mu}\equiv-\tfrac{1}{2}\big(H^{\mu}+\tilde{g}^{\rho\sigma} \Gamma^{\mu}_{\rho\sigma}\big) =0\,,
\end{equation}
where $H^\mu$ are the source functions, which can be freely chosen. These choices determine the gauge in that formulation, and amount to specifying evolution equations for the lapse and shift, which are derived below.

Applying the ADM $d+1$ decomposition presented in Chapter \ref{C:NR} to the vector of constraints \eqref{eq:Zmu}, we obtain the following constraints \cite{Alic:2011gg,Alic:2013xsa},
\begin{subequations}\label{constraints_C}
\begin{align}
\Theta\equiv\tfrac{1}{2}\Big[H^{\perp}+K+\frac{1}{\alpha^2}(1+a(x))\partial_{\perp}\alpha\Big]&=0\,, \label{eq:def_theta}\\
Z_i\equiv-\tfrac{1}{2}\Big[H_i + \Gamma_i - \frac{1+a(x)}{\alpha}\left(D_i\alpha +\frac{\gamma_{ij}}{\alpha}\partial_{\perp}\beta^j\right)\Big]&=0\,, \label{eq:def_Zi} 
\end{align}
\end{subequations}
where $H^{\perp}=n^{\mu}H_{\mu}$, $H_i=\gamma_i^{~\mu}H_{\mu}$ and $\Gamma_i\equiv \gamma_{ij}\gamma^{kl}\Gamma^{j}_{kl}$, with $\Gamma^{k}_{ij}$ being the Christoffel symbols of the spatial metric $\gamma_{ij}$. Therefore, we see that in the standard puncture gauge (i.e. when $a(x)=b(x)=0$), by choosing $H_i$ and $H^{\perp}$ as
\begin{subequations}\label{sources_choice}
\begin{align}
H^{\perp}=&\,(2\Theta-K)\Big(1-\tfrac{2}{\alpha} \Big)\,, \\ 
H_i =&\, \tfrac{D_i\alpha}{\alpha}+\tfrac{d-2}{2}\partial_i\chi+\hat{\Gamma}_i\Big(\tfrac{d}{2(d-1)\alpha^2}-\chi \Big)\,,
\end{align}
\end{subequations}
the conditions in \eqref{constraints_C} lead to the usual  $1+\log$ slicing and the (integrated) Gamma-driver evolution equations in \eqref{gauge}. 

However, the gauge coming from the evolution equations \eqref{gauge} is not adequate for our purposes since it does not have any dependency on the function $a(x)$ and hence it does not take advantage of the corresponding auxiliary metric that we have introduced. A suitably modified version of the $1+\log$ slicing and Gamma-driver equations \eqref{gauge} can be found instead by introducing the same choice of source functions as in \eqref{sources_choice} to the constraints \eqref{constraints_C} with $a(x)\neq 0$. 
The resulting modified gauge evolution equations become
\begin{subequations}\label{mgauge}
\begin{align}
&\partial_{\perp}\alpha = -\frac{2\alpha}{1+a(x)}(K-2\Theta)\,, \\
&\partial_{\perp}\beta^i = \frac{d}{2(d-1)} \frac{\hat{\Gamma}^i}{1+a(x)}-\frac{a(x)\,\alpha}{1+a(x)}\,D^i\alpha-\eta\beta^i\,.
\end{align} 
\end{subequations}

\section{Constraint damping}\label{sec:damping}

In this section we analyse the stability bounds for the damping terms that we have included in the Einstein's field equations in \eqref{ESFE} in a similar way as other NR formulations \cite{Gundlach:2005eh,Pretorius:2004jg,Alic:2011gg,Alic:2013xsa}. By analysing the propagation of the constraint violating modes around Minkowski space following the same procedure as in \cite{Gundlach:2005eh}, we obtain the bounds
\begin{equation}
    \kappa_1>0\,,\quad  \kappa_2>-\tfrac{2}{2+b(x)}\,,
\end{equation} 
which guarantee that constraint violating modes are exponentially suppressed (around a Minkowski background). Next, we give the details of the calculation of these bounds. 

Taking the divergence of \eqref{ESFE}, one gets
\begin{align}
\Box Z_{\mu}+R_{\mu\nu}Z^{\nu}-\kappa_1\nabla^{\nu}\left(2n_{(\mu}Z_{\nu)}+\hat{\kappa}_2g_{\mu\nu}n^{\rho}Z_{\rho}\right)\nonumber\\=\nabla^{\nu}\left[b(x)\left(2n_{\beta}n_{(\mu}\delta^{\alpha}_{\nu)}\nabla^{\beta}Z_{\alpha}-n_{\mu}n_{\nu}\nabla^{\rho}Z_{\rho}\right)\right]\,,
\end{align}
where we have defined $\hat{\kappa}_2=\frac{d-3}{2}+\frac{d-1}{2}\kappa_2$ as a shorthand notation. 

Now, linearising around a Minkowski background solution $g_{\mu\nu}^{(0)}$, hence with $R_{\mu\nu}^{(0)}=0$ and $Z_{\mu}^{(0)}=0$, and going to a frame where $n^{\mu}=(1,0,...,0)$, one gets without loss of generality that \footnote{Note that we have neglected the derivatives of the function $b(x)$, since we can assume that they have a small contribution in comparison to the other terms, especially given that we are setting $b(x)$ to be constant in all our simulations.}
\begin{subequations}
\begin{align}
\big(\Box-b(x)\partial_t^2\big) Z_0 -\kappa_1\big[(2+\hat{\kappa}_2)\partial_tZ_0-\partial^iZ_i \big]&= 0\,, \hspace{0.6cm} \\
\big(\Box-b(x)\partial_t^2\big) Z_i -\kappa_1\big(\partial_tZ_i+\hat{\kappa}_2\partial_iZ_0 \big)&= 0\,.
\end{align}
\end{subequations}

Then, using a plane-wave ansatz $Z_{\mu}=e^{st+i\,k_ix^i}\hat{Z}_{\mu}$, we are led to the following eigenvalue problem,
\begin{align}\label{mat_damping}
\begin{pmatrix}
\xi-\kappa_1(1+\hat{\kappa}_2)s && i\,\kappa_1\,k && 0 \\
-i\,\kappa_1\hat{\kappa}_2\,k && -\xi && 0 \\
0 && 0 && -\xi
\end{pmatrix}
\begin{pmatrix}
\hat{Z}_0 \\ \hat{Z}_n \\ \hat{Z}_A
\end{pmatrix}
=0\,,
\end{align}
where $\xi=-s^2(1+b(x))-k^2-\kappa_1s$, $\hat{Z}_n$ is the component of $\hat{Z}_i$ in the direction of $k_i$ and $\hat{Z}_A$ are the components orthogonal to $k_i$.

The eigenvalues for $\hat{Z}_A$ are given by
\begin{align}
s=\tfrac{-\kappa_1}{2(1+b(x))}\pm\sqrt{\Big(\tfrac{\kappa_1}{2(1+b(x))}\Big)^2-\tfrac{k^2}{1+b(x)}}\,,
\end{align}
while the corresponding eigenvalues for $\hat{Z}_0$ and  $\hat{Z}_n$ are more complicated and can be found by setting to zero the determinant of the upper-left block of the matrix in \eqref{mat_damping}, which yields the following quartic polynomial equation,
\begin{align}
    \left((1+b(x))s^2+k^2\right)^2 +\kappa_1^2\left(-k^2\hat{\kappa}_2+s^2(2+\hat{\kappa}_2)\right) \nonumber\\+ \kappa_1\,s \left((1+b(x))s^2+k^2\right)(3+\hat{\kappa}_2)=0\,.
\end{align}
For the special case $\hat{\kappa}_2=0$ they take the simple form
\begin{equation}
s=-\tfrac{\kappa_1}{1+b(x)}\pm\sqrt{\Big(\tfrac{\kappa_1}{1+b(x)}\Big)^2-\tfrac{k^2}{1+b(x)}}\,.
\label{eq:eigen_ZA}
\end{equation}
In this case one has that for large wavenumbers, $k\gg \kappa_1$,
\begin{equation}
s\approx -\tfrac{\kappa_1}{1+b(x)}\pm \tfrac{i\,k}{\sqrt{1+b(x)}}\,,\,  s\approx -\tfrac{\kappa_1}{2(1+b(x))}\pm\tfrac{i\,k}{\sqrt{1+b(x)}}\,,
\end{equation}
while for small wavenumbers, $k\ll \kappa_1$, we get
\begin{align}
s\approx -\tfrac{\kappa_1}{1+b(x)}\,,~ -\tfrac{k^2}{\kappa_1}\,,~ -\tfrac{2\kappa_1}{1+b(x)}\,,~ -\tfrac{k^2}{2\kappa_1}\,.
\end{align}
Clearly from \eqref{eq:eigen_ZA}, the real part is always negative, which implies that these modes are always damped. We have verified that the eigenvalues for $\hat{Z}_0$ and $\hat{Z}_n$ are undamped for $\hat{\kappa}_2<-\frac{2}{2+b(x)}$; for $\hat{\kappa}_2=-\frac{2}{2+b(x)}$, they also have a simple form, namely
\begin{subequations}
\begin{align}
s=&\pm\tfrac{i\,k}{\sqrt{1+b(x)}}, \\
s=&-\tfrac{(4+3b(x))\kappa_1}{2(1+b(x))(2+b(x))}\pm\sqrt{\Big(\tfrac{b(x)\kappa_1}{2(2+b(x))(1+b(x))}\Big)^2-\tfrac{k^2}{1+b(x)}}\,,
\end{align}
\end{subequations}
and hence they are undamped for all values of $k_i$. 

Therefore we conclude that damping occurs for $\kappa_1>0$ and $\hat{\kappa}_2>-\frac{2}{2+b(x)}$, which implies that $\kappa_2>-\frac{2}{2+b(x)}$.

\section{Equations of motion in $d+1$ form} 
\label{sec:3p1}

In this section we present the $(d+1)$-dimensional equations of Einstein gravity (i.e., GR without modifications), with a general matter source $T_{\mu\nu}$, in the modified CCZ4 (mCCZ4) formalism (hence derived from \eqref{ESFE}). 

We start considering the usual decomposition of the energy momentum tensor of the matter, 
\begin{equation}
\rho=n^{\mu}n^{\nu}T_{\mu\nu}\,,\quad J_i=-n^{\mu}\gamma_i ^{\phantom{i}\nu}T_{\mu\nu}\,,\quad S_{ij}=\gamma_i^{\phantom{i}\mu}\gamma_j^{\phantom{j}\nu}T_{\mu\nu}\,.
\end{equation}  
In the case of a massless scalar field with stress-energy tensor as defined in \eqref{eq:scalar_st}, we get
\begin{subequations}
\begin{align}
   \hspace{-0.5cm}\rho^{\phi}=&~\tfrac{1}{2}\big(K_{\phi}^2+(\partial\phi)^2\big)\,,\\
   \hspace{-0.5cm} J_i^{\phi}=&~K_{\phi}\,\partial_i\phi\,,\\
   \hspace{-0.5cm} S_{ij}^{\phi}=&~(\partial_i\phi)(\partial_j\phi)+\tfrac{1}{2}\,\gamma_{ij}\big(K_{\phi}^2-(\partial\phi)^2\big)\,,
\end{align}
\end{subequations}
with $(\partial\phi)^2=\gamma^{ij}(\partial_i\phi)(\partial_j\phi)$ and $K_{\phi}=-\tfrac{1}{\alpha}\partial_{\perp}\phi$.

The resulting $d+1$ form of the Einstein field equations coupled to matter in the mCCZ4 formalism yields
\begin{subequations}\label{eqsccz4}
\begin{align}
\partial_{\perp}\tilde{\gamma}_{ij} =& -2\alpha\tilde{A}_{ij}+2\tilde\gamma_{k(i}\partial_{j)}\beta^k-{\frac{2}{d}}\tilde{\gamma}_{ij}\partial_k\beta^k\,, \\
\partial_{\perp}\chi =&~ {\frac{2}{d}}\,\chi\,\big(\alpha K - \partial_k\beta^k\big)\,, \\
\partial_{\perp}K=&-D^iD_i\alpha +\alpha\left[R+2\,D_iZ^i +K(K-2\Theta)\right] -d\,\kappa_1(1+\kappa_2)\,\alpha\,\Theta\nonumber\\
&+{\frac{\kappa\,\alpha}{d-1}}\big[S-d\,\rho\big]-{\frac{d\,\alpha\,b(x)}{2(d-1)(1+b(x))}}\Big[R-\tilde{A}_{ij}\tilde{A}^{ij}+{\frac{d-1}{d}}K^2-2\,\kappa\,\rho \nonumber\\
&\hspace{6.5cm}-(d-1)\kappa_1(2+\kappa_2)\,\Theta\Big]\,,\\
\partial_{\perp}\Theta =&~{\frac{\alpha}{2}}\,\Big[R-\tilde A_{ij}\,\tilde A^{ij}+{\frac{d-1}{d}}\,K^2 -2\,\kappa\,\rho
+2\,(D^iZ_i-\Theta\,K)\Big]\nonumber\\&-{\frac{\kappa_1}{2}}\big(d+1+(d-1)\kappa_2\big)\,\alpha\,\Theta-Z_i\,D^i\alpha\nonumber\\
&-{\frac{b(x)}{1+b(x)}}\Big\{ {\frac{\alpha}{2}}\,\Big[R-\tilde A_{ij}\,\tilde A^{ij}+{\frac{d-1}{d}}\,K^2- 2\kappa\,\rho \Big] \nonumber\\
&\hspace{2cm} -{\frac{\kappa_1}{2}}\Big[{\frac{d-3}{2+b(x)}}+d+1+(d-1)\kappa_2\Big]\,\alpha\,\Theta
\Big\}\,,\\
\partial_{\perp}\tilde{A}_{ij}=&~\alpha\,[\tilde{A}_{ij}(K-2\Theta)-2\,\tilde{A}_{ik}\tilde{A}^k_{~j}]+2\tilde{A}_{k(i}\partial_{j)}\beta^k- {\frac{2}{d}}(\partial_k\beta^k)\tilde{A}_{ij}\nonumber\\&+\chi\left[\alpha\left(R_{ij} + 2D_{(i}Z_{j)}-\kappa\,S_{ij}\right)-D_iD_j\alpha\right]^{\text{TF}} \,,\\
\partial_{\perp}\hat\Gamma^i
=&~2\,\alpha\,\Big[\tilde\Gamma^i_{\phantom i kl}\tilde A^{kl}-{\frac{d-1}{d}}\tilde\gamma^{ij}\partial_jK-{\frac{d}{2\,\chi}}\,\tilde A^{ij}\partial_j\chi\Big] 
-2\,\tilde A^{ij}\partial_j\alpha-\hat\Gamma^j\partial_j\beta^i + {\frac{2}{d}}\,\hat\Gamma^i\partial_j\beta^j \nonumber\\&+ {\frac{d-2}{d}}\,\tilde\gamma^{ik}\partial_k\partial_j\beta^j + \tilde\gamma^{jk}\partial_j\partial_k\beta^i 
+2\,\alpha\,\tilde\gamma^{ij}\Big[\partial_j\Theta- {\frac{1}{\alpha}}\,\Theta\,\partial_j\alpha - {\frac{2}{d}}\,K\,Z_j - \kappa\, J_j\Big]\nonumber\\
&-2\,\kappa_1\,\alpha\,\tilde\gamma^{ij}Z_j\,-{\frac{2\alpha\,b(x)}{1+b(x)}}\Big\{
\tilde D_j\tilde A^{ij}-{\frac{d-1}{d}}\tilde\gamma^{ij}\partial_jK-{\frac{d}{2\,\chi}}\tilde A^{ij}\partial_j\chi - \tilde A^{ij}Z_j\nonumber\\
&\hspace{4.5cm}+\tilde\gamma^{ij}\Big[\partial_j\Theta-{\frac{1}{d}}\,K\,Z_j  -\kappa_1\,Z_j-\kappa\,J_j\Big]
\Big\}\,.
\end{align}
\end{subequations}

Note that by setting $b(x)=0$ and $d=3$ in \eqref{eqsccz4} we recover the equations derived in \cite{Brown:2011qg}. 

In the case of a massless scalar field minimally coupled to GR, we would also have the corresponding equation of motion for the scalar field,
\begin{equation}
    \Box\phi=0\,.
\end{equation} 
This equation can be written as two  first order (in time) equations for the scalar field $\phi$ and its curvature $K_{\phi}$; in the  $d+1$ decomposition of the spacetime metric \eqref{eq:adm_metric}, they are given by
\begin{subequations}\label{eq:scalar_eqs}
    \begin{align}
        \partial_{\perp}\phi =& -\alpha\, K_\phi\,, \\
        \partial_{\perp}K_{\phi} =&~ \alpha\,(-D^iD_i\phi + KK_{\phi}) - (D^i\phi) D_i\alpha\,.
    \end{align}    
\end{subequations}
Equations \eqref{mgauge}, \eqref{eqsccz4} together with \eqref{eq:scalar_eqs} provide the closed system of evolution equations whose hyperbolicity we will analyse in next section.

Regarding the $(d+1)$-dimensional evolution equations of Einstein-Gauss-Bonnet gravity and the Four-Derivative Scalar-Tensor theory of gravity in the modified puncture formalism, they are presented in Appendix \ref{App:EoM}.

\section{Hyperbolicity analysis}\label{sec:hyp}

In this section we prove the strong hyperbolicity of the $d+1$ equations of Einstein-Gauss-Bonnet gravity and the Four-Derivative Scalar-Tensor theory of gravity in the weak coupling regime (which will be properly defined in both Subsections \ref{subsec:hyp_EGB} and \ref{subsec:hyp_4dST}) in the formalism we have just presented. 

To this aim we start by studying the Einstein-scalar-field equations, which is the background on which the studied modified gravity theories are defined (though we will omit the scalar field for the first one), in the same way as in \cite{Brown:2011qg}.

\subsection{Einstein-scalar-field equations}

We start by writing down the principal part of the equations. For this purpose we need to introduce an orthonormal $d$-bein (triad in $d=3$) consisting of a unit covector $\xi_i$, such that $\xi_i\gamma^{ij}\xi_j=1$, and unit vectors $e_A^i$ with $A=1,...,d-1$ such that $\xi_ie_A^i=0$ and $e_A^i\gamma_{ij}e_B^j=\delta_{AB}$. Then, keeping the highest derivative terms in the equations \eqref{mgauge}, \eqref{eqsccz4} and \eqref{eq:scalar_eqs}, and replacing $\partial_{\mu}\to i\xi_{\mu}\equiv i(\xi_0,\xi_i)$,\footnote{Note that this $i$ factor differs from the conventions in \cite{Brown:2011qg}; here we follow instead the conventions of \cite{Kovacs:2020pns,Kovacs:2020ywu}.} we obtain the system
\begin{align}
i\xi_0U={\mathbb M}(\xi_k)U, 
\label{eq:principle_part}
\end{align}
where $U$ is a $2(3d+2)$-dimensional vector accounting for the principal part of the CCZ4 variables plus the scalar field $\phi$ and its curvature $K_{\phi}$, where we have taken into account the constraints $\det(\tilde{\gamma}_{ij})=1$ and $\text{Tr}(\tilde{A}_{ij})=0$, and ${\mathbb M}(\xi_k)$ is a $2(3d+2)\times2(3d+2)$-dimensional matrix dependent on the covector $\xi_i$ and the variables of the system. Explicitly, the principal part \eqref{eq:principle_part} for the Einstein-scalar-field system in our modified gauge is given by
\begin{subequations}\label{shypccz4}
\begin{align}
i\check{\xi}_0\hat{\tilde{\gamma}}_{ij} =&~ 2i\tilde{\gamma}_{k(i}\xi_{j)}\hat{\beta}^k-2\alpha\hat{\tilde{A}}_{ij}-\tfrac{2i}{d}\tilde{\gamma}_{ij}\xi_k\hat{\beta}^k\,, \\
i\check{\xi}_0\hat{\chi} =&~\tfrac{2}{d}\,\chi\,\big(\alpha\hat{K}-i\xi_k\hat{\beta}^k \big)\,, \\
i\check{\xi}_0\hat{\phi} =&-\alpha \hat{K}_{\phi}\,, \\
i\check{\xi}_0\hat{K} =&~\hat{\alpha} + i\alpha\chi \xi_i\hat{\hat{\Gamma}}^i-\alpha\Big[\tfrac{d-1}{\chi}\hat{\chi}-\tfrac{1}{2}\tilde{\gamma}^{jk}\hat{\tilde{\gamma}}_{jk}\Big]\nonumber\\&+\tfrac{b(x)d\alpha}{2\chi(d-1)(1+b(x))}\big(\xi^l\xi^k\hat{\tilde{\gamma}}_{kl}-\tilde{\gamma}^{jk}\hat{\tilde{\gamma}}_{jk}+(d-1)\hat{\chi} \big) \,, \\
i\check{\xi}_0\hat{K}_{\phi} =&~ \alpha\hat{\phi}\,, \\
i\check{\xi}_0\hat{\Theta} =& -\tfrac{\alpha}{2}\Big[\frac{d-1}{\chi}\hat{\chi}-\tfrac{1}{2}\tilde{\gamma}^{ij}\hat{\tilde{\gamma}}_{ij} \Big]+ \tfrac{i\alpha\chi \xi_i\hat{\hat{\Gamma}}^i}{2} \nonumber\\& +\tfrac{b(x)\alpha}{2\chi(1+b(x))}\big(\xi^l\xi^k\hat{\tilde{\gamma}}_{kl}-\chi\tilde{\gamma}^{jk}\hat{\tilde{\gamma}}_{jk}+(d-1)\hat{\chi} \big)\,, \\
i\check{\xi}_0\hat{\tilde{A}}_{ij} =&~ \Big[\xi_i\xi_j - \tfrac{1}{d}\tfrac{\tilde{\gamma}_{ij}}{\chi}\Big]\Big[\chi\hat{\alpha} - \tfrac{(d-2)\alpha}{2}\hat{\chi}\Big]+i\alpha\chi\Big[\tilde{\gamma}_{k(i}\xi_{j)}\hat{\hat{\Gamma}}^k-\tfrac{1}{d}\tilde{\gamma}_{ij}\xi_k\hat{\hat{\Gamma}}^k \Big]\nonumber\\&+\tfrac{1}{2}\alpha \Big[\hat{\tilde{\gamma}}_{ij}-\tfrac{1}{d}\tilde{\gamma}_{ij}\tilde{\gamma}^{kl}\hat{\tilde{\gamma}}_{kl} \Big]\,, \\
i\check{\xi}_0\hat{\hat{\Gamma}}^i =&~ \tfrac{2i\alpha\xi^i}{\chi}\Big[\hat{\Theta}-\tfrac{d-1}{d}\hat{K}\Big]- \tfrac{1}{\chi}\Big[\hat{\beta}^i + \tfrac{d-2}{d}\xi^i\xi_l\hat{\beta}^l \Big]\nonumber\\&-\tfrac{2i\alpha b(x)}{1+b(x)}\Big\{\tfrac{\xi^i}{\chi}\Big[\hat{\Theta}-\tfrac{d-1}{d}\hat{K}\Big]
+\xi_j\hat{\tilde{A}}^{ij} \Big\}\,,\\
i\check{\xi}_0\hat{\alpha} =& -\tfrac{2\alpha}{1+a(x)} (\hat{K}-2\hat{\Theta})\,, \\
i\check{\xi}_0\hat{\beta}^i =&~ \tfrac{d}{2(d-1)}\hat{\hat{\Gamma}}^i+\tfrac{a(x)}{1+a(x)}\Big[\tfrac{d}{2(d-1)}\hat{\hat{\Gamma}}^i-i\alpha\xi^i\hat{\alpha}\Big]\,, 
\end{align}
\end{subequations}
where $\check{\xi}_0=\xi_0-\beta^i\xi_i$ and the hat $\hat\,$ denotes the background values of the corresponding variables.

In the following,  we use the notation $\perp$ to denote the contraction of any tensor $T^i$ with the normal covector $\xi_i$, e.g., $T_\perp=T^i\xi_i$; therefore,  $\hat{\gamma}_{\perp\perp}=\hat{\gamma}_{ij}\xi^i\xi^j$ and so on. Similarly, upper case Latin indices are defined by contractions with the components of the $d$-bein; for instance, $\hat{\gamma}_{AB}=\hat{\gamma}_{ij}e_A^ie_B^j$ and analogously for the other variables. Having introduced the notation, we can now decompose the principal part of the equations \eqref{shypccz4} into a scalar, vector and traceless tensor blocks depending on how they transform with respect to transformations of the $d$-bein vectors $e_A^i$.  The tensor block  is given by
\begin{subequations}\label{hyp_tensor}
\begin{align}
i\check{\xi}_0\hat{\tilde{\gamma}}_{AB}^{\text{TF}} =& -2\alpha\hat{\tilde{A}}_{AB}^{\text{TF}}\,, \\ i\check{\xi}_0\hat{\tilde{A}}_{AB}^{\text{TF}} =&~ \tfrac{\alpha}{2}\hat{\tilde{\gamma}}_{AB}^{\text{TF}}\, ,
\end{align}
\end{subequations}
with eigenvalues $\check{\xi}_0=\pm\alpha$. Note that this block is unchanged with respect to the GR case in standard puncture gauge.%

The vector block is
\begin{subequations}\label{hyp_vector}
\begin{align}
i\check{\xi}_0\hat{\tilde{\gamma}}_{\perp A}=&~i\chi\hat{\beta}_A-2\alpha\hat{\tilde{A}}_{\perp A}\,, \\ i\check{\xi}_0\hat{\tilde{A}}_{\perp A} =&~ \tfrac{\alpha}{2}\hat{\tilde{\gamma}}_{\perp A} + \tfrac{i\alpha\chi^2}{2}\hat{\hat{\Gamma}}_A\,, \\ i\check{\xi}_0\hat{\beta}_A =&~ \tfrac{d}{2(d-1)(1+a(x))}\hat{\hat{\Gamma}}_A\,, \\ i\check{\xi}_0\hat{\hat{\Gamma}}_A =& -\tfrac{1}{\chi}\hat{\beta}_A -\tfrac{2ib(x)\alpha}{\chi^2(1+b(x))}\hat{\tilde{A}}_ {\perp A}\,,
\end{align}
\end{subequations}
with eigenvalues $\check{\xi}_0=\pm\frac{\alpha}{\sqrt{1+b(x)}}$ and $~\pm\sqrt{\frac{d}{2(d-1)\chi(1+a(x))}}$. These eigenvalues are degenerate for $\alpha^2\chi=\frac{d}{2(d-1)}\frac{1+b(x)}{1+a(x)}$,  which can be avoided if we choose $b(x)>\frac{d-2}{d}+\frac{2(d-1)a(x)}{d}$ given the ranges of $\alpha$ and $\chi$. This degeneracy reduces to the one already present in the standard CCZ4 formulation of GR when $a(x)=b(x)=0$, which does not cause problems in practical applications. The same appears to happen in our new formulation. Therefore, in practice we can replace this constraint by $b(x)\neq\frac{d-2}{d}+\frac{2(d-1)a(x)}{d}$ so as to avoid the degeneracy at spatial infinity.

Finally, the scalar block is
\begin{subequations}\label{hyp_scalar}
\begin{align}
i\check{\xi}_0\hat{\tilde{\gamma}}_{\perp\perp} =&~ \tfrac{2i(d-1)}{d}\chi\hat{\beta}^{\perp} - 2\alpha\hat{\tilde{A}}_{\perp\perp}\,,\\
i\check{\xi}_0\hat{\chi} =&~ \tfrac{2}{d}\chi(\alpha\hat{K} - i\hat{\beta}^{\perp} )\,, \\
i\check{\xi}_0\hat{\phi} =& -\alpha\hat{K}_{\phi}\,,\\
i\check{\xi}_0\hat{K} =& \hat{\alpha}+i\alpha\chi\hat{\hat{\Gamma}}^{\perp}+\tfrac{\alpha}{2\chi}\big(\hat{\tilde{\gamma}}_{\perp\perp}+\hat{\tilde{\gamma}}_{AB}\delta^{AB}\big)-\tfrac{\alpha}{d-1}\tfrac{\hat{\chi}}{\chi}\nonumber\\&
-\tfrac{b(x)d\alpha}{2\chi(1+b(x))}\Big[\tfrac{1}{d-1}\hat{\tilde{\gamma}}_{AB}\delta^{AB}-\hat{\chi}\Big]\,, \\
i\check{\xi}_0\hat{K}_{\phi} =&~ \alpha\hat{\phi}\,, \\
i\check{\xi}_0\hat{\Theta} =&~ \tfrac{i}{2}\alpha\chi\hat{\hat{\Gamma}}^{\perp} +\tfrac{\alpha}{4\chi}\big(\hat{\tilde{\gamma}}_{\perp\perp}+\hat{\tilde{\gamma}}_{AB}\delta^{AB}\big)- \tfrac{(d-1)\alpha}{2}\tfrac{\hat{\chi}}{\chi}\nonumber\\&-\tfrac{\alpha b(x)}{2(1+b(x))\chi}\big(\hat{\tilde{\gamma}}_{AB}\delta^{AB}-(d-1)\hat{\chi}\big)\,,\\
i\check{\xi}_0\hat{\tilde{A}}_{\perp\perp}=&~\tfrac{d-1}{d}\chi\hat{\alpha} - \tfrac{(d-1)(d-2)\alpha}{2d}\hat{\chi}+i\tfrac{(d-1)\alpha}{d}\chi^2\hat{\hat{\Gamma}}^{\perp}\nonumber\\&  - \tfrac{\alpha}{2d}\big(\hat{\tilde{\gamma}}_{AB}\delta^{AB}-(d-1)\hat{\tilde{\gamma}}_{\perp\perp})\big)\,,\\
i\check{\xi}_0\hat{\alpha} =& -\tfrac{2\alpha}{1+a(x)} (\hat{K}-2\hat{\Theta})\,,\\
i\check{\xi}_0\hat{\beta}^{\perp}=&~\tfrac{d}{2(d-1)(1+a(x))}\hat{\hat{\Gamma}}^{\perp}-\tfrac{ia(x)}{1+a(x)}\alpha \hat{\alpha}\,,\\
i\check{\xi}_0\hat{\hat{\Gamma}}^{\perp} =&~ \tfrac{2i\alpha}{\chi}\Big[\hat{\Theta}-\tfrac{d-1}{d}\hat{K}\Big]-\tfrac{2(d-1)}{d}\tfrac{1}{\chi}\hat{\beta}^{\perp}\nonumber\\& -\tfrac{2i\alpha b(x)}{\chi(1+b(x))}\Big[\hat{\Theta}-\tfrac{d-1}{d}\hat{K}+\tfrac{1}{\chi}\hat{\tilde{A}}_{\perp\perp}\Big]\,,
\end{align}
\end{subequations}
with eigenvalues $\check{\xi}_0 = \pm\frac{1}{\sqrt{\chi(1+a(x))}}\,,~\pm\sqrt{\frac{2\alpha}{1+a(x)}}\,,~\pm\alpha$ and $\pm\frac{\alpha}{\sqrt{1+b(x)}}$, with the last pair of multiplicity $2$. The eigenvalues are degenerate for $\alpha=\frac{1}{2\chi}$, $\alpha^2=\frac{1+b(x)}{\chi(1+a(x))}$ and $\alpha=\frac{2(1+b(x))}{1+a(x)}$,  which do not spoil the hyperbolicity of the system as long as $a(x)\neq b(x)$, so that again we avoid degeneracy at spatial infinity.\footnote{We note that while for $\alpha^2=\frac{1}{\chi(1+a(x))}$ some of the eigenvalues coincide, the corresponding eigenvectors remain distinct and hence there is no degeneracy in this case.}

To summarise, the constraints on the functions $a(x)$ and $b(x)$ that guarantee the hyperbolicity of the system are 
\begin{equation}
\begin{aligned}
&0<b(x)\neq\tfrac{d-2}{d}+\tfrac{2(d-1)a(x)}{d} \text{ and}\,\Bigg\{\hspace{-0.5cm}\begin{array}{cc}-1<a(x)<0 \\ \text{or} \\ \hspace{0.7cm}0 < a(x) < b(x)\end{array}\\
&\qquad\qquad\qquad\qquad\text{or}\\
&a(x)\neq 1+2b(x) \text{ and } 0 < b(x) < a(x)\,.
\end{aligned}
\end{equation}

Following \cite{Kovacs:2020pns,Kovacs:2020ywu}, we can classify the eigenvalues that we have computed into three types:\footnote{Note that the plus and minus signs of $\check{\xi}_0$ correspond to the ongoing and outgoing modes.}
\begin{itemize}
    \item Physical eigenvalues: $\check{\xi}_0=\pm\alpha$ with multiplicity $d$, consisting of the $2(d-1)$ polarisations of the gravitational field plus the additional two polarisations from the scalar field.\footnote{Note that the physical eigenvalues are the only ones not dependent on the functions $a(x)$ and $b(x)$ given that they are not gauge-dependent.}
    \item ``Gauge-condition violating'' eigenvalues: $~\check{\xi}_0=\pm\sqrt{\frac{2\alpha}{1+a(x)}}$, $~\pm\frac{1}{\sqrt{\chi(1+a(x))}}$ and $~\pm\sqrt{\frac{d}{2(d-1)\chi(1+a(x))}}$, with the last pair of multiplicity $d-1$.
    \item ``Pure-gauge'' eigenvalues: $\check{\xi}_0=\pm\frac{\alpha}{\sqrt{1+b(x)}}$ with multiplicity $d+1$.
\end{itemize}
Their corresponding eigenvectors have been explicitly written in Appendix \ref{app:eigenvectors}. Clearly the eigenvalues are real (recall that in all cases $a(x)>-1$ and $b(x)>0$), they smoothly depend on $\xi_i$ and so do the corresponding eigenvectors, which are linearly independent with each other.

Hence, we conclude that the hyperbolicity matrix $\mathbb M$ is diagonalisable. Moreover, the propagation of the constraints of the system (see Appendix \ref{app:constraints}) is strongly hyperbolic, showing that if they are satisfied at the initial time they will continue to be satisfied at future times. 

Therefore, we have proved that the system is strongly hyperbolic and, thus, well-posed. In the following two subsections we extend this well-posedness result to Einstein-Gauss-Bonnet gravity and the Four-Derivative Scalar-Tensor theory of gravity, which were introduced in Chapter \ref{C:MG}.

\subsection{Einstein-Gauss-Bonnet gravity}\label{subsec:hyp_EGB}

Here we show that the EGB gravity coming from the action \eqref{eq:action_Lovelock} with $d=4$ spatial dimensions \footnote{We believe, as for the result in \cite{Kovacs:2020ywu}, that well-posedness holds in our modified CCZ4 formulation for any EGB theory of gravity with arbitrary dimensions, but the method we have employed for such proof makes computations more difficult to cope with when the number of dimensions is not specified. This is the reason why we have used the minimum value of $d$ such that the modified gravity contribution is non-trivial.} is well-posed in our modified CCZ4 formulation. To this purpose, we need to find the principal part of the full theory, described by the evolution equations in Appendix \ref{App:EoMEGB} together with the modified puncture equations in  \eqref{mgauge}, which can be written down as
\begin{align}
    \mathbb{M}=\mathbb{M}_0+\delta\mathbb{M}\,,
    \label{eq:M_full}
\end{align}
where ${\mathbb M}_0$ is the principal part of the Einstein theory, already computed in \eqref{hyp_tensor}--\eqref{hyp_scalar} (without the contributions of the scalar field) and $\delta\mathbb{M}=\lambda^{\text{GB}}\mathbb{M}^{\text{GB}}$ are the contributions from the higher derivative terms, which are small compared to $\mathbb M_0$ in the weakly coupled regime, namely when
\begin{align}
    \left|\lambda^{\text{GB}}\right|\ll L^2\,,
\end{align}
where $L$ is the physical length scale of the spacetime, which can be heuristically defined as $L^{-1}=\max\{|R_{\mu\nu\rho\sigma}|^{1/2}\}$. The explicit form of ${\mathbb M}^{\text{GB}}$ can be found in the Appendix \ref{app:matEGB}.

Therefore, in order to prove that the full theory is well-posed in an open neighbourhood around the Einstein theory, we shall proceed by explicitly computing the eigenvalues and eigenvectors of \eqref{eq:M_full} perturbatively and showing that $\mathbb M$ has real eigenvalues and is diagonalisable.

Consider one of the eigenvalues\footnote{Here we suppress the subscript $0$ on $\xi_0$ to simplify the notation.} of the unperturbed principal part $\mathbb{M}_0$, namely $\xi$ with multiplicity $N^\xi$; let their associated right and left eigenvectors be $\{{\mathbf v}^{\xi}_{\text{\tiny R},i}\}_{i=1}^{N^{\xi}}$ and $\{{\mathbf v}^{\xi}_{\text{\tiny L},i}\}_{i=1}^{N^{\xi}}$ respectively, such that ${\mathbb M}_0{\mathbf v}^{\xi}_{\text{\tiny R},i}=\xi {\mathbf v}^{\xi}_{\text{\tiny R},i}$ and ${\mathbf v}^{\xi}_{\text{\tiny L},i}{\mathbb M_0}=\xi {\mathbf v}^{\xi}_{\text{\tiny L},i}$. Recall that the explicit form of the background right eigenvectors has been written down in Appendix \ref{app:eigenvectors}.

Then, the expressions of the perturbed eigenvalues $\left\{\xi+\delta\zeta^{\xi}_i\right\}_{i=1}^{N^{\xi}}$ and eigenvectors $\left\{\alpha^{\xi}_i\cdot{\mathbf v}_{\text{\tiny R}}^{\xi} + \delta{\mathbf w}^{\xi}_i \right\}_{i=1}^{N^{\xi}}$ of the principal part ${\mathbb{M}}$ can be obtained by solving the following eigenvalue problem \cite{hinch},
\begin{subequations}
\begin{align}
{\mathcal T}^{\xi}\alpha^{\xi}_i &= i\delta\zeta^{\xi}_i\alpha^{\xi}_i \,, \label{eq:system_hinch1}\\ 
\left(\mathbb{M}_0-i\xi{\mathbb I} \right)\delta{\mathbf w}^{\xi}_i &=\left(i\delta\zeta^{\xi}_i{\mathbb I}-\delta\mathbb{M} \right)(\alpha^{\xi}_i\cdot{\mathbf v}_{\text{\tiny R}}^{\xi})\,, \label{eq:system_hinch2}
\end{align}    
\end{subequations}
where 
\begin{equation} \label{eq:Tij}
    {\mathcal T}^{\xi}_{ij}= \frac{{\mathbf v}_{\text{\tiny L},i}^{\xi\dagger}\delta\mathbb{M}\,{\mathbf v}_{\text{\tiny R},j}^{\xi}}{{\mathbf v}_{\text{\tiny L},i}^{\xi\dagger}{\mathbf v}_{\text{\tiny R},i}^{\xi}}\,.
\end{equation}
Note that \eqref{eq:system_hinch1} ensures that the r.h.s. of \eqref{eq:system_hinch2} has no components parallel to $\xi$, which implies that the matrix  ${\mathbb M}_0-i\xi{\mathbb I}$ on the l.h.s. of \eqref{eq:system_hinch2} is invertible \cite{hinch}. Also the denominator in equation \eqref{eq:Tij} is always non-zero or can be chosen so in case of degeneracy with the suitable linear combination between the eigenvectors (see Chapter 11 of \cite{recipes} for the proof).

In order to prove well-posedness we need to verify that the matrices $\left\{{\mathcal T}^{\xi}\right\}_{\xi\in\text{Spec}({\mathbb M}_0)}$ are diagonalisable and that the perturbed eigenvectors depend smoothly on the covector $\xi_i$.

After computing for the studied system of equations the projection matrices ${\mathcal T}^{\xi}$ corresponding to each type of eigenvalues, one can see that the only non-trivial contributions  occur for the physical eigenvalues. In this case, the explicit form of those projection matrices yields 
\begin{align}
{\mathcal T}^{\pm\alpha}\hspace{-0.1cm}=\hspace{-0.1cm}\pm\hspace{-0.1cm}\begin{pmatrix}
2P_{00} & -2P_{01} & -2P_{02} & -2P_{12} & -2P_{12} \\
-2P_{01} & 2P_{11} & -2P_{12} & 0 & 2P_{02} \\
-2P_{02} & -2P_{12} & 2P_{22} & 2P_{01} & 0 \\
-2P_{12} & 0 & 2P_{01} & 2P_{11} &  \text{\footnotesize $\begin{matrix} P_{11}+P_{22}\\-P_{00} \end{matrix}$}\\
-2P_{12} & 2P_{02} & 0 & \text{\footnotesize $\begin{matrix} P_{11}+P_{22} \\ -P_{00} \end{matrix}$} & 2P_{22}
\end{pmatrix},\hspace{0.2cm}
\end{align}
with
\begin{equation}
\begin{aligned}
    P_{AB}=\lambda^{\text{GB}}\,e_A^ie_B^j(&{\mathcal L}_nK_{ij}+\frac{1}{\alpha}D_iD_j\alpha+K_{ik}K^k_{~j}\\
    &-2\xi^kN_{ikj}+\xi^k\xi^lM_{ikjl})\,,
\end{aligned}
\end{equation}
where $M_{ijkl}$ and $N_{ijk}$ are defined in Eq. \eqref{eq:MijNi}. Finding explicit expressions of the first order corrections to the physical eigenvalues is not necessary since we know that they exist and that they are real given that ${\mathcal T}^{\pm\alpha}$ are real and  symmetric. 

Therefore, this fact together with the smoothness of all the coefficients in ${\mathbb M}^{\text{GB}}$ ensures the well-posedness of the weakly coupled EGB gravity in the $4+1$ modified CCZ4 formulation that we have developed.

\subsection{Four-Derivative Scalar-Tensor theory of gravity}\label{subsec:hyp_4dST}

In order to show well-posedness for the $4\partial$ST theory of gravity coming from the action \eqref{eq:4dST} in our modified CCZ4 gauge, we proceed with the same perturbation analysis done for the previous case. Here we write down again the principal part of the theory, namely the evolution equations in Appendix \ref{App:EoMEsGB} together with the modified puncture gauge equations in \eqref{mgauge}, as
\begin{align}
    \mathbb{M}=\mathbb{M}_0+\delta\mathbb{M}\,,
\end{align}
where in this case ${\mathbb M}_0$ is the principal part of the Einstein-scalar-field theory (here also including the scalar field part) and $\delta\mathbb{M}=\lambda^{\text{GB}}\mathbb{M}^{\text{GB}}+g_2\mathbb{M}^{X}$ are the contributions from the higher derivative terms, which are small compared to $\mathbb M_0$ in the weakly coupled regime, namely when
\begin{align}
    L\gg\sqrt{\left|\lambda'(\phi)\right|}, \ \sqrt{\left|g_2(\phi)\right|}\,,
\end{align}
where $L$ is again the physical length scale of the spacetime, which now needs to take into account as well the gradients of the scalar field ($L^{-1}=\max\{|R_{ij}|^{1/2},|\nabla_{\mu}\phi|,|\nabla_{\mu}\nabla_{\nu}\phi|^{1/2},|{\mathcal L}^{\text{GB}}|^{1/4}\}$), which will be defined again in Chapters \ref{C:BBHin4dST} and \ref{C:Hyp}. ${\mathbb M}^{\text{GB}}$ is also written down in the Appendix \ref{app:matEsGB} and, as for ${\mathbb M}^{X}$, its only contribution comes from
\begin{align}
{\mathbb M}^X\hat{K}_{\phi}=2\big[&K_{\phi}^2-\xi^i\xi^j(D_i\phi)(D_j\phi)\big]\hat{\phi}+2iK_{\phi}(\xi^iD_i\phi)\,\hat{K}_{\phi}\,.
\end{align}

Here again the only non trivial contributions to the eigenvalues occur for the physical eigenvalues. Setting $\epsilon=\pm1$, we have that the corresponding projection matrices are
\begin{align}
{\mathcal T}^{\epsilon\alpha}=\begin{pmatrix}2\sigma_{\epsilon}&\frac{2\epsilon}{\chi}\psi_{01}&-\frac{\epsilon}{\chi}\big(\psi_{00}-\psi_{11}\big)\\
\frac{\epsilon\chi}{2}\psi_{01}&2\eta_{\epsilon}&0\\
-\frac{\epsilon\chi}{4}\big(\psi_{00}-\psi_{11}\big)&0&2\eta_{\epsilon}\end{pmatrix},
\end{align}
where
\begin{align}
     \eta_{\epsilon}=&\left[2\xi_i\gamma^i_{\mu}n_{\nu}-\epsilon\left(n_{\mu}n_{\nu} + \xi_i\xi_j\gamma^i_{\mu}\gamma^j_{\nu} \right)\right]{\mathcal C}^{\mu\nu}\,,\\
     \sigma_{\epsilon}=&~\frac{g_2}{2}\left[\xi_i(D^i\phi) K_{\phi}+\epsilon\left(K_{\phi}^2-\xi_i\xi_j(D^i\phi)(D^j\phi) \right) \right] \,,\\
     \psi_{AB}=&~\lambda^{\text{GB}}e_A^ie_B^j\Big[{{\mathcal L}_nK_{ij}+\frac{1}{\alpha}D_iD_j\alpha}\nonumber\ \\
     &\hspace{1.8cm}+R_{ij}+KK_{ij}-K_i^{~k}K_{jk}
     +2\xi_k\big(D^kK_{ij}-D_{(i}K_{j)}^{~k} \big)\Big]\,,
\end{align} 
where ${\mathcal C}_{\mu\nu}=\nabla_{\mu}\nabla_{\nu}\lambda(\phi)$.

Apart from proving that ${\mathcal T}^{\pm\alpha}$ diagonalises, we can explicitly compute the six physical eigenvalues of the $4\partial$ST theory in mCCZ4 perturbatively up to first order; the two corresponding to the purely gravitational sector\footnote{These eigenvectors are null with respect to the effective metric $C^{\mu\nu}=g^{\mu\nu}-4{\mathcal C}^{\mu\nu}$ as described in \cite{Reall:2021voz}, so their actual exact expression can be computed, as we will see in Chapter \ref{C:Hyp}.} are given by
 \begin{align}
 \label{eq:eigen_phys1}
 \xi_0=&~\alpha\left(\epsilon+2\eta_{\epsilon}\right)\,,
 \end{align} 
 and the four corresponding to the mixed gravitational-scalar field polarisations are 
 \begin{align}
 \label{eq:eigen_phys2}
 \xi_0=&~\alpha\bigg(\epsilon+\eta_{\epsilon}+\sigma_{\epsilon}\pm\sqrt{\left(\eta_{\epsilon}-\sigma_{\epsilon}\right)^2+\psi_{12}^2+\left(\frac{\psi_{11}-\psi_{22}}{2}\right)^2}\bigg)\,,
 \end{align}
where for simplicity we have shifted $\xi_0-\beta^k\xi_k\to \xi_0$. Furthermore, it is straightforward to see that the smoothness conditions are satisfied. 

Hence, this proves well-posedness in the weakly coupled $4\partial$ST theory of gravity in the $3+1$ modified CCZ4 formulation that we have developed.

\addtocounter{page}{1}%
\blankpage

\chapter{Binary Black Holes in Four-Derivative Scalar-Tensor theories of gravity}
\label{C:BBHin4dST}

In this chapter we present the numerical results in \cite{AresteSalo:2022hua,AresteSalo:2023mmd} regarding Binary Black Holes in the Four-Derivative Scalar-Tensor theory of gravity. These were obtained through the implementation of the equations in Appendix \ref{App:EoMEsGB} (coming from the action \eqref{eq:4dST}) in our recently released open-source NR code \texttt{GRFolres} \cite{AresteSalo:2023hcp}.


\section{Technical details}

In this section we discuss the main technical details of the simulations that we have run for the results presented in this chapter.

\subsection{Gauge parameters}
We have chosen the functions appearing in our modified CCZ4 gauge to be spatial constants, with $a(x)=0.2$ and $b(x)=0.4$ in all our simulations.  This choice gives reasonable results but our initial investigations suggest that tuning these values or choosing metric dependent functions may give better constraint conservation.\footnote{Making $a(x)$ and $b(x)$ functions of the evolution variable $\chi$ so that they interpolate between zero in the asymptotic region and the values quoted above near black holes does not seem to make any significant difference in the numerical stability of our simulations.} This has been studied in \cite{Doneva:2023oww}, which will be discussed in the next chapter.

\subsection{Excision of the EsGB terms}

As in \cite{AresteSalo:2022hua,Figueras:2020dzx,Figueras:2021abd} we smoothly switch off some of the higher derivative terms in the eoms well inside the apparent horizon (AH) \footnote{In Numerical Relativity, the event horizon (the boundary of ``no escape'' of the black hole) is very hard to track, as it relies on knowing the full time evolution. This is why we usually deal with the apparent horizon, which is not gauge-invariant, but implies the existence of an event horizon outside of it.} \footnote{We also note that characteristic hypersurfaces (and in particular the event horizon) are generally non-null in Horndeski theories \cite{Reall:2021voz} and, hence, for sufficiently strong couplings we could observe the AH that we compute lying outside of the event horizon. However, we have seen no numerical sign of this behaviour while remaining in the strongly hyperbolic regime.} by replacing $\lambda^{\text{GB}}\to\frac{\lambda^{\text{GB}}}{1+e^{-100(\chi-\chi_0)}}$ with $\chi_0=0.15$ for spinless black holes (BHs). The specific value of $\chi_0$ needs to be adjusted (smaller) for higher spin, with the value chosen to be within the AH in our chosen coordinates (see Figure C1 in \cite{Radia:2021smk} for the values in typical puncture gauge simulations). In Binary Black Hole (BBH) merger simulations we have found that it helps to change the value of $\chi_0$ after the merger, since the final remnant has a dimensionless spin of the order of $a/M\gtrsim 0.7$. 

Therefore, we are solving GR near the singularity (within the horizon), which helps us to deal better with the numerical simulations given that we could encounter physical problems near the singularity due to the modified gravity contributions, which cause a breakdown of the hyperbolicity at these curvatures. This is valid given that what happens inside the AH cannot physically affect the exterior of it.  Hence, as long as the excision happens well within the AH, this should not change the physical behaviour outside it.

\subsection{Constraint damping parameters}

We have also noted that the values of the damping constraint coefficients $\kappa_1$ and $\kappa_2$ play an essential role for keeping the violation of the Hamiltonian and momentum constraints of the system under control and in particular the best values appear to depend on the final spin of the stationary BH solution that the system evolves to. Therefore, we also increase the values after the remnant is formed -- more details are given in the following subsections. We use the usual rescaling   $\kappa_1\to\kappa_1/\alpha$ that allows for stable evolutions of BHs as in \cite{Alic:2013xsa}.

\subsection{Numerical set-up}

For the runs with single BHs we use a computational domain of $L = 256M$ with the BH situated at the centre of the grid, and $N=128$ grid points on the coarsest level. We use $6$ levels of refinement, which results in a finest resolution of $dx_{finest} = M/16$ on the finest grid, giving $\sim 35$ grid points across the BH horizon in the quasi-isotropic Kerr coordinates \cite{Liu:2009al} that we use to set the initial conditions for the metric.  These coordinates are a generalisation of the wormhole-like isotropic Schwarzschild coordinates, and similarly evolve into a trumpet-like solution for the (modified) puncture gauge within the first $\sim 10M$ of the simulation. At this point the BH horizon is located at $r\sim 0.98M$ in the zero spin case, which is similar to the GR puncture gauge value \cite{Radia:2021smk}.

For the BBH mergers we have chosen $L=512M$, with $N=128$ grid points on the coarsest level. We use $9$ levels of refinement, which results in a resolution of $dx_{finest} = M/64$ on the finest grid, which gives roughly 60 points across the horizon of each BH prior to their merger. We anticipate that higher resolutions would be required for detailed waveform templates, but for this study we are mainly interested in the overall phenomenology. For both type of simulations we use $6^{\text{th}}$ order finite differences to discretise the spatial derivatives and a standard RK4 time integrator to step forward in time.  We have checked convergence for these parameters, as shown in the Appendix \ref{App:convergence}.

We study two cases for the BBH mergers:
\begin{itemize}
    \item Case 1: The BHs have equal masses $m_{(1)} = m_{(2)} = 0.49M$, initial separation $11M$, initial velocities $v_{(i)} = (0, \pm0.09, 0)$ and are initially non-spinning. These initial conditions were tuned to have roughly circular initial orbits in GR such that the two black holes merge in approximately seven orbits. For this case we superpose the solutions for two boosted black holes as described in \cite{Baumgarte:2010ndz,Bowen:1980yu}, using a perturbative solution for the conformal factor that is accurate up to order $(P^i P_i)^2$.
    \item Case 2: An equal-mass binary where the component BHs each have non-zero initial (dimensionless) spin of $a_0/M=0.4$ aligned with the orbital axis. In this case we use a standalone version of the TwoPunctures code \cite{Ansorg:2004ds} to generate Bowen-York initial data \cite{Bowen:1980yu} with a separation of $11M$, initial velocities $v_{(i)}=(0,\pm 0.08, 0)$, equal masses of $m_1=m_2=0.31$ (so that the total ADM mass \footnote{The ADM mass is a gauge-invariant notion of total mass contained in asymptotically flat spacetimes, which is defined as a surface integral over a sphere with infinite radius.} is approximately $1$) and angular momentum $J_{(i)}/m_{(i)}^2=(0,0,0.4)$. In this case the orbits are only roughly circular and we have around eight orbits prior to merger in the GR case.
\end{itemize}

Note that in both cases we use GR initial data, which remains a solution of the constraint equations only in the case in which the additional scalar degree of freedom is zero. In some cases below we add a small perturbation in the field to source its growth after the merger. In these cases, where the constraints are initially violated, the violations are small and quickly damped away by the damping terms in the eoms. A generalisation of the initial data solver of \cite{Aurrekoetxea:2022mpw} to the $4\partial$ST theory is already available \cite{Brady:2023dgu} and will be used in the future for this type of simulation.

\section{Black hole solutions in $4\partial$ST}

The black hole solutions in a $4\partial$ST theory of the form \eqref{eq:4dST} can have scalar hair, i.e., a non-trivial configuration of scalar field, which depends mainly on the form of the coupling $\lambda(\phi)$. Previous works have divided the classes of coupling functions into the two following cases \cite{Elley:2022ept,R:2022hlf}:
\begin{itemize}
    \item Type I: $\lambda(\phi) \sim \phi + O(\phi^2)$. In this case the scalar field is always sourced by the presence of curvature and so the Kerr family of black holes are not stationary solutions of the theory. Since all the stationary black hole spacetimes in the theory must have hair, this case is strongly constrained by observations of astrophysical BHs. This case includes the so-called shift-symmetric and dilatonic couplings.
    \item Type II: $\lambda(\phi) \sim \phi^2 + O(\phi^3)$. In this case Kerr black holes can be stationary solutions of the theory, but in certain regions of the parameter space there can also exist black holes with non-trivial scalar configurations, i.e., hairy black holes. This means that astrophysical black holes may be on either the hairy or non-hairy branches, which makes them more difficult to constrain.
\end{itemize}

Analogously as we have considered when deriving the eoms in the $d+1$ form, we consider for simplicity a $4\partial$ST theory with no potential for the scalar field and with the coupling functions being $\lambda(\phi)=\frac{\lambda^{\text{GB}}}{4}f(\phi)$ and $g_2(\phi)=g_2$, where $f(\phi)$ is an arbitrary function (which is either linear, quadratic or exponential in our simulations) and $\lambda^{\text{GB}}$ and $g_2$ are constants that we assume to satisfy the weak coupling conditions, namely 
\begin{equation}
    L\gg \sqrt{|\lambda'(\phi)}|\,,\sqrt{|g_2|}\,,
\end{equation}
where $L$ accounts here as well for any characteristic length scale of the system associated to the spacetime curvature and the gradients of the scalar field.

As an initial test, we set the initial conditions to be the single Kerr BH with mass parameter $M=1$ as described above, and set the scalar field to zero initially. The values of the constraint damping coefficients have been chosen to be $\kappa_1=0.35-1.7$ (higher values for this parameter are found to be required for higher spins in order to stabilise the final state) and $\kappa_2=-0.5$. 

\begin{figure}[H]
    \centering
    \includegraphics[scale=.6]{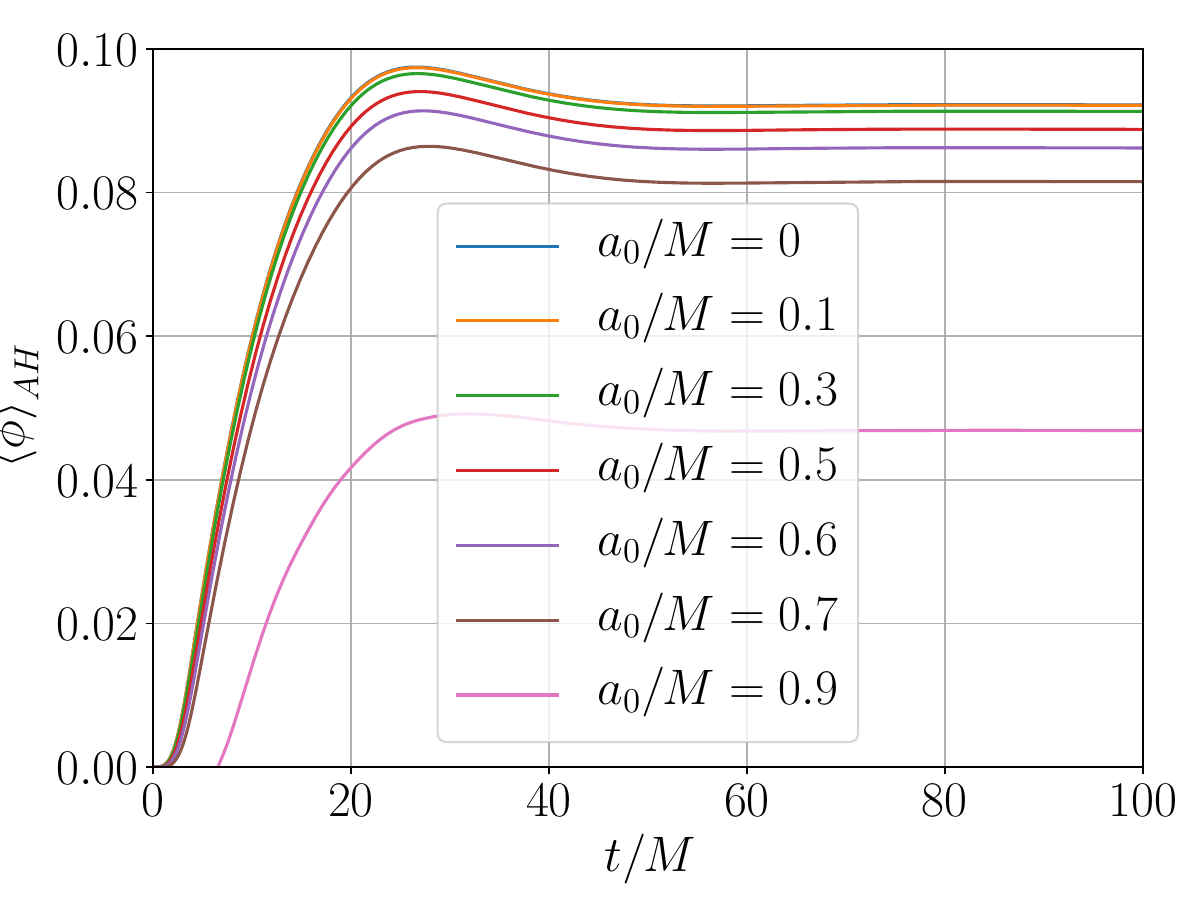}
    \caption[Evolution of hairy BHs in shift-symmetric EsGB: the average scalar field.]{Simulations of single BHs with spin in (Type I) shift-symmetric Einstein-scalar-Gauss-Bonnet theory for $\lambda^{\text{GB}}/M^2=0.2$. From an initial zero value of the scalar field, the curvature sources a growth of the scalar hair to the stationary state, with higher spins sourcing smaller average field values as expected. The energy for the scalar hair is extracted from the BH, which results in a decrease in its AH area (which is permitted in these modified theories of gravity) and mass. In cases with spin, angular momentum may also be extracted by the scalar field.
    Here we plot the average value of the scalar field at the AH for different values of the initial dimensionless spin $a_0/M$.}
    \label{F:kerr}
\end{figure}

\begin{figure}[hbt]
    \centering
    \includegraphics[scale=.6]{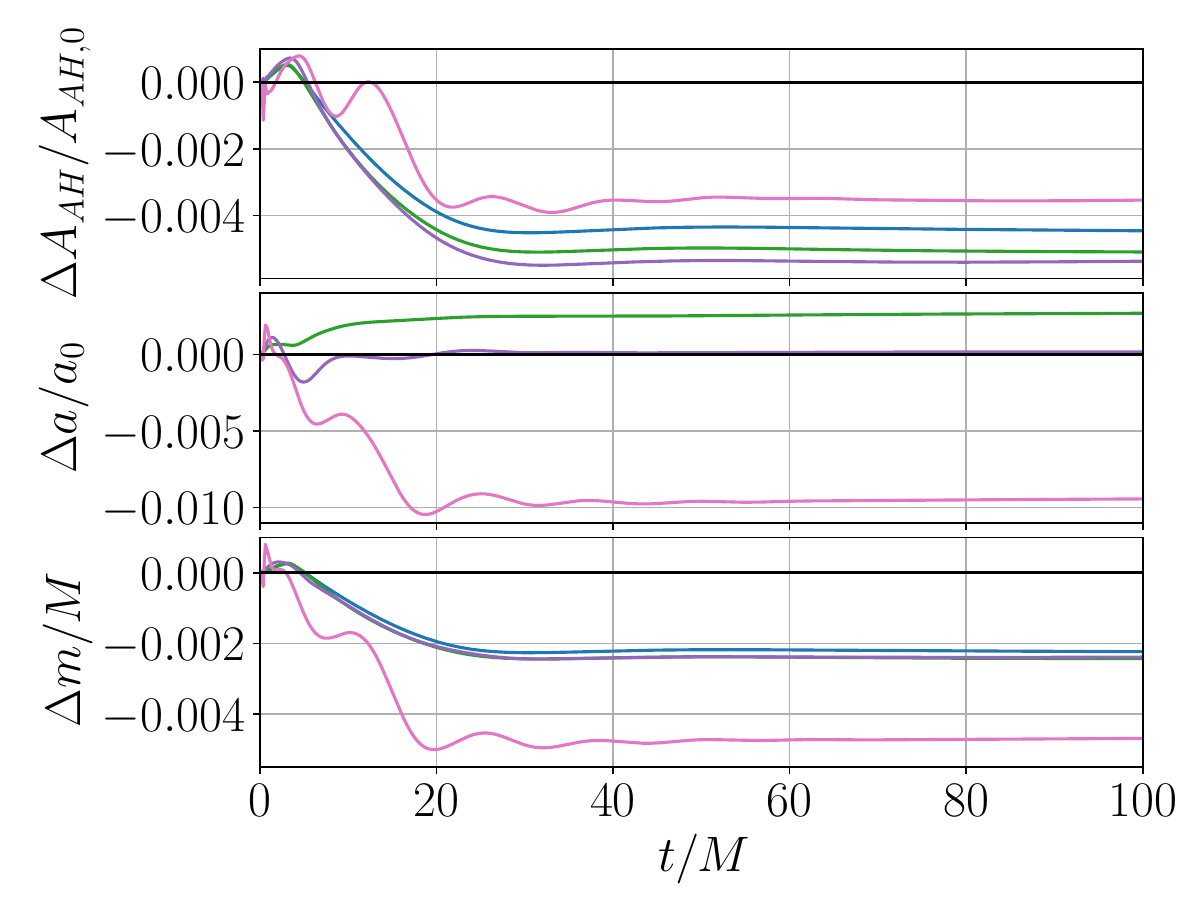}
    \caption[Evolution of hairy BHs in shift-symmetric EsGB: change in the AH area, spin and mass.]{Change in the AH area, spin and mass relative to their initial values for the same single BHs depicted in Figure \ref{F:kerr}.}
    \label{F:kerr_others}
\end{figure}

Figures \ref{F:kerr} and \ref{F:kerr_others} \footnote{In this and subsequent figures, we present the average value of certain quantities across the apparent horizon. We denote 
\begin{align}
    \langle\psi\rangle_{\text{AH}}=\tfrac{\int\tfrac{1}{\chi}\,\psi\,r^2(\theta,\phi)\sin\theta d\theta d\phi}{\int\tfrac{1}{\chi}\,r^2(\theta,\phi)\sin\theta drd\theta d\phi}\,,
\end{align}
where $r,\theta,\phi$ account for the spherical coordinates, $r=r(\theta,\phi)$ is the apparent horizon and $1/\chi$ is a factor coming from the determinant of the induced metric on a $2$-dimensional surface, which we find is a good approximation to the exact value since the determinant of the (Cartesian) conformal metric in our formulation is $1$.} show that a stationary hairy BH solution in shift-symmetric Einstein-scalar-Gauss-Bonnet theory, namely $4\partial$ST with $f(\phi)=\phi$ and $g_2=0$, is obtained for all the values of the dimensionless spin parameter $a_0/M$ after an initial transient period of growth of the scalar hair. These final stationary states are consistent with the results in \cite{East:2020hgw}.

\section{Binary Black Holes}
Next, we present our results in \cite{AresteSalo:2022hua,AresteSalo:2023mmd,unequal} in the context of Binary Black Holes mergers.

\subsection{Type I coupling -- shift-symmetric $4\partial$ST}

We start by considering the simplest case of scalarisation in the $4\partial$ST theory by adding a linear coupling $f(\phi)=\phi$, which is often referred to as shift-symmetric Einstein-scalar-Gauss-Bonnet (EsGB) theory (usually in the absence of the $g_2$ term, although this term also respects the shift-symmetry). As discussed above, due to the curvature sourcing the scalar field, BH solutions in this theory differ from the Kerr solution in that they possess a non-trivial scalar configuration, that is, they have scalar hair.

\begin{figure}[H]
\centering
\includegraphics[scale=.74]{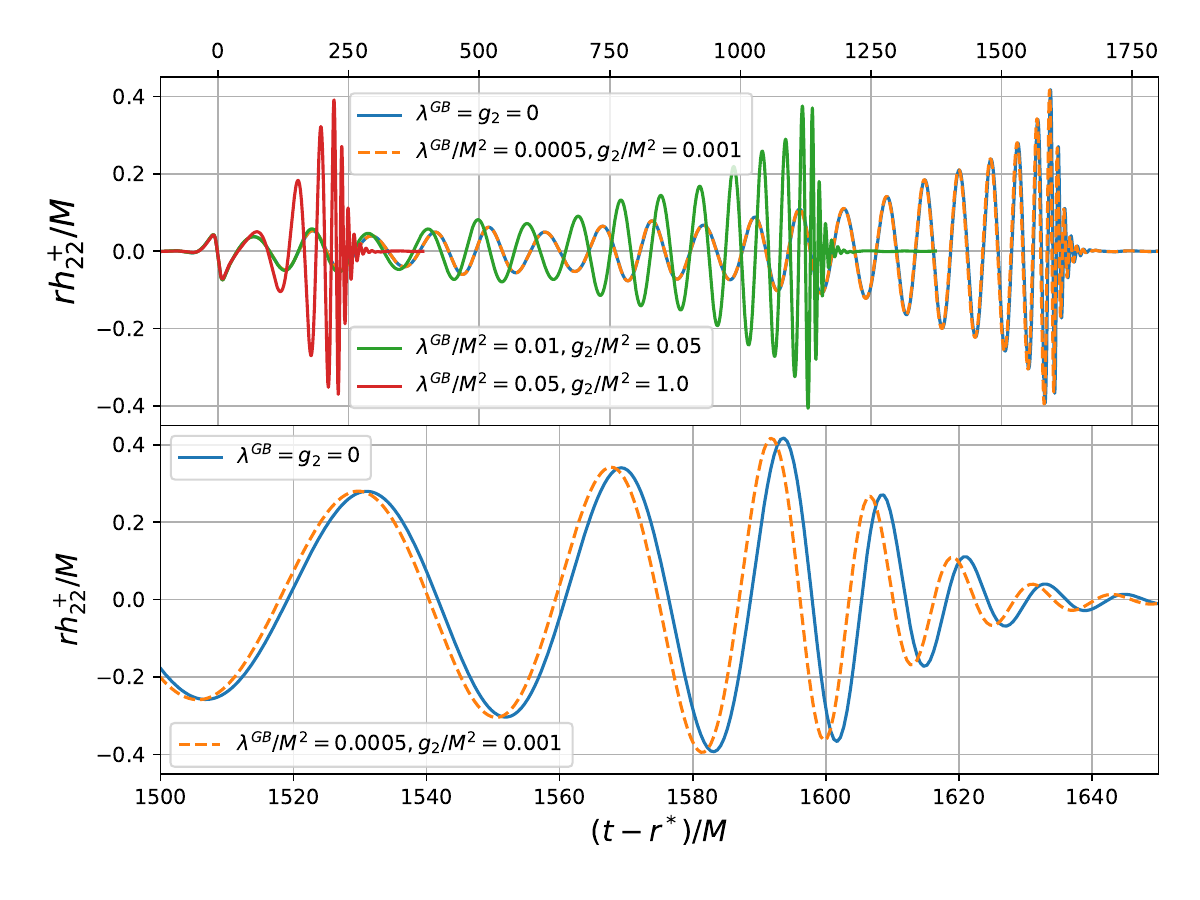}
\caption[Comparison of the (2,2) mode of the gravitational strain between GR and $4\partial$ST for initially non-spinning BBHs.]{\textit{Top}: Comparison of the (2,2) mode of the gravitational wave strain between GR (blue) and $4\partial$ST in retarded time, $u=t-r^*$, where $r^*$ is the tortoise coordinate, for different values of the couplings, namely small (orange dashed), medium (green) and large (red). \textit{Bottom}: Zoom in of the merger region for the small coupling case.}
\label{F:all_waves}
\end{figure}

\begin{figure}[H]
\centering
\includegraphics[width=14.8cm]{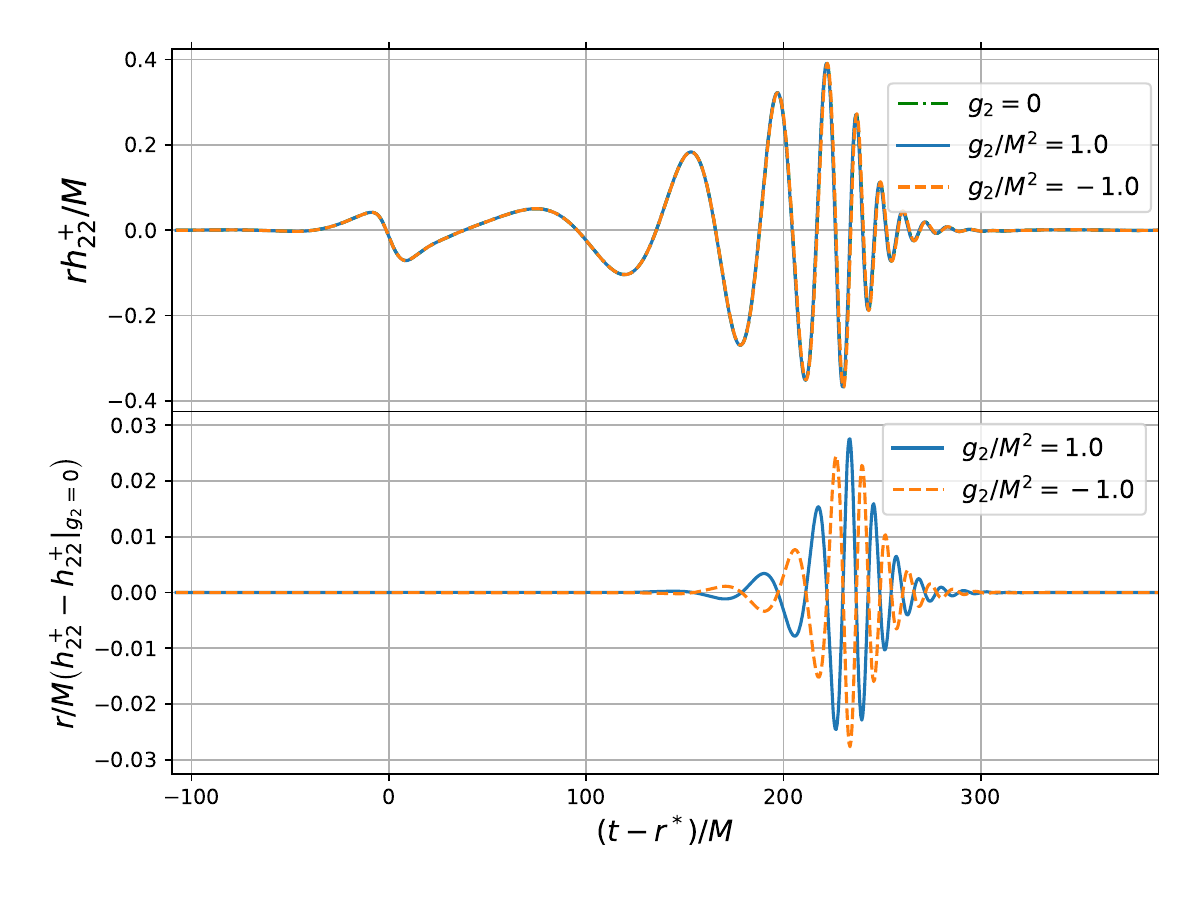}
\caption[Shift-symmetric $4\partial$ST BBHs waveforms with different values of the Horndeski coupling $g_2$.]{\textit{Top}: Waveforms for fixed $\lambda^{\text{GB}}/M^2=0.05$, with different values of the Horndeski coupling $g_2$. We see that changing $g_2$ has a small effect since the dynamics are mainly controlled by the Gauss-Bonnet coupling $\lambda^{\text{GB}}$. \textit{Bottom}: The lower plot shows the difference between the strains for the large couplings $g_2/M^2=\pm 1$ compared to the $g_2=0$ case.}
\label{F:large_coupling}
\end{figure}

Figures \ref{F:all_waves} and \ref{F:large_coupling} yield our main results in \cite{AresteSalo:2022hua}. We use equal-mass, non-spinning  BHs (Case 1 described above). The constraint damping coefficients are set to $\kappa_1=0.35/M$ and $\kappa_2=-0.1$. Since in the theory that we consider the Gauss-Bonnet term sources the scalar field, it is the associated coupling constant $\lambda^{\text{GB}}$ that plays the most prominent role and effectively controls the regime of validity of the EFT. For very small values of the couplings, the waveforms in $4\partial$ST tend to GR, as expected since the scalar field is effectively perturbative.\footnote{We emphasise that we still treat the system non-perturbatively, which allows us to avoid potential issues with secular effects.} The differences only become noticeable towards the late inspiral and merger (see the bottom panel in Fig. \ref{F:all_waves}) and they appear as an accumulation of phase shift in the waveform. This is expected since the spacetime curvature outside black holes is largest during this phase. 

The situation is drastically different for large values of the couplings.  For $\lambda^{\text{GB}}/M^2=0.05$, the binary merges in only 3 orbits as opposed to the 7 orbits that the black holes describe in GR with the same initial conditions. In the $4\partial$ST theory with large couplings the system can radiate strongly in scalar waves and hence shed energy and angular momentum more efficiently than in GR, so the larger the $\lambda^{\text{GB}}/M^2$ coupling, the sooner the binary merges. However, this case needs to be revisited, given that from a post-Newtonian approach the leading order effect in the inspiral is at $-1PN$, which is not noticeable with few orbits \cite{Shiralilou:2021mfl}.  The formation of the scalar cloud from an initial zero state may have an effect on the circular orbits which needs to be carefully quantified. Furthermore, we would also need to study whether the orbits become non-quasicircular and take into account a thorough alignment in time. 
We were able to increase the coupling to $\lambda^{\text{GB}}/M^2=0.1$ without major difficulties where we find that the binary merges even quicker. It seems possible to increase this coupling even further, but each increase necessitates a tuning of the damping parameters $\kappa_1$ and $\kappa_2$ to keep the truncation errors under control, and so we leave a full exploration of the limit to future work.

In Fig. \ref{F:large_coupling} we compare the effect of varying $g_2$ for a fixed (large) value of $\lambda^{\text{GB}}$. Even if the values of $g_2$ are large compared to the values used in \cite{Figueras:2021abd}, the effect is small. The reason is that the typical energy densities of the scalar field that result from the scalarisation process in the 4$\partial$ST are much smaller than in \cite{Figueras:2021abd}. 
It is interesting to note that changing the sign of $g_2$ gives rise to a phase shift of roughly $\pi$ rather than an advancement or delay of the wave as in \cite{Figueras:2021abd}.

\begin{figure}[hbt]
    \centering
    \includegraphics[scale=.65]{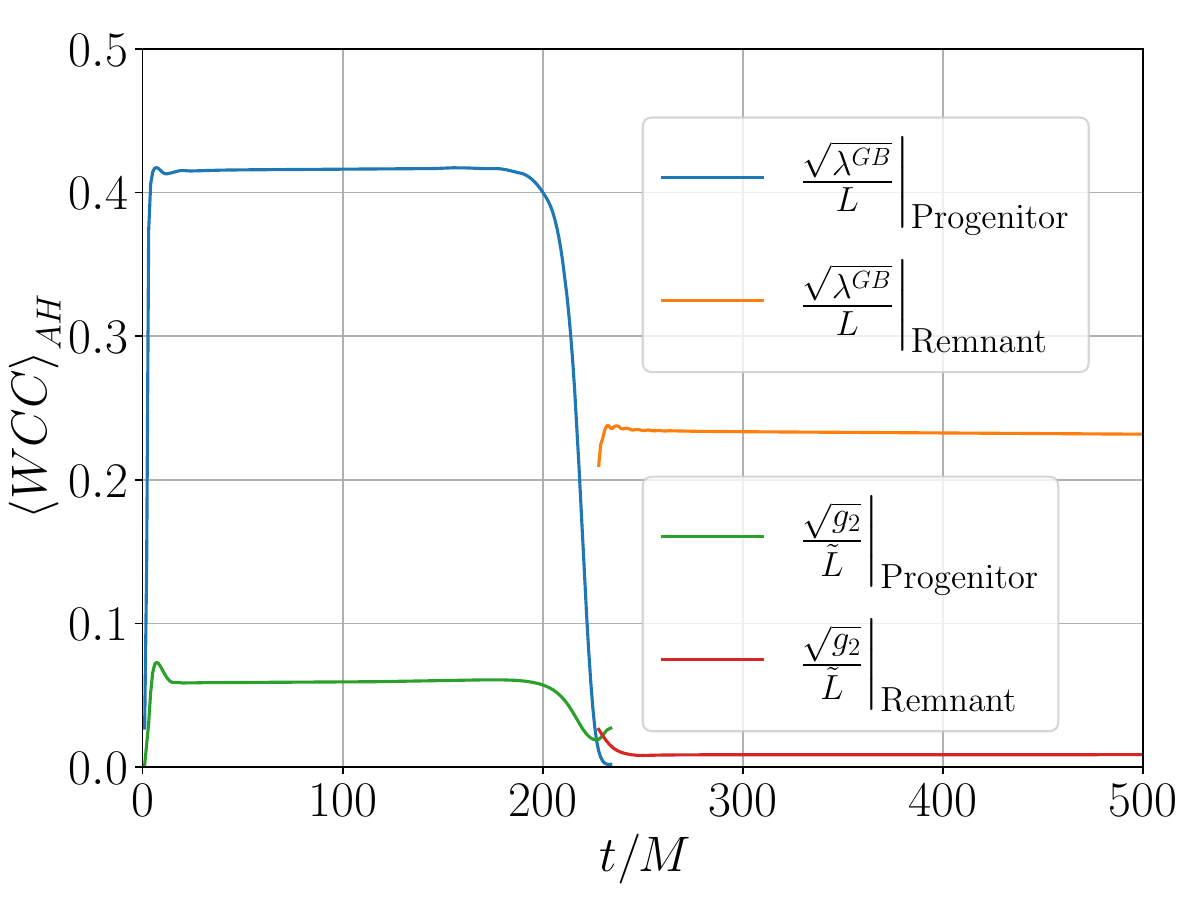}
    \caption[Evolution of the weak coupling condition in a shift-symmetric $4\partial$ST BBH.]{Here we show that the weak coupling regime holds throughout the BBH merger simulation in shift-symmetric $4\partial$ST theory with $\lambda^{\text{GB}}=0.05M^2$ and $g_2=M^2$, namely the one with the highest coupling constants considered in \cite{AresteSalo:2022hua}. 
    We depict the evolution of the WCCs in Equation \eqref{eq:wfc} for both the progenitor black holes throughout the inspiral and the remnant during the ringdown, seeing that they are not violated for these values of the couplings. As expected, the highest values are before the merger, due to the smaller curvature scales of the initial black holes compared to the final remnant.}
    \label{F:wfc}
\end{figure}

Next, we study whether the weak coupling condition (WCC) still holds for the case with the highest values of the couplings, namely $\lambda^{\text{GB}}=0.05M^2$ and $g_2=M^2$. We observe in Figure \ref{F:wfc} that the WCC
\begin{align}\label{eq:wfc}
    \sqrt{\lambda^{\text{GB}}}/L\ll1\,, \qquad \sqrt{g_2}/\tilde{L}\ll1\,,
\end{align}
still holds (even though it is close to the limit).
Here the relevant length scales that represent the curvature quantities of the metric and scalar sectors are
\begin{align}
    &L^{-1}=\max\{|R_{ij}|^{1/2},|\nabla_{\mu}\phi|,|\nabla_{\mu}\nabla_{\nu}\phi|^{1/2},|{\mathcal L}^{\text{GB}}|^{1/4}\} \,,\nonumber\\
    &\tilde{L}^{-1}=\max\{|K_{\phi}|,|D_i\phi D^i\phi|^{1/2}\} \,.
\end{align}

As expected, the highest values of the weak coupling conditions occur right before the merger, given that the curvature scales are larger near the initial BHs in comparison to the final remnant. Therefore if the WCC is not breached during the inspiral, it appears to be safe during merger and ringdown phases. Note that the WCC is not a well-defined mathematical condition, but is however a heuristic condition that helps us identify that we are in the regime of validity of the theory where the eigenvalues of the principal symbol do not differ significantly from GR.

For this same coupling function, we also test our ability to stably evolve equal-mass BBH cases with non-zero initial component spins (Case 2 above).
We used the following values of the constraint damping coefficients, $\kappa_1=1.4/M$ and $\kappa_2=-0.1$, which we changed to $\kappa_1=1.7/M$ and $\kappa_2=0$ after merger. We also decrease the value of $\chi_0$ from $\chi_0=0.15$ \footnote{Note that this value would need to be changed if the initial spins were higher than the ones used here (see Figure C1 in \cite{Radia:2021smk}).} to $\chi_0=0.05$ after the merger.

\begin{figure}[H]
\centering
\includegraphics[scale=.7]{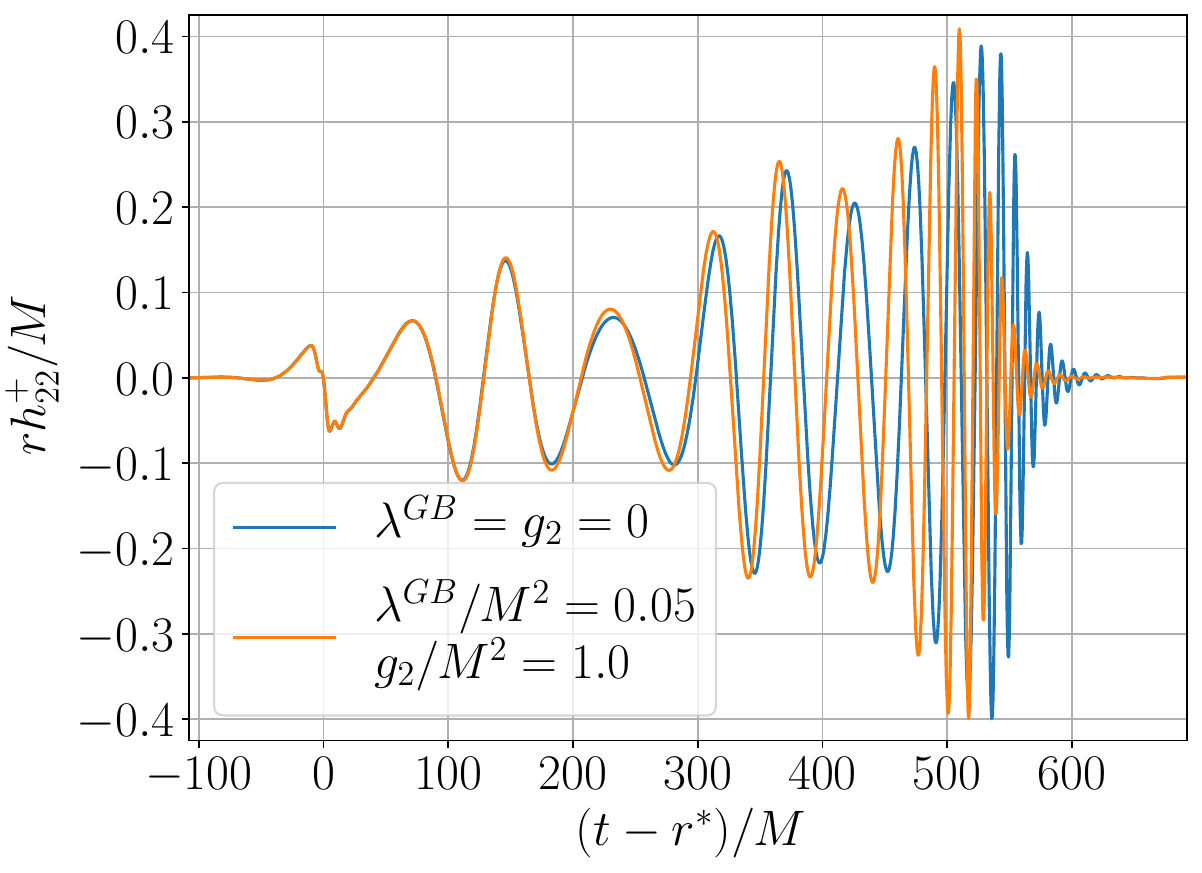}
\caption[Comparison of the (2,2) mode of the gravitational strain between GR and $4\partial$ST for initially spinning BBHs.]{Initially spinning Binary Black Hole mergers (with spins initially aligned along the orbital momentum, $a_0^{\pm}/M=0.4$) in GR (blue) and shift-symmetric $4\partial$ST (orange), for the following values of the coupling constants, $\lambda^{\text{GB}}=0.05M^2$ and $g_2=M^2$. We show the $(2,2)$ mode of the gravitational strain in retarded time, $u=t-r^*$ (where $r^*$ is the tortoise coordinate), observing that the additional extraction and radiation of energy via the scalar channel induces the merger to happen sooner compared to GR.}
\label{F:spin}
\end{figure}

The result is shown in Figure \ref{F:spin}, where we compare the $(2,2)$ mode of the gravitational strain with GR by extracting the gravitational waves at $r=100M$.\footnote{For the accuracy purposes of this work, we only considered the extraction at this radius but in a number of cases we checked that this result is essentially the same as the one obtained by extrapolating to null infinity.}  We find that for the chosen parameters the final spin reduces from $\sim0.85$ in GR to $\sim0.84$ in shift-symmetric $4\partial$ST theory, as expected from the extraction of spin caused by the non-trivial scalar field.

\subsection{Type II Coupling -- tachyonic growth and stealth scalarisation}

At this point we turn to the second class of coupling function, Type II, in which the coupling results in a (spatially dependent) mass term. These admit both scalarised and non-scalarised BH solutions. 

In this case we study the binary case directly, and we choose the coupling parameters so that the scalar hair is generated as a result of the merger. We use Case 1 of the BBH configurations described above throughout this section.

The simplest case of a Type II coupling is a quadratic coupling, namely $f(\phi)=\phi^2$. As studied in \cite{Silva:2020omi,Elley:2022ept}, this coupling function can have a tachyonic instability which leads to a spin-induced scalarisation or descalarisation. We study the case in which the remnant scalarises after the merger due to its spin for a sufficiently high negative value of the coupling $\lambda^{\text{GB}}$. 

\begin{figure}[H]
    \centering
    \includegraphics[scale=.7]{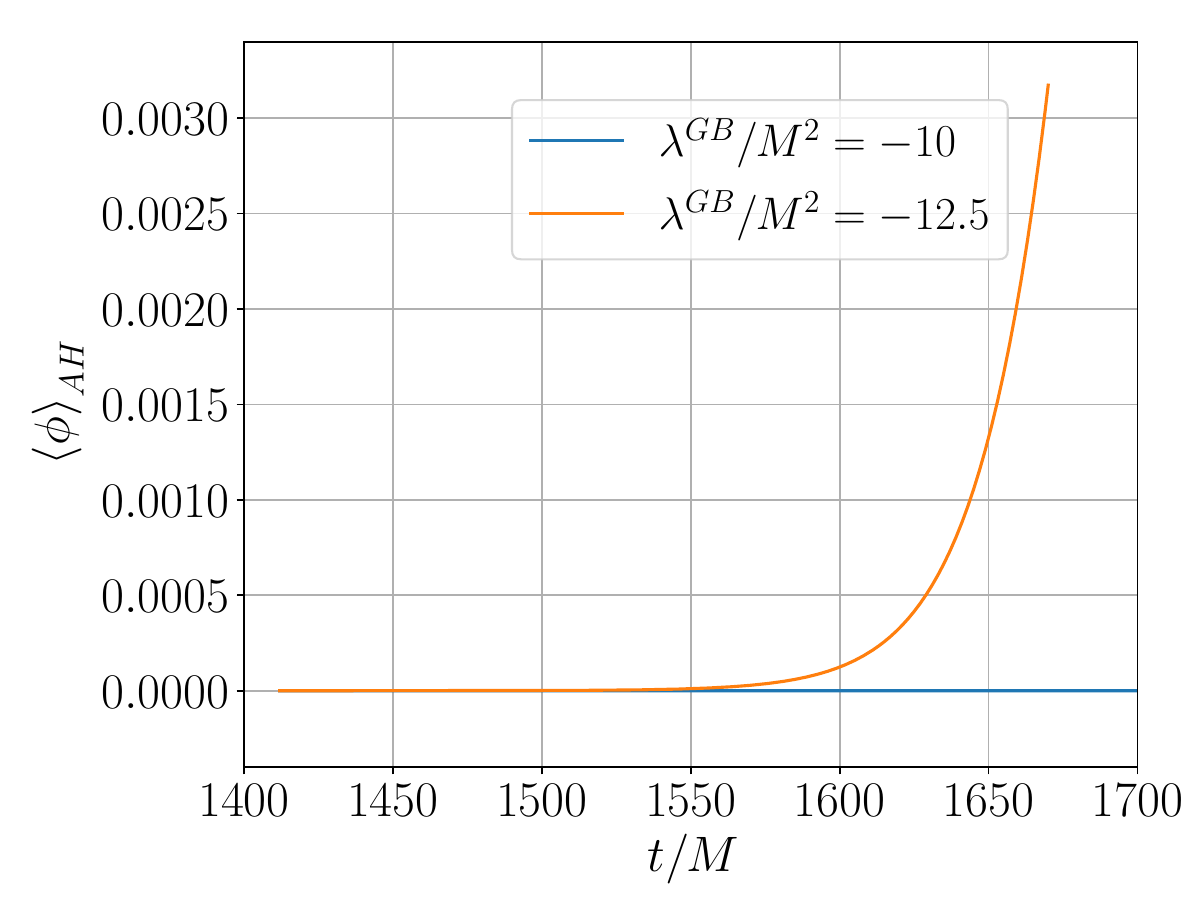}
    \caption[Evolution of the average value of the scalar field across the AH of the remnant in a BBH simulation for EsGB theory with quadratic coupling.]{Evolution of the average value of the scalar field across the AH, after the merger occurred in a Binary Black Hole simulation for Einstein-scalar-Gauss-Bonnet theory with quadratic coupling for two different values of the coupling $\lambda^{\text{GB}}$. We see that for a critical coupling value of around $\lambda^{\text{GB}} = -10 M^2$ the remnant (with $a/M\sim 0.7$) scalarises and the value of the scalar field grows exponentially. Eventually the simulation breaks down, since the weak field condition (and hence well-posedness) does not hold anymore.
    }
    \label{fig:phiquad}
\end{figure}

We used as constraint damping coefficients $\kappa_1=0.35/M$ and $\kappa_2=-0.1$ initially, but after merger changed them to $\kappa_1=1.7/M$ and $\kappa_2=0$, together with reducing the initial value of $\chi_0=0.15$ to $\chi_0=0.05$.
We also needed to add an initial  perturbation in the scalar field to seed the instability, for which we choose the (arbitrary) form $\phi(r)=10^{-3}(1+0.01r^2e^{-r^2})$.

Given that the initial BHs have zero spin, there is initially no scalarisation for this sign in the coupling and the scalar field dissipates. Only after the merger does the scalar field have a non-trivial evolution. In Figure \ref{fig:phiquad} we show the two possible behaviours -- exponential growth or zero growth, and find that the critical value of $\lambda^{\text{GB}}$ for which the transition occurs happens at around $10 M^2$. For the values of the coupling that induce exponential growth we observe that the weak coupling condition is eventually violated and, thus, at some point along the evolution the theory ceases to be well-posed, which results in the breakdown of the simulation (see also Chapter \ref{C:Hyp} for further results).

\begin{figure}[H]
\centering
\includegraphics[scale=.24]{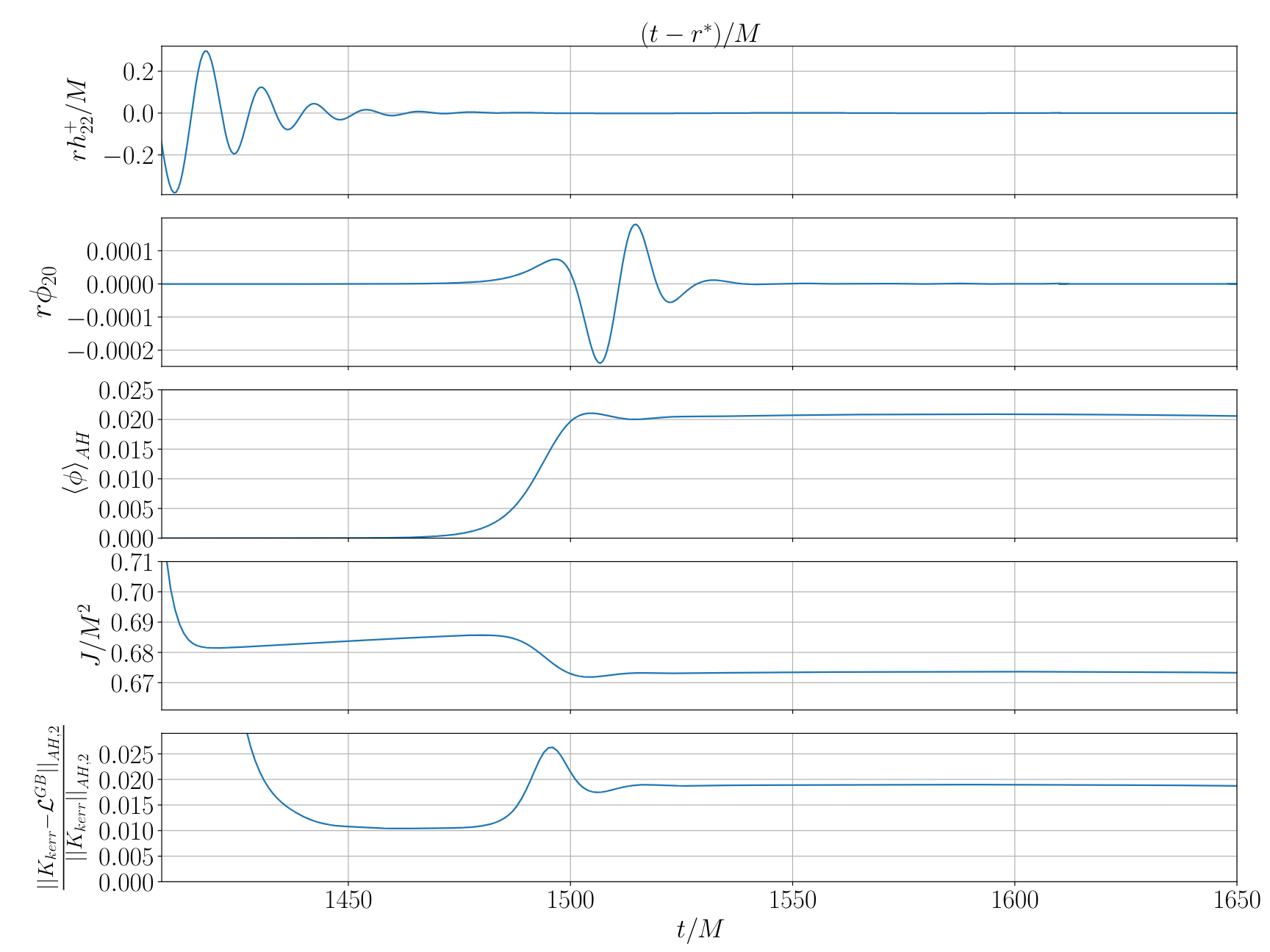}
\caption[Post-merger phase of a spin-induced scalarisation in EsGB with exponential quadratic coupling.]{Here we summarise the key results from the post-merger phase of spin-induced scalarisation in the EsGB theory with exponential quadratic coupling. 
    We see that the spin of the remnant following merger generates a tachyonic mass, by which the scalar field acquires a non-trivial configuration. This happens late in the ringdown of the tensor modes. It is accompanied by a burst of radiation in the scalar mode $(2,0)$, which coincides with the extraction of spin from the merger and the highest deviation of the Gauss-Bonnet curvature with respect to the Kretschmann scalar of a Kerr black hole (note that the initial deviation in this quantity is due to the merger state being far from Kerr).
    \emph{From top to bottom}: $(2,2)$ mode of the gravitational strain, $(0,2)$ scalar mode in retarded time, average value of the scalar field at the AH, evolution of the spin and $L^2$ norm of the Gauss-Bonnet curvature relative to the Kerr Kretschmann scalar.}
\label{F:expphi}
\end{figure}

\begin{figure}[H]
\centering
\includegraphics[scale=.63,trim={3.2cm 7cm 1cm 7cm},clip]{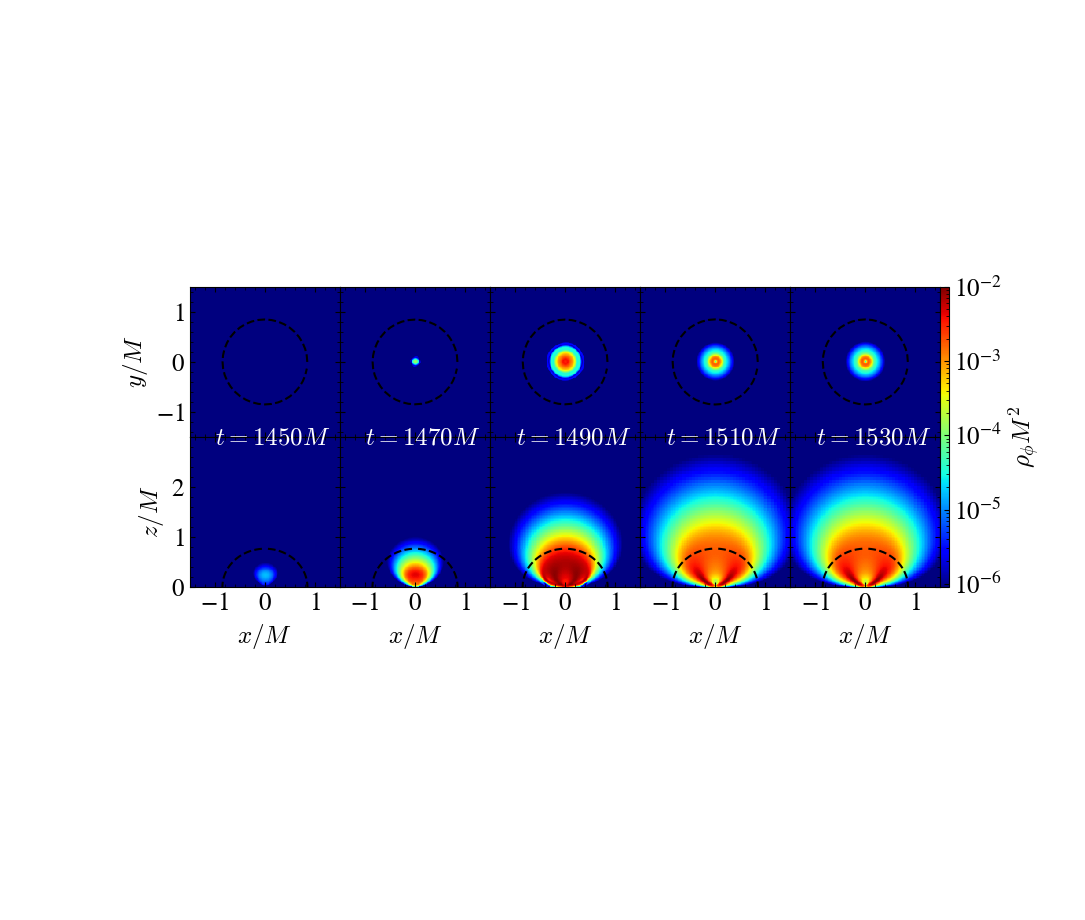}
\caption[Time evolution of the scalar cloud after the merger in EsGB with exponential quadratic coupling.]{Fully non-linear Stealth Scalarisation: 
    Here we show the time evolution of the scalar cloud after the merger for the Einstein-scalar-Gauss-Bonnet theory with exponential quadratic coupling (see eq. \eqref{eq:exp_quad_coupling}) on the rotation plane (upper row) and on a section orthogonal to it (lower row). The colour indicates the contribution of the kinetic and gradient terms to the energy density of the scalar. The dotted black lines denote the location of the apparent horizon. We see that the scalar cloud grows by extracting spin from the remnant, and stabilises with a density that is high compared to the curvature scale of the BH.}
\label{F:cloud}
\end{figure}

A more phenomenologically interesting class of Type II coupling functions was proposed in \cite{Doneva:2022byd,Doneva:2022yqu}, with the form
\begin{equation}
    f(\phi)=\omega^{-1}(1-e^{-\omega\phi^2})\,, \label{eq:exp_quad_coupling}
\end{equation} 
which we refer to as exponential quadratic. This type of coupling has the same initial behaviour as the quadratic one, but the tachyonic instability is saturated by the non-linearities at larger amplitudes, meaning that one can follow the growth of the scalar hair and settling of the solution into a steady hairy BH state after the merger while the theory remains weakly coupled throughout the evolution. This is the case referred to as ``stealth scalarisation'' in previous works \cite{East:2021bqk,Elley:2022ept}. Here we used again the same set-up as in the quadratic coupling case with $\omega=200$ \footnote{The motivation for this large value of $\omega$ is that it leads to $\omega\phi^2\sim1$ at the apparent horizon when the black hole has scalarised, which is where we expect the theory to start to break down. A further study of the impact of different values of this parameter has been carried out in Chapter \ref{C:Hyp}.}   and $\lambda^{\text{GB}}/M^2=-20$. The results are depicted in Figures  \ref{F:expphi} and \ref{F:cloud}. 

Figure \ref{F:expphi} shows that the single BH that results from the merger scalarises after the ringdown of the tensor modes, which coincides with a burst of radiation in the scalar mode $(2,0)$. At this point we observe the largest deviation of the Gauss-Bonnet curvature scalar with respect to the Kretschmann scalar of a Kerr BH (with the same angular momentum and mass as measured from the quasilocal quantities at the AH). This scalarisation process extracts spin from the remnant  BH, which decreases its intrinsic spin before settling into an equilibrium state. The end result is a stable hairy BH, but an observation of the effect would rely on the scalar mode being detectable as a secondary signal, since the tensor modes are emitted during a period in which the theory cannot be distinguished from GR and thus are unaffected -- at least to the precision to which we are able to measure the quasinormal modes (QNMs) here. This is consistent with the behaviour observed for the scalarisation of isolated Kerr BHs in \cite{East:2021bqk}.

\begin{figure}[H]
    \centering
    \includegraphics[width=11cm,trim={0.9cm 3.5cm 0.2cm 3cm},clip]{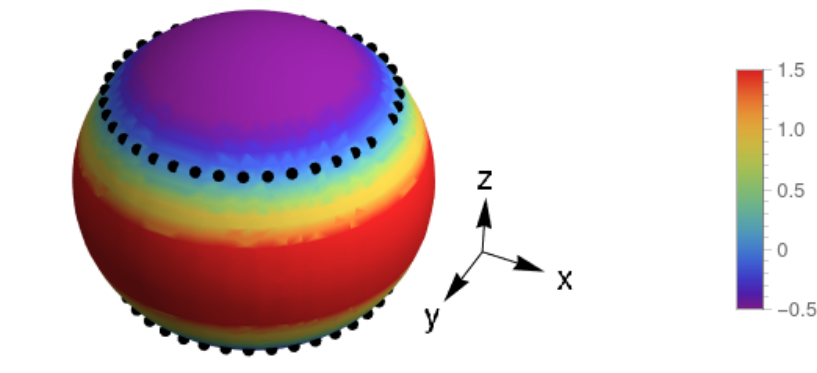}
    \caption[Value of the Gauss-Bonnet curvature around the AH of the final remnant of our BBH merger simulation in the exponential quadratic EsGB theory.]{The effective mass of the scalar field is proportional to the Gauss-Bonnet curvature ${\mathcal L}^{\text{GB}}$ (see \cite{Elley:2022ept} for a discussion). Hence, a change of sign (which occurs for high spins) gives rise to the spin-induced scalarisation in the Type II couplings. Here we show the value of the Gauss-Bonnet curvature around the AH of the final BH of our BBH merger simulation in the exponential quadratic EsGB theory at $t=1530M$ (when the value of the scalar field has already settled down). The dotted points denote the region where ${\mathcal L}^{\text{GB}}=0$. We observe that the negative regions coincide with those where the scalar field has a non-trivial contribution as from Figure \ref{F:cloud}.
    }
    \label{F:sphere}
\end{figure}

\begin{figure}[H]
\centering
\includegraphics[scale=.64,trim={3.2cm 9.5cm 1cm 9.5cm},clip]{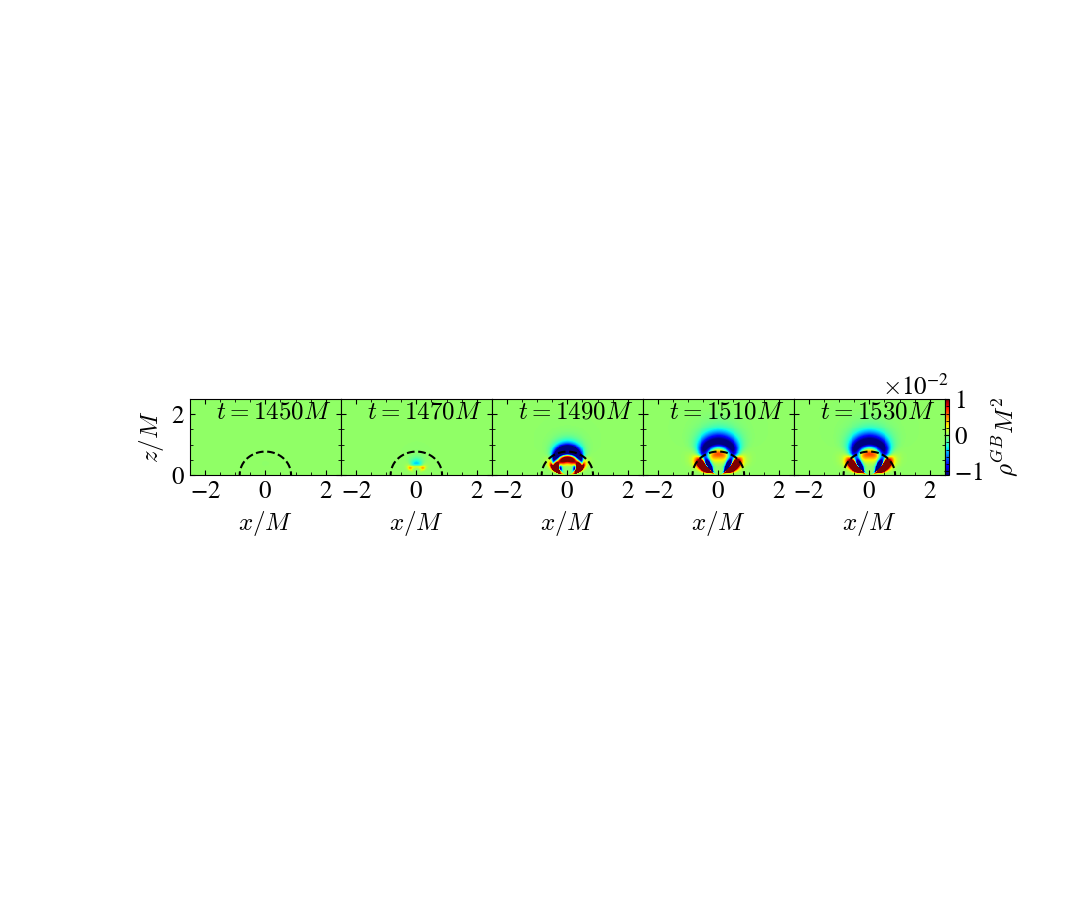}
\caption[Spatial configuration of the Gauss-Bonnet contribution to the effective energy density throughout the ringdown of an equal-mass BBH merger in the EsGB theory with exponential quadratic coupling.]{Spatial configuration of the Gauss-Bonnet contribution to the effective energy density $\rho^{\text{GB}}$ throughout the ringdown of an equal-mass BBH merger in the EsGB theory with an exponential quadratic coupling. The profile is shown in a section orthogonal to the rotation plane. We observe that in some regions around the AH $\rho^{\text{GB}}$  becomes negative, which explains the violation of the Null Curvature Condition (NCC) in this theory of gravity.}
\label{F:rhoGB}
\end{figure}

In Figure \ref{F:cloud} we see that the scalar field is localised around the poles of the AH, which is consistent with  the Gauss-Bonnet curvature acting as the source term for the scalar, as depicted in Figure \ref{F:sphere}. We also show in Figure \ref{F:rhoGB} the contribution of the Gauss-Bonnet term to the energy density, namely $\rho^{\text{GB}}$, from which we can see that it gives rise to a negative contribution to the total energy density in some regions around the AH. This permits a violation of the Null Curvature Condition (NCC) in this modified theory of gravity \cite{R:2022hlf}.

We note that for the chosen coupling function, the absolute value of the overall coupling constant $\lambda^{\text{GB}}$ could be increased beyond the value that we have used in order to increase the speed of growth of the scalar hair. This would push the field growth closer to the ringdown, potentially having an impact on the emission of tensor modes in this phase. However, in order to avoid breaking the hyperbolicity of the equations and weak coupling conditions during the evolution, the value of $\omega$ in the coupling function \eqref{eq:exp_quad_coupling} must also be increased in proportion to $\lambda^{\text{GB}}$ (i.e., keeping $\lambda^{\text{GB}}/(M^2\omega^{1/2})$ constant). As a result, the final maximum scalar field value will be smaller, and whilst the $\rho^{\text{GB}}$ values at maximum should remain the same, as in Fig. \ref{F:rhoGB}, the usual kinetic contribution of the field to the energy density, as shown in Fig. \ref{F:cloud}, will be reduced.  
Due to this trade off, there should exist optimum values of $\omega$ and $\lambda^{\text{GB}}$ that maximise the overlap of the growth of  the scalar hair and the ringdown of the BH, thus resulting in the largest modification of the tensor QNMs. We leave a full analysis of this to future work.

\subsection{Unequal-mass binaries}

Finally we look at unequal-mass binaries, which will be presented shortly in \cite{unequal}. In comparison to the capabilities of the modified harmonic gauge, as shown in \cite{Corman:2022xqg}, we have been able to run Binary Black Holes in the $4\partial$ST theory through merger for strong values of the couplings with mass ratios $1:2$ and $1:3$.

\begin{figure}[H]
\centering
\includegraphics[scale=.5]{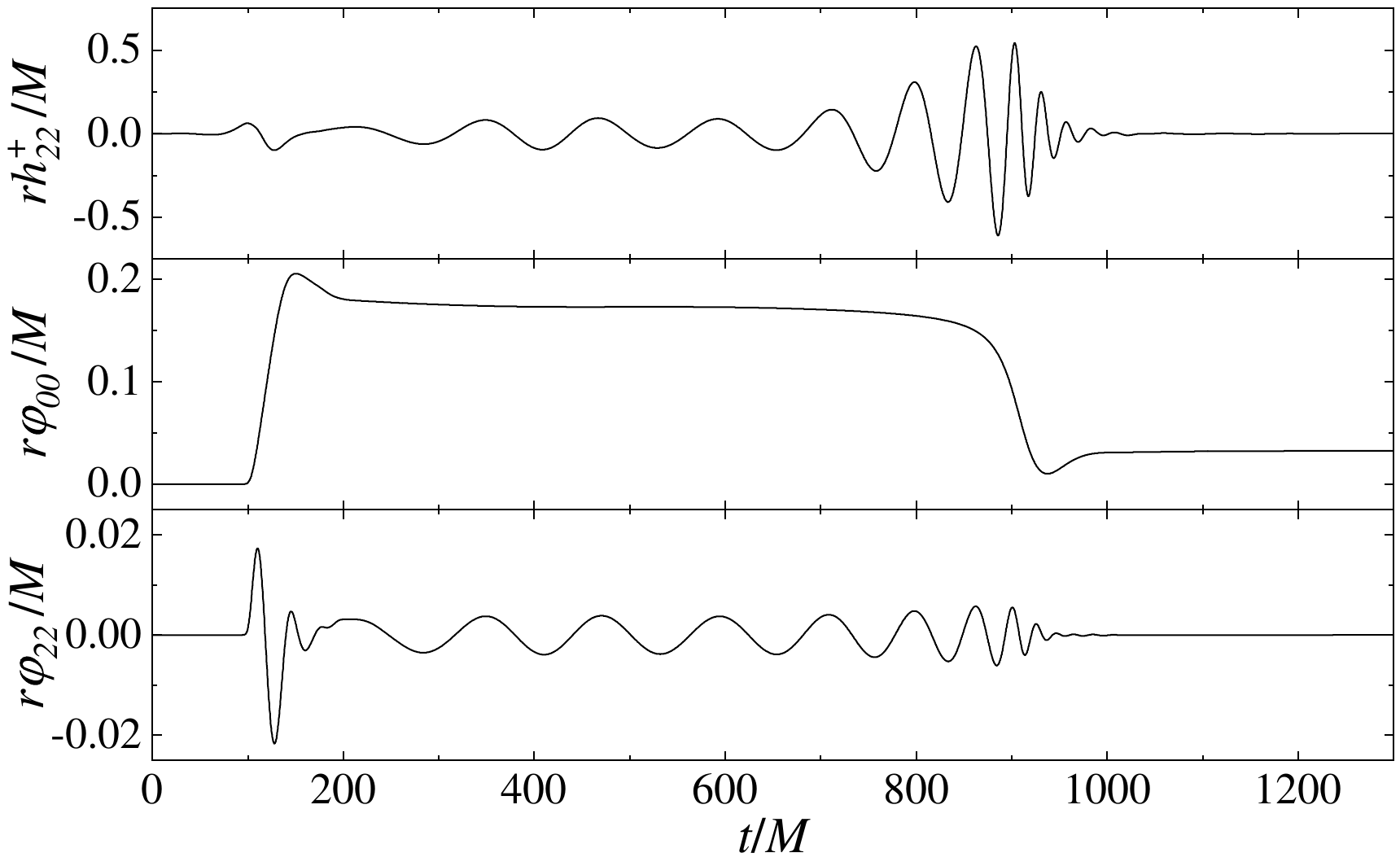}
\caption[BBH merger in shift-symmetric EsGB with mass ratio $1:3$.]{Binary Black Hole merger in shift-symmetric EsGB with mass ratio $1:3$. We show the evolution of the $(2,2)$ mode of the gravitational strain and the $(0,0)$ and $(2,2)$ scalar modes extracted at $r=100M$.}
\label{F:unequal}
\end{figure}

Figure \ref{F:unequal} shows the $(2,2)$ mode of the gravitational strain and the $(0,0)$ and $(2,2)$ scalar modes extracted at $r=100M$ of a $1:3$ Binary Black Hole merger in the shift-symmetric EsGB theory. We see that black holes scalarise during the inspiral, which coincides with a burst of radiation of the $(2,2)$ scalar mode. Further research needs to be carried out in order to find out whether this burst has a significant effect on the system. In future we will use our initial condition solver \cite{Brady:2023dgu} to start our simulation with hairy black holes as initial data. 

\addtocounter{page}{1}%
\blankpage

\chapter{Loss of hyperbolicity}
\label{C:Hyp}

In this chapter we review the regime in which the equations of motion of the Four-Derivative Scalar-Tensor theory become ill-posed, work that is based on \cite{Doneva:2023oww}. This study reinforces the idea that the theories should only be applied within their regime of validity, and not treated as complete theories in their own right.

The formulation of \cite{AresteSalo:2022hua,AresteSalo:2023mmd} has been proven to be strongly hyperbolic (and thus well-posed) for the Einstein-scalar-Gauss-Bonnet (EsGB) theory of gravity in the weakly coupled regime, where the contributions of the Gauss-Bonnet term to the field equations, measured by the coupling $\sqrt{\lambda\,f'(\varphi)}$ \footnote{In order to match with the notation used in \cite{Doneva:2023oww} we will consider throughout this chapter $\lambda^{\text{GB}}$ to be $\lambda$ and the scalar field $\phi$ to be $\varphi$.}, are smaller than the two-derivative Einstein-scalar field terms. This yields the following \emph{Weak Coupling Condition},
    \begin{align}\label{wfc}
        \sqrt{|\lambda\,f'(\varphi)|}/L\ll 1\,,
    \end{align}
    where $L^{-1}=\max\{|R_{ij}|^{1/2}, |\nabla_{\mu}\varphi|, |\nabla_{\mu}\nabla_{\nu}\varphi|^{1/2},|{\mathcal L}^{GB}|^{1/4}\}$ is the inverse of the shortest physical length scale characterizing the system, i.e., the maximum curvature scale.

    Here we explore the validity of the effective field theory (EFT) by monitoring the condition \eqref{wfc} during the evolution. This condition is, in a sense, more important than hyperbolicity, since once breached one can no longer trust the EFT. Nevertheless, it is interesting to explore the interplay between the two conditions and how closely they coincide in practice for generic classes of initial conditions.

\section{Breakdown of strong hyperbolicity}

 It has been shown in Chapter \ref{C:Wellposedness} that the contribution from the Gauss-Bonnet sector to the principal part of the evolution equations only affects the physical modes (and not the gauge modes). These can be separated into purely gravitational modes and mixed scalar-gravitational ones \cite{Reall:2021voz}. Therefore, strong hyperbolicity (and thus, well-posedness) fails when the eigenvalues corresponding to those modes become imaginary.

    In \cite{AresteSalo:2022hua,AresteSalo:2023mmd} all the physical eigenvalues were computed perturbatively. However, the eigenvalues from the purely gravitational sector can be derived exactly in the full theory, given that they lie on the null cone of the effective metric \cite{Reall:2021voz},
    \begin{align}
        g_{\text{eff}}^{\mu\nu}=g^{\mu\nu}-4{\mathcal C}^{\mu\nu}\,,
    \end{align}
    where (as defined previously) ${\mathcal C}^{\mu\nu}=\nabla^{\mu}\nabla^{\nu}\lambda(\phi)$. Hence, one can find that the determinant of the effective metric (normalised to its value in pure GR) is given by
    \begin{align}\label{disc}
        \frac{\det(g^{\mu\nu}_{\text{eff}})}{\det(g^{\mu\nu})}=&~\tfrac{1}{(1+\Omega^{\perp\perp})^2}\det\left\{\tfrac{1}{\chi}\Big[(\gamma^{ij} - \Omega^{ij})(1 + \Omega^{\perp\perp}) -\tfrac{2}{\alpha}\Omega^{\perp(i} \beta^{j)}\right. \nonumber\\&\hspace{5cm} \left.-\Omega^{\perp\perp} \tfrac{\beta^i \beta^j}{\alpha^2}+ \Omega^{\perp i} \Omega^{\perp j}\Big]\right\}\,,
    \end{align}
    where $\Omega^{ij}=4\,\gamma^i_{\mu}\gamma^j_{\nu}{\mathcal C}^{\mu\nu}$, $\Omega^{\perp i}=-4\,n_{\mu}\gamma^i_{\nu}{\mathcal C}^{\mu\nu}$ and $\Omega^{\perp\perp}=4\,n_{\mu}n_{\nu}{\mathcal C}^{\mu\nu}$. \footnote{In the results shown later in the chapter we will consider the normalised determinant 
    \begin{equation} \label{eq:Geff}
            G_{\textrm{eff}} =  \frac{\det(g^{\mu\nu}_{\text{eff}})}{\det(g^{\mu\nu})} \left(1+\Omega^{\perp\perp}\right)^2 ~,
    \end{equation}
    which typically has values of order one and is normalised to unity in the absence of any scalar field.}    
    When the value of the ratio in \eqref{disc} becomes negative, this tells us that strong hyperbolicity no longer holds. The eigenvalues of these modes of the system have become imaginary, and when this occurs outside of the apparent horizon the evolution cannot continue. It is also possible that in the strongly coupled regime strong hyperbolicity could be violated in the mixed scalar-gravitational sector; in this case the diagnosis of the problem is less easy to formulate
    \footnote{We only identify when loss of hyperbolicity is observed in the purely gravitational modes. Since the loss of hyperbolicity in these modes appears to coincide with the breakdown of the simulation, we do not investigate further the mixed scalar-gravitational ones. Whilst the latter are the ``fastest'' modes \cite{Reall:2021voz}, it is not necessarily the case that hyperbolicity loss should occur first in their sector. Further work is needed to understand at what level they contribute to the loss of hyperbolicity we observe.}.

    When hyperbolicity is lost, as discussed in \cite{Bernard:2019fjb} (see also \cite{Figueras:2020dzx}), the equations change character in some regions of the spacetime from hyperbolic to parabolic or elliptic. This behaviour can be described by analogy with two model equations, namely the Tricomi equation,
    \begin{align}
        \partial_y^2u(x,y)+y\partial_x^2u(x,y)=0\,,
    \end{align}
    where the characteristic speeds go to zero at $y=0$ and the equations become parabolic, or the Keldysh equation,
    \begin{align}
        \partial_y^2u(x,y)+\tfrac{1}{y}\partial_x^2u(x,y)=0\,,
    \end{align}
    where the characteristic speeds diverge at the transition line $y=0$. Finding out if loss of hyperbolicity happens due to a Tricomi or Keldysh-type transition is of interest since the cure is different for each type. For instance, one can choose a different gauge for a Keldysh-type transition or add derivative self-interactions for a Tricomi-type transition, see the discussion in \cite{Barausse:2022rvg} for more details. 

    \section{Numerical set-up}

    In order to study the loss of hyperbolicity and its relation to the violation of the weak coupling condition we have performed $3+1$ non-linear evolutions of spin-induced scalarisation for isolated rotating black holes. We use the \texttt{GRChombo} code \cite{Andrade:2021rbd,Radia:2021smk} and more specifically its modification to include EsGB gravity with singularity avoiding coordinates \cite{AresteSalo:2022hua,AresteSalo:2023mmd}, namely \texttt{GRFolres} \cite{AresteSalo:2023hcp}, in contrast to other recent studies that make use of mGHC \cite{East:2021bqk}. Using different gauge formulations helps us to explore the possibility that these are physical breakdowns in the theory and not simply gauge issues.

    The size of the computational domain is $L=256M$ along each of the coordinate directions and we use 6 refinement levels (so 7 levels in total), with a refinement ratio of $2:1$. The rest of the parameters are fixed to $\kappa_1=2.0/M$, $\kappa_2=-0.1$, while the Kreiss-Oliger numerical dissipation coefficient is set to  $\sigma=2.0$ (see \cite{Radia:2021smk} for the precise definition of these parameters).
    
    
    Our initial conditions are isolated Kerr black holes in quasi-isotropic coordinates (see \cite{Liu:2009al} for details) with initial angular momentum parameters $a_0/M=0.6$ and $a_0/M=0.8$ respectively. On top of the GR background, we add a scalar field perturbation in the form of a small Gaussian pulse located at a distance of $30M$ from the centre of the black hole with an amplitude of $10^{-5}$ and zero initial momentum. Due to the use of the puncture gauge, our initial data is not a stationary solution of the 3+1 evolution equations and the variables need a characteristic time of $\sim 10-20M$ to settle to a nearly stationary state. 
    We note that the addition of the scalar pulse violates the constraints by a small amount, and we rely on the constraint damping terms in the equations of motion to get rid of this error during the evolution.  However, for our choice of scalar amplitude the initial constraint violations are small compared to the truncation errors introduced by the discretisation of the equations. 
    By the time the scalar pulse ``hits'' the black hole, the initial period of gauge adjustment is over and the constraints are well satisfied. 
	
	In our simulations, $|\lambda/M^2|$ is chosen large enough such that the corresponding Kerr black hole undergoes spin-induced scalarisation.
    The small scalar field perturbation grows exponentially until it settles into an equilibrium distribution. In an astrophysical set-up, this exponential growth would happen during stellar core collapse to a black hole. In that case, the curvature quickly grows as the stellar core compactifies and this would trigger the scalar field development \cite{Kuan:2021lol}. Thus, whilst the set-up we consider is somewhat artificial, it is a reasonable proxy for the astrophysical processes that could produce scalarised black holes.
    As we will see below, the strongest violation of the hyperbolicity and weak coupling condition happens during the exponential growth of the scalar field in the intermediate stages of evolution, so this is the phase that should constrain theories most strongly.
	
	As already observed in $1+1$ non-linear simulations in the spherically symmetric case, the non-hyperbolic region first develops inside the black hole horizon. As the evolution proceeds it can emerge above the horizon leading to an ill-posed initial value problem \cite{Corelli:2022phw}. Even though a non-hyperbolic region would not, in principle, be a problem from a physical point of view if it is hidden inside the apparent horizon, from a numerical point of view it still leads to problems continuing the evolution in a horizon-penetrating puncture gauge (in the mGHC approach one excises the region inside the black hole horizon from the evolution domain \cite{Ripley:2019hxt,Ripley:2020vpk}). We fix this by changing the equations of motion inside the apparent horizon, smoothly switching off the coupling constant as in \cite{Figueras:2020dzx,Figueras:2021abd, AresteSalo:2022hua,AresteSalo:2023mmd}. This means that our interior is effectively Kerr and thus hyperbolic, while outside the horizon the full EsGB system of equations is evolved. This approach is justified as long as we turn off the Gauss-Bonnet term fully inside the horizon and the details of how we switch off this term do not influence the physics of the system on and outside the horizon. 
	To achieve this, the coupling function that we have employed in practice has the form
	\begin{equation} \label{eq:excision}
		f(\varphi) =  f_{\textrm{orig}}(\varphi) /\left(1 + e^{-\beta_{\textrm{ex}} (r - r_{\textrm{ex}})}\right)\,,
	\end{equation}  
    where $f_{\textrm{orig}}$ is the exponential quadratic coupling that we used in Chapter \ref{C:BBHin4dST}, namely
    \begin{equation}\label{eq:coupling_function}
        f_{\textrm{orig}}(\varphi)=\frac{1}{2\beta}(1-e^{-\beta\varphi^2})\,.
    \end{equation}
    In our simulations, we have set  $\beta_{\textrm{ex}}=400$ while $r_{\textrm{ex}}$ is a parameter smaller than the calculated black hole apparent horizon radius. For relatively weak scalar fields, the value of $r_{\textrm{ex}}$ would depend mostly on the initial spin $a_0/M$.  We have checked that the scalar field evolution outside the apparent horizon remains unchanged (if hyperbolic) with the decrease of  $r_{\textrm{ex}}$. Since, as discussed above, unstable regions often develop within the horizon and gradually extend beyond it, then in order to determine the limiting parameters for loss of hyperbolicity in the exterior of the BH we have to choose the maximum possible $r_{\textrm{ex}}$ inside the apparent horizon. Thus, for initial spin $a_0/M=0.6$ we have worked with $r_{\textrm{ex}}=0.75M$ while for $a_0/M=0.8$, the excision radius $r_{\textrm{ex}}=0.54M$.
	
    A final comment on $r_{\textrm{ex}}$ concerns the fact that the apparent horizon radius for the final rotating black holes states may not be exactly spherical after the gauge evolution. 
    Thus, using a spherical excision radius might introduce an error in the measured values. Our aim is to make an order of magnitude estimate of the threshold for hyperbolicity loss and the validity of the weak coupling condition, and so we consider our approach to be appropriate for this goal.

\section{Hyperbolicity loss threshold}\label{sec:wfc}

In order to determine the hyperbolicity loss threshold, for each value of $a_0/M$ we have performed a series of simulations where we vary the parameters $\lambda/M^2$ and $\beta$ in turn.
    The goal is to determine the threshold of $\beta$ for each combination of $a_0/M$ and $\lambda/M^2$ where loss of hyperbolicity is observed. The results are summarised in Table \ref{tab:threshold}. We see that $\beta_{\textrm{threshold}}$ is typically of the order of $10^2-10^3$ for values of $|\lambda/M^2|$ close to the minimum value that ensures scalarisation. The value of $\beta_{\textrm{threshold}}$  increases rapidly for higher $|\lambda/M^2|$ since larger $|\lambda/M^2|$ leads to faster and stronger development of the scalar field. Thus, one has to increase $\beta$ in order to saturate the scalar field at a lower value and keep the system in the hyperbolic region. One can also verify that for each $a_0/M$ the ratio $\lambda/(M^2\sqrt{\beta_{\textrm{threshold}}})$ is roughly a constant as seen in the last column of Table \ref{tab:threshold}. This empirical trend can be understood by considering the coupling function used, for which the maximum value of $f'(\varphi)$ (for any $\varphi$) is limited from above for positive $\beta$ and scales as $f'(\varphi)_{\textrm{max}}\sim 1/\sqrt{\beta}$. Thus the quantity $\lambda/(M^2\sqrt{\beta_{\textrm{threshold}}})$ acts as an effective coupling at the threshold. We see that $\lambda/(M^2\sqrt{\beta_{\textrm{threshold}}})$ decreases for larger $a_0/M$, which is expected since the absolute value of ${\mathcal L}^{\text{GB}}$ at the pole increases (recall that the negative ${\mathcal L}^{\text{GB}}$ around the poles of the apparent horizon is the source of the spin-induced scalarisation).

	\begin{table}[H]
		\centering
			\begin{tabular}{|c|c|c|c|}
			\hline
				$a_0/M$ & $\lambda/M^2$ & $\beta_{\textrm{threshold}}$ & $\lambda/(M^2\sqrt{\beta_{\textrm{threshold}}})$ \\
			\hline
				0.6 & -200 & 1000 & -6.3\\ 
				\hline
				0.6 & -400 & 5000 & -5.7\\
				\hline
				0.6 & -800 & 20950 & -5.5\\
				\hline
				\hline
				0.8 & -50 & 240 & -3.2\\
				\hline
				0.8 & -100 & 1150 & -3.0\\
				\hline
				0.8 & -200 & 4650 & -2.9\\
			\hline
					
			\end{tabular}
   \caption[Threshold for loss of hyperbolicity for several values of the initial angular momentum and the Gauss-Bonnet coupling]{Threshold for loss of hyperbolicity calculated for several combinations of the initial angular momentum $a_0/M$ and $\lambda/M^2$. The third column represents the minimum $\beta_{\textrm{threshold}}$ for which the simulations are still hyperbolic while the last one is the ratio $\lambda/(M^2\sqrt{\beta_{\textrm{threshold}}})$ at the threshold, which is roughly a constant for every value of $a_0/M$.}
		\label{tab:threshold}
	\end{table}

\begin{figure}[hbt]
            \centering
		\includegraphics[width=1.0\textwidth]{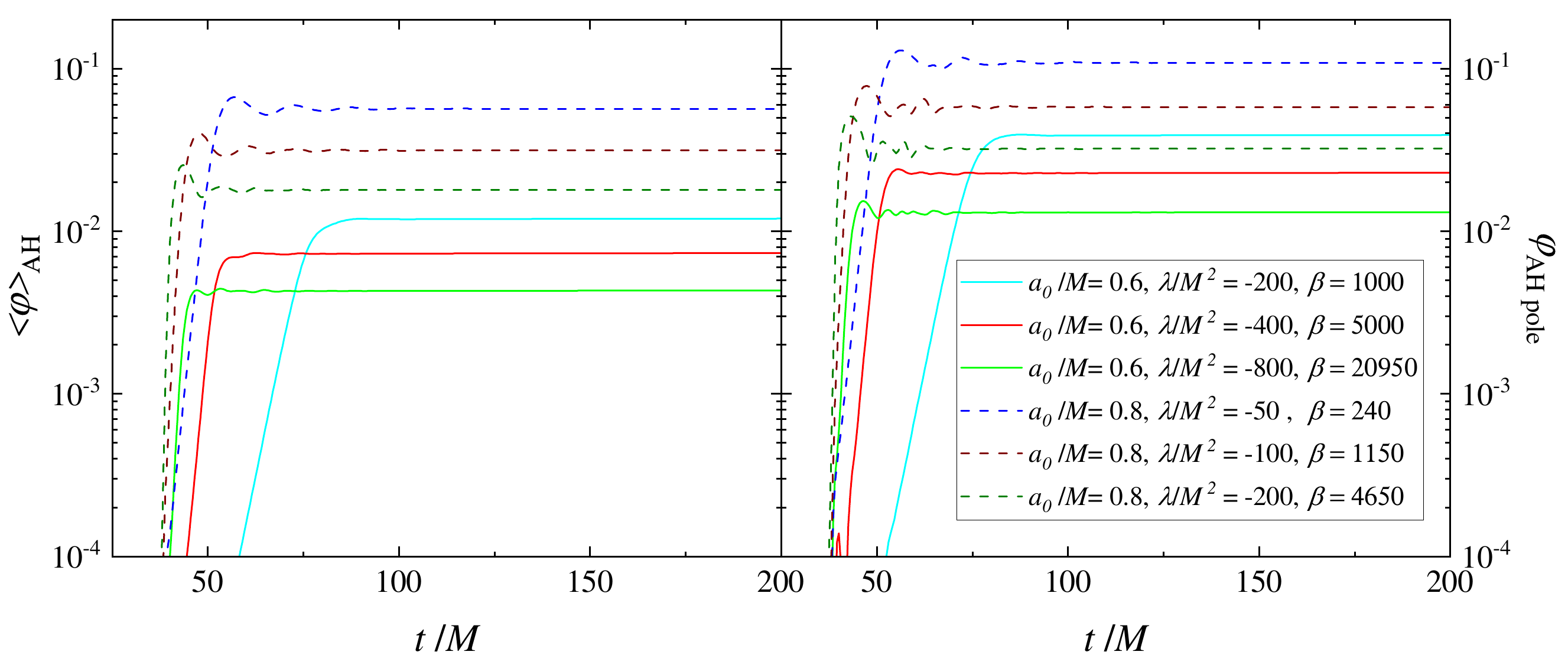}
		\includegraphics[width=1.0\textwidth]{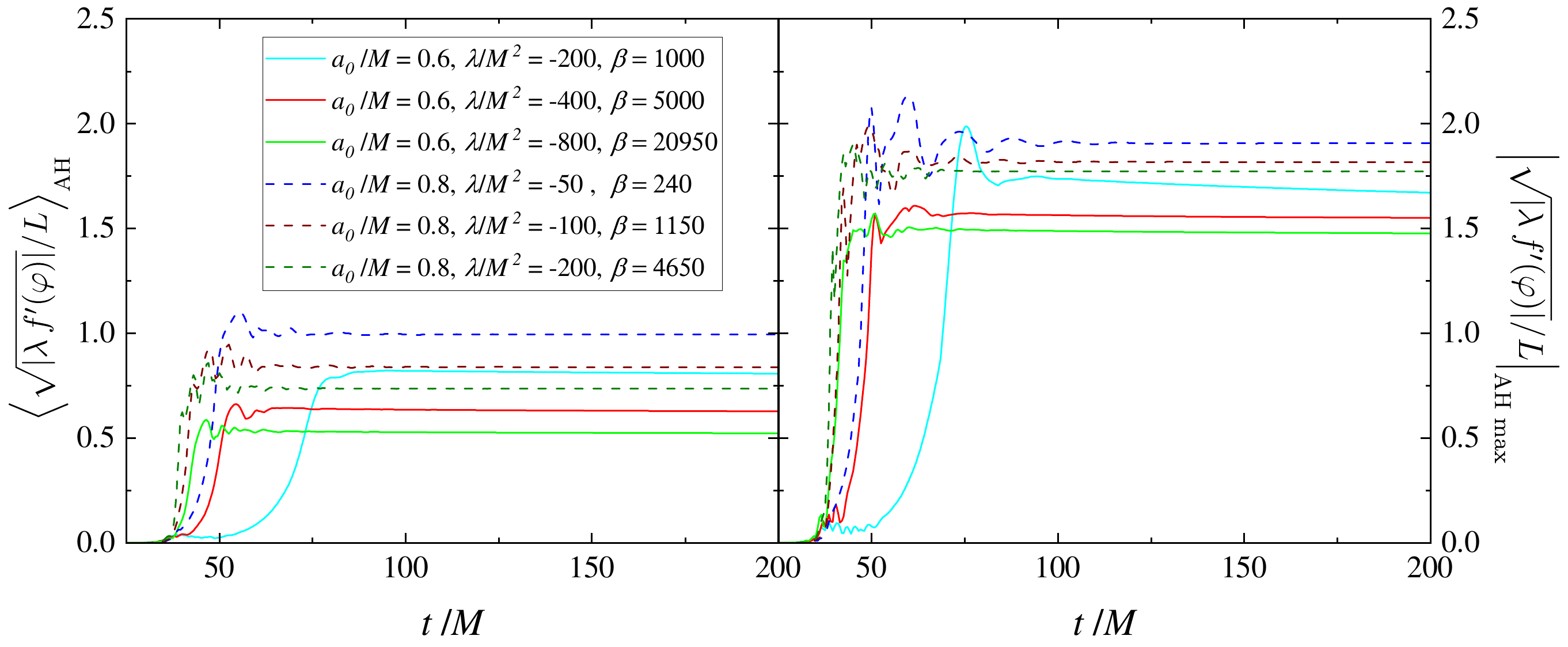}
		\caption[Time evolution of the scalar field and the weak coupling condition for hairy BHs in exponential quadratic coupling EsGB just above the hyperbolicity loss threshold.]{Time evolution of the models from Table \ref{tab:threshold} that are just above the hyperbolicity loss threshold. \textit{Top:} The scalar field at the apparent horizon taken either as the $L^2$ norm (\textit{left}) or at the pole of the apparent horizon, where it has its maximum (\textit{right}). \textit{Bottom:} The weak coupling condition (given by Eq. \eqref{wfc}) at the apparent horizon taken  as the $L^2$ norm (\textit{left}) and its maximum value at the apparent horizon (\textit{right}).}
		\label{fig:Evol_threshold}
	\end{figure}

    For the threshold models listed in Table \ref{tab:threshold}, in Fig. \ref{fig:Evol_threshold} we show the evolution of the scalar field on the horizon as well as the weak coupling condition defined in Eq. \eqref{wfc}. Both quantities are displayed as their $L^2$ norm at the apparent horizon in the left panel. In the right panel, we depict the time evolution of the maximum value of the scalar field and the weak coupling condition. For the scalar field $\varphi$, the maximum typically occurs at the pole of the apparent horizon where the Gauss-Bonnet invariant is most negative. We, therefore, plot $\varphi$ at the pole. The extremum of the weak coupling condition happens away from the pole and its position slightly varies with time; we plot its maximum value on the apparent horizon as a function of time. The profiles of these quantities outside the black hole horizon will be shown in more detail later in this section.
         
    Let us first consider the scalar field evolution in the top panels of Fig. \ref{fig:Evol_threshold}. We see that the scalar field behaviour (specifically its $L^2$ norm across the apparent horizon and at the pole of the apparent horizon) is very similar for all cases -- after a phase of exponential growth, small amplitude oscillations are observed until the scalar field settles to a constant profile. The value of the scalar field at the pole of the horizon $S^2$, which corresponds to the maximum, is roughly twice the size of the averaged  $L^2$ norm. 
        
    The differences between the plots in the left and the right panel are more pronounced for the weak coupling condition. In order for the EFT approach to be justified, the weak coupling condition defined in \eqref{wfc} should be much less than unity. As one can see in the left panel, the weak coupling condition given by eq. \eqref{wfc} is less than unity, i.e. $\sqrt{\lambda f'(\varphi)}/L \lesssim 1$, for all threshold models if one looks at the $L^2$ norm. Its maximum value can reach much larger values though, more than twice the average ones. In addition, at intermediate times, as the scalar field is still rapidly evolving, we observe spikes in the weak coupling condition, especially in its maximum value on horizon. 

\begin{figure}[!hbt]
            \centering
		\includegraphics[width=1.0\textwidth]{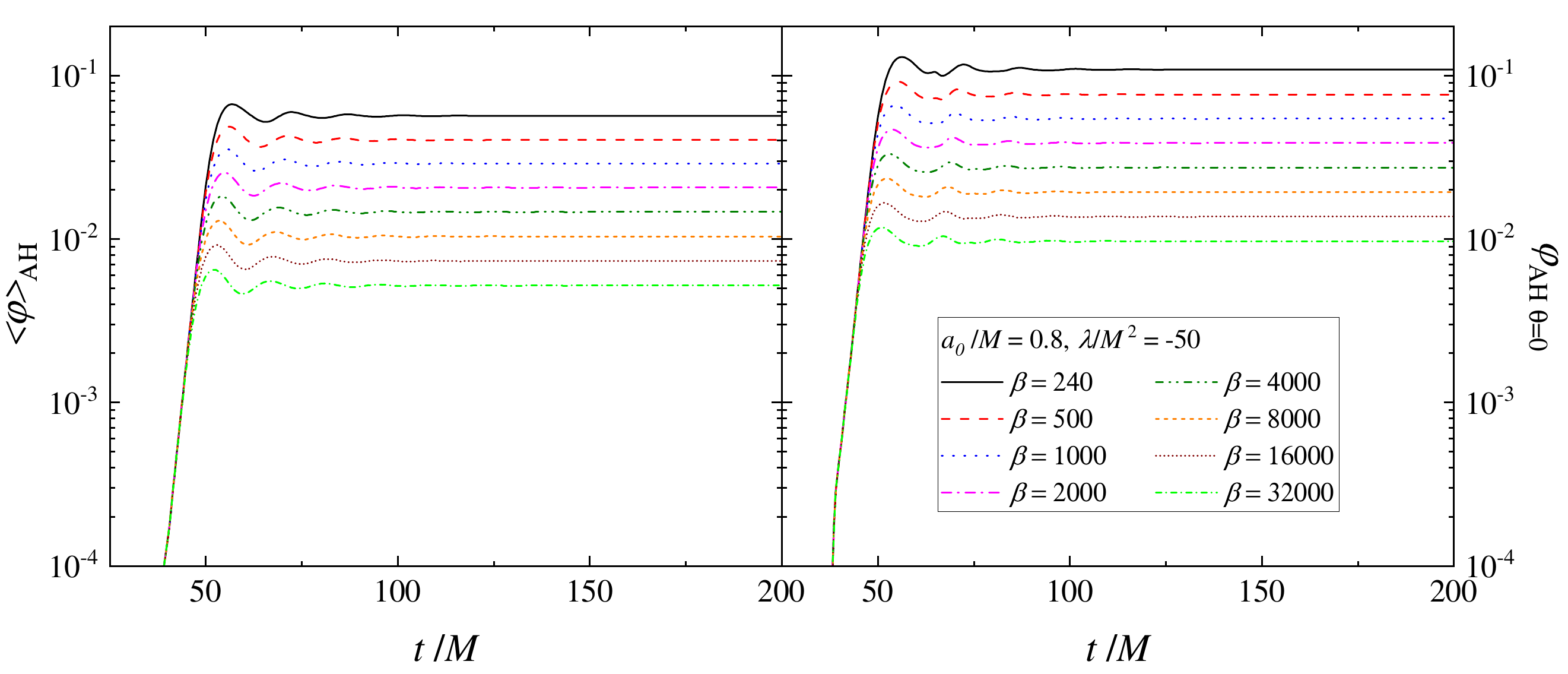}
		\includegraphics[width=1.0\textwidth]{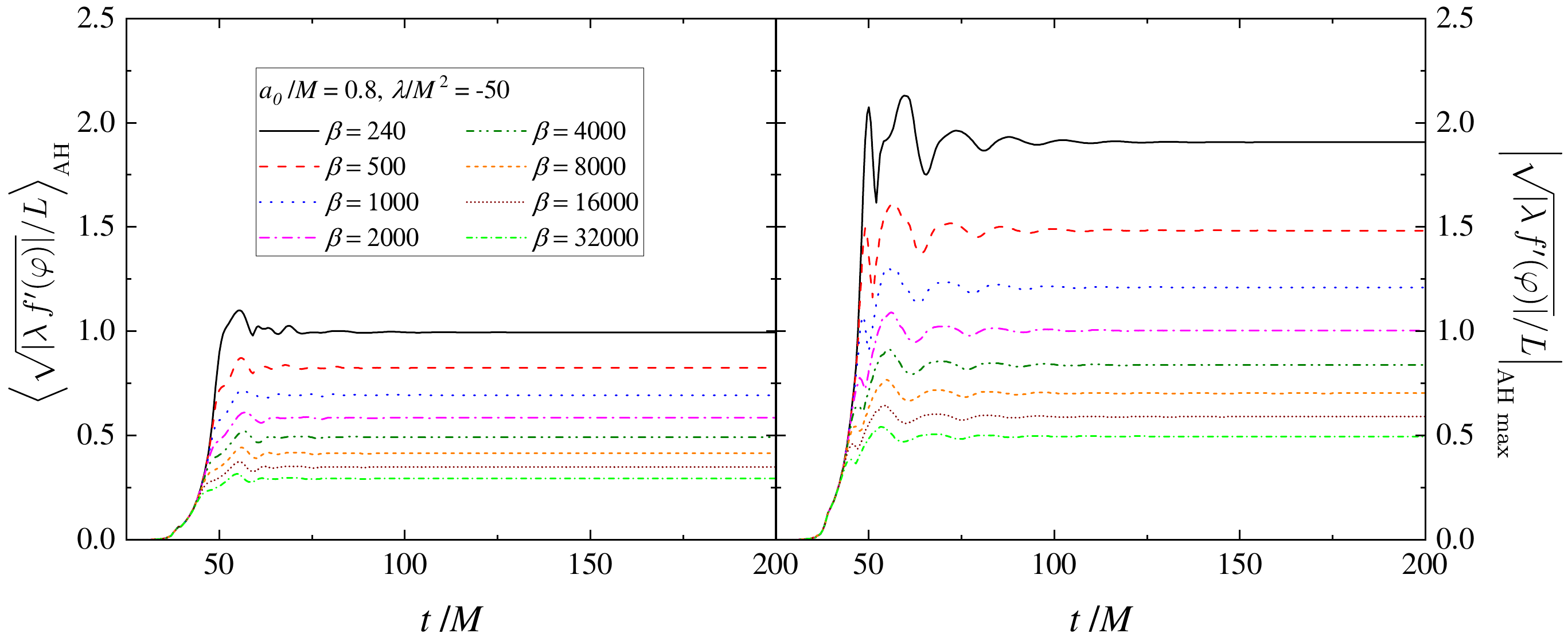}
		\caption[Time evolution of the scalar field and the weak coupling condition for more hairy BHs in exponential quadratic EsGB above the hyperbolicity loss threshold.]{Time evolution of models with $a_0/M=0.8$, $\lambda/M^2=-50$ and varying $\beta$ starting from the threshold for hyperbolicity loss in Table \ref{tab:threshold} and reaching values for which the weak coupling condition is much less than unity. \textit{Top:} The scalar field at the apparent horizon taken either as the $L^2$ norm (\textit{left}) or at the pole, where it has a maximum (\textit{right}). \textit{Bottom:} The weak coupling condition (given by eq. \eqref{wfc}) at the apparent horizon taken  as the $L^2$ norm (\textit{left}) and its maximum value at the apparent horizon (\textit{right}).}
		\label{fig:Evol_Fixed_a_beta}
	\end{figure}

    We conclude that the range of parameters where hyperbolicity breaks down is actually well into the regime where the weak coupling condition is violated and the effective field theory treatment is no longer justified. We can explore how large $\beta$ should be for fixed $\lambda/M^2$ and $a_0/M$ in order to maintain the maximum weak coupling condition value less than unity. For that purpose, we have chosen one combination of parameters, namely $a_0/M=0.8$ and $\lambda/M^2=-50$, and increased $\beta$ starting from its threshold value given in Table \ref{tab:threshold}. The results from the corresponding time evolutions are depicted in Fig. \ref{fig:Evol_Fixed_a_beta}. As one can see, in order to maintain $\sqrt{\lambda f'(\varphi)}/L \ll 1$, $\beta$ needs to be at least two orders of magnitude larger than the minimum one that preserves hyperbolicity. The effect on the scalar field magnitude is less pronounced, and the two limiting cases in the figure, $\beta=240$ and $\beta=32000$, have roughly one order of magnitude difference in the equilibrium value of the scalar field after saturation. 

\begin{figure}[!hbt]
        \centering
	\includegraphics[width=1.0\textwidth]{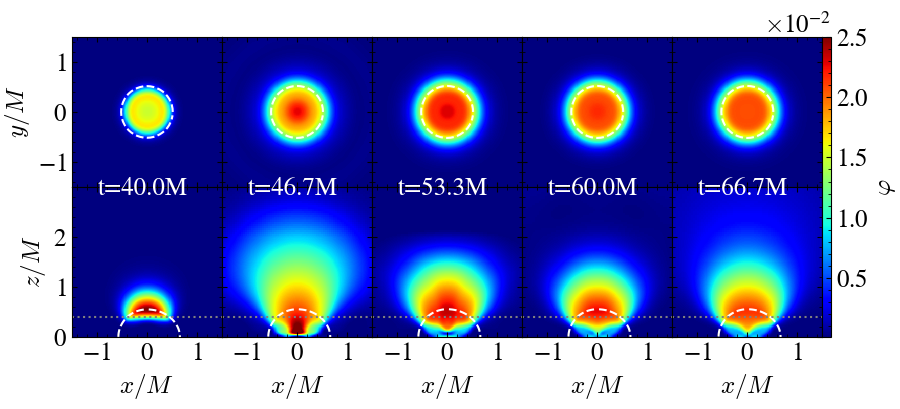}
	\caption[Time evolution of the scalar field profile in a hairy BHs in exponential quadratic EsGB.]{The scalar field profile at several equally spaced coordinate times during the scalarisation. In the upper panels we have represented the scalar field taken on a $x-y$ slice, while in the lower panels $x-z$ slices are depicted. The $x-z$ slices cross the centre of the BH while the $x-y$ slices are taken at $z=0.4$ above the centre (marked with grey dotted lines in the lower panels), where the scalar field is strongest. The dashed white line represents the position of the apparent horizon. Note that in the black hole interior the Gauss-Bonnet term is completely turned off, thus the inner region does not represent a solution of the EsGB field equations. The values of the employed parameters are $a_0/M=0.8$, $\lambda/M^2=-200$, and $\beta=10000$.}
	\label{fig:2D_Slice_phi}
    \end{figure}

\begin{figure}[!hbt]
    \centering
	   \includegraphics[width=1.0\textwidth]{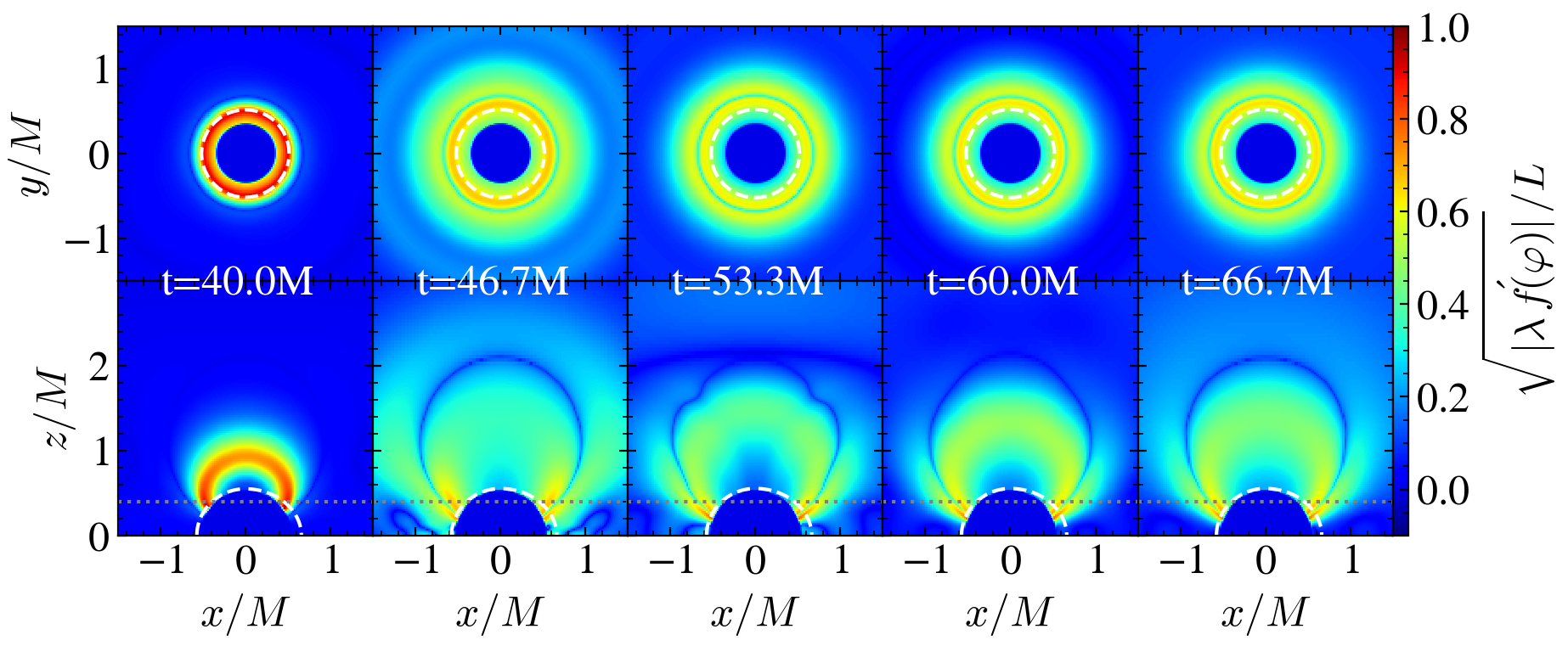}
	   \caption[Time evolution of the weak coupling condition profile in a hairy BH in exponential quadratic EsGB.]{The same configuration as in Fig. \ref{fig:2D_Slice_phi} but illustrating the evolution of the weak coupling condition given by Eq. \eqref{wfc}.}
	   \label{fig:2D_Slice_WFC}
    \end{figure}

\begin{figure}[H]
   \centering
	   \includegraphics[width=1.0\textwidth]{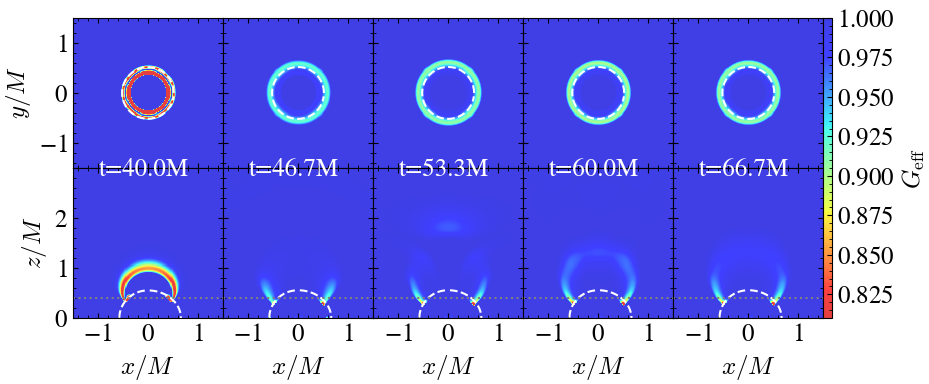}
	   \caption[Time evolution of the profile of the normalised determinant of the effective metric in a hairy BH in exponential quadratic EsGB.]{The same configuration as in Fig. \ref{fig:2D_Slice_phi} but illustrating the evolution of the normalised determinant of the effective metric $G_{\textrm{eff}}$, as defined in Eq. \eqref{eq:Geff}.}
	   \label{fig:2D_Slice_discr}
    \end{figure}

   In Figs. \ref{fig:2D_Slice_phi}, \ref{fig:2D_Slice_WFC}, and \ref{fig:2D_Slice_discr} we have depicted $x-y$ and $x-z$ slices of three main diagnostic quantities -- the scalar field, the weak coupling condition and the determinant of the effective metric, in the form of 2D slice plots. This provides further context on the hyperbolicity loss and weak coupling condition violation. Five equally spaced characteristic times during the evolution are represented starting from the moment when the initial scalar field seed pulse has arrived at the black hole (when the scalar hair starts growing), until it settles into an equilibrium profile. On each figure, the position of the apparent horizon is indicated with a dashed white line. The simulations shown correspond to the case of initial angular momentum $a_0/M=0.8$, $\lambda/M^2=-200$, and $\beta=10000$.

The scalar field depicted in Fig. \ref{fig:2D_Slice_phi} has a maximum on the rotational axis and it forms a bulb shape there, while close to the equator it has a minimum. The weak coupling condition has a different distribution as seen in Fig. \ref{fig:2D_Slice_WFC} -- it has a maximum at a ring-like structure around the pole. This is because the maximum of the weak coupling condition is influenced not only by the scalar field maximum but also by its derivatives.

    Let us now consider the normalised determinant of the effective metric depicted in Fig. \ref{fig:2D_Slice_discr}. As discussed above, the moment when hyperbolicity is lost is associated with the determinant becoming zero.
    Since inside the black hole interior we turn off the Gauss-Bonnet coupling term, the value of the determinant is unity there. The values of $a_0/M$, $\lambda/M^2$, and $\beta$ are chosen in such a way that the models we evolve are above the threshold for loss of hyperbolicity, so the determinant remains positive outside the apparent horizon.  It reaches a minimum value (closer to hyperbolicity loss) at intermediate times, when the scalar field grows fastest. At that point of evolution, the weak coupling condition is also at its maximum. At late times the determinant approaches larger values (close to unity), that is, further away from hyperbolicity loss. This effect is expected since both the determinant and the weak coupling condition are influenced by the scalar field time and spatial derivatives. Thus, the loss of hyperbolicity will depend not only on the final stationary solution but also on the dynamical evolution that leads to it.

    In our simulations of models with smaller $\beta$, in a regime for which the evolution is stable but closer to the threshold for loss of hyperbolicity, we observe a transient period where hyperbolicity is lost within the horizon. Specifically, inside the apparent horizon but still outside the region where the Gauss-Bonnet coupling is turned off, the determinant has negative values at intermediate times. Since this happens inside the apparent horizon for a relatively short time, the evolution manages to continue. This is consistent with the observations made in the $1+1$ non-linear evolution of \cite{Corelli:2022phw}.

    \section{Gauge dependence of hyperbolicity loss}\label{subsec:gauge}
    
    As discussed in the previous sections, the breakdown in hyperbolicity that we observe appears to be linked to the physical modes.  These are gauge invariant \cite{Reall:2021voz}, and so we do not expect a change in gauge to prevent the breakdown observed. However, it is possible that in the strongly coupled regime the gauge modes themselves may become problematic. It is also interesting to explore the effect of the modified puncture gauge choice (which is newly developed and little explored) on the accuracy and resolution of the numerical simulations. For that purpose, in this section we explore the impact of modifying the gauge, by adjusting the free functions $a(x)$ and $b(x)$ that determine the auxiliary metrics (see eq. \eqref{abfunc}). 
        
    In previous sections we have worked with constant $a(x)=0.2$ and $b(x)=0.4$, values similar to those used in \cite{East:2021bqk}. This choice is by no means unique, and we have tried several other combinations of constant $a(x)$ and $b(x)$, keeping always $b(x)>a(x)$ and $\kappa_2>-\frac{2}{2+b(x)}$ as required in the modified puncture gauge. We focus on simulations of black holes with $a_0/M=0.8$ and $\lambda/M^2=-50$. The combinations of $\{a(x),b(x)\}$ we have tested are $\{0.05,0.1\}$, $\{0.1,0.2\}$, $\{0.2,0.4\}$, $\dots$, $\{1.6,3.2\}$, as well as $\{0.2,0.3\}$ and $\{0.2,0.6\}$. For all cases, the threshold for loss of hyperbolicity is found consistently to be at $\beta=240$, with any deviations lying within our numerical uncertainties. A practical observation is that, despite the wide range of $\{a(x),b(x)\}$ that we have considered, we were able to perform evolutions for all of the combinations mentioned above, keeping the same value of the damping parameters $\kappa_1$ and $\kappa_2$ as for the $\{0.2,0.4\}$ case.
             
    \begin{figure}[H]
    \centering
	   \includegraphics[scale=0.8]{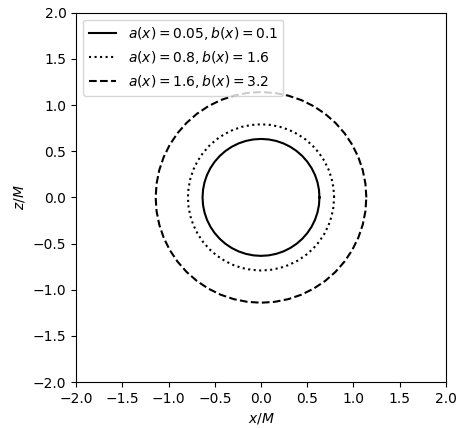}
	   \caption[A slice of the BH apparent horizon for different combinations of the gauge parameters $a(x)$ and $b(x)$.]{An $x-z$ slice of the apparent horizon at $t=400M$ once the BH has settled to a stationary gauge. The simulations are performed for  $a_0/M=0.8$, $\lambda/M^2=-50$, $\beta=2000$ and different combinations of the gauge parameters $a(x)$ and $b(x)$. }
		\label{fig:horizon_gauge}
	\end{figure}
 
    We now highlight another interesting observation not directly related to the loss of hyperbolicity. While experimenting with the values of $a(x)$ and $b(x)$ we have noticed that changing their values in turn changes significantly the coordinate size of the apparent horizon radius, as we can see in Fig. \ref{fig:horizon_gauge}. For example the difference between the two limiting cases we have considered, namely $\{a(x),b(x)\}=\{0.05,0.1\}$ and $\{a(x),b(x)\}=\{1.6,3.2\}$ is roughly twice, as evident from Fig. \ref{fig:horizon_gauge}. The coordinate radius of the apparent horizon increases further with the increase of $a(x)$ and $b(x)$. This might be helpful for some simulations, as it may provide a way to improve the resolution at the apparent horizon. Another positive outcome concerns the violation of the Hamiltonian constraint, that always occurs near the puncture in singularity avoiding coordinates. When we increase $a(x)$ and $b(x)$ the size of this constraint violating region grows slower compared to the growth of the apparent horizon radius. As a result, in the simulation with  $\{a(x),b(x)\}=\{1.6,3.2\}$, the constraint violating region is further away from the black hole apparent horizon compared to the standard $\{a(x),b(x)\}=\{0.2,0.4\}$ case. This should improve the accuracy of the evolution outside the horizon for a fixed resolution, since the separation (in terms of grid points) between the apparent horizon and the region with sizeable constraint violations is increased. 

    These results, whilst not definitive, suggest that the hyperbolicity loss that we observe does not strongly depend on the choice of the functions $a(x)$ and $b(x)$, and is dominated by the physical sector of the eigenmodes (rather than the gauge modes). This also follows from our observation that the breakdown of the simulations is usually preceded by the determinant of the effective metric turning negative, which indicates that some of the eigenvalues from the physical sector have become complex. Thus, the results are consistent with the theoretical investigations in \cite{Reall:2021voz}.

        \section{Reasons for loss of hyperbolicity}

 In this section we investigate whether the hyperbolicity loss happens because of a Tricomi or a Keldysh-type of transition. Since the pure gravitational sector of the physical eigenmodes (whose eigenvalues we can directly compute) lie at the null cone of the effective metric, one can show that their propagation speeds will only diverge if the quantity $1+\Omega^{\perp\perp}$ vanishes, thus this indicates a Keldysh-type hyperbolicity loss. If instead the determinant of the effective metric vanishes with the quantity $1+\Omega^{\perp\perp}$ remaining positive, the propagation speeds remain finite but they become equal, which indicates a Tricomi-type hyperbolicity loss (see \cite{Lara:2021piy} for a more detailed discussion) \footnote{Note that in Tricomi equation the propagation speeds go to zero as explained in Section \ref{sec:wfc}. However, the determinant of the effective metric can also vanish for the more general case in which both propagation speeds become equal with an analogous behaviour to the Tricomi-type loss of hyperbolicity.}.

    In order to test the type of transition, we have performed two simulations that lose hyperbolicity at a certain point with fixed $a_0/M=0.8$, $\lambda/M^2=-50$, and varying $\beta=\{100,200\}$. While  $\beta=200$ is only slightly below the threshold given in Table \ref{tab:threshold}, for $\beta=100$ the hyperbolicity loss happens faster with the formation of a larger region of a negative determinant of the effective metric.

\begin{figure}[!hbt]
    \centering
		\includegraphics[scale=.45]{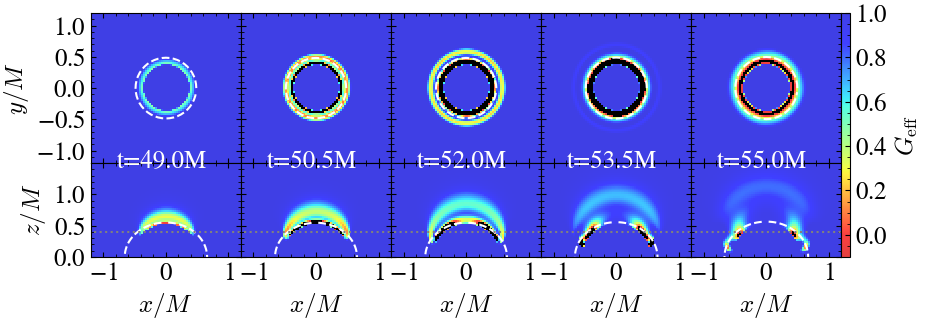}
		\includegraphics[scale=.45]{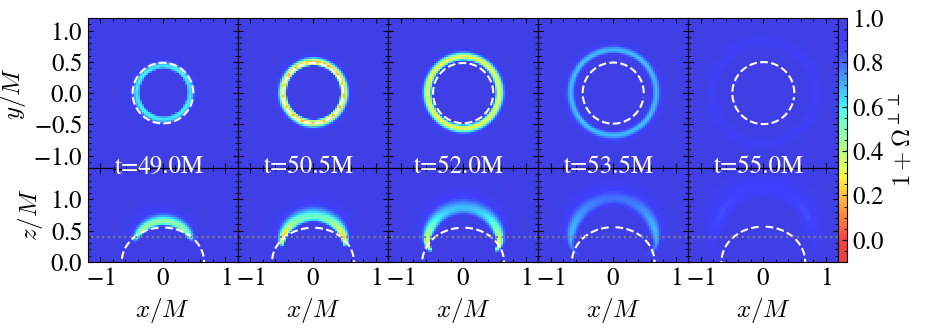}
		\caption[Time evolution of the normalised determinant of the effective metric and the quantity $1+\Omega^{\perp\perp}$ in a hairy BH in exponential quadratic EsGB just below the threshold of hyperbolicity loss.]{Time evolution of a model just below the threshold of hyperbolicity loss with $a_0/M=0.8$, $\lambda/M=-50$, and $\beta=200$. Several equally spaced coordinate times during the scalarisation are plotted, capturing the evolution just before the code breaks down due to hyperbolicity loss. In each figure, both $x-y$ and $x-z$ slices are depicted, as in the figures above. The apparent horizon is plotted as a white dashed line. \textit{(top)} Time evolution of the normalised determinant of the effective metric $G_{\textrm{eff}}$ defined in Eq. \eqref{eq:Geff}. Negative values are depicted in black. \textit{(bottom)} Time evolution of $1+\Omega^{\perp\perp}$.}
		\label{F:discriminant_beta200}
	\end{figure}

    The quantity $1+\Omega^{\perp\perp}$  behaves differently in the two cases. For $\beta=200$, it is always positive as seen in Fig. \ref{F:discriminant_beta200}. Thus, the hyperbolicity loss is caused by a Tricomi-type transition, as was observed in other simulations in EsGB gravity \cite{Ripley:2019hxt}.  Note that the normalised determinant of the effective metric, depicted in the same figure, is negative only in a small region outside the apparent horizon. As may be expected, shortly after this region forms, the numerical evolution cannot be continued.  

\begin{figure}[!ht]
    \centering
		\includegraphics[scale=.62]{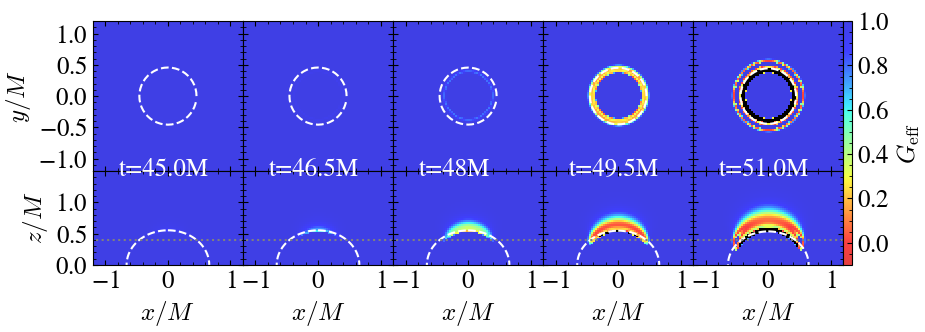}
		\includegraphics[scale=.62]{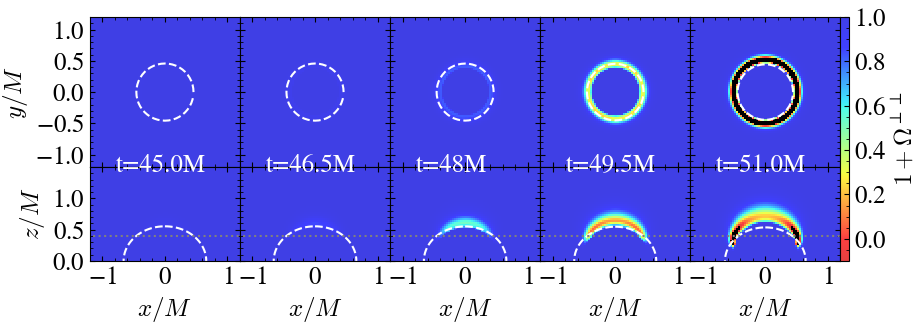}
		\caption[Time evolution of the normalised determinant of the effective metric and the quantity $1+\Omega^{\perp\perp}$ in a hairy BH in exponential quadratic EsGB far beyond the threshold of hyperbolicity loss.]{As in Fig. \ref{F:discriminant_beta200} but for a model far beyond the boundary of hyperbolicity loss with $a_0/M=0.8$, $\lambda/M=-50$, and $\beta=100$. Negative values of the determinant of the effective metric and the quantity  $1+\Omega^{\perp\perp}$ are depicted in black.}
		\label{F:discriminant_beta100}
	\end{figure}

      As depicted in Fig. \ref{F:discriminant_beta100}, for $\beta=100$ the quantity $1+\Omega^{\perp\perp}$ can become negative outside the apparent horizon, which means that the propagation speeds diverge. Interestingly, in the region where this happens, the determinant is positive. Thus, one can speculate that in the region with a negative determinant (in black in the upper panel of Fig. \ref{F:discriminant_beta100}, around the pole of the apparent horizon), the hyperbolicity loss is again of the Tricomi-type since the propagation speeds are finite there. However, the appearance of another region with diverging speeds (in black in the bottom panel of Fig. \ref{F:discriminant_beta100}) implies that a second ``problematic'' region develops away from the pole, causing a loss of hyperbolicity of the Keldysh-type. Further investigation is required to confirm that the loss of hyperbolicity happens because Tricomi-type and Keldysh-type-of transitions occur simultaneously in different regions of the spacetime, or whether one is dominant.

\addtocounter{page}{1}%
\blankpage

\chapter{GRFolres}
\label{C:GRFolres}

This chapter is based on the overview in \cite{AresteSalo:2023hcp}, which is part of the submission of the code to the Journal of Open Source Software, as well as the complementary information that can be found in the wiki page in \href{https://github.com/GRTLCollaboration/GRFolres}{\texttt{https://github.com/GRTLCollaboration/GRFolres}}.

\section{Capabilities}

\texttt{GRFolres} uses a lot of the features and follows a similar design philosophy to \texttt{GRChombo}. The key difference is that it includes the contributions to the Einstein's equations coming from certain modified gravity theories.

In particular, it currently implements the Four-Derivative Scalar-Tensor theory of gravity ($4\partial$ST) coming from the action
\begin{equation}
    S=\int d^4x\sqrt{-g}\Big(\frac{R}{16\pi G}+X+g_2(\phi)X^2+f(\phi){\mathcal L}^{\text{GB}} \Big)\,,
\end{equation}
whose equations are implemented in our modified puncture CCZ4 formalism presented in Chapter \ref{C:Wellposedness}.

The code also implements cubic Horndeski, whose results are in \cite{Figueras:2020dzx,Figueras:2021abd}, as well as the $4\partial$ST without back-reaction, which was used in \cite{Evstafyeva:2022rve}. The code has sufficient flexibility that new modified theories of gravity that are well-posed in our formulation can be added straightforwardly.

\texttt{GRFolres} contains some additional features, which have not yet been made publicly available but that are being used in ongoing projects.
\begin{itemize}
    \item $4\partial$ST initial conditions - The CTTK approach \cite{Aurrekoetxea:2022mpw} has been adapted to treat the additional $4\partial$ST terms for general initial configurations of the scalar field \cite{Brady:2023dgu}.
    \item Cosmic inflation - \texttt{GRFolres} has been adapted to cosmological spacetimes in order to understand how $4\partial$ST modifications to GR could affect the start of inflation and any observational consequences.
    \item Higher dimensional Gauss-Bonnet gravity - The equations for Einstein-Gauss-Bonnet theory in 4+1 dimensions have been implemented as in \texttt{GRFolres} using the Cartoon formalism.
\end{itemize}

 \section{General features}

 \texttt{GRFolres} has been implemented in an analogous way as the scalar field example in \texttt{GRChombo} with some additional features:
 \begin{itemize}
    \item Modified puncture gauge - The equations are written in a formalism that differs from the standard CCZ4 approach by adding extra spatially dependent functions $a(x)$ and $b(x)$, which modify the speed of propagation of the gauge modes and ensure well-posedness when adding the modified gravity contributions.

 \item Matrix inversion - For $4\partial$ST
 (and many other modified gravity theories) there are terms in the time derivatives (aka, the right hand side or RHS) of the 
 variables that depend (linearly) on the time derivatives of other variables. Therefore, in order to solve the full non-linear equations one needs to express the evolution of those variables at each spatial point as a linear system, and invert an $8\times8$ matrix.
\item Coupling and potential - In the same way as the ScalarField class is templated over the potential, we template the class associated to the theory we are solving (the FourDerivScalarTensor class in the case of $4\partial$ST) over the coupling to the modified gravity terms and the potential of the scalar field. Therefore, in the case of $4\partial$ST the user needs to create a class to compute the following quantities:
\begin{itemize}
    \item \texttt{dfdphi}: the derivative of the coupling to the Gauss-Bonnet curvature $f(\phi)$ with respect to $\phi$.
 \item \texttt{d2fdphi2}: the second derivative of the coupling $f(\phi)$ with respect to $\phi$.
 \item \texttt{g2}: the coupling to the square of the kinetic term $X^2$ of the scalar field, $g_2(\phi)$.
 \item \texttt{V\_of\_phi}: the potential of the scalar field $\phi$.
 \item \texttt{dVdphi}: the derivative of the potential with respect to $\phi$.
\end{itemize}
Several examples of such classes are provided, for shift-symmetric and exponential couplings, which should be easy to adapt to other cases.
\item Excision - In the puncture approach, numerical excision of the region inside the apparent horizon is not needed (in contrast to Generalised Harmonic Coordinates). However, the equations can in principle be changed in this region without affecting the rest of the spacetime, and this is usually necessary since close to the singularity EsGB type theories develop non hyperbolic regions. Therefore the code allows the user to modify the coupling 
 so that it vanishes in the singularity and then smoothly transitions to its chosen value before reaching the apparent horizon. This has been implemented in two ways,
\begin{itemize}
    \item using the radial coordinate 
 that measures the distance to the black hole: $f(\phi)\to f(\phi)/(1+e^{-\tau(r-r_0)})$, or
\item using the conformal factor $\chi$: $f(\phi)\to f(\phi)/(1+e^{-\tau(\chi-\chi_0)})$,
\end{itemize}
where $\tau$, $r_0$ and $\chi_0$ have to be chosen so that the function is smooth enough and only affects the region inside the apparent horizon.
\end{itemize}

\section{Structure of the code}

The code is structured in three main folders:
\begin{itemize}
    \item Source, where all the additional features of \texttt{GRFolres} are implemented.
    
    The main class where the equations of motion of the theory are encoded is called the ModifiedCCZ4 class (in analogy to the MatterCCZ4 class), which inherits from the \texttt{GRChombo} CCZ4 class. The ModifiedCCZ4 class is templated over the theory (which can be $4\partial$ST, Cubic Horndeski...), the gauge (the modified puncture gauge, which includes the functions $a(x)$ and $b(x)$) and the derivatives (which can be $4^{\text{th}}$ or $6^{\text{th}}$ order stencils).

The algorithm that the ModifiedCCZ4 class follows in order to compute the equations of motion of a given theory is the following:
\begin{itemize}
    \item Compute the Einstein GR equations in the puncture CCZ4 formalism.
    \item Add the $a(x)$ and $b(x)$ terms from the modified puncture gauge.
    \item Add the modified gravity contribution to the RHS of the 3+1 variables.
    \item Add the equations for the additional fields (if any) that the theory incorporates (phi, Pi...)
    \item Compute the LHS matrix of the equations of motion and solve the linear system, if required for the given theory.
\end{itemize}
The functions in charge of computing the specific form of those terms for the given theory (which are required in order to execute the algorithm above) are included in the class corresponding to that theory, which is for example the FourDerivScalarTensor class in the case of the $4\partial$ST.
    \item Examples, where we include some examples with either single Kerr Black Holes or Binary Black Holes in the modified theories of gravity that we have implemented ($4\partial$ST, Cubic Horndeski).
    \item Tests, where one can find the tests that we have carried out in order to ensure that the equations of motions have been correctly implemented.
\end{itemize}

 \section{Parameters}

 When running an example in \texttt{GRFolres}, the additional parameters that need to be specified apart from the ones needed in any \texttt{GRChombo} run are the following ones:
 \begin{itemize}
     \item  Modified puncture gauge sector:
\begin{itemize}
    \item \texttt{a0} is the value of the $a(x)$ function, which we set to be a constant (default value: \texttt{a0}$=0$).
 \item \texttt{b0} is the value of the $b(x)$ function, which we set to be a constant (default value: \texttt{b0}$=0$).
\end{itemize}

\item Coupling and potential sector, which includes in the case of the $4\partial$ST:
\begin{itemize}
    \item \texttt{lambda\_GB} is the coupling constant in the Gauss-Bonnet coupling $f(\phi)$ (default value: \texttt{lambda\_GB}$=0$).
    \item \texttt{cutoff\_GB} is the value of $r_0$ or $\chi_0$ for the excision of the Gauss-Bonnet terms (default value: \texttt{cutoff\_GB}$=0.15$).
 \item \texttt{factor\_GB} $=\tau$ is the parameter in the excision function (default value: \texttt{factor\_GB}$=100$).
\item \texttt{g2} is the value of the $g_2(\phi)$ coupling, which we set to be a constant (default value: \texttt{g2}$=0$).
\end{itemize}
 \end{itemize}
 Note that when the theory contains additional variables apart from the CCZ4 ones (e.g. the scalar field and its conjugate momentum), one may need to take into account as well the parameters that concern their initial conditions in the same way as is done in the \texttt{GRChombo} scalar field example.

.

\chapter{Conclusions and further work}
\label{C:Conclusions}

\section{Summary}

It is likely that it will be in the highly dynamical, strong field regime where the fundamental nature of gravity is revealed, and the key tool to access it is Numerical Relativity. Motivated originally by
the prospect of detecting GWs from compact object mergers, NR has expanded beyond its original
premise. Whilst the most well-known example of its vital role in cutting-edge research is the characterisation of BH merger signals from the Advanced LIGO/Virgo network of GW detectors, in recent
years this tool has started to have a significant impact in studies of fundamental gravitational theory
and cosmology. This thesis shows how we are pioneering the application of NR tools to scenarios beyond GR and the Standard Model.

In Chapter \ref{C:Wellposedness} we have successfully developed a modified version of the CCZ4 formulation of the Einstein equations based on the recently proposed modified harmonic gauge \cite{Kovacs:2020ywu,Kovacs:2020pns}, together with a modification of the puncture gauge extensively used in numerical relativity to simulate black hole binary mergers. With these modifications we have proved the well-posedness in the weak coupling regime for the Einstein-Gauss-Bonnet theory of gravity and the Four-Derivative Scalar-Tensor theory of gravity. It seems plausible that one can also extend the well-posedness results in
singularity avoiding coordinates to the general Horndeski and Lovelock theories - see the argument in \cite{Kovacs:2020ywu,Kovacs:2020pns}, which avoids the explicit computation of the eigenvalues and eigenvectors
of the theory.

We have implemented the equations of motion of the Four-Derivative Scalar-Tensor theory of gravity in \texttt{GRFolres}, an extension of our NR code \texttt{GRChombo}, which has recently been made publicly available in order to assist other NR groups working in modified gravity and whose capabilities and details are reviewed in Chapter \ref{C:GRFolres}. This has enabled us to study equal and non-equal mass Binary Black Hole mergers and extract its waveforms, which we present in Chapter \ref{C:BBHin4dST}. 
These studies have demonstrated the robustness of our formulation, which allows us to treat the theory fully non-linearly, thus avoiding the secular effects of order reduced formulations while capturing the non-perturbative physics.

We have studied both Type I and Type II couplings. For the first one, we have observed that for very small values of the couplings the differences of the waveforms with respect to GR only become
noticeable in the merger phase and they appear as a phase shift, while for larger values the system can radiate strongly in scalar waves and hence shed energy and
angular momentum more efficiently than in GR, so the larger the couplings, the sooner the binary merges. For the Type II coupling we have studied the so-called ``stealth-scalarisation''
effect where the scalar cloud arises due to the spin of the
remnant after merger. 
As in previous studies using alternative gauges we found that the scalarisation generically occurs too late after merger to impact on the tensor waveform. Too large values of the coupling – to accelerate the
growth of the scalar hair – result in a breakdown of the theory as it is pushed into the strongly coupled regime in which well-posedness is no longer assured. However,
we point out that this can be compensated in our chosen coupling function by tuning the values of the higher order interactions. Without such tuning, observation will rely
on detection of the scalar GWs that we show accompany the scalarisation post-merger.

In Chapter \ref{C:Hyp} we have investigated, for spin-induced scalarisation of EsGB rotating black holes, the threshold and dynamics of the loss of hyperbolicity, the influence of our
chosen gauge, and how these issues connect to the regime of validity of the model as an effective
field theory. We have found that the determinant of the effective metric turns negative just before the breakdown of the simulations, meaning that hyperbolicity is typically lost because of the purely gravitational physical modes and, thus, is not dependent on gauge choices. We have also examined the type of loss of hyperbolicity, finding that it
is of Tricomi-type for models close to the scalarisation threshold while the transition appears to change to a Keldysh-type or a mixture of both developing in different regions of the spacetime for models deep inside the strongly coupled regime. Finally, we have studied the weak coupling condition, which is violated when the theory exits the domain of validity of the EFT, i.e, when higher-derivative
corrections cannot be neglected.  Our results show that we were outside of the weak coupling regime in all those cases in which loss of hyperbolicity was observed.

\section{Further work}

The results that have been obtained in the thesis open up many possibilities for further research. 
The two most promising directions that we are pursuing are waveform modelling and cosmological spacetimes:

\begin{itemize}
    \item \textbf{Waveform modelling:} 
The studies in this thesis are highly timely, with demand for beyond GR waveforms driven by the large amounts of data that are now being
generated by the GW detector network LIGO-Virgo-KAGRA, and the anticipation of a further explosion in such data as new detectors come online in the future, including the space based mission LISA
that will launch in around 2035. Such data will open up new frequency ranges for study, and push the
limits of what can be detected to the earliest moments of the universe.

Whilst beyond GR effects are currently searched for in detector data, these are primarily tests based
on parameterised deviations, without any underlying specific model. Such tests have the significant
advantage of being fully general, but it is not yet clear that for specific, physically motivated models,
they would indeed capture the resulting changes in the signal, rather than, say, attributing them to the
system as having a larger mass ratio or different component spins. It is also not clear to what extent
they can recover model parameters in specific cases.

Testing this, with truly well modelled signals, is an extremely challenging task that required the development of significant new tools and techniques in NR. After the work carried out in this thesis, which represents a proof of principle of the methods that we have developed, we are now in a position to apply what we have developed to actual data pipelines.

Our main goal now is to provide waveform templates to be used to test the LIGO backward inference pipeline. To
discover and interpret hints of physics beyond the realms of GR requires a detailed forward model
that predicts the observable features of new fundamental physics. Whilst our simulations provide
exactly this forward model, the limitation is that they only generate single cases at a time in a vast
parameter space of modified theory parameters and couplings, and each case is costly to obtain at
the required quality. Therefore the construction of a full waveform template bank against which to
match signals, whilst a potential longer term goal, should wait until we obtain a better understanding
of the phenomenology of the signals, and can focus on regions of particular interest.

Instead, we focus on the more modest (but still cutting-edge) goal of testing the robustness of
the backward models that are used by groups like LIGO to identify and explain new observations \cite{Yunes:2013dva,Berti:2015itd}. A growing body of work has been developed in backward models for GWs \cite{LIGOScientific:2021sio}, from 
non-parameterised tests such as inspiral-merger ringdown tests that check consistency with GR to highly-tuned parameterised modelled methods that seek to identify variations. 
By providing example waveforms from our exact forward modelling, we can probe the ability of such pipelines to identify new
physics or quantify how it shifts the inference of the system parameters.

We plan to use our simulations to perform
the first end-to-end validation of current-generation tests of General Relativity. Adding the simulated
signals to real data observed by the LIGO-Virgo-KAGRA ground-based detectors we will then process them through the standard ``pipeline'', checking how and if they will be found by modelled and
unmodelled search algorithms, understanding any tell-tale signature the new physics imprints on inference methods that assume GR, and finally investigating the capability of tests of GR to identify
and characterise the new physics. By considering the holistic impact across the pipeline, we will be
able to predict the pipeline performance and identify areas where new ideas are needed to ensure
full sensitivity. In the event of a future detection of beyond-GR physics, the outcomes of this project
will uniquely help to validate and understand the results.

\item \textbf{Cosmological spacetimes:}

Whilst GW data drives the research, the tools that we have developed in NR for studying beyond GR
waveforms has applications in more fundamental aspects of gravitational theory. In such studies, NR
simulations provide a kind of ``experimental test'' of different theories, ruling out scenarios that are
incompatible with our observed universe or physics that we know. 

In the understandable excitement surrounding the detection of GWs produced by mergers of compact
objects, it can often be forgotten that there are other opportunities to test GR in the strong field
regime. Cosmology, and the early Universe in particular, also belong to the strong field regime of
Einstein’s theory of gravity. Therefore, modifications of GR, which are
expected to generically arise from effective theories of quantum gravity, could play a role in the
gravitational dynamics of the early Universe, when the energy scale of GR theory was surpassed.

Cosmic inflation is the leading paradigm for the early Universe, which explains the current homogeneous and spatially flat state. However, one important open question is how generic inflation is and
under which initial conditions it can occur. The problem of initial conditions for inflation has been extensively studied, but only recently fully non-perturbatively for initial configurations in the strong field
regime using NR \cite{Clough:2016ymm, Clough:2017efm,Aurrekoetxea:2019fhr}. No studies exist of the fully non-linear dynamics of gravity in the
early Universe beyond GR in the most general $4\partial$ST theory, which is well-motivated by effective field theory (EFT) whilst retaining second order equations
of motion.

In an ongoing project we are exploring the impact of theories of gravity with
higher derivative corrections on inflationary spacetimes in order to understand how these modifications could affect the start of inflation
and any observational consequences, extending the previous results in pure GR ,
which showed that three broad classes of inflationary potentials (``convex'', ``hilltop''
and ``plateau'') exemplify the possible inhomogeneous inflationary dynamics. In each case, our goal will be
to identify the consistent regime where the initial conditions and the couplings are such that inflation
can occur and the EFT remains in its regime of validity at all times. This will provide new information
for model building in inflation and quantum gravity. For cases that remain within the valid EFT regime,
another goal will be to study the dynamics and compare them with standard GR minimally coupled to a scalar
field.




\end{itemize}

There are many other exciting future directions, some of these that form part of currently ongoing projects are listed below:

\begin{itemize}

\item \textbf{Massive scalar-Gauss-Bonnet:}

One of the the main difficulties of testing the $4\partial$ST theory of gravity is that its degrees of freedom are not coupling constants but arbitrary functions. Therefore, some of the observational constraints that have been deduced for a shift-symmetric EsGB theory may become weaker or stronger when one considers other coupling functions or a mass term. The introduction of a scalar field mass leads to a suppression of the scalar field in black hole solutions and this can have interesting dynamics in Binary Black Hole systems.

\item \textbf{High dimensions:}

As we have motivated throughout this thesis, Einstein-Gauss-Bonnet gravity in high dimensions is especially relevant since it is an effective field theory of gravity and can arise in the low energy limit of string theory when the theory is compactified from 11D to 5D. Furthermore, black hole solutions are also particularly interesting and the well-posedness studies in this thesis lay the ground work for a numerical implementation. In a project led by Shunhui Yao and Pau Figueras, the equations of motions of the 5-dimensional Einstein-Gauss-Bonnet gravity are implemented using the Cartoon formalism (a dimensional reduction method which makes use of the symmetries of the problem) and applied to study the Gregory-Laflamme instability for black strings, which was recently studied in GR in \cite{Figueras:2022zkg}.

\item \textbf{Ricci-scalar-Gauss-Bonnet:}

On the other hand, a lower order term in the EsGB theory containing a coupling between the Ricci scalar and the scalar field was recently proposed in \cite{Thaalba:2023fmq}. This stabilises the scalar field in cases when a tachyonic instability in EsGB leads to ill-posedness of the equations. We are currently looking at 3+1 simulations in this context in order to study the resemblance of adding this additional Ricci coupling with changing the EsGB coupling function.

\item \textbf{Hydrodynamics:}

The numerical studies comprised in this thesis have all been obtained in the framework of black hole solutions. However, there are other compact objects, such as Neutron Stars (NS), which can lead to interesting dynamics in the context of modified gravity. In particular, Neutron Stars are expected to probe modifications of GR at smaller curvature scales than for BBHs and the presence of matter can generate additional
non-linear effects. The significant challenge here is that the matter in the star is traditionally modelled effectively as a fluid and that the star can carry a
magnetic field, so one needs to couple the existing equations of motion to magneto-hydrodynamics. This strongly motivates to extend our results in this context in order to be able to simulate both Binary Neutron Stars and mixed systems of BH-NS.

Therefore, on the one hand, we are implementing the equations of motion of the $4\partial$ST in our formalism in \texttt{MHDuet} through the platform \texttt{Simflowny}, which already contains all the necessary tools for hydrodynamics. On the other hand, we are currently working on implementing hydrodynamics in \texttt{GRChombo}, which will enable us to study astrophysical systems including matter sources in a similar way as other NR codes do.

\item \textbf{Anti de Sitter spacetimes:}

Finally, another topic of interest in NR is the study of asymptotically Anti de Sitter (AdS) spacetimes. The main motivation to understand AdS is the deep connection between gravity in AdS to certain conformal field theories (CFT), known as the AdS/CFT correspondence, which provides an important window into the real-time dynamics of strongly
interacting quantum field theories far from equilibrium. AdS with reflective boundary conditions plays the role of a box that naturally keeps propagating waves confined to its interior, where they are perpetually interacting. Thus, even the smallest perturbations in AdS can enter the strong field regime, where qualitatively new gravitational phenomena emerge.

Evolution in AdS is notoriously hard for several reasons. On the one hand, it is an initial-boundary value
which requires data to be prescribed
not only at an initial space-like hypersurface, but also
at spatial and null infinity which constitute the timelike
boundary of an asymptotically AdS spacetime. On the other hand, the most interesting phenomena involve spacetimes that
have very little or no symmetry and there is, as well, a variety of physical scales that must be adequately resolved to correctly
capture the relevant physics.

Recent works have used the Generalised Harmonic Coordinates approach in order to numerically study those spacetimes \cite{Bantilan:2012vu,Bantilan:2017kok,Bantilan:2020xas,Figueras:2023ihz}. We believe that an adaptation to the puncture gauge approach in a similar way as we have done for the modified harmonic gauge would improve the robustness of the formalism and definitely help in treating singularities. This could lead us to study critical collapse and the behaviour of black hole solutions in asymptotically AdS spacetimes among other topics of interest.

\end{itemize}

In summary, the new methods and techniques developed in this thesis open new avenues for the exploration of modified gravity theories and their effect on gravitational waves, which we believe that will improve our understanding of gravity.


\appendix 

\part{Extra material}

\addtocounter{page}{1}%
\blankpage

\renewcommand{\chaptername}{Appendix}
\chapter{Equations of motion in $d+1$ form}
\label{App:EoM}

In this appendix we present the equations of motion of the EGB and the $4\partial$ST theory in the modified puncture gauge coming respectively from the actions \eqref{eq:action_Lovelock} and \eqref{eq:4dST}. We neglect the potential and take the coupling functions to be $g_2(\phi)=g_2$ and $\lambda(\phi)=\frac{1}{4}\lambda^{\text{GB}}f(\phi)$.

\section{Einstein-Gauss-Bonnet gravity}\label{App:EoMEGB}

The tensor ${\mathcal H}_{\mu\nu}$ defined in \eqref{eq:Hmunu_EGB} that appears on the r.h.s. of the equations of motion that result from varying the action \eqref{eq:action_Lovelock} with respect to the metric plays the role of an effective stress-energy tensor. Therefore, its $d+1$ decomposition gives\footnote{The signs have been chosen so that the quantities in the $d+1$ decomposition of $\mathcal{H}_{\mu\nu}$ enter the equations of motion with the same signs as the analogous quantities for a standard stress-energy tensor.}
\begin{subequations}
    \begin{align}
        \kappa\,\rho =&~n^\mu\,n^\nu\,\mathcal{H}_{\mu\nu}\,,\\
        \kappa\,J_i =&-n^{\mu}\gamma_i^{\phantom{i}\nu}{\mathcal H}_{\mu\nu}\,,\\
        \kappa\,S_{ij} = &~\gamma_i^{\phantom{i}\mu}\gamma_j^{\phantom{j}\nu}\mathcal{H}_{\mu\nu}\,,
    \end{align}
\end{subequations}
where 
\begin{subequations}\label{egbcomp}
\begin{align}
\kappa\,\rho=&-\tfrac{\lambda^{\text{GB}}}{2}\big(M^2-4M_{ij}M^{ij}+M_{ijkl}M^{ijkl} \big), \\
\kappa\,J_i=&-2\lambda^{\text{GB}}\big(MN_i-2M_i^{~j}N_j+2M^{jk}N_{ijk} -M_i^{~ljk}N_{jkl} \big)\,, \\
\kappa\,S_{ij}=&~ 2\lambda^{\text{GB}}\Big[4M^k_{(i}F_{j)k}+2M_{i~j}^{~k~l}F_{kl}-MF_{ij}-2M_{ij}F+ 2N_iN_j\nonumber
        \\ &  -4N^kN_{k(ij)}- N_{kli}N^{kl}_{~~j}-2N_{ikl}N_j^{~kl}+ MM_{ij} + M_{iklm}M_j^{~klm}\nonumber
        \\&-2\big(M_{ik}M^k_{~j}+M^{kl}M_{ikjl}\big) +\gamma_{ij}\Big(MF-2M^{kl}F_{kl}+N_{klm}N^{klm}\nonumber
        \\&\hspace{2cm}-2N_kN^k-\tfrac{1}{4}\big(M^2-4M_{kl}M^{kl}+M_{klmn}M^{klmn}\big) \Big)\Big]\,,
\end{align}
\end{subequations}
with
\begin{subequations}
\begin{align}\label{eq:MijNi}
M_{ijkl}=&~R_{ijkl}+K_{ik}K_{jl}-K_{il}K_{jk}\,, \\
N_{ijk}=&~D_iK_{jk}-D_{j}K_{ik}\,, \\
F_{ij}=&~{\mathcal L}_nK_{ij}+\frac{D_iD_j\alpha}{\alpha}+K_{ik}K_j^{~k}\,,
\end{align}
\end{subequations}
where $M_{ij}=\gamma^{kl}M_{ikjl}$, $M=\gamma^{ij}M_{ij}$ is the GR Hamiltonian constraint and $N_i=\gamma^{jk}N_{jik}$ is the GR momentum constraint.

The $d+1$ equations are obtained by inserting the above quantities in Eqs. \eqref{eqsccz4} except for $\tilde{A}_{ij}$ and $K$, whose evolution equations are given by the following coupled system,
\begin{align}\label{mat_egb}
\begin{pmatrix} X^{kl}_{ij} & Y_{ij} \\ X^{kl}_K & Y_K \end{pmatrix}
\begin{pmatrix} \partial_t \tilde{A}_{kl} \\ \partial_tK  \end{pmatrix}=
\begin{pmatrix} Z_{ij}^{\tilde{A}} \\ Z^K  \end{pmatrix}\,,
\end{align}
where the elements of the matrix are
\begin{subequations}
\begin{align}
X_{ij}^{kl}=&~\gamma_i^{~k}\gamma_j^{~l}+2\lambda^{\text{GB}}\big[M\gamma_i^{~k}\gamma_j^{~l}+\tfrac{6}{d}\gamma_{ij}M^{kl}-2\big(M_{i~j}^{~k~l}+2M_{(i}^{~k}\gamma_{j)}^{~l} \big) \big]\,, \\
X_K^{kl}=&-\tfrac{4(d-3)}{(d-1)\chi}\lambda^{\text{GB}}M^{kl}\,, \\
Y_{ij}=&~\tfrac{4(d-3)}{d}\lambda^{\text{GB}}\chi\Big(M_{ij}-\tfrac{1}{d}\gamma_{ij}M\Big)\,, \\
Y_K =&~1+\tfrac{2(d-2)(d-3)}{d(d-1)}\lambda^{\text{GB}}M\,,
\end{align}
\end{subequations}
whereas the r.h.s. terms are
\begin{subequations}
\begin{align}
Z_{ij}^{\tilde{A}}=&~{\mathcal L}_{\beta}\tilde{A}_{ij}-2\alpha\tilde{A}_{il}\tilde{A}^l_{~j}+ \chi\big[\alpha\big(R_{ij} + 2D_{(i}Z_{j)} -\kappa\,\bar{S}_{ij}\big)-D_iD_j\alpha\big]^{\text{TF}}  \nonumber\\&+\alpha\tilde{A}_{ij}(K-2\Theta) - \tfrac{2}{d}\partial_k\beta^k\tilde{A}_{ij}\,, \\
Z^K=&~\beta^i\partial_iK-D^iD_i\alpha +\alpha\left[R+2\,D_iZ^i +K(K-2\Theta)\right]-d\,\kappa_1(1+\kappa_2)\,\alpha\,\Theta \nonumber\\
&+\tfrac{\kappa\,\alpha}{d-1}\big(\bar{S}-d\rho\big)-\frac{d\,\alpha\,b(x)}{2(d-1)(1+b(x))}\Big[R-\tilde{A}_{ij}\tilde{A}^{ij}+\frac{d-1}{d}K^2\nonumber\\
&\hspace{5.5cm} -(d-1)\kappa_1(2+\kappa_2)\,\Theta -2\,\kappa\,\rho\Big]\,,
\end{align}
\end{subequations}
with $\bar{S}_{ij}$ and $\bar{S}$, which are obtained by subtracting the time derivatives of $\tilde{A}_{ij}$ and $K$ from $S_{ij}$ (which are in turn computed from Eq. \eqref{mat_egb}), given by
\begin{subequations}
\begin{align}
\kappa\,\bar{S}_{ij}^{\text{TF}}=&~\lambda^{\text{GB}}\Big\{\tfrac{8}{\chi}M_{(i}^{~k}\hat{\mathcal O}_{j)k}+\tfrac{4}{\chi}M_{i~j}^{~k~l}\hat{\mathcal O}_{kl}-4\hat{\mathcal O} M_{ij}-\tfrac{2}{\chi}M\hat{\mathcal O}_{ij}\nonumber\\
&\hspace{0.6cm}+\tfrac{12}{d}\tfrac{\tilde{\gamma}_{ij}}{\chi}\Big(\tfrac{M\hat{\mathcal O}}{2}-\tfrac{M^{kl}\hat{\mathcal O}_{kl}}{\chi}+N_kN^k-\tfrac{N_{klm}N^{klm}}{2} \Big)\Big\}-\tfrac{4\,\kappa}{d}\tfrac{\tilde{\gamma}_{ij}}{\chi}\rho\nonumber\\
&\hspace{-0.5cm}-2\lambda^{\text{GB}}\big[M_{iklm}M_ j^{~klm}-2(M_{ik}M^k_{~j}+M^{kl}M_{ikjl})+MM_{ij}\nonumber\\&\hspace{0.7cm}+2\big(N_iN_j-2N^kN_{k(ij)} -\tfrac{1}{2}N_{kli}N^{kl}_{~~j}-N_{ikl}N_j^{~kl}\big)\big] \,, \\
\kappa\,\bar{S}=&~ 2\lambda^{\text{GB}}(d-3)\Big(M\hat{\mathcal O} - \tfrac{2}{\chi}M^{ij}\hat{\mathcal O}_{ij}+2N_iN^i-N_{ijk}N^{ijk} \Big)+(4-d)\,\kappa\,\rho\,,
\end{align}
\end{subequations}
where $\hat{\mathcal O}_{ij}=\frac{1}{\alpha}{\mathcal L}_{\beta}\tilde{A}_{ij}-\tilde{A}_{ik}\tilde{A}_j^{~k}+\frac{2}{d}K\tilde{A}_{ij}+\tilde{\gamma}_{ij}\left(\frac{K}{d}\right)^2-\frac{\chi}{\alpha} D_iD_j\alpha-\frac{2}{d}\tilde{A}_{ij}\frac{\partial_k\beta^k}{\alpha}+\frac{\tilde{\gamma}_{ij}}{d}\frac{1}{\alpha}{\mathcal L}_{\beta}K$ and $\hat{\mathcal O}=\frac{1}{\alpha}{\mathcal L}_{\beta}K+\tilde{A}_{ij}\tilde{A}^{ij}+\frac{K^2}{d}-\frac{1}{\alpha}D_kD^k\alpha$\,.

\section{Four-Derivative Scalar-Tensor theory of gravity}\label{App:EoMEsGB}

We start writing the $3+1$ decomposition of $T_{\mu\nu}^X$ appearing in Eq. \eqref{eq:X_tensor},
\begin{subequations}
\begin{align}
    \rho^X=&~\tfrac{1}{8}\big(K_{\phi}^2-(\partial\phi)^2\big)\big(3K_{\phi}^2+(\partial\phi)^2\big)\,,\\
    J_i^X=&~\tfrac{1}{2}K_{\phi}\partial_i\phi\big(K_{\phi}^2-(\partial\phi)^2\big)\,,\\
    S_{ij}^X=&~\tfrac{1}{2}\big(K_{\phi}^2-(\partial\phi)^2\big)\Big[(D_i\phi) D_j\phi+\tfrac{1}{4}\gamma_{ij}\big(K_{\phi}^2-(\partial\phi)^2\big)\Big]\,,
\end{align}
\end{subequations}
and from ${\mathcal H}_{\mu\nu}$ in \eqref{eq:H_tensor} as well,
\begin{subequations}\label{edgbcomp}
\begin{align}
\rho^{\text{GB}}=&~\tfrac{\Omega M}{2} - M_{kl}\Omega^{kl}, \, \\
J^{\text{GB}}_i=&~\tfrac{\Omega_iM}{2}-M_{ij}\Omega^j - 2\Big(\Omega^j_ {~[i}N_{j]}-\Omega^{jk}D_{[i}K_{j]k}\Big),\, \\
S^{\text{GB}}_{ij}=&-\tfrac{\Omega}{3}\Big({\mathcal L}_nA_{ij}+\tfrac{1}{\alpha}(D_iD_j\alpha)^{\text{TF}}+A_{im}A^m_{~j} \Big)-\Omega_{nn}M_{ij}+N_{(i}\Omega_{j)}-2\epsilon_{(i}^{~kl}B_{j)k}\Omega_l\nonumber\\
&+\gamma_{ij}\Big[\rho^{\text{GB}}-N^k\Omega_k+\tfrac{M}{6}\Big(\Omega_{nn}+\tfrac{\Omega}{3}\Big)-\Omega^{\text{TF},kl}\Big({\mathcal L}_nA_{kl}+\tfrac{1}{\alpha}(D_kD_l\alpha)^{\text{TF}}+A_{km}A^m_{~l}\Big)\nonumber\\
&\hspace{1cm}-\tfrac{1}{3}\Omega^{\text{TF},kl}M_{kl}+\tfrac{2\Omega}{9}\Big({\mathcal L}_nK+\tfrac{D^kD_k\alpha}{\alpha}-\tfrac{3}{2}A_{kl}A^{kl}-\tfrac{K^2}{3}\Big)\Big]\nonumber\\
&+2\gamma^k_{~(i}\Omega_{j)}^{\text{TF},l}\Big({\mathcal L}_nA_{kl}+\tfrac{1}{\alpha}(D_kD_l\alpha)^{\text{TF}}+A_{km}A^m_{~l}\Big)\nonumber\\
&-\Omega_{ij}^{\text{TF}}\Big({\mathcal L}_nK+\tfrac{1}{\alpha}D^kD_k\alpha-3A_{kl}A^{kl}-\tfrac{K^2}{3}\Big)\,,
\end{align}
\end{subequations}
with
\begin{subequations}
\begin{align}\label{MNeq}
M_{ij}=&~R_{ij}+\tfrac{1}{\chi}\left(\tfrac{2}{9}\tilde{\gamma}_{ij}K^2+\tfrac{1}{3}K\tilde{A}_{ij}-\tilde{A}_{ik}\tilde{A}_j^{~k} \right)\,, \\
N_i=&~\tilde{D}_j\tilde{A}_i^{~j}-\tfrac{3}{2\chi}\tilde{A}_i^{~j}\partial_j\chi-\tfrac{2}{3}\partial_iK\,, \\
B_{ij}=&~\epsilon_{(i}^{~kl}D_kA_{j)l}\,,\\
\Omega_i\equiv&-\tfrac{4}{\lambda^{\text{GB}}}\,\gamma_{~i}^{\mu}n^{\nu}{\mathcal C}_{\mu\nu}=f'\big(\partial_iK_{\phi}-\tilde{A}^j_{~i}\partial_j\phi-\tfrac{K}{3}\partial_i\phi \big)+f''K_{\phi}\partial_i\phi\,,\\
\Omega_{ij}\equiv&~\tfrac{4}{\lambda^{\text{GB}}}\,\gamma_{~i}^{\mu}\gamma_{j}^{~\nu}{\mathcal C}_{\mu\nu}=f'\left(D_iD_j\phi-K_{\phi}K_{ij}\right)+f''(\partial_i\phi) \partial_j\phi\,,\\
\Omega_{nn}\equiv&~\tfrac{4}{\lambda^{\text{GB}}}\,n^{\mu}n^{\nu}{\mathcal C}_{\mu\nu}=f''K_{\phi}^2-\tfrac{f'}{\alpha}D^k\alpha D_k\phi-\tfrac{f'}{\alpha}\partial_{\perp}K_{\phi}\,,
\end{align}
\end{subequations}
where $M_{ij}$ and $N_i$ are as defined in \eqref{eq:MijNi}, $B_{ij}$ is the magnetic part of the Weyl tensor and $\Omega_i$, $\Omega_{ij}$ and $\Omega_{nn}$ come from the $3+1$ decomposition of ${\mathcal C}_{\mu\nu}$ in Eq. \eqref{Cmunu}. In addition, we have 
\begin{subequations}
\begin{align}
    M^{\text{TF}}_{ij} &\equiv M_{ij}-\tfrac{1}{3}\gamma_{ij}M\,,\\
    \Omega^{\text{TF}}_{ij} &\equiv \Omega_{ij}-\tfrac{1}{3}\gamma_{ij}\Omega\,,
\end{align}
\end{subequations}
where $M=\gamma^{kl}M_{kl}$ and $\Omega=\gamma^{kl}\Omega_{kl}$. 

So, using that 
\begin{subequations}
\begin{align}
\kappa\,\rho=&~\tfrac{1}{2}\rho^{\phi}+g_2\rho^X+\lambda^{\text{GB}}\rho^{\text{GB}}\,,\\ \kappa\,J_i=&~\tfrac{1}{2}J_i^{\phi}+g_2J_i^X+\lambda^{\text{GB}}J_i^{\text{GB}}\,,\\
\kappa\,\bar{S}_{ij}=&~\tfrac{1}{2}S_{ij}^{\phi}+g_2S_{ij}^X+\lambda^{\text{GB}}\bar{S}_{ij}^{\text{GB}}\,,
\end{align}
\end{subequations}
where the bar denotes again that the terms depending on the time derivatives of ${\tilde{A}}_{ij}$ and $K$ are substracted since they are taken into account in the matrix on the l.h.s. in Eq. \eqref{mat_esgb}, we can obtain the equations of motion in the $3+1$ form by replacing those quantities in Eqs. \eqref{eqsccz4} and \eqref{eq:scalar_eqs} with $d=3$, except for $K$, $\tilde{A}_{ij}$ and $K_{\phi}$, which satisfy the following system of coupled partial differential equations:
\begin{align}\label{mat_esgb}
\begin{pmatrix} X^{kl}_{ij} & Y_{ij} & 0 \\ X^{kl}_K & Y_K & 0 \\ X^{kl}_{K_{\phi}} & Y_{K_{\phi}} & I \end{pmatrix}
\begin{pmatrix} \partial_t \tilde{A}_{kl} \\ \partial_tK \\ \partial_tK_{\phi} \end{pmatrix}=
\begin{pmatrix} Z_{ij}^{\tilde{A}} \\ Z^K \\ Z^{K_{\phi}} \end{pmatrix},
\end{align}
where the elements of the matrix are defined as follows,
\begin{subequations}
\begin{align}
X_{ij}^{kl}=&~ \gamma_i^k\gamma_j^l\,\Big(1-\tfrac{\lambda^{\text{GB}}}{3}\Omega \Big)+2\lambda^{\text{GB}}\Big(\gamma_{(i}^k\Omega_{j)}^{\text{TF},l}
-\tfrac{\gamma_{ij}}{3}\Omega^{\text{TF},kl}-\tfrac{\lambda^{\text{GB}}}{\Sigma} 
{f'}^2 M_{ij}^{\text{TF}} M^{\text{TF},kl}\Big)\,,\\
X_K^{kl}=&~\tfrac{\lambda^{\text{GB}}}{2\chi}\Big(\Omega^{\text{TF},kl}-\tfrac{\lambda^{\text{GB}}}{\Sigma}{f'}^2M\,M^{\text{TF},kl}\Big)\,, \\
X_{K_{\phi}}^{kl}=&~\tfrac{\lambda^{\text{GB}}}{2\chi}f'M^{\text{TF},kl}\,, \\
Y_{ij}=&~\tfrac{\lambda^{\text{GB}}}{3}\chi\Big(-\Omega^{\text{TF}}_{ij}+\tfrac{\lambda^{\text{GB}}}{\Sigma}{f'}^2M\,M^{\text{TF}}_{ij} \Big)\,, \\
Y_K=&~1+\tfrac{\lambda^{\text{GB}}}{3}\Big(-\Omega+\tfrac{\lambda^{\text{GB}}}{4\Sigma}{f'}^2M^2 \Big)\,,\\
Y_{K_{\phi}}=&-\tfrac{\lambda^{\text{GB}}}{12}f'M\,, \\
I=&~\Sigma\,,
\end{align}
\end{subequations}
where $\Sigma=1+g_2(3K_{\phi}^2-(\partial\phi)^2)$, while the terms of the r.h.s. are
\begin{subequations}
\begin{align}
Z_{ij}^{\tilde{A}}=&~\chi\big[-D_iD_j\alpha+\alpha\left(R_{ij} + 2D_{(i}Z_{j)} -\kappa\,\bar{S}_{ij}\right) \big]^{\text{TF}}+\alpha\left[\tilde{A}_{ij}(K-2\Theta)-2\tilde{A}_{il}\tilde{A}^l_{~j}\right]\nonumber\\
&+\beta^k\partial_k\tilde{A}_{ij}+2\,\tilde A_{k(i}\partial_{j)}\beta^k-\tfrac{2}{3}\tilde{A}_{ij}(\partial_k\beta^k)\,,\\
Z^K=&~\beta^i\partial_iK +\alpha\left[R+2\,D_iZ^i +K(K-2\Theta)\right] -3\,\kappa_1(1+\kappa_2)\,\alpha\,\Theta+\tfrac{\kappa\,\alpha}{2}(\bar{S}-3\rho)\nonumber\\
&-D^iD_i\alpha-\frac{3\,\alpha\,b(x)}{4(1+b(x))}\Big[R-\tilde{A}_{ij}\tilde{A}^{ij}+\frac{2}{3}K^2 -2\kappa_1(2+\kappa_2)\,\Theta -2\,\kappa\,\rho\Big]\,,\\
Z^{K_{\phi}}=&~ \Sigma\,\big[\beta^i\partial_i K_\phi+\alpha\,\big(-D^iD_i\phi+K\,K_\phi\big)-(D^i\phi)D_i\alpha\big]
+\alpha\,g_2\,Z^{g_2}-\tfrac{\lambda^{\text{GB}}}{4}\,\alpha\,f'\,\bar{\mathcal L}^{\text{GB}}\,,
\end{align}
\end{subequations}
where
\begin{align}
    Z^{g_2}=&~2K_{\phi}^2(D^iD_i\phi-KK_{\phi})+2 D^i\phi \left[(D^j\phi) D_iD_j\phi\right.\nonumber \\&\left.
    -K_{\phi}\big(2D_iK_{\phi}-D^j\phi\tfrac{1}{\chi}\tilde{A}_{ij}-\tfrac{1}{3}KD_i\phi\big)\right]\,,
\end{align}
with $\bar{\mathcal L}^{\text{GB}}$ also denoting that we are subtracting the terms with time derivatives, which are take into account in the elements of the matrix above. Finally the expression of these remaining quantities yields
\begin{subequations}
\begin{align}
\bar{S}_{ij}^{\text{GB,TF}}=&-\tfrac{1}{3}\left(\Omega^{TF}_{ij}-\tfrac{\lambda^{\text{GB}}}{\Sigma}{f'}^2MM^{\text{TF}}_{ij}\right)\Big[-\tfrac{1}{\alpha}\beta^i\partial_iK
+\tfrac{1}{\alpha}D_iD^i\alpha-\tilde{A}_{kl}\tilde{A}^{kl}-\tfrac{K^2}{3} \Big]\nonumber\\
&-2\,\left(\tfrac{1}{3}\,\gamma_{ij}\,\Omega^{\text{TF},kl}+\tfrac{\lambda^{\text{GB}}}{\Sigma}{f'}^2M_{ij}^{\text{TF}}M^{\text{TF},kl} \right)\left[\tfrac{1}{\alpha}D_kD_l\alpha
+\tfrac{1}{\chi}\big(\tilde{A}_{km}\tilde{A}^m_{~l}-\hat{\Theta}_{kl}\big) \right]\nonumber\\
&-\tfrac{1}{3}\,\Omega\left[\tfrac{1}{\alpha}D_iD_j\alpha+\tfrac{1}{\chi}\big(\tilde{A}_{im}\tilde{A}^m_{~j}-\hat{\Theta}_{ij}\big) \right]^{\text{TF}} -\tfrac{2}{3}\,\Omega_{ij}^{\text{TF}}\left(\tfrac{1}{\alpha}D_kD^k\alpha-\tilde{A}_{kl}\tilde{A}^{kl} \right)\nonumber\\
&+2\,\Omega_{(i}^{\text{TF},k}\left[\tfrac{1}{\alpha}D_{j)}D_k\alpha+\tfrac{1}{\chi}\big(\tilde{A}_{j)m}\tilde{A}_k^{~m}-\hat{\Theta}_{j)k}\big) \right]+\left[N_{(i}\Omega_{j)}\right]^{\text{TF}} \nonumber\\
&-M_{ij}^{TF}\left[\Omega+f''(K_{\phi}^2-(\partial\phi)^2)-\tfrac{g_2}{\Sigma}f'Z^{g_2}-\tfrac{\lambda^{\text{GB}}}{\Sigma}{f'}^2H \right]\nonumber\\
&-2\left(D_kA_{ij}-D_{(i}A_{j)k}\right)\Omega^k-\gamma_ {ij}\,(D^kA_{kl})\,\Omega^l+\Omega_{(i}D^kA_{j)k} \,,\\
\bar{S}^{\text{GB}}=&~\tfrac{2}{3}\left(\Omega-\tfrac{\lambda^{\text{GB}}}{4\Sigma}{f'}^2M^2\right)\left[-\tfrac{1}{\alpha}\beta^i\partial_iK+\tfrac{1}{\alpha}D_iD^i\alpha-\tilde{A}_{ij}\tilde{A}^{ij}-\tfrac{K^2}{3} \right]\nonumber\\
&+2\, M\left(\tfrac{1}{4}\,f''(K_{\phi}^2-(\partial\phi)^2)-\tfrac{g_2}{4\Sigma}f'Z^{g_2}-\tfrac{\lambda^{\text{GB}}}{4\Sigma}{f'}^2H+\tfrac{1}{3}\,\Omega\right)\nonumber\\
&+\big(\Omega^{\text{TF},kl}-\tfrac{\lambda^{\text{GB}}}{\Sigma}{f'}^2MM^{\text{TF},kl}\big)\left(\tfrac{1}{\alpha}D_kD_l\alpha+\tfrac{1}{\chi}\tilde{A}_{km}\tilde{A}_{~l}^m -\tfrac{\hat{\Theta}_{kl}}{\chi} \right)\nonumber\\
&-2\Omega^iN_i-\Omega^{\text{TF},ij}\,M^{\text{TF}}_{ij} -\rho^{\text{GB}}\,,  \\
\bar{\mathcal L}^{\text{GB}}=&-\tfrac{4}{3}\,M\left[-\tfrac{1}{\alpha}\beta^i\partial_iK + \tfrac{1}{\alpha}D_iD^i\alpha-\tilde{A}_{ij}\tilde{A}^{ij}-\tfrac{K^2}{3} \right] -4\,H\nonumber\\
&+8\,M^{\text{TF},kl}\left[\tfrac{1}{\alpha}D_kD_l\alpha+\tfrac{1}{\chi}\left(\tilde{A}_{kj}\tilde{A}^j_{~l}-\hat{\Theta}_{kl}\right) \right]\, ,
\end{align}
\end{subequations}
where we have used $\hat{\Theta}_{kl}=\tfrac{1}{\alpha}{\mathcal L}_{\beta}\tilde{A}_{kl}+\tfrac{2}{3}\left(K-\tfrac{1}{\alpha}\partial_i\beta^i \right)\tilde{A}_{kl}$ with ${\mathcal L}_{\beta}\tilde{A}_{ij}=\beta^k\partial_k \tilde A_{ij}+2\tilde A_{k(i}\partial_{j)}\beta^k$ and 
\begin{equation}
    \begin{aligned}
    H=~2\,B_{ij}B^{ij}-N_i&N^i 
    =
    -\tfrac{4}{3}D_iK\big(N^i+\tfrac{D^iK}{3}\big)+2\,D_iA_{jk}\big(D^iA^{jk}-D^jA^{ik} \big)-2N_iN^i\,.
    \end{aligned}
\end{equation}

\addtocounter{page}{1}%
\blankpage

\chapter{Eigenvectors of the Einstein-scalar-field principal part}
\label{app:eigenvectors}

In this appendix we display the expression of the eigenvectors of the Einstein-scalar-field principal part in the modified CCZ4 gauge in $d+1$ spacetime dimensions in Tables \ref{tab:phys}, \ref{tab:viol} and \ref{tab:pure}, corresponding respectively to the physical, ``gauge-condition violating'' and ``pure-gauge'' categories:

\begin{table}[ht]
\centering
\makebox[0pt][c]{\parbox{1.0\textwidth}{%
    \begin{minipage}[c]{0.42\hsize}\centering\small
    
    \scalebox{0.75}{\begin{tabular}{c|c|c|c c}
        $\hat{\tilde{\gamma}}_{ij}$ &  $\hat{\tilde{A}}_{ij}$ & $\hat{\phi}$ & $\hat{K}_{\phi}$  \\
        \cline{1-4}
        0 & 0 & $\mp1$ & 1 \\
        $\mp2e_A^ie_B^j$ & $e_A^ie_B^j$ & 0 & 0  \\
        \vspace{-0.3cm}
        & & & & \hspace{-0.3cm}$\forall A\neq B$ \\
            \vspace{-0.2cm}
        $\pm2(e_A^ie_A^j-e_B^ie_B^j)$ & $-e_A^ie_A^j+e_B^ie_B^j$ & 0 & 0 
    \end{tabular}}
    \caption[Physical eigenvectors of the Einstein-scalar-field principal part.]{Physical eigenvectors of the Einstein-scalar-field principal part.}
    \label{tab:phys}
    \end{minipage}
    \hfill
    \begin{minipage}[c]{0.52\hsize}\centering\small
        \scalebox{0.75}{
        \begin{tabular}{c|c|c|c}
        $\hat{\tilde{\gamma}}_{ij}$ & $\hat{\chi}$ & $\hat{\beta}^i$ & $\hat{\alpha}$  \\
        $\hat{\tilde{A}}_{ij}$ & $\hat{K}$ & $\hat{\hat{\Gamma}}^i$  \\
        \hline
        $-\frac{\chi^2}{d-1}e_A^ie_B^i\delta^{AB}$ & $-\frac{\chi^2}{d-1}$ &  $\pm\frac{d\sqrt{\chi}}{2(d-1)\sqrt{1+a(x)}}\xi_i$ & 0 \\
        0 & 0 & $\xi_i$\\
        \hline
        $\chi^2e_A^i\xi^j$ & 0 & $\pm\sqrt{\frac{d\chi}{2(d-1)(1+a(x))}}e^A_i$ & 0\\
        0 & 0 & $e^A_i$ \\
        \hline
         $-\frac{\chi^2}{d-1}e_A^ie_B^j\delta^{AB}$ & $-\frac{\chi^2}{d-1}$ & $\pm\frac{d\sqrt{2}\left(1+a(x)(1-2\alpha\chi)\right)}{4(d-1)\sqrt{\alpha(1+a(x))}}\xi_i$ & $\frac{d(-1+2\alpha\chi)}{2(d-1)\alpha}$ \\
         $\mp\sigma e_A^ie_B^j\delta^{AB}$  & $\pm \tfrac{d\sigma}{\chi}$  & $\xi_i$ 
    \end{tabular}}
    \caption[``Gauge-condition violating'' eigenvectors of the Einstein-scalar-field principal part.]{``Gauge-condition violating'' eigenvectors of the Einstein-scalar-field principal part, with $\sigma=\frac{\chi\sqrt{2(1+a(x))}(1-2\alpha\chi)}{4(d-1)\alpha^{3/2}}$.}
    \label{tab:viol}
    \end{minipage}%
}}
\end{table}

    \begin{table}[H]
    \centering
    \small
    \scalebox{0.77}{
    \begin{tabular}{c|c|c}
        $\hat{\tilde{\gamma}}_{ij}$ & $\hat{\chi}$ & $\hat{\beta}^i$ \\ 
        $\hat{\tilde{A}}_{ij}$ &$\hat{K}$ & $\hat{\hat{\Gamma}}^i$ \\
        $\hat{\Theta}$ & $\hat{\alpha}$ \\
        \hline
       $\frac{\chi(1+b(x))}{b(x)}\left(-\frac{d}{2(d-1)(1+a(x))\alpha^2}+\chi \right)e_A^i\xi_j$ & 0 & $\pm\frac{d\sqrt{1+b(x)}}{2(d-1)\alpha(1+a(x))}e^A_i$ \\ $\pm\frac{\chi\sqrt{1+b(x)}}{2b(x)\alpha^2}\left(\frac{d}{2(d-1)}\frac{1+b(x)}{1+a(x)}-\alpha^2\chi\right)e_A^i\xi^j$ & 0 & $e^A_i$ \\
       0 & 0 \\
       \hline
        $-\frac{\chi^2(1+b(x))}{d-2+2(d-1)b(x)}\left(\frac{d\left(2-\alpha(1+a(x))\right)}{4\alpha^2\chi a(x)b(x)}+d-1\right)e_A^ie_B^j\delta^{AB}$ & $\chi\frac{1+b(x)}{2a(x)\alpha^2}\frac{4\left(d-(d-2)a(x)\alpha^2\chi\right)-\frac{d^2}{d-1}\alpha\frac{1+a(x)}{1+b(x)}}{(d-2)(d-2+2(d-1)b(x))}$ & $\frac{d}{2(d-1)a(x)\alpha}\xi_i$  \\
         $\pm\chi^2\frac{\sqrt{1+b(x)}}{d-2+2(d-1)b(x)}\left(\frac{d\left(2-\alpha(1+a(x))\right)}{4(d-1)\alpha^2\chi a(x)b(x)}+1 \right)e_A^ie_B^j\delta^{AB}$  &  $\pm\frac{d\sqrt{1+b(x)}}{4a(x)\alpha^2}\frac{4(d-(d-2)a(x)\alpha^2\chi)-\frac{d^2}{d-1}\alpha\frac{1+a(x)}{1+b(x)}}{(d-2)(d-2+2(d-1)b(x))}$ & $\xi_i$ \\
       $\pm\frac{d\sqrt{1+b(x)}}{4a(x)\alpha^2}\frac{2(d-(d-2)a(x)\alpha^2\chi)-\alpha(2-(d-2)b(x))\frac{1+a(x)}{1+b(x)}}{(d-2)(d-2+2(d-1)b(x))}$ & 0 \\
        \hline
        $\pm\frac{\chi\sqrt{1+b(x)}}{a(x)b(x)\alpha}\frac{2(1+a(x)(1+b(x)))-\alpha(1+a(x))^2}{d-2+2(d-1)b(x)}e_A^ie_B^j\delta^{AB}$ & $\mp\frac{\sqrt{1+b(x)}\chi}{a(x)\alpha}\frac{2(2(d-1)+da(x))-d\alpha\frac{(1+a(x))^2}{1+b(x)}}{(d-2)(d-2+2(d-1)b(x))}$ & $\xi_i$\\ $-\chi\frac{1+a(x)\left(1+2\frac{d-1}{d}b(x)(1+b(x))\right)-\frac{\alpha}{2}(1+a(x))^2}{a(x)b(x)(d-2+2(d-1)b(x))\alpha}e_A^ie_B^j\delta^{AB}$ & $\frac{4(d-1)\left(a(x)(1+b(x))-\frac{d}{d-2}(1+a(x))\right)+d^2\alpha\frac{(1+a(x)^2}{1+b(x)}}{2\alpha a(x)(d-2+2(d-1)b(x))}$ & 0 \\
        $\frac{d-1}{2a(x)\alpha}\frac{\frac{(1+a(x))^2}{1+b(x)}(2-(d-2)b(x))\alpha-2\left(d+a(x)(2-(d-2)b(x))\right)}{(d-2)(d-2+2(d-1)b(x))}$ & $\mp\frac{1+a(x)}{\alpha\sqrt{1+b(x)}}$
    \end{tabular}}
    \caption[{``Pure-gauge'' eigenvectors of the Einstein-scalar-field principal part.}]{``Pure-gauge'' eigenvectors of the Einstein-scalar-field principal part.}
    \label{tab:pure}
\end{table}

\addtocounter{page}{1}%
\blankpage

\chapter{Propagation of the constraints}\label{app:constraints}

Below we consider the propagation of the constraints in the modified puncture formulation in $d+1$ spacetime dimensions. Let the Hamiltonian and momentum constraints be denoted by  ${\mathcal H}$ and ${\mathcal M}_i$ respectively. Then, we find that the constraints obey the following evolution equations: 
\begin{subequations}\label{eq_constraints}
\begin{align}
\partial_{\perp}{\mathcal H}=&~\tfrac{2+b(x)}{1+b(x)}\alpha\,K\,{\mathcal H}-\tfrac{2}{\alpha}D^i\big(\alpha^2{\mathcal M}_i\big)+4\alpha\big(K\gamma^{ij}-K^{ij}\big)\big(D_iZ_j-\Theta\,K_{ij}\big)\nonumber\\
&-\tfrac{2(d-1)}{1+b(x)}\kappa_1\alpha\Big[1+\tfrac{\kappa_2}{2}(2+b(x))\Big]K\Theta\,,\\
\partial_{\perp}{\mathcal M}_i=&~\alpha\,K\,{\mathcal M}_i-\tfrac{1}{2\alpha}D_i(\alpha^2{\mathcal H})-2D^j\big[\alpha\big(D^kZ_k\gamma_{ij}-D_{(i}Z_{j)}+\Theta(K_{ij}-K\gamma_{ij})\big)\big]\nonumber\\
&+\tfrac{b(x)}{2(1+b(x))}D_i(\alpha{\mathcal H})+\tfrac{d-1}{1+b(x)}\kappa_1\Big[1+\tfrac{\kappa_2}{2}(2+b(x))\Big]D_i(\alpha\Theta)\,,\\
\partial_{\perp}\Theta =& \tfrac{\alpha}{2(1+b(x))}{\mathcal H}+\alpha(D_iZ^i-K\Theta)-
Z^iD_i\alpha\nonumber\\&-\tfrac{\alpha\kappa_1}{1+b(x)}\Big(\tfrac{d+1+2b(x)}{2+b(x)}+\tfrac{d-1}{2}\kappa_2\Big)\Theta\,, \\
\partial_{\perp}Z_i =& -\Theta D_i\alpha+\tfrac{\alpha}{1+b(x)}\big(D_i\Theta+{\mathcal M}_i-Z_jK_i^{~j}(2+b(x))-\kappa_1Z_i\big)\,.
\end{align}
\end{subequations}

We consider the principal part of \eqref{eq_constraints} and decompose it into its scalar and vector sectors respectively, as in Section \ref{sec:hyp}. The scalar sector is given by
\begin{subequations}
  \begin{align}
    \check{\xi}_0\hat{\mathcal H}=&-2\alpha\,\hat{\mathcal M}_{\perp}\,,\\
    \check{\xi}_0\hat{\mathcal M}_{\perp}=&-\tfrac{\alpha}{2(1+b(x))}\hat{\mathcal H}\,,\\
    \check{\xi}_0\hat{Z}_{\perp}=&~\tfrac{\alpha}{1+b(x)}(\hat{\mathcal M}_{\perp}+\hat{\Theta})\,,\\
    \check{\xi}_0\Theta=&~\tfrac{\alpha}{2(1+b(x))}\hat{\mathcal H}+\alpha\hat{Z}_{\perp}\,.
\end{align}  
\end{subequations}
The respective eigenvalues are $\xi_0=\beta^{\perp}\pm\frac{\alpha}{\sqrt{1+b(x)}}$, each of them with multiplicity $2$ but there is no degeneracy in the corresponding eigenvectors. The vector sector of the principal part is given by
\begin{subequations}
    \begin{align}
        \check{\xi}_0\hat{\mathcal M}_A=&~\alpha\hat{Z}_A\,,\\
        \check{\xi}_0\hat{Z}_A=&~\tfrac{\alpha}{1+b(x)}{\mathcal M}_A\,,
    \end{align}
\end{subequations}
with eigenvalues $\xi_0=\beta^{\perp}\pm\frac{\alpha}{\sqrt{1+b(x)}}$. Therefore, the system is strongly hyperbolic and, thus, it follows that if the constraints are satisfied initially then they continue to hold throughout the evolution.

\chapter{Perturbative hyperbolicity matrices}

Here we compute the perturbation hyperbolicity matrices ${\mathbb M}^{\text{GB}}$, both in the $4+1$-dimensional Einstein-scalar Gauss-Bonnet theory of gravity and in the the $3+1$-dimensional Four-Derivative Scalar Tensor theory of gravity, that have been defined in Section \ref{sec:hyp}.

\section{Einstein-Gauss-Bonnet gravity}\label{app:matEGB}
\begin{table}[ht]
\noindent
\centering
\small
\tabcolsep=0.02cm
\scalebox{0.77}{
\begin{tabular}{c|c c c c c c c c c  c c c c c c c  c c c c c  c c c c  c c c c}
& $i\hat{\tilde{\gamma}}_{00}$ & $i\hat{\tilde{\gamma}}_{11}$ & $i\hat{\tilde{\gamma}}_{22}$ & $i\hat{\tilde{\gamma}}_{01}$ & $i\hat{\tilde{\gamma}}_{02}$ & $i\hat{\tilde{\gamma}}_{12}$ & $i\hat{\tilde{\gamma}}_{\perp0}$ & $i\hat{\tilde{\gamma}}_{\perp1}$ & $i\hat{\tilde{\gamma}}_{\perp2}$ & $i\hat{\chi}$ & $i\hat{{\tilde{A}}}_{00}$ & $i\hat{{\tilde{A}}}_{11}$ & $i\hat{{\tilde{A}}}_{22}$ & $i\hat{{\tilde{A}}}_{01}$ & $i\hat{{\tilde{A}}}_{02}$ & $i\hat{{\tilde{A}}}_{12}$ & $i\hat{{\tilde{A}}}_{\perp0}$ & $i\hat{{\tilde{A}}}_{\perp1}$ & $i\hat{{\tilde{A}}}_{\perp2}$ & $i\hat{K}$ 
& $i\hat{\hat{\Gamma}}_{\perp}$ & $i\hat{\hat{\Gamma}}_0$ & $i\hat{\hat{\Gamma}}_1$ & $i\hat{\hat{\Gamma}}_2$  \\
\hline
$i\check{\xi}_0\hat{{\tilde{A}}}_{00}$ & $ia_{00}^{\tilde{\gamma}}$ & $ia_{11}^{\tilde{\gamma}}$ & $ia_{22}^{\tilde{\gamma}}$ & $ia_{01}^{\tilde{\gamma}}$ & $ia_{02}^{\tilde{\gamma}}$ & $ia_{12}^{\tilde{\gamma}}$ & $ia_{\perp0}^{\tilde{\gamma}}$ & $ia_{\perp1}^{\tilde{\gamma}}$ & $ia_{\perp2}^{\tilde{\gamma}}$ & $ia^{\chi}$ & $a_{00}^{\tilde{A}}$ & $a_{11}^{\tilde{A}}$ & $a_{22}^{\tilde{A}}$ & $a_{01}^{\tilde{A}}$ & $a_{02}^{\tilde{A}}$ & $a_{12}^{\tilde{A}}$ & $a_{\perp0}^{\tilde{A}}$ & $a_{\perp1}^{\tilde{A}}$ & $a_{\perp2}^{\tilde{A}}$ & $a^K$ & $a_{\perp}^{\hat{\Gamma}}$ & $a_0^{\hat{\Gamma}}$ & $a_1^{\hat{\Gamma}}$ & $a_2^{\hat{\Gamma}}$\\
$i\check{\xi}_0\hat{{\tilde{A}}}_{11}$ & $ib_{00}^{\tilde{\gamma}}$ & $ib_{11}^{\tilde{\gamma}}$ & $ib_{22}^{\tilde{\gamma}}$ & $ib_{01}^{\tilde{\gamma}}$ & $ib_{02}^{\tilde{\gamma}}$ & $ib_{12}^{\tilde{\gamma}}$ & $ib_{\perp0}^{\tilde{\gamma}}$ & $ib_{\perp1}^{\tilde{\gamma}}$ & $ib_{\perp2}^{\tilde{\gamma}}$ & $ib^{\chi}$ & $b_{00}^{\tilde{A}}$ & $b_{11}^{\tilde{A}}$ & $b_{22}^{\tilde{A}}$ & $b_{01}^{\tilde{A}}$ & $b_{02}^{\tilde{A}}$ & $b_{12}^{\tilde{A}}$ & $b_{\perp0}^{\tilde{A}}$ & $b_{\perp1}^{\tilde{A}}$ & $b_{\perp2}^{\tilde{A}}$ & $b^K$ & $b_{\perp}^{\hat{\Gamma}}$ & $b_0^{\hat{\Gamma}}$ & $b_1^{\hat{\Gamma}}$ & $b_2^{\hat{\Gamma}}$\\
$i\check{\xi}_0\hat{{\tilde{A}}}_{22}$ & $ic_{00}^{\tilde{\gamma}}$ & $ic_{11}^{\tilde{\gamma}}$ & $ic_{22}^{\tilde{\gamma}}$ & $ic_{01}^{\tilde{\gamma}}$ & $ic_{02}^{\tilde{\gamma}}$ & $ic_{12}^{\tilde{\gamma}}$ & $ic_{\perp0}^{\tilde{\gamma}}$ & $ic_{\perp1}^{\tilde{\gamma}}$ & $ic_{\perp2}^{\tilde{\gamma}}$ & $ic^{\chi}$ & $c_{00}^{\tilde{A}}$ & $c_{11}^{\tilde{A}}$ & $c_{22}^{\tilde{A}}$ & $c_{01}^{\tilde{A}}$ & $c_{02}^{\tilde{A}}$ & $c_{12}^{\tilde{A}}$ & $c_{\perp0}^{\tilde{A}}$ & $c_{\perp1}^{\tilde{A}}$ & $c_{\perp2}^{\tilde{A}}$ & $c^K$  & $c_{\perp}^{\hat{\Gamma}}$ & $c_0^{\hat{\Gamma}}$ & $c_1^{\hat{\Gamma}}$ & $c_2^{\hat{\Gamma}}$\\
$i\check{\xi}_0\hat{{\tilde{A}}}_{01}$ & $id_{00}^{\tilde{\gamma}}$ & $id_{11}^{\tilde{\gamma}}$ & $id_{22}^{\tilde{\gamma}}$ & $id_{01}^{\tilde{\gamma}}$ & $id_{02}^{\tilde{\gamma}}$ & $id_{12}^{\tilde{\gamma}}$ & $id_{\perp0}^{\tilde{\gamma}}$ & $id_{\perp1}^{\tilde{\gamma}}$ & $id_{\perp2}^{\tilde{\gamma}}$ & $id^{\chi}$ & $d_{00}^{\tilde{A}}$ & $d_{11}^{\tilde{A}}$ & $d_{22}^{\tilde{A}}$ & $d_{01}^{\tilde{A}}$ & $d_{02}^{\tilde{A}}$ & $d_{12}^{\tilde{A}}$ & $d_{\perp0}^{\tilde{A}}$ & $d_{\perp1}^{\tilde{A}}$ & $d_{\perp2}^{\tilde{A}}$ & $d^K$ & $d_{\perp}^{\hat{\Gamma}}$ & $d_0^{\hat{\Gamma}}$ & $d_1^{\hat{\Gamma}}$ & $d_2^{\hat{\Gamma}}$\\
$i\check{\xi}_0\hat{{\tilde{A}}}_{02}$ & $ie_{00}^{\tilde{\gamma}}$ & $ie_{11}^{\tilde{\gamma}}$ & $ie_{22}^{\tilde{\gamma}}$ & $ie_{01}^{\tilde{\gamma}}$ & $ie_{02}^{\tilde{\gamma}}$ & $ie_{12}^{\tilde{\gamma}}$ & $ie_{\perp0}^{\tilde{\gamma}}$ & $ie_{\perp1}^{\tilde{\gamma}}$ & $ie_{\perp2}^{\tilde{\gamma}}$ & $ie^{\chi}$ & $e_{00}^{\tilde{A}}$ & $e_{11}^{\tilde{A}}$ & $e_{22}^{\tilde{A}}$ & $e_{01}^{\tilde{A}}$ & $e_{02}^{\tilde{A}}$ & $e_{12}^{\tilde{A}}$ & $e_{\perp0}^{\tilde{A}}$ & $e_{\perp1}^{\tilde{A}}$ & $e_{\perp2}^{\tilde{A}}$ & $e^K$ & $e_{\perp}^{\hat{\Gamma}}$ & $e_0^{\hat{\Gamma}}$ & $e_1^{\hat{\Gamma}}$ & $e_2^{\hat{\Gamma}}$\\
$i\check{\xi}_0\hat{{\tilde{A}}}_{12}$ & $if_{00}^{\tilde{\gamma}}$ & $if_{11}^{\tilde{\gamma}}$ & $if_{22}^{\tilde{\gamma}}$ & $if_{01}^{\tilde{\gamma}}$ & $if_{02}^{\tilde{\gamma}}$ & $if_{12}^{\tilde{\gamma}}$ & $if_{\perp0}^{\tilde{\gamma}}$ & $if_{\perp1}^{\tilde{\gamma}}$ & $if_{\perp2}^{\tilde{\gamma}}$ & $if^{\chi}$ & $f_{00}^{\tilde{A}}$ & $f_{11}^{\tilde{A}}$ & $f_{22}^{\tilde{A}}$ & $f_{01}^{\tilde{A}}$ & $f_{02}^{\tilde{A}}$ & $f_{12}^{\tilde{A}}$ & $f_{\perp0}^{\tilde{A}}$ & $f_{\perp1}^{\tilde{A}}$ & $f_{\perp2}^{\tilde{A}}$ & $f^K$  & $f_{\perp}^{\hat{\Gamma}}$ & $f_0^{\hat{\Gamma}}$ & $f_1^{\hat{\Gamma}}$ & $f_2^{\hat{\Gamma}}$\\
$i\check{\xi}_0\hat{A}_{\perp0}$ & $ig_{00}^{\tilde{\gamma}}$ & $ig_{11}^{\tilde{\gamma}}$ & $ig_{22}^{\tilde{\gamma}}$ & $ig_{01}^{\tilde{\gamma}}$ & $ig_{02}^{\tilde{\gamma}}$ & $ig_{12}^{\tilde{\gamma}}$ & $ig_{\perp0}^{\tilde{\gamma}}$ & $ig_{\perp1}^{\tilde{\gamma}}$ & $ig_{\perp2}^{\tilde{\gamma}}$ & $ig^{\chi}$ & $g_{00}^{\tilde{A}}$ & $g_{11}^{\tilde{A}}$ & $g_{22}^{\tilde{A}}$ & $g_{01}^{\tilde{A}}$ & $g_{02}^{\tilde{A}}$ & $g_{12}^{\tilde{A}}$ & $g_{\perp0}^{\tilde{A}}$ & $g_{\perp1}^{\tilde{A}}$ & $g_{\perp2}^{\tilde{A}}$ & $g^K$ & $g_{\perp}^{\hat{\Gamma}}$ & $g_0^{\hat{\Gamma}}$ & $g_1^{\hat{\Gamma}}$ & $g_2^{\hat{\Gamma}}$\\
$i\check{\xi}_0\hat{A}_{\perp1}$ & $ih_{00}^{\tilde{\gamma}}$ & $ih_{11}^{\tilde{\gamma}}$ & $ih_{22}^{\tilde{\gamma}}$ & $ih_{01}^{\tilde{\gamma}}$ & $ih_{02}^{\tilde{\gamma}}$ & $ih_{12}^{\tilde{\gamma}}$ & $ih_{\perp0}^{\tilde{\gamma}}$ & $ih_{\perp1}^{\tilde{\gamma}}$ & $ih_{\perp2}^{\tilde{\gamma}}$ & $ih^{\chi}$ & $h_{00}^{\tilde{A}}$ & $h_{11}^{\tilde{A}}$ & $h_{22}^{\tilde{A}}$ & $h_{01}^{\tilde{A}}$ & $h_{02}^{\tilde{A}}$ & $h_{12}^{\tilde{A}}$ & $h_{\perp0}^{\tilde{A}}$ & $h_{\perp1}^{\tilde{A}}$ & $h_{\perp2}^{\tilde{A}}$ & $h^K$  & $h_{\perp}^{\hat{\Gamma}}$ & $h_0^{\hat{\Gamma}}$ & $h_1^{\hat{\Gamma}}$ & $h_2^{\hat{\Gamma}}$\\
$i\check{\xi}_0\hat{A}_{\perp2}$ & $ii_{00}^{\tilde{\gamma}}$ & $ii_{11}^{\tilde{\gamma}}$ & $ii_{22}^{\tilde{\gamma}}$ & $ii_{01}^{\tilde{\gamma}}$ & $ii_{02}^{\tilde{\gamma}}$ & $ii_{12}^{\tilde{\gamma}}$ & $ii_{\perp0}^{\tilde{\gamma}}$ & $ii_{\perp1}^{\tilde{\gamma}}$ & $ii_{\perp2}^{\tilde{\gamma}}$ & $ii^{\chi}$ & $i_{00}^{\tilde{A}}$ & $i_{11}^{\tilde{A}}$ & $i_{22}^{\tilde{A}}$ & $i_{01}^{\tilde{A}}$ & $i_{02}^{\tilde{A}}$ & $i_{12}^{\tilde{A}}$ & $i_{\perp0}^{\tilde{A}}$ & $i_{\perp1}^{\tilde{A}}$ & $i_{\perp2}^{\tilde{A}}$ & $i^K$ & $i_{\perp}^{\hat{\Gamma}}$ & $i_0^{\hat{\Gamma}}$ & $i_1^{\hat{\Gamma}}$ & $i_2^{\hat{\Gamma}}$\\
$i\check{\xi}_0\hat{K}$ & $ij_{00}^{\tilde{\gamma}}$ & $ij_{11}^{\tilde{\gamma}}$ & $ij_{22}^{\tilde{\gamma}}$ & $ij_{01}^{\tilde{\gamma}}$ & $ij_{02}^{\tilde{\gamma}}$ & $ij_{12}^{\tilde{\gamma}}$ & $ij_{\perp0}^{\tilde{\gamma}}$ & $ij_{\perp1}^{\tilde{\gamma}}$ & $ij_{\perp2}^{\tilde{\gamma}}$ & $ij^{\chi}$ & $j_{00}^{\tilde{A}}$ & $j_{11}^{\tilde{A}}$ & $j_{22}^{\tilde{A}}$ & $j_{01}^{\tilde{A}}$ & $j_{02}^{\tilde{A}}$ & $j_{12}^{\tilde{A}}$ & $j_{\perp0}^{\tilde{A}}$ & $j_{\perp1}^{\tilde{A}}$ & $j_{\perp2}^{\tilde{A}}$ & $j^K$  & $j_{\perp}^{\hat{\Gamma}}$ & $j_0^{\hat{\Gamma}}$ & $j_1^{\hat{\Gamma}}$ & $j_2^{\hat{\Gamma}}$\\
$i\check{\xi}_0\hat{\Theta}$  & $ik_{00}^{\tilde{\gamma}}$ & $ik_{11}^{\tilde{\gamma}}$ & $ik_{22}^{\tilde{\gamma}}$ & $ik_{01}^{\tilde{\gamma}}$ & $ik_{02}^{\tilde{\gamma}}$ & $ik_{12}^{\tilde{\gamma}}$ & $ik_{\perp0}^{\tilde{\gamma}}$ & $ik_{\perp1}^{\tilde{\gamma}}$ & $ik_{\perp2}^{\tilde{\gamma}}$ & $ik^{\chi}$ & 0 & 0 &0 & 0 & 0 & 0 & 0 & 0 & 0 & 0 & 0 & 0 & 0 & 0 \\
$i\check{\xi}_0\hat{\hat{\Gamma}}_{\perp}$ & $im_{00}^{\tilde{\gamma}}$ & $im_{11}^{\tilde{\gamma}}$ & $im_{22}^{\tilde{\gamma}}$ & $im_{01}^{\tilde{\gamma}}$ & $im_{02}^{\tilde{\gamma}}$ & $im_{12}^{\tilde{\gamma}}$ & $im_{\perp0}^{\tilde{\gamma}}$ & $im_{\perp1}^{\tilde{\gamma}}$ & $im_{\perp2}^{\tilde{\gamma}}$ & $im^{\chi}$ & $m_{00}^{\tilde{A}}$ & $m_{11}^{\tilde{A}}$ & $m_{22}^{\tilde{A}}$ & $m_{01}^{\tilde{A}}$ & $m_{02}^{\tilde{A}}$ & $m_{12}^{\tilde{A}}$ & $m_{\perp0}^{\tilde{A}}$ & $m_{\perp1}^{\tilde{A}}$ & $m_{\perp2}^{\tilde{A}}$ & $m^K$  & 0 & 0 & 0 & 0\\
$i\check{\xi}_0\hat{\hat{\Gamma}}_0$ &  $in_{00}^{\tilde{\gamma}}$ & $in_{11}^{\tilde{\gamma}}$ & $in_{22}^{\tilde{\gamma}}$ & $in_{01}^{\tilde{\gamma}}$ & $in_{02}^{\tilde{\gamma}}$ & $in_{12}^{\tilde{\gamma}}$ & $in_{\perp0}^{\tilde{\gamma}}$ & $in_{\perp1}^{\tilde{\gamma}}$ & $in_{\perp2}^{\tilde{\gamma}}$ & $in^{\chi}$ & $n_{00}^{\tilde{A}}$ & $n_{11}^{\tilde{A}}$ & $n_{22}^{\tilde{A}}$ & $n_{01}^{\tilde{A}}$ & $n_{02}^{\tilde{A}}$ & $n_{12}^{\tilde{A}}$ & $n_{\perp0}^{\tilde{A}}$ & $n_{\perp1}^{\tilde{A}}$ & $n_{\perp2}^{\tilde{A}}$ & $n^K$ & 0 & 0 & 0 & 0 \\
$i\check{\xi}_0\hat{\hat{\Gamma}}_1$ &  $ip_{00}^{\tilde{\gamma}}$ & $ip_{11}^{\tilde{\gamma}}$ & $ip_{22}^{\tilde{\gamma}}$ & $ip_{01}^{\tilde{\gamma}}$ & $ip_{02}^{\tilde{\gamma}}$ & $ip_{12}^{\tilde{\gamma}}$ & $ip_{\perp0}^{\tilde{\gamma}}$ & $ip_{\perp1}^{\tilde{\gamma}}$ & $ip_{\perp2}^{\tilde{\gamma}}$ & $ip^{\chi}$ & $p_{00}^{\tilde{A}}$ & $p_{11}^{\tilde{A}}$ & $p_{22}^{\tilde{A}}$ & $p_{01}^{\tilde{A}}$ & $p_{02}^{\tilde{A}}$ & $p_{12}^{\tilde{A}}$ & $p_{\perp0}^{\tilde{A}}$ & $p_{\perp1}^{\tilde{A}}$ & $p_{\perp2}^{\tilde{A}}$ & $p^K$  & 0 & 0 & 0 & 0\\
$i\check{\xi}_0\hat{\hat{\Gamma}}_2$ &  $iq_{00}^{\tilde{\gamma}}$ & $iq_{11}^{\tilde{\gamma}}$ & $iq_{22}^{\tilde{\gamma}}$ & $iq_{01}^{\tilde{\gamma}}$ & $iq_{02}^{\tilde{\gamma}}$ & $iq_{12}^{\tilde{\gamma}}$ & $iq_{\perp0}^{\tilde{\gamma}}$ & $iq_{\perp1}^{\tilde{\gamma}}$ & $iq_{\perp2}^{\tilde{\gamma}}$ & $iq^{\chi}$ & $q_{00}^{\tilde{A}}$ & $q_{11}^{\tilde{A}}$ & $q_{22}^{\tilde{A}}$ & $q_{01}^{\tilde{A}}$ & $q_{02}^{\tilde{A}}$ & $q_{12}^{\tilde{A}}$ & $q_{\perp0}^{\tilde{A}}$ & $q_{\perp1}^{\tilde{A}}$ & $q_{\perp2}^{\tilde{A}}$ & $q^K$   & 0 & 0 & 0 & 0\\
\end{tabular}}
\caption[Hyperbolicity matrix in Einstein-Gauss-Bonnet gravity.]{Perturbative hyperbolicity matrix ${\mathbb M}^{\text{GB}}$ in $4+1$ Einstein-Gauss-Bonnet gravity.}\label{tab:matEGB}
\end{table}
In this case those are the values of the elements of the matrix in \ref{tab:matEGB}, where ${\mathcal M}_{\perp A\perp B}=e_A^i\xi^je_B^k\xi^lM_{ijkl}$, ${\mathcal M}_{\perp ABC}=\xi^ie_A^je_B^ke_C^lM_{ijkl}$, ${\mathcal M}_{ABCD}=e_A^ie_B^je_C^ke_D^lM_{ijkl}$, ${\mathcal N}_{A\perp B}=e_A^i\xi^je_B^kN_{ijk}$, ${\mathcal N}_{ABC}=e_A^ie_B^je_C^kN_{ijk}$ and $F_{AB}=e_A^ie_B^j({\mathcal L}_nK_{ij}+\frac{1}{\alpha}D_iD_j\alpha+K_{ik}K^k_{~j})$:
\begin{subequations}
\begin{align}
a^{\tilde{\gamma}}_{00}=&~\tfrac{1}{6(1+b(x))}\big[2b(x)\big({\mathcal M}_{0101}+{\mathcal M}_{0202}\big)+3(1-b(x)){\mathcal M}_{1212}+6b(x){\mathcal M}_{\perp0\perp0} \nonumber\\&+(3+b(x))\big({\mathcal M}_{\perp1\perp1}+{\mathcal M}_{\perp2\perp2}\big)\big]+\tfrac{F_{11}+F_{22}}{2}\,,\nonumber\\
a^{\tilde{\gamma}}_{11}=&~\tfrac{F_{00}-3F_{22}}{2}+\tfrac{1}{6(1+b(x))}\big[3(1+3b(x)){\mathcal M}_{\perp0\perp0}-2b(x)({\mathcal M}_{\perp1\perp1}-{\mathcal M}_{0101}) \nonumber\\&-(9+7b(x)){\mathcal M}_{0202} -(9+11b(x)){\mathcal M}_{\perp2\perp2}+6(2+b(x)){\mathcal M}_{1212}\big]\,, \nonumber\\
a^{\tilde{\gamma}}_{22}=&~\tfrac{F_{00}-3F_{11}}{2}+\tfrac{1}{6(1+b(x))}\big[3(1+3b(x)){\mathcal M}_{\perp0\perp0}-2b(x)({\mathcal M}_{\perp2\perp2}-{\mathcal M}_{0202})\nonumber\\&-(9+7b(x)){\mathcal M}_{0101}  -(9+11b(x)){\mathcal M}_{\perp1\perp1}+6(2+b(x)){\mathcal M}_{1212}\big]\,, \nonumber\\
a^{\chi}=&~\tfrac{3+5b(x)}{2(1+b(x))}\big[3{\mathcal M}_{1212}-\big({\mathcal M}_{0101}+{\mathcal M}_{0202} \big)\big]-F_{00}+F_{11}+F_{22}
\nonumber\\&+\tfrac{1}{1+b(x)}\big[(1+2b(x))({\mathcal M}_{\perp1\perp1}+{\mathcal M}_{\perp2\perp2})-(1+4b(x)){\mathcal M}_{\perp0\perp0}\big]\,,\nonumber\\
a^{\tilde{\gamma}}_{\perp0}=&-{\mathcal M}_{\perp101}-{\mathcal M}_{\perp202}\,, \ \ a^{\tilde{\gamma}}_{\perp1}=3{\mathcal M}_{\perp212}-{\mathcal M}_{\perp010}\,, \ \
a^{\tilde{\gamma}}_{\perp2}=3{\mathcal M}_{\perp121}-{\mathcal M}_{\perp020}\,, \nonumber\\ 
a^{\tilde{A}}_{00}=&-2({\mathcal N}_{1\perp1}+{\mathcal N}_{2\perp2})\,, \ \ a^{\tilde{A}}_{11}=6{\mathcal N}_{2\perp2}-2{\mathcal N}_{0\perp0}\,, \ \
a^{\tilde{A}}_{22}=6{\mathcal N}_{1\perp1}-2{\mathcal N}_{0\perp0}\,, \nonumber\\ a^K=&~\chi({\mathcal N}_{1\perp1}+{\mathcal N}_{2\perp2}-{\mathcal N}_{0\perp0})\,, \ \
a^{\hat{\Gamma}}_{\perp}=\chi^2(3{\mathcal M}_{1212}-{\mathcal M}_{0101}-{\mathcal M}_{0202})\,, \nonumber\\ a^{\hat{\Gamma}}_0=&~\chi^2({\mathcal M}_{\perp101}+{\mathcal M}_{\perp202})\,, \ \ a^{\hat{\Gamma}}_1=\chi^2({\mathcal M}_{\perp010}-3{\mathcal M}_{\perp212})\,,\nonumber\\
a^{\hat{\Gamma}}_2=&~\chi^2({\mathcal M}_{\perp020}-3{\mathcal M}_{\perp121})\,,\\
b^{\tilde{\gamma}}_{00}=&~\tfrac{F_{11}-3F_{22}}{2}+\tfrac{1}{6(1+b(x))}\big[6(2+b(x)){\mathcal M}_{0202}-2b(x)({\mathcal M}_{\perp0\perp0}-{\mathcal M}_{0101})\nonumber\\&-(9+7b(x)){\mathcal M}_{1212} -(9+11b(x)){\mathcal M}_{\perp2\perp2}+3(1+3b(x)){\mathcal M}_{\perp1\perp1}\big]\,, \nonumber\\
b^{\tilde{\gamma}}_{11}=&~\tfrac{1}{6(1+b(x))}\big[2b(x)\left({\mathcal M}_{0101}+{\mathcal M}_{1212}\right)+3(1-b(x)){\mathcal M}_{0202}+6b(x){\mathcal M}_{\perp1\perp1} \nonumber\\&+(3+b(x))\left({\mathcal M}_{\perp0\perp0}+{\mathcal M}_{\perp2\perp2}\right)\big]+\tfrac{F_{00}+F_{22}}{2}\,,\nonumber\\
b^{\tilde{\gamma}}_{22}=&~\tfrac{F_{11}-3F_{00}}{2}+\tfrac{1}{6(1+b(x))}\big[3(1+3b(x)){\mathcal M}_{\perp1\perp1}-2b(x)({\mathcal M}_{\perp2\perp2}-{\mathcal M}_{1212}) \nonumber\\&-(9+7b(x)){\mathcal M}_{0101}-(9+11b(x)){\mathcal M}_{\perp0\perp0}+6(2+b(x)){\mathcal M}_{0202}\big]\,, \nonumber\\
b^{\chi}=&~\tfrac{3+5b(x)}{2(1+b(x))}\big[3{\mathcal M}_{0202}-\left({\mathcal M}_{0101}+{\mathcal M}_{1212} \right)\big]+F_{00}-F_{11}+F_{22}
\nonumber\\&+\tfrac{1}{1+b(x)}\big[(1+2b(x))({\mathcal M}_{\perp0\perp0}+{\mathcal M}_{\perp2\perp2})-(1+4b(x)){\mathcal M}_{\perp1\perp1}\big]\,,\nonumber\\
b^{\tilde{\gamma}}_{\perp0}=&~3{\mathcal M}_{\perp202}-{\mathcal M}_{\perp101}\,, \ \
b^{\tilde{\gamma}}_{\perp1}=-{\mathcal M}_{\perp010}-{\mathcal M}_{\perp212}\,, \ \ b^{\tilde{\gamma}}_{\perp2}=3{\mathcal M}_{\perp020}-{\mathcal M}_{\perp121}, \nonumber \\
b^{\tilde{A}}_{00}=&~6{\mathcal N}_{2\perp2}-2{\mathcal N}_{1\perp1}\,, \ \ b^{\tilde{A}}_{11}=-2({\mathcal N}_{0\perp0}+{\mathcal N}_{2\perp2})\,, \ \
b^{\tilde{A}}_{22}=6{\mathcal N}_{0\perp0}-2{\mathcal N}_{1\perp1}\,, \nonumber\\ b^K=&~\chi({\mathcal N}_{0\perp0}-{\mathcal N}_{1\perp1}+{\mathcal N}_{2\perp2})\,, \ \
b^{\hat{\Gamma}}_{\perp}=\chi^2(3{\mathcal M}_{0202}-{\mathcal M}_{0101}-{\mathcal M}_{1212})\,, \nonumber\\ b^{\hat{\Gamma}}_0=&~\chi^2({\mathcal M}_{\perp101}-3{\mathcal M}_{\perp202})\,, \ \ b^{\hat{\Gamma}}_1=\chi^2({\mathcal M}_{\perp010}+{\mathcal M}_{\perp212})\,, \nonumber\\
b^{\hat{\Gamma}}_2=&~\chi^2({\mathcal M}_{\perp121}-3{\mathcal M}_{\perp020})\,, \\
c^{\tilde{\gamma}}_{00}=&~\tfrac{F_{22}-3F_{11}}{2}+\tfrac{1}{6(1+b(x))}\big[6(2+b(x)){\mathcal M}_{0101}-2b(x)({\mathcal M}_{\perp0\perp0}-{\mathcal M}_{0202})\nonumber\\&-(9+7b(x)){\mathcal M}_{1212} -(9+11b(x)){\mathcal M}_{\perp1\perp1}+3(1+3b(x)){\mathcal M}_{\perp2\perp2}\big]\,, \nonumber\\
c^{\tilde{\gamma}}_{11}=&~\tfrac{F_{22}-3F_{00}}{2}+\tfrac{1}{6(1+b(x))}\big[3(1+3b(x)){\mathcal M}_{\perp2\perp2}-2b(x)({\mathcal M}_{\perp1\perp1}-{\mathcal M}_{1212})\nonumber\\&-(9+7b(x)){\mathcal M}_{0202}  -(9+11b(x)){\mathcal M}_{\perp0\perp0}+6(2+b(x)){\mathcal M}_{0101}\big]\,, \nonumber\\
c^{\tilde{\gamma}}_{22}=&~\tfrac{1}{6(1+b(x))}\big[2b(x)\left({\mathcal M}_{0202}+{\mathcal M}_{1212}\right)+3(1-b(x)){\mathcal M}_{0101}+6b(x){\mathcal M}_{\perp2\perp2} \nonumber\\&+(3+b(x))\left({\mathcal M}_{\perp0\perp0}+{\mathcal M}_{\perp1\perp1}\right)\big]+\tfrac{F_{00}+F_{11}}{2}\,,\nonumber\\
c^{\chi}=&~\tfrac{3+5b(x)}{2(1+b(x))}\big[3{\mathcal M}_{0101}-\big({\mathcal M}_{0202}+{\mathcal M}_{1212} \big)\big]+F_{00}+F_{11}-F_{22}
\nonumber\\&+\tfrac{1}{1+b(x)}\big[(1+2b(x))({\mathcal M}_{\perp0\perp0}+{\mathcal M}_{\perp1\perp1})-(1+4b(x)){\mathcal M}_{\perp2\perp2}\big]\,,\nonumber\\
c^{\tilde{\gamma}}_{\perp0}=&~3{\mathcal M}_{\perp101}-{\mathcal M}_{\perp202}\,, \ \ c^{\tilde{\gamma}}_{\perp1}=3{\mathcal M}_{\perp010}-{\mathcal M}_{\perp212}\,, \ \
c^{\tilde{\gamma}}_{\perp2}=-{\mathcal M}_{\perp020}-{\mathcal M}_{\perp121}\,, \nonumber \\
c^{\tilde{A}}_{00}=&~6{\mathcal N}_{1\perp1}-2{\mathcal N}_{2\perp2}\,, \ \ c^{\tilde{A}}_{11}=6{\mathcal N}_{0\perp0}-2{\mathcal N}_{2\perp2}\,, \ \
c^{\tilde{A}}_{22}=-2({\mathcal N}_{0\perp0}+{\mathcal N}_{1\perp1})\,, \nonumber\\ c^K=&~\chi({\mathcal N}_{0\perp0}+{\mathcal N}_{1\perp1}-{\mathcal N}_{2\perp2})\,, \ \
c^{\hat{\Gamma}}_{\perp}=\chi^2(3{\mathcal M}_{0101}-{\mathcal M}_{0202}-{\mathcal M}_{1212})\,, \nonumber\\ c^{\hat{\Gamma}}_0=&~\chi^2({\mathcal M}_{\perp202}-3{\mathcal M}_{\perp101})\,, \ \ c^{\hat{\Gamma}}_1=\chi^2({\mathcal M}_{\perp212}-3{\mathcal M}_{\perp010})\,, \nonumber\\
c^{\hat{\Gamma}}_2=&~\chi^2({\mathcal M}_{\perp020}+{\mathcal M}_{\perp121})\,, \\
d^{\tilde{\gamma}}_{00}=&~d^{\tilde{\gamma}}_{11}=d^{\tilde{\gamma}}_{22}-2({\mathcal M}_{\perp0\perp1}+F_{01})\,, \ \ 
d^{\tilde{\gamma}}_{01}=2({\mathcal M}_{\perp2\perp2}+F_{22})\,, \nonumber\\ d^{\tilde{\gamma}}_{02}=&-2({\mathcal M}_{\perp1\perp2}+F_{12})\,, \ \
d^{\tilde{\gamma}}_{12}=-2({\mathcal M}_{\perp0\perp2}+F_{02})\,, \nonumber\\
d^{\tilde{A}}_{00}=&~d^{\tilde{A}}_{11}=0\,, \ \ d^{\tilde{A}}_{22}=-8{\mathcal N}_{(0|\perp|1)}\,, \ \
d^K=-2\chi{\mathcal N}_{(0|\perp|1))}\,, \ \ d^{\hat{\Gamma}}_{\perp}=-4\chi^2{\mathcal M}_{0212}\,, \nonumber \\
d^{\hat{\Gamma}}_0=&~2\chi^2{\mathcal M}_{\perp212}\,, \ \
d^{\hat{\Gamma}}_1=2\chi^2{\mathcal M}_{\perp202}\,, \ \ d^{\hat{\Gamma}}_2=4\chi^2{\mathcal M}_{\perp(1|2|0)}\,,\\
e^{\tilde{\gamma}}_{00}=&~e^{\tilde{\gamma}}_{22}=e^{\tilde{\gamma}}_{11}-2({\mathcal M}_{\perp0\perp2}+F_{02})\,, \ \ 
e^{\tilde{\gamma}}_{01}=-2({\mathcal M}_{\perp1\perp2}+F_{12})\,, \nonumber\\ e^{\tilde{\gamma}}_{02}=&~2({\mathcal M}_{\perp1\perp1}+F_{11})\,, \ \
e^{\tilde{\gamma}}_{12}=-2({\mathcal M}_{\perp0\perp1}+F_{01})\,, \nonumber\\
e^{\tilde{A}}_{00}=&~e^{\tilde{A}}_{22}=0\,, \ \ e^{\tilde{A}}_{11}=-8{\mathcal N}_{(0|\perp|2)}\,, \ \
e^K=-2\chi{\mathcal N}_{(0|\perp|2)}\,, \ \
e_{\perp}^{\hat{\Gamma}}=-4\chi^2{\mathcal M}_{0121}\,, \nonumber\\
e^{\hat{\Gamma}}_0=&~2\chi^2{\mathcal M}_{\perp121}\,, \ \
e^{\hat{\Gamma}}_1=4\chi^2{\mathcal M}_{\perp(2|1|0)}\,, \ \ e^{\hat{\Gamma}}_2=2\chi^2{\mathcal M}_{\perp101}\,, \\
f^{\tilde{\gamma}}_{11}=&~f^{\tilde{\gamma}}_{22}=f^{\tilde{\gamma}}_{22}-2({\mathcal M}_{\perp1\perp2}+F_{12})\,, \ \ 
f^{\tilde{\gamma}}_{01}=-2({\mathcal M}_{\perp0\perp2}+F_{02})\,, \nonumber\\ f^{\tilde{\gamma}}_{02}=&-2({\mathcal M}_{\perp0\perp1}+F_{01})\,, \ \
f^{\tilde{\gamma}}_{12}=2({\mathcal M}_{\perp0\perp0}+F_{00})\,, \nonumber\\
f^{\tilde{A}}_{11}=&~f^{\tilde{A}}_{22}=0\,, \ \
f^{\tilde{A}}_{00}=-8{\mathcal N}_{(1|\perp|2)}\,, \ \
f^K=-2\chi{\mathcal N}_{(1|\perp|2)}\,, \ \
f^{\hat{\Gamma}}_ {\perp}=-4\chi^2{\mathcal M}_{0102}\,, \nonumber\\ f^{\hat{\Gamma}}_0=&~4\chi^2{\mathcal M}_{\perp(1|0|2)}\,, \ \
f^{\hat{\Gamma}}_1=2\chi^2{\mathcal M}_{\perp020}\,, \ \ f^{\hat{\Gamma}}_2=2\chi^2{\mathcal M}_{\perp010}\,, \\
g^{\tilde{\gamma}}_{00}=&-\tfrac{2b(x)}{3(1+b(x))}({\mathcal M}_{\perp101}+{\mathcal M}_{\perp202})\,, \ \
g^{\tilde{\gamma}}_{11}=g^{\tilde{\gamma}}_{00}+2{\mathcal M}_{\perp202}\,, \nonumber\\ g^{\tilde{\gamma}}_{22}=&~g^{\tilde{\gamma}}_{00}+2{\mathcal M}_{\perp101}\,, \ \
g^{\chi}=-\tfrac{2}{1+b(x)}({\mathcal M}_{\perp101}+{\mathcal M}_{\perp202})\,, \nonumber \\
g^{\tilde{\gamma}}_{\perp0}=&~2{\mathcal M}_{1212}\,, \ \ g^{\tilde{\gamma}}_{\perp1}=-2{\mathcal M}_{0212}\,, \ \
g^{\tilde{\gamma}}_{\perp2}=-2{\mathcal M}_{0121}\,, \nonumber\\
g^{\tilde{A}}_{00}=&~0\,, \ \ g^{\tilde{A}}_{11}=-4{\mathcal N}_{202}\,, \ \ g^{\tilde{A}}_{22}=-4{\mathcal N}_{101}\,, \ \
g^{\tilde{A}}_{01}=4{\mathcal N}_{212}\,, \ \ g^{\tilde{A}}_{02}=4{\mathcal N}_{121}\,, \nonumber\\ g^{\tilde{A}}_{12}=&-8{\mathcal N}_{0(12)}\,, \ \
g^K=-\chi({\mathcal N}_{202}+{\mathcal N}_{101})\,, \ \
g^{\tilde{A}}_{\perp A}=g^{\hat{\Gamma}}_{\perp}=0 \ \ \forall A=0,1,2\,,\nonumber\\ g^{\hat{\Gamma}}_0=&-2\chi^2{\mathcal M}_{1212}\,, \ \ 
g^{\hat{\Gamma}}_1=2\chi^2{\mathcal M}_{0212}\,, \ \ g^{\hat{\Gamma}}_2=2\chi^2{\mathcal M}_{0121}\,, \\
h^{\tilde{\gamma}}_{11}=&-\tfrac{2b(x)}{3(1+b(x))}({\mathcal M}_{\perp010}+{\mathcal M}_{\perp212})\,, \ \
h^{\tilde{\gamma}}_{00}=h^{\tilde{\gamma}}_{11}+2{\mathcal M}_{\perp212}\,,\nonumber \\ h^{\tilde{\gamma}}_{22}=&~h^{\tilde{\gamma}}_{11}+2{\mathcal M}_{\perp010}\,, \ \
h^{\chi}=-\tfrac{2}{1+b(x)}({\mathcal M}_{\perp010}+{\mathcal M}_{\perp212})\,, \nonumber \\
h^{\tilde{\gamma}}_{\perp0}=&-2{\mathcal M}_{0212}\,, \ \ h^{\tilde{\gamma}}_{\perp1}=2{\mathcal M}_{0202}\,, \ \
h^{\tilde{\gamma}}_{\perp2}=-2{\mathcal M}_{0102}\,,\nonumber\\
h^{\tilde{A}}_{00}=&-4{\mathcal N}_{212}\,, \ \ h^{\tilde{A}}_{11}=0\,, \ \ h^{\tilde{A}}_{22}=-4{\mathcal N}_{010}\,, \ \
h^{\tilde{A}}_{01}=4{\mathcal N}_{202}\,, \ \ h^{\tilde{A}}_{02}=-8{\mathcal N}_{1(02)}\,, \nonumber \\ h^{\tilde{A}}_{12}=&~4{\mathcal N}_{020}\,, \ \
h^K=-\chi({\mathcal N}_{212}+{\mathcal N}_{010})\,, \ \
h^{\tilde{A}}_{\perp A}=h^{\hat{\Gamma}}_{\perp}=0 \ \ \forall A=0,1,2\,,\nonumber\\ h^{\hat{\Gamma}}_0=&~2\chi^2{\mathcal M}_{0212}\,,\ \
h^{\hat{\Gamma}}_1=-2\chi^2{\mathcal M}_{0202}\,, \ \ h^{\hat{\Gamma}}_2=2\chi^2{\mathcal M}_{0102}\,, \\
i^{\tilde{\gamma}}_{22}=&-\tfrac{2b(x)}{3(1+b(x))}({\mathcal M}_{\perp020}+{\mathcal M}_{\perp121})\,, \ \
i^{\tilde{\gamma}}_{00}=i^{\tilde{\gamma}}_{22}+2{\mathcal M}_{\perp121}\,, \nonumber\\ i^{\tilde{\gamma}}_{11}=&~i^{\tilde{\gamma}}_{22}+2{\mathcal M}_{\perp020}\,, \ \
i^{\chi}=-\tfrac{2}{1+b(x)}({\mathcal M}_{\perp020}+{\mathcal M}_{\perp121})\,, \nonumber \\
i^{\tilde{\gamma}}_{\perp0}=&-2{\mathcal M}_{0121}\,, \ \
i^{\tilde{\gamma}}_{\perp1}=-2{\mathcal M}_{0102}\,, \ \ i^{\tilde{\gamma}}_{\perp2}=2{\mathcal M}_{0101}\,, \nonumber \\
i^{\tilde{A}}_{00}=&-4{\mathcal N}_{121}\,, \ \ i^{\tilde{A}}_{11}=-4{\mathcal N}_{020} \,, \ \ i^{\tilde{A}}_{22}=0\,, \ \
i^{\tilde{A}}_{01}=-8{\mathcal N}_{2(10)}\,, \ \ i^{\tilde{A}}_{02}=4{\mathcal N}_{101}\,, \nonumber\\ i^{\tilde{A}}_{12}=&~4{\mathcal N}_{010}\,, \ \
i^K=-\chi({\mathcal N}_{020}+{\mathcal N}_{121})\,, \ \
i^{\tilde{A}}_{\perp A}=i^{\hat{\Gamma}}_{\perp}=0 \ \ \forall A=0,1,2\,, \nonumber\\
i^{\hat{\Gamma}}_0=&~2\chi^2{\mathcal M}_{0121}\,, \ \ i^{\hat{\Gamma}}_1=2\chi^2{\mathcal M}_{0102}\,, \ \ i^{\hat{\Gamma}}_2=-2\chi^2{\mathcal M}_{0101}\,,\\
j^{\tilde{\gamma}}_{00}=&~\tfrac{2(F_{11}+F_{22})}{3\chi}-\tfrac{1}{9\chi(1+b(x))}\big[4b(x)\left({\mathcal M}_{\perp0\perp0}+{\mathcal M}_{0101}+{\mathcal M}_{0202}\right)\nonumber\\&+2(9-b(x)){\mathcal M}_{1212}-2(3+b(x))({\mathcal M}_{\perp1\perp1}+{\mathcal M}_{\perp2\perp2})
\big]\,,\nonumber\\
j^{\tilde{\gamma}}_{11}=&~\tfrac{2(F_{00}+F_{22})}{3\chi}-\tfrac{1}{9\chi(1+b(x))}\big[4b(x)\left({\mathcal M}_{\perp1\perp1}+{\mathcal M}_{0101}+{\mathcal M}_{1212}\right)\nonumber\\&+2(9-b(x)){\mathcal M}_{0202}-2(3+b(x))({\mathcal M}_{\perp0\perp0}+{\mathcal M}_{\perp2\perp2})
\big]\,,\nonumber\\
j^{\tilde{\gamma}}_{22}=&~\tfrac{2(F_{00}+F_{11})}{3\chi}-\tfrac{1}{9\chi(1+b(x))}\big[4b(x)\left({\mathcal M}_{\perp2\perp2}+{\mathcal M}_{0202}+{\mathcal M}_{1212}\right)\nonumber\\&+2(9-b(x)){\mathcal M}_{0101}-2(3+b(x))({\mathcal M}_{\perp0\perp0}+{\mathcal M}_{\perp1\perp1})
\big]\,,\nonumber\\
j^{\chi}=&-\tfrac{4}{3\chi}\big(\tfrac{{\mathcal M}_{\perp0\perp0}+{\mathcal M}_{\perp1\perp1}+{\mathcal M}_{\perp2\perp2}}{1+b(x)}+F_{00}+F_{11}+F_{22}\big)\nonumber\\&-\tfrac{2(3+5b(x))({\mathcal M}_{0101}+{\mathcal M}_{0202}+{\mathcal M}_{1212})}{3\chi(1+b(x))}\,,\nonumber\\
j^{\tilde{\gamma}}_{\perp0}=&-\tfrac{4({\mathcal M}_{\perp101}+{\mathcal M}_{\perp202})}{3\chi}\,, \ \ j^{\tilde{\gamma}}_{\perp1}=-\tfrac{4({\mathcal M}_{\perp010}+{\mathcal M}_{\perp212})}{3\chi}\,, \nonumber\\
j^{\tilde{\gamma}}_{\perp2}=&-\tfrac{4({\mathcal M}_{\perp020}+{\mathcal M}_{\perp121})}{3\chi}\,, \ \
j^{\tilde{A}}_{00}=-\tfrac{8({\mathcal N}_{1\perp1}+{\mathcal N}_{2\perp2})}{3\chi}\,, \nonumber\\ j^{\tilde{A}}_{11}=&-\tfrac{8({\mathcal N}_{0\perp0}+{\mathcal N}_{2\perp2})}{3\chi}\,, \ \
j^{\tilde{A}}_{22}=-\tfrac{8({\mathcal N}_{0\perp0}+{\mathcal N}_{1\perp1})}{3\chi}, \ \ j^{\tilde{A}}_{01}=\tfrac{16{\mathcal N}_{(0|\perp|1)}}{3\chi}\,, \nonumber\\
j^{\tilde{A}}_{02}=&~\tfrac{16}{3\chi}{\mathcal N}_{(0|\perp|1)}\,, \ \ j^{\tilde{A}}_{12}=\tfrac{16}{3\chi}{\mathcal N}_{(1|\perp|2)}\,, \ \
j^{\tilde{A}}_{\perp0}=\tfrac{8}{3\chi}({\mathcal N}_{101}+{\mathcal N}_{202})\,, \nonumber\\
j^{\tilde{A}}_{\perp1}=&~\tfrac{8}{3\chi}({\mathcal N}_{010}+{\mathcal N}_{212})\,, \ \
j^{\tilde{A}}_{\perp2}=\tfrac{8({\mathcal N}_{020}+{\mathcal N}_{121})}{3\chi}\,,  \nonumber\\
j^K=&-\tfrac{4({\mathcal N}_{0\perp0}+{\mathcal N}_{1\perp1}+{\mathcal N}_{2\perp2})}{3}\,, \ \
j^{\hat{\Gamma}}_{\perp}=-\tfrac{4}{3}\chi({\mathcal M}_{0101}+{\mathcal M}_{0202}+{\mathcal M}_{1212})\,,  \nonumber\\
j^{\hat{\Gamma}}_0=&~\tfrac{4\chi({\mathcal M}_{\perp101}+{\mathcal M}_{\perp202})}{3}\,, \ \ 
j^{\hat{\Gamma}}_1=\tfrac{4}{3}\chi({\mathcal M}_{\perp010}+{\mathcal M}_{\perp212})\,, 
\nonumber\\
j^{\hat{\Gamma}}_2=&~\tfrac{4}{3}\chi({\mathcal M}_{\perp121}+{\mathcal M}_{\perp020})\,, \\
k^{\tilde{\gamma}}_{00}=& -\tfrac{2{\mathcal M}_{1212}}{\chi(1+b(x))}\,, \ \ k^{\tilde{\gamma}}_{11}=-\tfrac{2{\mathcal M}_{0202}}{\chi(1+b(x))}\, \ \
k^{\tilde{\gamma}}_{22}= -\tfrac{2{\mathcal M}_{0101}}{\chi(1+b(x))}\,, \nonumber\\ k^{\tilde{\gamma}}_{01}=&~\tfrac{4{\mathcal M}_{0212}}{\chi(1+b(x))}\,, \ \
k^{\tilde{\gamma}}_{02}= \tfrac{4{\mathcal M}_{0121}}{\chi(1+b(x))}\,, \ \ k^{\tilde{\gamma}}_{12}=\tfrac{4{\mathcal M}_{0102}}{\chi(1+b(x))}\,, \nonumber \\
k^{\chi}=&~2\tfrac{{\mathcal M}_{0101}+{\mathcal M}_{0202}+{\mathcal M}_{1212}}{\chi(1+b(x))}\,, \ \ k^{\tilde{\gamma}}_{\perp A}=0 \ \ \forall A=0,1,2\,, \\
m^{\tilde{\gamma}}_{AB}=&~m^{\tilde{\gamma}}_{\perp A}=m^{\chi}=0 \ \ \forall A,B=0,1,2\,, \ \
m^{\tilde{A}}_{\perp A}=0 \ \ \forall A=0,1,2\,,  \nonumber \\
m^{\tilde{A}}_{00}=&-\tfrac{8{\mathcal M}_{1212}}{\chi^2(1+b(x))}\,, \ \ m^{\tilde{A}}_{11}=-\tfrac{8{\mathcal M}_{0202}}{\chi^2(1+b(x))}\,, \ \
m^{\tilde{A}}_{22}=-\tfrac{8{\mathcal M}_{0101}}{\chi^2(1+b(x))}\,, \nonumber \\
 m^{\tilde{A}}_{01}=&~\tfrac{16{\mathcal M}_{0212}}{\chi^2(1+b(x))}\,, \ \
m^{\tilde{A}}_{02}=\tfrac{16{\mathcal M}_{0121}}{\chi^2(1+b(x))}\,, \nonumber \\
 m^{\tilde{A}}_{12}=&~\tfrac{16{\mathcal M}_{0102}}{\chi^2(1+b(x))}\,, \ \ 
 m^K=-\tfrac{{\mathcal M}_{0101}+{\mathcal M}_{0202}+{\mathcal M}_{1212}}{4\chi(1+b(x))}\,, \\
n^{\tilde{\gamma}}_{\perp A}=&~n^{\tilde{\gamma}}_{00}=0 \ \ \forall A =0,1,2\,, \ \
n^{\tilde{\gamma}}_{11}=-\tfrac{4{\mathcal N}_{202}}{\chi^2(1+b(x))}\,, \ \ n^{\tilde{\gamma}}_{22}=-\tfrac{4{\mathcal N}_{101}}{\chi^2(1+b(x))}\,, \nonumber\\
 n^{\tilde{\gamma}}_{01}=&~\tfrac{4{\mathcal N}_{212}}{\chi^2(1+b(x))}\,, \ \ n^{\tilde{\gamma}}_{02}=\tfrac{4{\mathcal N}_{121}}{\chi^2(1+b(x))}\,, \ \ 
 n^{\tilde{\gamma}}_{12}=-\tfrac{8{\mathcal N}_{0(12)}}{\chi^2(1+b(x))}\,, \nonumber \\
n^{\chi}=&~\tfrac{4({\mathcal N}_{101}+{\mathcal N}_{202})}{\chi^2(1+b(x))}\,,  \ \
n^{\tilde{A}}_{00}=0\,, \ \ n^{\tilde{A}}_{11}=\tfrac{8{\mathcal M}_{\perp202}}{\chi^2(1+b(x))}\,, \ \ 
n^{\tilde{A}}_{22}=\tfrac{8{\mathcal M}_{\perp101}}{\chi^2(1+b(x))}\,, \nonumber\\ 
n^{\tilde{A}}_{01}=&-\tfrac{8{\mathcal M}_{\perp212}}{\chi^2(1+b(x))}\,, \ \
n^{\tilde{A}}_{02}=-\tfrac{8{\mathcal M}_{\perp121}}{\chi^2(1+b(x))}\,, \ \
n^{\tilde{A}}_{12}=\tfrac{16{\mathcal M}_{\perp(12)0}}{\chi^2(1+b(x))}\,, \nonumber \\
n^{\tilde{A}}_{\perp0}=&~\tfrac{8{\mathcal M}_{1212}}{\chi^2(1+b(x))}\,, \ \ n^{\tilde{A}}_{\perp1}=-\tfrac{8{\mathcal M}_{0212}}{\chi^2(1+b(x))}\,, \nonumber\\
n^{\tilde{A}}_{\perp2}=&-\tfrac{8{\mathcal M}_{0121}}{\chi^2(1+b(x))}\,, \ \ n^K=\tfrac{2({\mathcal M}_{\perp101}+{\mathcal M}_{\perp202})}{\chi^2(1+b(x))}\,, \\
p^{\tilde{\gamma}}_{\perp A}=&~p^{\tilde{\gamma}}_{11}=0 \ \ \forall A =0,1,2\,, \ \
 p^{\tilde{\gamma}}_{00}=-\tfrac{4{\mathcal N}_{212}}{\chi^2(1+b(x))}\,, \ \ p^{\tilde{\gamma}}_{22}=-\tfrac{4{\mathcal N}_{010}}{\chi^2(1+b(x))}\,, \nonumber\\
p^{\tilde{\gamma}}_{01}=&~\tfrac{4{\mathcal N}_{202}}{\chi^2(1+b(x))}\,, \ \ p^{\tilde{\gamma}}_{02}=-\tfrac{8{\mathcal N}_{1(02)}}{\chi^2(1+b(x))}\,, \ \
p^{\tilde{\gamma}}_{12}=\tfrac{4{\mathcal N}_{020}}{\chi^2(1+b(x))}\,, \nonumber \\
p^{\chi}=&~\tfrac{4({\mathcal N}_{010}+{\mathcal N}_{212})}{\chi^2(1+b(x))}\,, \ \
p^{\tilde{A}}_{00}=\tfrac{8{\mathcal M}_{\perp212}}{\chi^2(1+b(x))}\,, \ \ p^{\tilde{A}}_{11}=0\,, \ \ 
p^{\tilde{A}}_{22}=\tfrac{8{\mathcal M}_{\perp010}}{\chi^2(1+b(x))}\,, \nonumber\\
p^{\tilde{A}}_{01}=&-\tfrac{8{\mathcal M}_{\perp202}}{\chi^2(1+b(x))}\,, \ \ p^{\tilde{A}}_{02}=\tfrac{16{\mathcal M}_{\perp(02)1}}{\chi^2(1+b(x))}\,, \ \
p^{\tilde{A}}_{12}=-\tfrac{8{\mathcal M}_{\perp020}}{\chi^2(1+b(x))}\,,  \nonumber\\
p^{\tilde{A}}_{\perp0}=&-\tfrac{8{\mathcal M}_{0212}}{\chi^2(1+b(x))}\,, \ \ p^{\tilde{A}}_{\perp1}=\tfrac{8{\mathcal M}_{0202}}{\chi^2(1+b(x))}\,, \nonumber\\
p^{\tilde{A}}_{\perp2}=&-\tfrac{8{\mathcal M}_{0102}}{\chi^2(1+b(x))}\,, \ \
p^K=\tfrac{2({\mathcal M}_{\perp010}+{\mathcal M}_{\perp212})}{\chi^2(1+b(x))}\,, \\ 
q^{\tilde{\gamma}}_{\perp A}=&~q^{\tilde{\gamma}}_{22}=0 \ \ \forall A =0,1,2\,, \ \
q^{\tilde{\gamma}}_{00}=-\tfrac{4{\mathcal N}_{121}}{\chi^2(1+b(x))}\,, \ \ q^{\tilde{\gamma}}_{11}=-\tfrac{4{\mathcal N}_{020}}{\chi^2(1+b(x))}\,, \nonumber \\
q^{\tilde{\gamma}}_{01}=&-\tfrac{8{\mathcal N}_{2(01)}}{\chi^2(1+b(x))}\,, \ \ q^{\tilde{\gamma}}_{02}=\tfrac{4{\mathcal N}_{101}}{\chi^2(1+b(x))}\,, \ \
q^{\tilde{\gamma}}_{12}=\tfrac{4{\mathcal N}_{010}}{\chi^2(1+b(x))}\,, \nonumber \\
q^{\chi}=&~\tfrac{4({\mathcal N}_{020}+{\mathcal N}_{121})}{\chi^2(1+b(x))}\,, \ \
q^{\tilde{A}}_{00}=\tfrac{8{\mathcal M}_{\perp121}}{\chi^2(1+b(x))}\,, \ \
q^{\tilde{A}}_{11}=\tfrac{8{\mathcal M}_{\perp020}}{\chi^2(1+b(x))}\,, \ \ q^{\tilde{A}}_{22}=0\,, \nonumber\\
q^{\tilde{A}}_{01}=&~\tfrac{16{\mathcal M}_{\perp(10)2}}{\chi^2(1+b(x))}\,, \ \ q^{\tilde{A}}_{02}=-\tfrac{8{\mathcal M}_{\perp101}}{\chi^2(1+b(x))}\,, \ \ q^{\tilde{A}}_{12}=-\tfrac{8{\mathcal M}_{\perp010}}{\chi^2(1+b(x))}\,,  \nonumber\\ 
q^{\tilde{A}}_{\perp0}=&-\tfrac{8{\mathcal M}_{0121}}{\chi^2(1+b(x))}\,, \ \ q^{\tilde{A}}_{\perp1}=-\tfrac{8{\mathcal M}_{0102}}{\chi^2(1+b(x))}\,, \nonumber \\
q^{\tilde{A}}_{\perp2}=&~\tfrac{8{\mathcal M}_{0101}}{\chi^2(1+b(x))}\,, \ \ q^K=\tfrac{2({\mathcal M}_{\perp020}+{\mathcal M}_{\perp121})}{\chi^2(1+b(x))}\,.
\end{align}
\end{subequations}

\section{Four-Derivative Scalar Tensor theory of gravity}\label{app:matEsGB}

\begin{table}[ht]
\centering
\small
\tabcolsep=0.07cm
\scalebox{0.8}{\begin{tabular}{c|c c c c c c c c c  c c c c c c c  c c c c c  c c c c }
& $i\hat{\tilde{\gamma}}_{00}$ & $i\hat{\tilde{\gamma}}_{11}$ & $i\hat{\tilde{\gamma}}_{01}$ & $i\hat{\tilde{\gamma}}_{\perp0}$ & $i\hat{\tilde{\gamma}}_{\perp1}$ & $i\hat{\chi}$& $i\hat{{\tilde{A}}}_{00}$ & $i\hat{{\tilde{A}}}_{11}$ & $i\hat{{\tilde{A}}}_{01}$ & $i\hat{{\tilde{A}}}_{\perp0}$ & $i\hat{{\tilde{A}}}_{\perp1}$ & $i\hat{K}$ 
& $i\hat{\hat{\Gamma}}_{\perp}$ & $i\hat{\hat{\Gamma}}_0$ & $i\hat{\hat{\Gamma}}_1$ & $i\hat{\phi}$ & $i\hat{K}_{\phi}$ \\
\hline
$i\check{\xi}_0\hat{{\tilde{A}}}_{00}$ & $im_{00}^{\tilde{\gamma}}$ & $im_{11}^{\tilde{\gamma}}$ & $im_{01}^{\tilde{\gamma}}$ & $im_{\perp0}^{\tilde{\gamma}}$ & $im_{\perp1}^{\tilde{\gamma}}$ & $im^{\chi}$ & $m_{00}^{\tilde{A}}$ & $m_{11}^{\tilde{A}}$ & $m_{01}^{\tilde{A}}$ & $m_{\perp0}^{\tilde{A}}$ & $m_{\perp1}^{\tilde{A}}$ & $m^K$ &  $m_{\perp}^{\hat{\Gamma}}$ & $m_0^{\hat{\Gamma}}$ & $m_1^{\hat{\Gamma}}$ & $if'm^{\phi}$ & $f'm^{K_{\phi}}$\\
$i\check{\xi}_0\hat{{\tilde{A}}}_{11}$ & $in_{00}^{\tilde{\gamma}}$ & $in_{11}^{\tilde{\gamma}}$ & $in_{01}^{\tilde{\gamma}}$ & $in_{\perp0}^{\tilde{\gamma}}$ & $in_{\perp1}^{\tilde{\gamma}}$ & $in^{\chi}$ & $n_{00}^{\tilde{{\tilde{A}}}}$ & $n_{11}^{\tilde{A}}$ & $n_{01}^{\tilde{A}}$ & $n_{\perp0}^{\tilde{A}}$ & $n_{\perp1}^{\tilde{A}}$  & $n^K$  & $n_{\perp}^{\hat{\Gamma}}$ & $n_0^{\hat{\Gamma}}$ & $n_1^{\hat{\Gamma}}$ & $if'n^{\phi}$ & $f'n^{K_{\phi}}$\\
$i\check{\xi}_0\hat{{\tilde{A}}}_{01}$ & $ip^{\tilde{\gamma}}_{00}$ & $ip^{\tilde{\gamma}}_{11}$ & $ip^{\tilde{\gamma}}_{01}$ & $ip^{\tilde{\gamma}}_{\perp0}$ & $ip^{\tilde{\gamma}}_{\perp1}$ & $ip^{\chi}$ & $p^{\tilde{A}}_{00}$ & $p^{\tilde{A}}_{11}$ & $p^{\tilde{A}}_{01}$ & $p^{\tilde{A}}_{\perp0}$ & $p^K_{\perp1}$  & $p^K$  & $p^{\hat{\Gamma}}_{\perp}$ & $p^{\hat{\Gamma}}_0$ & $p^{\hat{\Gamma}}_1$ & $if'p^{\phi}$ & $f'p^{K_{\phi}}$\\
$i\check{\xi}_0\hat{A}_{\perp0}$ & $iq_{00}^{\tilde{\gamma}}$ & $iq_{11}^{\tilde{\gamma}}$ & $iq_{01}^{\tilde{\gamma}}$ & $iq_{\perp0}^{\tilde{\gamma}}$ & $iq_{\perp1}^{\tilde{\gamma}}$ & $iq^{\chi}$ & $q_{00}^{\tilde{A}}$ & $q_{11}^{\tilde{A}}$ & $q_{01}^{\tilde{A}}$ & $q_{\perp0}^{\tilde{A}}$ & $q_{\perp1}^{\tilde{A}}$ & $q^K$   & $q_{\perp}^{\hat{\Gamma}}$ & $q_0^{\hat{\Gamma}}$ & $q_1^{\hat{\Gamma}}$ & $if'q^{\phi}$ & $f'q^{K_{\phi}}$\\
$i\check{\xi}_0\hat{A}_{\perp1}$ & $ir_{00}^{\tilde{\gamma}}$ & $ir_{11}^{\tilde{\gamma}}$ & $ir_{01}^{\tilde{\gamma}}$ & $ir_{\perp0}^{\tilde{\gamma}}$ & $ir_{\perp1}^{\tilde{\gamma}}$ & $ir^{\chi}$ & $r_{00}^{\tilde{A}}$ & $r_{11}^{\tilde{A}}$ & $r_{01}^{\tilde{A}}$ & $r_{\perp0}^{\tilde{A}}$ & $r_{\perp1}^{\tilde{A}}$ & $r^K$   & $r_{\perp}^{\hat{\Gamma}}$ & $r_0^{\hat{\Gamma}}$ & $r_1^{\hat{\Gamma}}$ & $if'r^{\phi}$ & $f'r^{K_{\phi}}$\\
$i\check{\xi}_0\hat{K}$ & $is_{00}^{\tilde{\gamma}}$ & $is_{11}^{\tilde{\gamma}}$ & $is_{01}^{\tilde{\gamma}}$ & $is_{\perp0}^{\tilde{\gamma}}$ & $is_{\perp1}^{\tilde{\gamma}}$ & $is^{\chi}$ & $s_{00}^{\tilde{A}}$ & $s_{11}^{\tilde{A}}$ & $s_{01}^{\tilde{A}}$ & $s_{\perp0}^{\tilde{A}}$ & $s_{\perp1}^{\tilde{A}}$ & $s^K$  & $s_{\perp}^{\hat{\Gamma}}$ & $s_0^{\hat{\Gamma}}$ & $s_1^{\hat{\Gamma}}$ & $if's^{\phi}$ & $f's^{K_{\phi}}$\\
$i\check{\xi}_0\hat{\Theta}$  & $it_{00}^{\tilde{\gamma}}$ & $it_{11}^{\tilde{\gamma}}$ & $it_{01}^{\tilde{\gamma}}$ & $it_{\perp0}^{\tilde{\gamma}}$ & $it_{\perp1}^{\tilde{\gamma}}$ & $it^{\chi}$ & 0 & 0 & 0 & 0 & 0 & 0 & 0 & 0 & 0 & $if't^{\phi}$ & 0 \\
$i\check{\xi}_0\hat{\hat{\Gamma}}_{\perp}$ & $iv_{00}^{\tilde{\gamma}}$ & $iv_{11}^{\tilde{\gamma}}$ & $iv_{01}^{\tilde{\gamma}}$ & $iv_{\perp0}^{\tilde{\gamma}}$ & $iv_{\perp1}^{\tilde{\gamma}}$ & $iv^{\chi}$ & $v_{00}^{\tilde{A}}$ & $v_{11}^{\tilde{A}}$ & $v_{01}^{\tilde{A}}$ & $v_{\perp0}^{\tilde{A}}$ & $v_{\perp1}^{\tilde{A}}$ & $v^K$  & 0 & 0 & 0 & $if'v^{\phi}$ & $f'v^{K_{\phi}}$ \\
$i\check{\xi}_0\hat{\hat{\Gamma}}_0$ & $iw_{00}^{\tilde{\gamma}}$ & $iw_{11}^{\tilde{\gamma}}$ & $iw_{01}^{\tilde{\gamma}}$ & $iw_{\perp0}^{\tilde{\gamma}}$ & $iw_{\perp1}^{\tilde{\gamma}}$ & $iw^{\chi}$ & $w_{00}^{\tilde{A}}$ & $w_{11}^{\tilde{A}}$ & $w_{01}^{\tilde{A}}$ & $w_{\perp0}^{\tilde{A}}$ & $w_{\perp1}^{\tilde{A}}$ & $w^K$  & 0 & 0 & 0 & $if'w^{\phi}$ & $f'w^{K_{\phi}}$\\
$i\check{\xi}_0\hat{\hat{\Gamma}}_1$ & $ix_{00}^{\tilde{\gamma}}$ & $ix_{11}^{\tilde{\gamma}}$ & $ix_{01}^{\tilde{\gamma}}$ & $ix_{\perp0}^{\tilde{\gamma}}$ & $ix_{\perp1}^{\tilde{\gamma}}$ & $ix^{\chi}$ & $x_{00}^{\tilde{A}}$ & $x_{11}^{\tilde{A}}$ & $x_{01}^{\tilde{A}}$ & $x_{\perp0}^{\tilde{A}}$ & $x_{\perp1}^{\tilde{A}}$ & $x^K$  & 0 & 0 & 0 & $if'x^{\phi}$ & $f'x^{K_{\phi}}$ \\
$i\check{\xi}_0\hat{K}_{\phi}$ & $if'y_{00}^{\tilde{\gamma}}$ & $if'y_{11}^{\tilde{\gamma}}$ & $if'y_{01}^{\tilde{\gamma}}$ & $if'y_{\perp0}^{\tilde{\gamma}}$ & $if'y_{\perp1}^{\tilde{\gamma}}$ & $if'y^{\chi}$ & $f'y_{00}^{\tilde{A}}$ & $f'y_{11}^{\tilde{A}}$ & $f'y_{01}^{\tilde{A}}$ & $f'y_{\perp0}^{\tilde{A}}$ & $f'y_{\perp1}^{\tilde{A}}$ & $f'y^K$ & $f'y_{\perp}^{\hat{\Gamma}}$ & $f'y_0^{\hat{\Gamma}}$ & $f'y_1^{\hat{\Gamma}}$ & 0 & 0  
\end{tabular}}
\caption[Hyperbolicity matrix in Einstein-scalar-Gauss-Bonnet theory of gravity.]{Perturbative hyperbolicity matrix ${\mathbb M}^{\text{GB}}$ in $3+1$ Einstein-scalar-Gauss-Bonnet theory of gravity.}\label{tab:matEsGB}
\end{table}
The elements of the matrix in Table \ref{tab:matEsGB} yield as follows,
 where we have used that $W_{\perp}=\xi^i\Omega_i$, $W_A=e_A^i\Omega_i$, $W_{\perp \perp}=\xi^i\xi^j\Omega_{ij}$, $W_{\perp A}=e_A^i\xi^j\Omega_{ij}$, $W_{AB}=e_A^ie_B^j\Omega_{ij}$, ${\mathcal M}_{\perp\perp}=\xi^i\xi^jM_{ij}$, ${\mathcal M}_{\perp A}=\xi^ie_A^jM_{ij}$, ${\mathcal M}_{AB}=e_A^ie_B^jM_{ij}$, ${\mathcal N}_{\perp}=\xi^iN_i$, ${\mathcal N}_A=e_A^iN_i$, ${\mathcal N}_{\perp AB}=\xi^ie_A^je_B^kN_{ijk}$) and $F_{AB}=e_A^ie_B^j\left({\mathcal L}_nK_{ij} +\frac{1}{\alpha}D_iD_j\alpha+K_{im}K^m_{~j}+M_{ij}\right)$:
\begin{subequations}
\begin{align}
m_{00}^{\tilde{\gamma}}=&~\tfrac{1+2b(x)}{6(1+b(x))}W_{00}-\tfrac{6+7b(x)}{12(1+b(x))}W_{11}
-\tfrac{2+3b(x)}{12(1+b(x))}W_{\perp\perp}-\tfrac{n^{\mu}n^{\nu}{\mathcal C}_{\mu\nu}}{6\lambda^{\text{GB}}}\,, \nonumber\\ m_{11}^{\tilde{\gamma}}=&~\tfrac{b(x)}{6(1+b(x))}W_{00}-\tfrac{4+5b(x)}{12(1+b(x))}W_{11}+\tfrac{4+3b(x)}{12(1+b(x))}W_{\perp\perp}+\tfrac{n^{\mu}n^{\nu}{\mathcal C}_{\mu\nu}}{3\lambda^{\text{GB}}}\,, \nonumber\\ m_{01}^{\tilde{\gamma}}=&~\tfrac{W_{01}}{3}\,, \ \
 m_{\perp0}^{\tilde{\gamma}}=\tfrac{W_{\perp0}}{3}\,, \ \ m_{\perp1}^{\tilde{\gamma}}=-\tfrac{2W_{\perp1}}{3}\,, \nonumber \\
m^{\chi}=&~\tfrac{3+b(x)}{6(1+b(x))}W_{00}-\tfrac{3+2b(x)}{6(1+b(x))}W_{11}-\tfrac{W_{\perp\perp}}{6(1+b(x))}-\tfrac{n^{\mu}n^{\nu}{\mathcal C}_{\mu\nu}}{6\lambda^{\text{GB}}}\,,\nonumber\\
m_{00}^{\tilde{A}}=&~\tfrac{2W_{\perp}}{3}\,, \ \ m_{11}^{\tilde{A}}=-\tfrac{4W_{\perp}}{3}\,, \ \ m_{01}^{\tilde{A}}=0\,, \ \
m_{\perp0}^{\tilde{A}}=-\tfrac{2W_0}{3}\,, \ \ m_{\perp1}^{\tilde{A}}=\tfrac{4W_1}{3}\,, \nonumber\\
 m^{K}=&-\tfrac{2\chi W_{\perp}}{9}\,,\ \ m_{\perp}^{\tilde{\Gamma}}=\tfrac{\chi^2(W_{00}-2W_{11})}{3}\,, \ \ m_0^{\tilde{\Gamma}}=-\tfrac{\chi^2W_{\perp0}}{3}\,, \ \ m_1^{\tilde{\Gamma}}=\tfrac{2\chi^2W_{\perp1}}{3}\,,\nonumber \\
 m^{\phi}=&~\xi+\chi F_{00}\,, \ \ m^{K_{\phi}}=\tfrac{\chi}{3}\left({\mathcal N}_{\perp}+6{\mathcal N}_{\perp00} \right)\,,\\ 
n_{00}^{\tilde{\gamma}}=&-\tfrac{4+5b(x)}{12(1+b(x))}W_{00}+\tfrac{b(x)}{6(1+b(x))}W_{11}+\tfrac{4+3b(x)}{12(1+b(x))}W_{\perp\perp}+\tfrac{n^{\mu}n^{\nu}{\mathcal C}_{\mu\nu}}{3\lambda^{\text{GB}}}\,, \nonumber\\ n_{11}^{\tilde{\gamma}}=&-\tfrac{6+7b(x)}{12(1+b(x))}W_{00}+\tfrac{1+2b(x)}{6(1+b(x))}W_{11}-\tfrac{2+3b(x)}{12(1+b(x))}W_{\perp\perp}-\tfrac{n^{\mu}n^{\nu}{\mathcal C}_{\mu\nu}}{6\lambda^{\text{GB}}}\,, \nonumber\\n_{01}^{\tilde{\gamma}}=&~\tfrac{W_{01}}{3}\,, \ \ n_{\perp0}^{\tilde{\gamma}}=-\tfrac{2W_{\perp0}}{3}\,, \ \ n_ {\perp1}^{\tilde{\gamma}}=\tfrac{W_{\perp1}}{3}\,, \nonumber \\
n^{\chi}=&-\tfrac{3+2b(x)}{6(1+b(x))}W_{00}+\tfrac{3+b(x)}{6(1+b(x))}W_{11}-\tfrac{W_{\perp\perp}}{6(1+b(x))}-\tfrac{n^{\mu}n^{\nu}{\mathcal C}_{\mu\nu}}{6\lambda^{\text{GB}}}\,,\nonumber\\
n_{00}^{\tilde{A}}=&-\tfrac{4W_{\perp}}{3}\,, \ \ n_{11}^{\tilde{A}}=\tfrac{2W_{\perp}}{3}\,, \ \ n_{01}^{\tilde{A}}=0\,, \ \
n_{\perp0}^{\tilde{A}}=\tfrac{4W_0}{3}\,, \ \ n_ {\perp1}^{\tilde{A}}=-\tfrac{2W_1}{3}\,, \nonumber\\
 n^{K}=&-\tfrac{2\chi W_{\perp}}{9}\,, \ \
 n_{\perp}^{\tilde{\Gamma}}=\tfrac{\chi^2(-2W_{00}+W_{11})}{3}\,, \ \ n_0^{\tilde{\Gamma}}=\tfrac{2\chi^2W_{\perp0}}{3}\,, \ \ n_1^{\tilde{\Gamma}}=-\tfrac{\chi^2W_{\perp1}}{3}\,,\nonumber \\
 n^{\phi}=&~\xi+\chi F_{11}\,, \ \ n^{K_{\phi}}=\tfrac{\chi}{3}\left({\mathcal N}_{\perp}+6{\mathcal N}_{\perp11} \right)\,,\\ 
p_{00}^{\tilde{\gamma}}=&~p_{11}^{\tilde{\gamma}}=\tfrac{b(x)W_{01}}{4(1+b(x))}, \ \ p_{01}^{\tilde{\gamma}}=-\tfrac{W_{\perp\perp}}{2}-\tfrac{n^{\mu}n^{\nu}{\mathcal C}_{\mu\nu}}{2\lambda^{\text{GB}}}\,, \ \ p_{\perp0}^{\tilde{\gamma}}=\tfrac{W_{\perp1}}{2}\,, \ \ p_{\perp1}^{\tilde{\gamma}}=\tfrac{W_{\perp0}}{2}\,, \nonumber\\ p^{\chi}=&~\tfrac{4+3b(x)}{2(1+b(x))}W_{01}\,, \ \
p_{00}^{\tilde{A}}=p_{11}^{\tilde{A}}=0\,, \ \ p_{01}^{\tilde{A}}=2W_{\perp}\,, \ \ p_{\perp0}^{\tilde{A}}=-W_1\,, \ \ p_{\perp1}^{\tilde{A}}=-W_0\,, \nonumber\\ p^K=&~0\,, \ \
p_{\perp}^{\tilde{\Gamma}}=\chi^2W_{01}\,, \ \ p_0^{\tilde{\Gamma}}=-\tfrac{\chi^2W_{\perp1}}{2}\,, \ \ p_1^{\tilde{\Gamma}}=-\tfrac{\chi^2W_{\perp0}}{2}\,, \nonumber\\
p^{\phi}=&~\chi F_{01}\,, \ \ p^{K_{\phi}}=2\chi{\mathcal N}_{\perp(01)}\,,\\
q_{00}^{\tilde{\gamma}}=&~\tfrac{b(x)W_{\perp0}}{4(1+b(x))} , \ \ q_{11}^{\tilde{\gamma}}=-\tfrac{2+b(x)}{4(1+b(x))}W_{\perp0}, \ \ q_{01}^{\tilde{\gamma}}=\tfrac{W_{\perp1}}{2} , \ \ q_{\perp0}^{\tilde{\gamma}}=-\tfrac{W_{11}}{2}\,,\nonumber \\ 
q_{\perp1}^{\tilde{\gamma}}=&~\tfrac{W_{01}}{2}\,, \ \ q^{\chi}=-\tfrac{W_{\perp0}}{2}\,, \ \
q_{00}^{\tilde{A}}=0\,, \ \ q_{11}^{\tilde{A}}=W_0, \ \ q_{01}^{\tilde{A}}=-W_1\,, \ \ q_{\perp0}^{\tilde{A}}=q_{\perp1}^{\tilde{A}}=0\,, \nonumber\\ q^K=&~\tfrac{\chi W_0}{3}\,, \ \
q_{\perp}^{\tilde{\Gamma}}=0\, \ \ q_0^{\tilde{\Gamma}}=\tfrac{\chi^2W_{11}}{2}\,, \ \ q_1^{\tilde{\Gamma}}=-\tfrac{\chi^2W_{01}}{2}\,, \nonumber\\
q^{\phi}=&~\chi {\mathcal M}_{\perp0}\,, \ \ q^{K_{\phi}}=-\chi\left({\mathcal N}_0-2{\mathcal N}_{\perp\perp0}\right)\,, \\
r_{00}^{\tilde{\gamma}}=&-\tfrac{2+b(x)}{4(1+b(x))}W_{\perp1}\,, \ \ r_{11}^{\tilde{\gamma}}=\tfrac{b(x)W_{\perp1}}{4(1+b(x))}\,, \ \ r_{01}^{\tilde{\gamma}}=\tfrac{W_{\perp0}}{2}\,, \ \ r_{\perp0}^{\tilde{\gamma}}=\tfrac{W_{01}}{2}\,, \nonumber\\ r_{\perp1}^{\tilde{\gamma}}=&-\tfrac{W_{00}}{2}\,, \ \ r^{\chi}=-\tfrac{W_{\perp1}}{2}\,, \ \
r_{00}^{\tilde{A}}=W_1\,, \ \ r_{11}^{\tilde{A}}=0\,, \ \ r_{01}^{\tilde{A}}=-W_0\,, \ \ r_{\perp0}^{\tilde{A}}=r_{\perp1}^{\tilde{A}}=0\,, \nonumber\\ r^K=&~\tfrac{\chi W_1}{3}\,, \ \
r_{\perp}^{\tilde{\Gamma}}=0\,, \ \ r_0^{\tilde{\Gamma}}=-\tfrac{\chi^2W_{01}}{2}\,, \ \ r_1^{\tilde{\Gamma}}=\tfrac{\chi^2W_{00}}{2}\,, \nonumber\\
r^{\phi}=&~\chi {\mathcal M}_{\perp1}\,, \ \ r^{K_{\phi}}=-\chi\left({\mathcal N}_1-2{\mathcal N}_{\perp\perp1}\right)\,, \\
s_{00}^{\tilde{\gamma}}=&~\tfrac{(1+2b(x))W_{00}+(3+b(x))W_{11}-W_{\perp\perp}}{4\chi(1+b(x))}-\tfrac{n^{\mu}n^{\nu}{\mathcal C}_{\mu\nu}}{4\lambda^{\text{GB}}\chi}\,, \ \ s_{\perp0}^{\tilde{A}}=-\tfrac{W_0}{\chi}\,,\nonumber\\ s_{11}^{\tilde{\gamma}}=&~\tfrac{(3+b(x))W_{00}+(1+2b(x))W_{11}-W_{\perp\perp}}{4\chi(1+b(x))}-\tfrac{n^{\mu}n^{\nu}{\mathcal C}_{\mu\nu}}{4\lambda^{\text{GB}}\chi}\,, \ \ s_{\perp1}^{\tilde{\gamma}}=\tfrac{W_{\perp1}}{2\chi}\,, \nonumber\\ s_{01}^{\tilde{\gamma}}=&~\tfrac{-2+b(x)}{2\chi(1+b(x))}W_{01}\,, \ \
s^{\chi}=\tfrac{b(x)\left(W_{00}+W_{11}\right)+2W_{\perp\perp}}{4\chi(1+b(x))}+\tfrac{n^{\mu}n^{\nu}{\mathcal C}_{\mu\nu}}{2\lambda^{\text{GB}}\chi}\,, \nonumber \\
s_{00}^{\tilde{A}}=&~s_{11}^{\tilde{A}}=\tfrac{W_{\perp}}{\chi}\,, \ \ s_{01}^{\tilde{A}}=0\,, \ \ s_{\perp0}^{\tilde{A}}=-\tfrac{W_0}{\chi}\,, \ \ s_{\perp1}^{\tilde{A}}=-\tfrac{W_1}{\chi}\,, \ \ s^K=\tfrac{2W_{\perp}}{3}\,,\nonumber\\
s_{\perp}^{\tilde{\Gamma}}=&~\tfrac{\chi}{2}\left(W_{00}+W_{11}\right)\,, \ \ s_0^{\tilde{\Gamma}}=-\tfrac{\chi W_{\perp0}}{2}\,, \ \ s_1^{\tilde{\Gamma}}=-\tfrac{\chi W_{\perp1}}{2}\,, \nonumber\\
s^{\phi}=&~\tfrac{F_{00}+F_{11}}{2}-\tfrac{(4+b(x))({\mathcal M}_{00}+{\mathcal M}_{11}-{\mathcal M}_{\perp\perp})}{4(1+b(x))}\,, \ \ s^{K_{\phi}}=-{\mathcal N}_{\perp}\,, \\
t_{00}^{\tilde{\gamma}}=&~\tfrac{W_{11}}{2\chi(1+b(x))}\,, \ \ t_{11}^{\tilde{\gamma}}=\tfrac{W_{00}}{2\chi(1+b(x))}\,, \ \ t_{01}^{\tilde{\gamma}}=-\tfrac{W_{01}}{\chi(1+b(x))}\,, \ \ t_{\perp0}^{\tilde{\gamma}}=t_{\perp1}^{\tilde{\gamma}}=0\,, \nonumber\\ t^{\chi}=&-\tfrac{W_{00}+W_{11}}{2\chi(1+b(x))},\ \
t^{\phi}=-\tfrac{{\mathcal M}_{00}+{\mathcal M}_{11}-{\mathcal M}_{\perp\perp}}{2(1+b(x))}\,, \\
v_{00}^{\tilde{\gamma}}=&~v_{11}^{\tilde{\gamma}}=v_{01}^{\tilde{\gamma}}=v_{\perp0}^{\tilde{\gamma}}=v_{\perp1}^{\tilde{\gamma}}=v^{\chi}=0\,, \ \
v_{00}^{\tilde{A}}=\tfrac{2W_{11}}{\chi^2(1+b(x))}\,, \ \ v_{11}^{\tilde{A}}=\tfrac{2W_{00}}{\chi^2(1+b(x))}\,, \nonumber\\ v_{01}^{\tilde{A}}=&-\tfrac{4W_{01}}{\chi^2(1+b(x))}\,, \ \ v_{\perp0}^{\tilde{A}}=v_{\perp1}^{\tilde{A}}=0\,, \ \ v^K=\tfrac{2(W_{00}+W_{11})}{3\chi(1+b(x))}\,, \nonumber\\
v^{\phi}=&~0\,, \ \ v^{K_{\phi}}=-\tfrac{{\mathcal M}_{00}+{\mathcal M}_{11}-{\mathcal M}_{\perp\perp}}{\chi(1+b(x))}\,, \\
w_{00}^{\tilde{\gamma}}=&~0\,, \ \ w_{11}^{\tilde{\gamma}}=\tfrac{W_0}{\chi^2(1+b(x))}\,, \ \ w_{01}^{\tilde{\gamma}}=-\tfrac{W_1}{\chi^2(1+b(x))}\,, \ \
w_{\perp0}^{\tilde{\gamma}}=w_{\perp1}^{\tilde{\gamma}}=0\,, \nonumber \\ w^{\chi}=&-\tfrac{W_0}{\chi^2(1+b(x))}\,, \ \
w_{00}^{\tilde{A}}=0\,, \ \ w_{11}^{\tilde{A}}=-\tfrac{2W_{\perp0}}{\chi^2(1+b(x))}\,, \ \ 
w_{01}^{\tilde{A}}=\tfrac{2W_{\perp1}}{\chi^2(1+b(x))}\,, \nonumber\\ w_{\perp0}^{\tilde{A}}=&~\tfrac{-2W_{11}}{\chi^2(1+b(x))}, \ \
w_{\perp1}=\tfrac{2W_{01}}{\chi^2(1+b(x))}\,, \ \ w^K=\tfrac{-2W_{\perp0}}{3\chi(1+b(x))}, \nonumber\\
w^{\phi}=&-\tfrac{{\mathcal N}_0}{\chi(1+b(x))}\,, \ \ w^{K_{\phi}}=\tfrac{2{\mathcal M}_{\perp0}}{\chi(1+b(x))}\,, \\
x_{00}^{\tilde{\gamma}}=&~\tfrac{W_1}{\chi^2(1+b(x))}\,, \ \ x_{11}^{\tilde{\gamma}}=0, \ \ x_{01}^{\tilde{\gamma}}=-\tfrac{W_0}{\chi^2(1+b(x))}\,, \ \
x_{\perp0}^{\tilde{\gamma}}=x_{\perp1}^{\tilde{\gamma}}=0\,, \nonumber\\ x^{\chi}=&-\tfrac{W_1}{\chi^2(1+b(x))}\,, \ \
x_{00}^{\tilde{A}}=-\tfrac{2W_{\perp1}}{\chi^2(1+b(x))}\,, \ \ x_{11}^{\tilde{A}}=0\,, \ \ 
x_{01}^{\tilde{A}}=\tfrac{2W_{\perp0}}{\chi^2(1+b(x))}\,, \nonumber\\ x_{\perp0}^{\tilde{A}}=&~\tfrac{2W_{01}}{\chi^2(1+b(x))}\,, \ \
x_{\perp1}^{\tilde{A}}=\tfrac{-2W_{00}}{\chi^2(1+b(x))}\,, \ \ x^K=\tfrac{-2W_{\perp1}}{3\chi(1+b(x))}\,, \nonumber\\
x^{\phi}=&-\tfrac{{\mathcal N}_1}{\chi(1+b(x))}\,, \ \ x^{K_{\phi}}=\tfrac{2{\mathcal M}_{\perp1}}{\chi(1+b(x))}\,, \\
y_{00}^{\tilde{\gamma}}-&y_{11}^{\tilde{\gamma}}=\tfrac{F_{00}-F_{11}}{\chi}\,, \ \ y_{01}^{\tilde{\gamma}}=\tfrac{2F_{01}}{\chi}\,, \nonumber\\
y_{00}^{\tilde{A}}-&y_{11}^{\tilde{A}}=\tfrac{4}{\chi}\left({\mathcal N}_{\perp00}-{\mathcal N}_{\perp11}  \right)\,, \ \ 
y_{01}^{\tilde{A}}=\tfrac{8}{\chi} {\mathcal N}_{\perp(01)}\,.
\end{align}
\end{subequations}

\chapter{Convergence} \label{App:convergence}

Here we present an example of our convergence tests for the phase difference between the two polarisations of the
$(2, 2)$ mode of the strain. This is a rather standard test in NR. We show the example of an equal-mass binary black holes merger in the $4\partial$ST theory
presented in Chapter \ref{C:BBHin4dST} with the coupling constants $\lambda^{\text{GB}}/M^2=0.05$ and $g_2/M^2=1$ respectively. This example is not in the weakly coupled regime and hence showing convergence in this case is non-trivial.

\begin{figure}[H]
\centering
\includegraphics[scale=0.7]{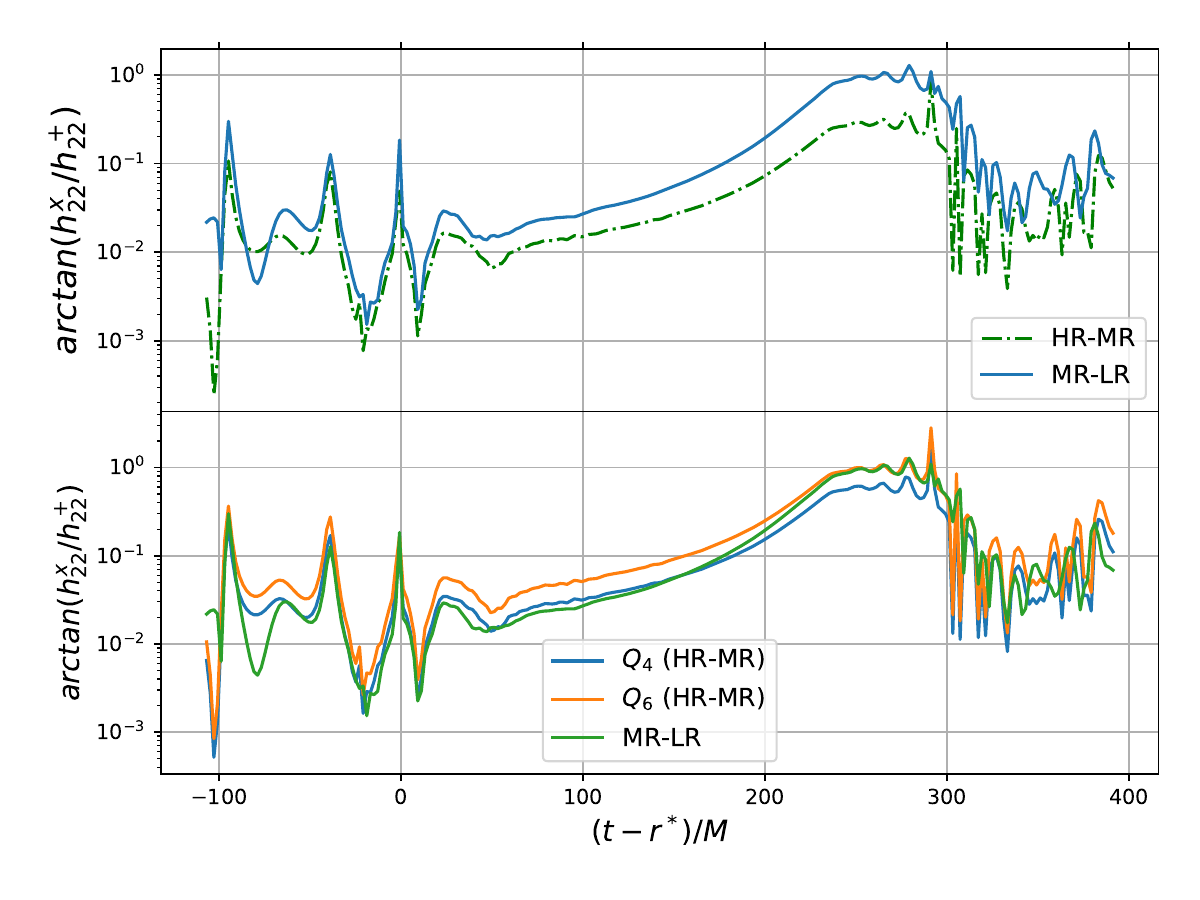}
\caption[Convergence test for an equal-mass BBH merger in the $4\partial$ST theory.]{\textit{Top:} Differences of across resolutions of the phase difference between the two polarisations of the $(l,m)=(2,2)$ mode of the strain. \textit{Bottom:} Same as in the panel above but now the difference between the high and the medium resolution runs has been rescaled by the convergence factor $Q_n\equiv\frac{\Delta_{HR}^n-\Delta_{MR}^n}{\Delta_{MR}^n-\Delta_{LR}^n}$ assuming fourth (blue) and sixth (orange) order convergence.}
\label{F:convergence}
\end{figure}

We consider three runs on a computational domain of fixed size $\Delta=512M$ and three different resolutions
on the coarsest level with grid spacings $h_{\text{LR}}=\Delta/96$, $h_{MR}=\Delta/128$ and $h_{HR}=\Delta/160$
respectively. For each
of these three runs, we added the same number of refinement levels, namely 8 (so the total number of levels is
9). The results presented in the main text were obtained
with the medium resolution.
In Fig. \ref{F:convergence} we show the differences between resolutions for
the phase difference between the two polarisations of the
$(l, m) = (2, 2)$ mode of the strain. This figure shows that
during the inspiral and merger phases of the binary, the
convergence order is around four and it increases to six
after the merger, which is consistent with the order of
the finite difference stencils used. This mild over-convergence that \texttt{GRChombo} exhibits in the phase was already observed in the detailed studies that \cite{Radia:2021smk} carried out. The
results of convergence analysis presented in this section
indicate that our simulations are stable and in the convergent regime.

\backmatter 

\renewcommand\bibname{References}
\bibliography{InitialFinalMaterials/Bibliography.bib}{}

\begin{thebibliography}{100}

\bibitem{apocrita}
Thomas King, Simon Butcher, and Lukasz Zalewski.
\newblock {\em {Apocrita - High Performance Computing Cluster for Queen Mary
  University of London}}, March 2017.

\bibitem{LIGOScientific:2016aoc}
B.~P. Abbott et~al.
\newblock {Observation of Gravitational Waves from a Binary Black Hole Merger}.
\newblock {\em Phys. Rev. Lett.}, 116(6):061102, 2016.

\bibitem{LIGOScientific:2016lio}
B.~P. Abbott et~al.
\newblock {Tests of general relativity with GW150914}.
\newblock {\em Phys. Rev. Lett.}, 116(22):221101, 2016.
\newblock [Erratum: Phys.Rev.Lett. 121, 129902 (2018)].

\bibitem{LIGOScientific:2016dsl}
B.~P. Abbott et~al.
\newblock {Binary Black Hole Mergers in the first Advanced LIGO Observing Run}.
\newblock {\em Phys. Rev. X}, 6(4):041015, 2016.
\newblock [Erratum: Phys.Rev.X 8, 039903 (2018)].

\bibitem{LIGOScientific:2017vwq}
B.~P. Abbott et~al.
\newblock {GW170817: Observation of Gravitational Waves from a Binary Neutron
  Star Inspiral}.
\newblock {\em Phys. Rev. Lett.}, 119(16):161101, 2017.

\bibitem{LIGOScientific:2018dkp}
B.~P. Abbott et~al.
\newblock {Tests of General Relativity with GW170817}.
\newblock {\em Phys. Rev. Lett.}, 123(1):011102, 2019.

\bibitem{Pretorius:2005gq}
Frans Pretorius.
\newblock {Evolution of binary black hole spacetimes}.
\newblock {\em Phys. Rev. Lett.}, 95:121101, 2005.

\bibitem{LIGOScientific:2021sio}
R.~Abbott et~al.
\newblock {Tests of General Relativity with GWTC-3}.
\newblock 12 2021.

\bibitem{Lovelock:1971yv}
D.~Lovelock.
\newblock {The Einstein tensor and its generalizations}.
\newblock {\em J. Math. Phys.}, 12:498--501, 1971.

\bibitem{Horndeski:1974wa}
Gregory~Walter Horndeski.
\newblock {Second-order scalar-tensor field equations in a four-dimensional
  space}.
\newblock {\em Int. J. Theor. Phys.}, 10:363--384, 1974.

\bibitem{Richards:2023xsr}
Chloe Richards, Alexandru Dima, and Helvi Witek.
\newblock {Black holes in massive dynamical Chern-Simons gravity: Scalar hair
  and quasibound states at decoupling}.
\newblock {\em Phys. Rev. D}, 108(4):044078, 2023.

\bibitem{R:2022tqa}
Abhishek Hegade~K. R, Elias~R. Most, Jorge Noronha, Helvi Witek, and Nicol\'as
  Yunes.
\newblock {How do axisymmetric black holes grow monopole and dipole hair?}
\newblock {\em Phys. Rev. D}, 107(10):104047, 2023.

\bibitem{Okounkova:2022grv}
Maria Okounkova, Maximiliano Isi, Katerina Chatziioannou, and Will~M. Farr.
\newblock {Gravitational wave inference on a numerical-relativity simulation of
  a black hole merger beyond general relativity}.
\newblock {\em Phys. Rev. D}, 107(2):024046, 2023.

\bibitem{Elley:2022ept}
Matthew Elley, Hector~O. Silva, Helvi Witek, and Nicol\'as Yunes.
\newblock {Spin-induced dynamical scalarization, descalarization, and
  stealthness in scalar-Gauss-Bonnet gravity during a black hole coalescence}.
\newblock {\em Phys. Rev. D}, 106(4):044018, 2022.

\bibitem{Doneva:2022byd}
Daniela~D. Doneva, Alex Va\~n\'o Vi\~nuales, and Stoytcho~S. Yazadjiev.
\newblock {Dynamical descalarization with a jump during a black hole merger}.
\newblock {\em Phys. Rev. D}, 106(6):L061502, 2022.

\bibitem{Okounkova:2020rqw}
Maria Okounkova.
\newblock {Numerical relativity simulation of GW150914 in Einstein dilaton
  Gauss-Bonnet gravity}.
\newblock {\em Phys. Rev. D}, 102(8):084046, 2020.

\bibitem{Silva:2020omi}
Hector~O. Silva, Helvi Witek, Matthew Elley, and Nicol\'as Yunes.
\newblock {Dynamical Descalarization in Binary Black Hole Mergers}.
\newblock {\em Phys. Rev. Lett.}, 127(3):031101, 2021.

\bibitem{Okounkova:2019zjf}
Maria Okounkova, Leo~C. Stein, Jordan Moxon, Mark~A. Scheel, and Saul~A.
  Teukolsky.
\newblock {Numerical relativity simulation of GW150914 beyond general
  relativity}.
\newblock {\em Phys. Rev. D}, 101(10):104016, 2020.

\bibitem{Okounkova:2019dfo}
Maria Okounkova, Leo~C. Stein, Mark~A. Scheel, and Saul~A. Teukolsky.
\newblock {Numerical binary black hole collisions in dynamical Chern-Simons
  gravity}.
\newblock {\em Phys. Rev. D}, 100(10):104026, 2019.

\bibitem{Witek:2018dmd}
Helvi Witek, Leonardo Gualtieri, Paolo Pani, and Thomas~P. Sotiriou.
\newblock {Black holes and binary mergers in scalar Gauss-Bonnet gravity:
  scalar field dynamics}.
\newblock {\em Phys. Rev. D}, 99(6):064035, 2019.

\bibitem{Evstafyeva:2022rve}
Tamara Evstafyeva, Michalis Agathos, and Justin~L. Ripley.
\newblock {Measuring the ringdown scalar polarization of gravitational waves in
  Einstein-scalar-Gauss-Bonnet gravity}.
\newblock {\em Phys. Rev. D}, 107(12):124010, 2023.

\bibitem{Franchini:2022ukz}
Nicola Franchini, Miguel Bezares, Enrico Barausse, and Luis Lehner.
\newblock {Fixing the dynamical evolution in scalar-Gauss-Bonnet gravity}.
\newblock {\em Phys. Rev. D}, 106(6):064061, 2022.

\bibitem{Cayuso:2023aht}
Ramiro Cayuso, Pau Figueras, Tiago Fran\c{c}a, and Luis Lehner.
\newblock {Modelling self-consistently beyond General Relativity}.
\newblock {\em Phys. Rev. Lett.}, 131:111403, 3 2023.

\bibitem{Lara:2024rwa}
Guillermo Lara, Harald~P. Pfeiffer, Nikolas~A. Wittek, Nils~L. Vu, Kyle~C.
  Nelli, Alexander Carpenter, Geoffrey Lovelace, Mark~A. Scheel, and William
  Throwe.
\newblock {Scalarization of isolated black holes in scalar Gauss-Bonnet theory
  in the fixing-the-equations approach}.
\newblock 3 2024.

\bibitem{Liu:2022fxy}
Yunqi Liu, Cheng-Yong Zhang, Qian Chen, Zhoujian Cao, Yu~Tian, and Bin Wang.
\newblock {Critical scalarization and descalarization of black holes in a
  generalized scalar-tensor theory}.
\newblock {\em Sci. China Phys. Mech. Astron.}, 66(10):100412, 2023.

\bibitem{Thaalba:2023fmq}
Farid Thaalba, Miguel Bezares, Nicola Franchini, and Thomas~P. Sotiriou.
\newblock {Spherical collapse in scalar-Gauss-Bonnet gravity: Taming
  ill-posedness with a Ricci coupling}.
\newblock {\em Phys. Rev. D}, 109(4):L041503, 2024.

\bibitem{Kovacs:2020pns}
\'Aron~D. Kov\'acs and Harvey~S. Reall.
\newblock {Well-Posed Formulation of Scalar-Tensor Effective Field Theory}.
\newblock {\em Phys. Rev. Lett.}, 124(22):221101, 2020.

\bibitem{Kovacs:2020ywu}
\'Aron~D. Kov\'acs and Harvey~S. Reall.
\newblock {Well-posed formulation of Lovelock and Horndeski theories}.
\newblock {\em Phys. Rev. D}, 101(12):124003, 2020.

\bibitem{East:2020hgw}
William~E. East and Justin~L. Ripley.
\newblock {Evolution of Einstein-scalar-Gauss-Bonnet gravity using a modified
  harmonic formulation}.
\newblock {\em Phys. Rev. D}, 103(4):044040, 2021.

\bibitem{East:2021bqk}
William~E. East and Justin~L. Ripley.
\newblock {Dynamics of Spontaneous Black Hole Scalarization and Mergers in
  Einstein-Scalar-Gauss-Bonnet Gravity}.
\newblock {\em Phys. Rev. Lett.}, 127(10):101102, 2021.

\bibitem{East:2022rqi}
William~E. East and Frans Pretorius.
\newblock {Binary neutron star mergers in Einstein-scalar-Gauss-Bonnet
  gravity}.
\newblock {\em Phys. Rev. D}, 106(10):104055, 2022.

\bibitem{Corman:2022xqg}
Maxence Corman, Justin~L. Ripley, and William~E. East.
\newblock {Nonlinear studies of binary black hole mergers in
  Einstein-scalar-Gauss-Bonnet gravity}.
\newblock {\em Phys. Rev. D}, 107(2):024014, 2023.

\bibitem{Nakamura:1987zz}
T.~Nakamura, K.~Oohara, and Y.~Kojima.
\newblock {General Relativistic Collapse to Black Holes and Gravitational Waves
  from Black Holes}.
\newblock {\em Prog. Theor. Phys. Suppl.}, 90:1--218, 1987.

\bibitem{Shibata:1995we}
Masaru Shibata and Takashi Nakamura.
\newblock {Evolution of three-dimensional gravitational waves: Harmonic slicing
  case}.
\newblock {\em Phys. Rev. D}, 52:5428--5444, 1995.

\bibitem{Baumgarte:1998te}
Thomas~W. Baumgarte and Stuart~L. Shapiro.
\newblock {On the numerical integration of Einstein's field equations}.
\newblock {\em Phys. Rev. D}, 59:024007, 1998.

\bibitem{Bona:2003fj}
C.~Bona, T.~Ledvinka, C.~Palenzuela, and M.~Zacek.
\newblock {General covariant evolution formalism for numerical relativity}.
\newblock {\em Phys. Rev. D}, 67:104005, 2003.

\bibitem{Bernuzzi:2009ex}
Sebastiano Bernuzzi and David Hilditch.
\newblock {Constraint violation in free evolution schemes: Comparing BSSNOK
  with a conformal decomposition of Z4}.
\newblock {\em Phys. Rev. D}, 81:084003, 2010.

\bibitem{Alic:2011gg}
Daniela Alic, Carles Bona-Casas, Carles Bona, Luciano Rezzolla, and Carlos
  Palenzuela.
\newblock {Conformal and covariant formulation of the Z4 system with
  constraint-violation damping}.
\newblock {\em Phys. Rev. D}, 85:064040, 2012.

\bibitem{Campanelli:2005dd}
Manuela Campanelli, C.~O. Lousto, P.~Marronetti, and Y.~Zlochower.
\newblock {Accurate evolutions of orbiting black-hole binaries without
  excision}.
\newblock {\em Phys. Rev. Lett.}, 96:111101, 2006.

\bibitem{Baker:2005vv}
John~G. Baker, Joan Centrella, Dae-Il Choi, Michael Koppitz, and James van
  Meter.
\newblock {Gravitational wave extraction from an inspiraling configuration of
  merging black holes}.
\newblock {\em Phys. Rev. Lett.}, 96:111102, 2006.

\bibitem{AresteSalo:2022hua}
Llibert Arest\'e~Sal\'o, Katy Clough, and Pau Figueras.
\newblock {Well-Posedness of the Four-Derivative Scalar-Tensor Theory of
  Gravity in Singularity Avoiding Coordinates}.
\newblock {\em Phys. Rev. Lett.}, 129(26):261104, 2022.

\bibitem{AresteSalo:2023mmd}
Llibert Arest\'e~Sal\'o, Katy Clough, and Pau Figueras.
\newblock {Puncture gauge formulation for Einstein-Gauss-Bonnet gravity and
  four-derivative scalar-tensor theories in d+1 spacetime dimensions}.
\newblock {\em Phys. Rev. D}, 108(8):084018, 2023.

\bibitem{Doneva:2023oww}
Daniela~D. Doneva, Llibert Arest\'e~Sal\'o, Katy Clough, Pau Figueras, and
  Stoytcho~S. Yazadjiev.
\newblock {Testing the limits of scalar-Gauss-Bonnet gravity through nonlinear
  evolutions of spin-induced scalarization}.
\newblock {\em Phys. Rev. D}, 108(8):084017, 2023.

\bibitem{AresteSalo:2023hcp}
Llibert Arest\'e~Sal\'o, Sam~E. Brady, Katy Clough, Daniela Doneva, Tamara
  Evstafyeva, Pau Figueras, Tiago Fran\c{c}a, Lorenzo Rossi, and Shunhui Yao.
\newblock {GRFolres: A code for modified gravity simulations in strong
  gravity}.
\newblock {\em J. Open Source Softw.}, 9(98):6369, 2024.

\bibitem{Wald:1984rg}
Robert~M. Wald.
\newblock {\em {General Relativity}}.
\newblock Chicago Univ. Pr., Chicago, USA, 1984.

\bibitem{Lanczos:1938sf}
Cornelius Lanczos.
\newblock {A Remarkable property of the Riemann-Christoffel tensor in four
  dimensions}.
\newblock {\em Annals Math.}, 39:842--850, 1938.

\bibitem{Lovelock:1972vz}
D.~Lovelock.
\newblock {The four-dimensionality of space and the einstein tensor}.
\newblock {\em J. Math. Phys.}, 13:874--876, 1972.

\bibitem{Ezquiaga:2018btd}
Jose~Mar\'\i{}a Ezquiaga and Miguel Zumalac\'arregui.
\newblock {Dark Energy in light of Multi-Messenger Gravitational-Wave
  astronomy}.
\newblock {\em Front. Astron. Space Sci.}, 5:44, 2018.

\bibitem{Burgess:2014lwa}
C.~P. Burgess and M.~Williams.
\newblock {Who You Gonna Call? Runaway Ghosts, Higher Derivatives and
  Time-Dependence in EFTs}.
\newblock {\em JHEP}, 08:074, 2014.

\bibitem{Solomon:2017nlh}
Adam~R. Solomon and Mark Trodden.
\newblock {Higher-derivative operators and effective field theory for general
  scalar-tensor theories}.
\newblock {\em JCAP}, 02:031, 2018.

\bibitem{Allwright:2018rut}
Gwyneth Allwright and Luis Lehner.
\newblock {Towards the nonlinear regime in extensions to GR: assessing possible
  options}.
\newblock {\em Class. Quant. Grav.}, 36(8):084001, 2019.

\bibitem{Reall:2021ebq}
Harvey~S. Reall and Claude~M. Warnick.
\newblock {Effective field theory and classical equations of motion}.
\newblock {\em J. Math. Phys.}, 63(4):042901, 2022.

\bibitem{Woodard:2015zca}
Richard~P. Woodard.
\newblock {Ostrogradsky's theorem on Hamiltonian instability}.
\newblock {\em Scholarpedia}, 10(8):32243, 2015.

\bibitem{Motohashi:2014opa}
Hayato Motohashi and Teruaki Suyama.
\newblock {Third order equations of motion and the Ostrogradsky instability}.
\newblock {\em Phys. Rev. D}, 91(8):085009, 2015.

\bibitem{Donoghue:2021eto}
John~F. Donoghue and Gabriel Menezes.
\newblock {Ostrogradsky instability can be overcome by quantum physics}.
\newblock {\em Phys. Rev. D}, 104(4):045010, 2021.

\bibitem{Camanho:2014apa}
Xian~O. Camanho, Jose~D. Edelstein, Juan Maldacena, and Alexander Zhiboedov.
\newblock {Causality Constraints on Corrections to the Graviton Three-Point
  Coupling}.
\newblock {\em JHEP}, 02:020, 2016.

\bibitem{Bueno:2024fzg}
Pablo Bueno, Pablo~A. Cano, and Robie~A. Hennigar.
\newblock {Kasner Eons}.
\newblock 2 2024.

\bibitem{Garraffo:2008hu}
Cecilia Garraffo and Gaston Giribet.
\newblock {The Lovelock Black Holes}.
\newblock {\em Mod. Phys. Lett. A}, 23:1801--1818, 2008.

\bibitem{Boulware:1985wk}
David~G. Boulware and Stanley Deser.
\newblock {String Generated Gravity Models}.
\newblock {\em Phys. Rev. Lett.}, 55:2656, 1985.

\bibitem{Reall:2014pwa}
Harvey Reall, Norihiro Tanahashi, and Benson Way.
\newblock {Causality and Hyperbolicity of Lovelock Theories}.
\newblock {\em Class. Quant. Grav.}, 31:205005, 2014.

\bibitem{Reall:2021voz}
Harvey~S. Reall.
\newblock {Causality in gravitational theories with second order equations of
  motion}.
\newblock {\em Phys. Rev. D}, 103(8):084027, 2021.

\bibitem{Burgess:2003jk}
C.~P. Burgess.
\newblock {Quantum gravity in everyday life: General relativity as an effective
  field theory}.
\newblock {\em Living Rev. Rel.}, 7:5--56, 2004.

\bibitem{Baker:2019gxo}
Tessa Baker et~al.
\newblock {Novel Probes Project: Tests of gravity on astrophysical scales}.
\newblock {\em Rev. Mod. Phys.}, 93(1):015003, 2021.

\bibitem{Baker:2020apq}
Tessa Baker and Ian Harrison.
\newblock {Constraining Scalar-Tensor Modified Gravity with Gravitational Waves
  and Large Scale Structure Surveys}.
\newblock {\em JCAP}, 01:068, 2021.

\bibitem{Bellini:2015xja}
Emilio Bellini, Antonio~J. Cuesta, Raul Jimenez, and Licia Verde.
\newblock {Constraints on deviations from \ensuremath{\Lambda}CDM within
  Horndeski gravity}.
\newblock {\em JCAP}, 02:053, 2016.
\newblock [Erratum: JCAP 06, E01 (2016)].

\bibitem{Kreisch:2017uet}
C.~D. Kreisch and E.~Komatsu.
\newblock {Cosmological Constraints on Horndeski Gravity in Light of GW170817}.
\newblock {\em JCAP}, 12:030, 2018.

\bibitem{Noller:2018wyv}
Johannes Noller and Andrina Nicola.
\newblock {Cosmological parameter constraints for Horndeski scalar-tensor
  gravity}.
\newblock {\em Phys. Rev. D}, 99(10):103502, 2019.

\bibitem{SpurioMancini:2019rxy}
A.~Spurio~Mancini, F.~K\"ohlinger, B.~Joachimi, V.~Pettorino, B.~M. Sch\"afer,
  R.~Reischke, Edo van Uitert, S.~Brieden, M.~Archidiacono, and J.~Lesgourgues.
\newblock {KiDS + GAMA: constraints on horndeski gravity from combined
  large-scale structure probes}.
\newblock {\em Mon. Not. Roy. Astron. Soc.}, 490(2):2155--2177, 2019.

\bibitem{LIGOScientific:2017ync}
B.~P. Abbott et~al.
\newblock {Multi-messenger Observations of a Binary Neutron Star Merger}.
\newblock {\em Astrophys. J. Lett.}, 848(2):L12, 2017.

\bibitem{Baker:2017hug}
T.~Baker, E.~Bellini, P.~G. Ferreira, M.~Lagos, J.~Noller, and I.~Sawicki.
\newblock {Strong constraints on cosmological gravity from GW170817 and GRB
  170817A}.
\newblock {\em Phys. Rev. Lett.}, 119(25):251301, 2017.

\bibitem{Ezquiaga:2017ekz}
Jose~Mar\'\i{}a Ezquiaga and Miguel Zumalac\'arregui.
\newblock {Dark Energy After GW170817: Dead Ends and the Road Ahead}.
\newblock {\em Phys. Rev. Lett.}, 119(25):251304, 2017.

\bibitem{Creminelli:2017sry}
Paolo Creminelli and Filippo Vernizzi.
\newblock {Dark Energy after GW170817 and GRB170817A}.
\newblock {\em Phys. Rev. Lett.}, 119(25):251302, 2017.

\bibitem{Sakstein:2017xjx}
Jeremy Sakstein and Bhuvnesh Jain.
\newblock {Implications of the Neutron Star Merger GW170817 for Cosmological
  Scalar-Tensor Theories}.
\newblock {\em Phys. Rev. Lett.}, 119(25):251303, 2017.

\bibitem{Boran:2017rdn}
S.~Boran, S.~Desai, E.~O. Kahya, and R.~P. Woodard.
\newblock {GW170817 Falsifies Dark Matter Emulators}.
\newblock {\em Phys. Rev. D}, 97(4):041501, 2018.

\bibitem{Mastrogiovanni:2020gua}
S.~Mastrogiovanni, D.~Steer, and M.~Barsuglia.
\newblock {Probing modified gravity theories and cosmology using
  gravitational-waves and associated electromagnetic counterparts}.
\newblock {\em Phys. Rev. D}, 102(4):044009, 2020.

\bibitem{deRham:2018red}
Claudia de~Rham and Scott Melville.
\newblock {Gravitational Rainbows: LIGO and Dark Energy at its Cutoff}.
\newblock {\em Phys. Rev. Lett.}, 121(22):221101, 2018.

\bibitem{Kobayashi:2011nu}
Tsutomu Kobayashi, Masahide Yamaguchi, and Jun'ichi Yokoyama.
\newblock {Generalized G-inflation: Inflation with the most general
  second-order field equations}.
\newblock {\em Prog. Theor. Phys.}, 126:511--529, 2011.

\bibitem{Weinberg:2008hq}
Steven Weinberg.
\newblock {Effective Field Theory for Inflation}.
\newblock {\em Phys. Rev. D}, 77:123541, 2008.

\bibitem{Lyu:2022gdr}
Zhenwei Lyu, Nan Jiang, and Kent Yagi.
\newblock {Constraints on Einstein-dilation-Gauss-Bonnet gravity from black
  hole-neutron star gravitational wave events}.
\newblock {\em Phys. Rev. D}, 105(6):064001, 2022.
\newblock [Erratum: Phys.Rev.D 106, 069901 (2022), Erratum: Phys.Rev.D 106,
  069901 (2022)].

\bibitem{Bezares:2020wkn}
Miguel Bezares, Marco Crisostomi, Carlos Palenzuela, and Enrico Barausse.
\newblock {K-dynamics: well-posed 1+1 evolutions in K-essence}.
\newblock {\em JCAP}, 03:072, 2021.

\bibitem{Lara:2022gof}
Guillermo Lara, Miguel Bezares, Marco Crisostomi, and Enrico Barausse.
\newblock {Robustness of kinetic screening against matter coupling}.
\newblock {\em Phys. Rev. D}, 107(4):044019, 2023.

\bibitem{Boskovic:2023dqk}
Mateja Bo\v{s}kovi\'c and Enrico Barausse.
\newblock {Two-body problem in theories with kinetic screening}.
\newblock {\em Phys. Rev. D}, 108(6):064033, 2023.

\bibitem{Will:1993hxu}
C.~M. Will.
\newblock {\em {Theory and Experiment in Gravitational Physics}}.
\newblock Cambridge University Press, 1993.

\bibitem{Will:2014kxa}
Clifford~M. Will.
\newblock {The Confrontation between General Relativity and Experiment}.
\newblock {\em Living Rev. Rel.}, 17:4, 2014.

\bibitem{Nicolis:2008in}
Alberto Nicolis, Riccardo Rattazzi, and Enrico Trincherini.
\newblock {The Galileon as a local modification of gravity}.
\newblock {\em Phys. Rev. D}, 79:064036, 2009.

\bibitem{Brans:1961sx}
C.~Brans and R.~H. Dicke.
\newblock {Mach's principle and a relativistic theory of gravitation}.
\newblock {\em Phys. Rev.}, 124:925--935, 1961.

\bibitem{Damour:1992kf}
Thibault Damour and Kenneth Nordtvedt.
\newblock {General relativity as a cosmological attractor of tensor scalar
  theories}.
\newblock {\em Phys. Rev. Lett.}, 70:2217--2219, 1993.

\bibitem{Damour:1993hw}
Thibault Damour and Gilles Esposito-Farese.
\newblock {Nonperturbative strong field effects in tensor - scalar theories of
  gravitation}.
\newblock {\em Phys. Rev. Lett.}, 70:2220--2223, 1993.

\bibitem{Franchini:2019npi}
Nicola Franchini and Thomas~P. Sotiriou.
\newblock {Cosmology with subdominant Horndeski scalar field}.
\newblock {\em Phys. Rev. D}, 101(6):064068, 2020.

\bibitem{Doneva:2022ewd}
Daniela~D. Doneva, Fethi~M. Ramazano\u{g}lu, Hector~O. Silva, Thomas~P.
  Sotiriou, and Stoytcho~S. Yazadjiev.
\newblock {Spontaneous scalarization}.
\newblock {\em Rev. Mod. Phys.}, 96(1):015004, 2024.

\bibitem{Glavan:2019inb}
Dra\v{z}en Glavan and Chunshan Lin.
\newblock {Einstein-Gauss-Bonnet Gravity in Four-Dimensional Spacetime}.
\newblock {\em Phys. Rev. Lett.}, 124(8):081301, 2020.

\bibitem{Fernandes:2021ysi}
Pedro G.~S. Fernandes, Pedro Carrilho, Timothy Clifton, and David~J. Mulryne.
\newblock {Black holes in the scalar-tensor formulation of 4D
  Einstein-Gauss-Bonnet gravity: Uniqueness of solutions, and a new candidate
  for dark matter}.
\newblock {\em Phys. Rev. D}, 104(4):044029, 2021.

\bibitem{Fernandes:2022zrq}
Pedro G.~S. Fernandes, Pedro Carrilho, Timothy Clifton, and David~J. Mulryne.
\newblock {The 4D Einstein\textendash{}Gauss\textendash{}Bonnet theory of
  gravity: a review}.
\newblock {\em Class. Quant. Grav.}, 39(6):063001, 2022.

\bibitem{Eling:2004dk}
Christopher Eling, Ted Jacobson, and David Mattingly.
\newblock {Einstein-Aether theory}.
\newblock In {\em {Deserfest: A Celebration of the Life and Works of Stanley
  Deser}}, pages 163--179, 10 2004.

\bibitem{Jacobson:2007veq}
Ted Jacobson.
\newblock {Einstein-aether gravity: A Status report}.
\newblock {\em PoS}, QG-PH:020, 2007.

\bibitem{Mattingly:2005re}
David Mattingly.
\newblock {Modern tests of Lorentz invariance}.
\newblock {\em Living Rev. Rel.}, 8:5, 2005.

\bibitem{Gupta:2021vdj}
Toral Gupta, Mario Herrero-Valea, Diego Blas, Enrico Barausse, Neil Cornish,
  Kent Yagi, and Nicol\'as Yunes.
\newblock {New binary pulsar constraints on Einstein-\ae{}ther theory after
  GW170817}.
\newblock {\em Class. Quant. Grav.}, 38(19):195003, 2021.

\bibitem{Horava:2008ih}
Petr Ho\v{r}ava.
\newblock {Membranes at Quantum Criticality}.
\newblock {\em JHEP}, 03:020, 2009.

\bibitem{Horava:2009uw}
Petr Ho\v{r}ava.
\newblock {Quantum Gravity at a Lifshitz Point}.
\newblock {\em Phys. Rev. D}, 79:084008, 2009.

\bibitem{Sotiriou:2010wn}
Thomas~P. Sotiriou.
\newblock {Horava-Lifshitz gravity: a status report}.
\newblock {\em J. Phys. Conf. Ser.}, 283:012034, 2011.

\bibitem{deRham:2010kj}
Claudia de~Rham, Gregory Gabadadze, and Andrew~J. Tolley.
\newblock {Resummation of Massive Gravity}.
\newblock {\em Phys. Rev. Lett.}, 106:231101, 2011.

\bibitem{deRham:2011qq}
Claudia de~Rham, Gregory Gabadadze, and Andrew~J. Tolley.
\newblock {Helicity decomposition of ghost-free massive gravity}.
\newblock {\em JHEP}, 11:093, 2011.

\bibitem{deRham:2023ngf}
Claudia de~Rham, Jan Ko\.zuszek, Andrew~J. Tolley, and Toby Wiseman.
\newblock {Dynamical formulation of ghost-free massive gravity}.
\newblock {\em Phys. Rev. D}, 108(8):084052, 2023.

\bibitem{Stelle:1976gc}
K.~S. Stelle.
\newblock {Renormalization of Higher Derivative Quantum Gravity}.
\newblock {\em Phys. Rev. D}, 16:953--969, 1977.

\bibitem{Held:2021pht}
Aaron Held and Hyun Lim.
\newblock {Nonlinear dynamics of quadratic gravity in spherical symmetry}.
\newblock {\em Phys. Rev. D}, 104(8):084075, 2021.

\bibitem{Held:2023aap}
Aaron Held and Hyun Lim.
\newblock {Nonlinear evolution of quadratic gravity in 3+1 dimensions}.
\newblock {\em Phys. Rev. D}, 108(10):104025, 2023.

\bibitem{Cayuso:2023dei}
Ramiro Cayuso.
\newblock {Gravitational collapse in quadratic gravity}.
\newblock {\em Phys. Rev. D}, 108(12):124066, 2023.

\bibitem{Endlich:2017tqa}
Solomon Endlich, Victor Gorbenko, Junwu Huang, and Leonardo Senatore.
\newblock {An effective formalism for testing extensions to General Relativity
  with gravitational waves}.
\newblock {\em JHEP}, 09:122, 2017.

\bibitem{Cayuso:2017iqc}
Juan Cayuso, N\'estor Ortiz, and Luis Lehner.
\newblock {Fixing extensions to general relativity in the nonlinear regime}.
\newblock {\em Phys. Rev. D}, 96(8):084043, 2017.

\bibitem{Israel:1976efz}
W.~Israel and J.~M. Stewart.
\newblock {Thermodynamics of nonstationary and transient effects in a
  relativistic gas}.
\newblock {\em Phys. Lett. A}, 58(4):213--215, 1976.

\bibitem{Israel:1976tn}
W.~Israel.
\newblock {Nonstationary irreversible thermodynamics: A Causal relativistic
  theory}.
\newblock {\em Annals Phys.}, 100:310--331, 1976.

\bibitem{Muller:1967zza}
Ingo Muller.
\newblock {Zum Paradoxon der Warmeleitungstheorie}.
\newblock {\em Z. Phys.}, 198:329--344, 1967.

\bibitem{Arnowitt:1959ah}
Richard~L. Arnowitt, Stanley Deser, and Charles~W. Misner.
\newblock {Dynamical Structure and Definition of Energy in General Relativity}.
\newblock {\em Phys. Rev.}, 116:1322--1330, 1959.

\bibitem{Baumgarte:2010ndz}
Thomas~W. Baumgarte and Stuart~L. Shapiro.
\newblock {\em {Numerical Relativity: Solving Einstein's Equations on the
  Computer}}.
\newblock Cambridge University Press, 2010.

\bibitem{Alcubierre:2008}
Miguel Alcubierre.
\newblock {\em {Introduction to 3+1 Numerical Relativity}}.
\newblock Oxford University Press, 04 2008.

\bibitem{Gundlach:2005eh}
Carsten Gundlach, Jos\'e~M. Mart\'in-Garc\'ia, Gioel Calabrese, and Ian Hinder.
\newblock {Constraint damping in the Z4 formulation and harmonic gauge}.
\newblock {\em Class. Quant. Grav.}, 22:3767--3774, 2005.

\bibitem{Bona:2003qn}
C.~Bona, T.~Ledvinka, C.~Palenzuela, and M.~Zacek.
\newblock {A Symmetry breaking mechanism for the Z4 general covariant evolution
  system}.
\newblock {\em Phys. Rev. D}, 69:064036, 2004.

\bibitem{Alic:2008pw}
Daniela Alic, Carles Bona, and Carles Bona-Casas.
\newblock {Towards a gauge-polyvalent Numerical Relativity code}.
\newblock {\em Phys. Rev. D}, 79:044026, 2009.

\bibitem{Alic:2013xsa}
Daniela Alic, Wolfgang Kastaun, and Luciano Rezzolla.
\newblock {Constraint damping of the conformal and covariant formulation of the
  Z4 system in simulations of binary neutron stars}.
\newblock {\em Phys. Rev. D}, 88(6):064049, 2013.

\bibitem{Pretorius:2004jg}
Frans Pretorius.
\newblock {Numerical relativity using a generalized harmonic decomposition}.
\newblock {\em Class. Quant. Grav.}, 22:425--452, 2005.

\bibitem{Szilagyi:2009qz}
Bela Szilagyi, Lee Lindblom, and Mark~A. Scheel.
\newblock {Simulations of Binary Black Hole Mergers Using Spectral Methods}.
\newblock {\em Phys. Rev. D}, 80:124010, 2009.

\bibitem{Brown:2011qg}
J.~David Brown.
\newblock {Generalized Harmonic Equations in 3+1 Form}.
\newblock {\em Phys. Rev. D}, 84:124012, 2011.

\bibitem{Bona:1994dr}
Carles Bona, Joan Mass\'o, Edward Seidel, and Joan Stela.
\newblock {A New formalism for numerical relativity}.
\newblock {\em Phys. Rev. Lett.}, 75:600--603, 1995.

\bibitem{Hannam:2008sg}
Mark Hannam, Sascha Husa, Frank Ohme, Bernd Br{\"u}gmann, and Niall
  O'Murchadha.
\newblock {Wormholes and trumpets: The Schwarzschild spacetime for the
  moving-puncture generation}.
\newblock {\em Phys. Rev. D}, 78:064020, 2008.

\bibitem{Figueras:2015hkb}
Pau Figueras, Markus Kunesch, and Saran Tunyasuvunakool.
\newblock {End Point of Black Ring Instabilities and the Weak Cosmic Censorship
  Conjecture}.
\newblock {\em Phys. Rev. Lett.}, 116(7):071102, 2016.

\bibitem{Figueras:2017zwa}
Pau Figueras, Markus Kunesch, Luis Lehner, and Saran Tunyasuvunakool.
\newblock {End Point of the Ultraspinning Instability and Violation of Cosmic
  Censorship}.
\newblock {\em Phys. Rev. Lett.}, 118(15):151103, 2017.

\bibitem{Alcubierre:2002kk}
Miguel Alcubierre, Bernd Br{\"u}gmann, Peter Diener, Michael Koppitz, Denis
  Pollney, Edward Seidel, and Ryoji Takahashi.
\newblock {Gauge conditions for long term numerical black hole evolutions
  without excision}.
\newblock {\em Phys. Rev. D}, 67:084023, 2003.

\bibitem{Shibata:2015}
Masaru Shibata.
\newblock {\em Numerical Relativity}.
\newblock World Scientific, 2015.

\bibitem{Baumgarte:2006ug}
Thomas~W. Baumgarte, Niall~O Murchadha, and Harald~P. Pfeiffer.
\newblock {The Einstein constraints: Uniqueness and non-uniqueness in the
  conformal thin sandwich approach}.
\newblock {\em Phys. Rev. D}, 75:044009, 2007.

\bibitem{Aurrekoetxea:2022mpw}
Josu~C. Aurrekoetxea, Katy Clough, and Eugene~A. Lim.
\newblock {CTTK: a new method to solve the initial data constraints in
  numerical relativity}.
\newblock {\em Class. Quant. Grav.}, 40(7):075003, 2023.

\bibitem{Brady:2023dgu}
Sam~E. Brady, Llibert Arest\'e~Sal\'o, Katy Clough, Pau Figueras, and
  Annamalai~P. S.
\newblock {Solving the initial conditions problem for modified gravity
  theories}.
\newblock {\em Phys. Rev. D}, 108(10):104022, 2023.

\bibitem{Kreiss2004InitialBoundaryVP}
Heinz-Otto Kreiss and Jens Lorenz.
\newblock {\em Initial-Boundary Value Problems and the Navier-Stokes
  Equations}.
\newblock Society for Industrial and Applied Mathematics, 2004.

\bibitem{Andrade:2021rbd}
Tom\'as Andrade et~al.
\newblock {GRChombo: An adaptable numerical relativity code for fundamental
  physics}.
\newblock {\em J. Open Source Softw.}, 6(68):3703, 2021.

\bibitem{Radia:2021smk}
Miren Radia, Ulrich Sperhake, Amelia Drew, Katy Clough, Pau Figueras, Eugene~A.
  Lim, Justin~L. Ripley, Josu~C. Aurrekoetxea, Tiago Fran\c{c}a, and Thomas
  Helfer.
\newblock {Lessons for adaptive mesh refinement in numerical relativity}.
\newblock {\em Class. Quant. Grav.}, 39(13):135006, 2022.

\bibitem{Croft:2022bxq}
Robin Croft, Thomas Helfer, Bo-Xuan Ge, Miren Radia, Tamara Evstafyeva,
  Eugene~A. Lim, Ulrich Sperhake, and Katy Clough.
\newblock {The gravitational afterglow of boson stars}.
\newblock {\em Class. Quant. Grav.}, 40(6):065001, 2023.

\bibitem{Evstafyeva:2022bpr}
Tamara Evstafyeva, Ulrich Sperhake, Thomas Helfer, Robin Croft, Miren Radia,
  Bo-Xuan Ge, and Eugene~A. Lim.
\newblock {Unequal-mass boson-star binaries: initial data and merger dynamics}.
\newblock {\em Class. Quant. Grav.}, 40(8):085009, 2023.

\bibitem{Aurrekoetxea:2019fhr}
Josu~C. Aurrekoetxea, Katy Clough, Raphael Flauger, and Eugene~A. Lim.
\newblock {The Effects of Potential Shape on Inhomogeneous Inflation}.
\newblock {\em JCAP}, 05:030, 2020.

\bibitem{Aurrekoetxea:2023jwd}
Josu~C. Aurrekoetxea, Katy Clough, and Francesco Muia.
\newblock {Oscillon formation during inflationary preheating with general
  relativity}.
\newblock {\em Phys. Rev. D}, 108(2):023501, 2023.

\bibitem{Bantilan:2019bvf}
Hans Bantilan, Pau Figueras, Markus Kunesch, and Rodrigo Panosso~Macedo.
\newblock {End point of nonaxisymmetric black hole instabilities in higher
  dimensions}.
\newblock {\em Phys. Rev. D}, 100(8):086014, 2019.

\bibitem{Andrade:2020dgc}
Tom\'as Andrade, Pau Figueras, and Ulrich Sperhake.
\newblock {Evidence for violations of Weak Cosmic Censorship in black hole
  collisions in higher dimensions}.
\newblock {\em JHEP}, 03:111, 2022.

\bibitem{deJong:2021bbo}
Eloy de~Jong, Josu~C. Aurrekoetxea, and Eugene~A. Lim.
\newblock {Primordial black hole formation with full numerical relativity}.
\newblock {\em JCAP}, 03(03):029, 2022.

\bibitem{deJong:2023gsx}
Eloy de~Jong, Josu~C. Aurrekoetxea, Eugene~A. Lim, and Tiago Fran\c{c}a.
\newblock {Spinning primordial black holes formed during a matter-dominated
  era}.
\newblock {\em JCAP}, 10:067, 2023.

\bibitem{Helfer:2018qgv}
Thomas Helfer, Josu~C. Aurrekoetxea, and Eugene~A. Lim.
\newblock {Cosmic String Loop Collapse in Full General Relativity}.
\newblock {\em Phys. Rev. D}, 99(10):104028, 2019.

\bibitem{Drew:2022iqz}
Amelia Drew and E.~P.~S. Shellard.
\newblock {Radiation from global topological strings using adaptive mesh
  refinement: Massive modes}.
\newblock {\em Phys. Rev. D}, 107(4):043507, 2023.

\bibitem{Drew:2023ptp}
Amelia Drew, Tomasz Kinowski, and E.~P.~S. Shellard.
\newblock {Axion String Source Modelling}.
\newblock 12 2023.

\bibitem{Papallo:2017qvl}
Giuseppe Papallo and Harvey~S. Reall.
\newblock {On the local well-posedness of Lovelock and Horndeski theories}.
\newblock {\em Phys. Rev. D}, 96(4):044019, 2017.

\bibitem{hinch}
E.~J.~Hinch.
\newblock {\em {Perturbation Methods}}.
\newblock {Cambridge University Press}, 1991.

\bibitem{recipes}
William~H. Press, Brian~P. Flannery, Saul~A. Teukolsky, and William~T.
  Vetterling.
\newblock {\em Numerical Recipes in C: The Art of Scientific Computing}.
\newblock Cambridge University Press, 1992.

\bibitem{Figueras:2020dzx}
Pau Figueras and Tiago Fran\c{c}a.
\newblock {Gravitational Collapse in Cubic Horndeski Theories}.
\newblock {\em Class. Quant. Grav.}, 37(22):225009, 2020.

\bibitem{Figueras:2021abd}
Pau Figueras and Tiago Fran\c{c}a.
\newblock {Black hole binaries in cubic Horndeski theories}.
\newblock {\em Phys. Rev. D}, 105(12):124004, 2022.

\bibitem{Liu:2009al}
Yuk~Tung Liu, Zachariah~B. Etienne, and Stuart~L. Shapiro.
\newblock {Evolution of near-extremal-spin black holes using the moving
  puncture technique}.
\newblock {\em Phys. Rev. D}, 80:121503, 2009.

\bibitem{Bowen:1980yu}
Jeffrey~M. Bowen and James~W. York, Jr.
\newblock {Time asymmetric initial data for black holes and black hole
  collisions}.
\newblock {\em Phys. Rev. D}, 21:2047--2056, 1980.

\bibitem{Ansorg:2004ds}
Marcus Ansorg, Bernd Br{\"u}gmann, and Wolfgang Tichy.
\newblock {A Single-domain spectral method for black hole puncture data}.
\newblock {\em Phys. Rev. D}, 70:064011, 2004.

\bibitem{R:2022hlf}
Abhishek Hegade~K. R, Justin~L. Ripley, and Nicol\'as Yunes.
\newblock {Where and why does Einstein-scalar-Gauss-Bonnet theory break down?}
\newblock {\em Phys. Rev. D}, 107(4):044044, 2023.

\bibitem{unequal}
Daniela~D. Doneva, Llibert Arest\'e~Sal\'o, Katy Clough, Pau Figueras, and
  Stoytcho~S. Yazadjiev.
\newblock {Unequal-mass binaries in the Four-Derivative Scalar-Tensor theory of
  gravity}.
\newblock 2024.
\newblock in preparation.

\bibitem{Shiralilou:2021mfl}
Banafsheh Shiralilou, Tanja Hinderer, Samaya~M. Nissanke, N\'estor Ortiz, and
  Helvi Witek.
\newblock {Post-Newtonian gravitational and scalar waves in
  scalar-Gauss\textendash{}Bonnet gravity}.
\newblock {\em Class. Quant. Grav.}, 39(3):035002, 2022.

\bibitem{Doneva:2022yqu}
Daniela~D. Doneva, Lucas~G. Collodel, and Stoytcho~S. Yazadjiev.
\newblock {Spontaneous nonlinear scalarization of Kerr black holes}.
\newblock {\em Phys. Rev. D}, 106(10):104027, 2022.

\bibitem{Bernard:2019fjb}
Laura Bernard, Luis Lehner, and Raimon Luna.
\newblock {Challenges to global solutions in Horndeski\textquoteright{}s
  theory}.
\newblock {\em Phys. Rev. D}, 100(2):024011, 2019.

\bibitem{Barausse:2022rvg}
Enrico Barausse, Miguel Bezares, Marco Crisostomi, and Guillermo Lara.
\newblock {The well-posedness of the Cauchy problem for self-interacting vector
  fields}.
\newblock {\em JCAP}, 11:050, 2022.

\bibitem{Kuan:2021lol}
Hao-Jui Kuan, Daniela~D. Doneva, and Stoytcho~S. Yazadjiev.
\newblock {Dynamical Formation of Scalarized Black Holes and Neutron Stars
  through Stellar Core Collapse}.
\newblock {\em Phys. Rev. Lett.}, 127(16):161103, 2021.

\bibitem{Corelli:2022phw}
Fabrizio Corelli, Marina De~Amicis, Taishi Ikeda, and Paolo Pani.
\newblock {Nonperturbative gedanken experiments in
  Einstein-dilaton-Gauss-Bonnet gravity: Nonlinear transitions and tests of the
  cosmic censorship beyond general relativity}.
\newblock {\em Phys. Rev. D}, 107(4):044061, 2023.

\bibitem{Ripley:2019hxt}
Justin~L. Ripley and Frans Pretorius.
\newblock {Hyperbolicity in Spherical Gravitational Collapse in a Horndeski
  Theory}.
\newblock {\em Phys. Rev. D}, 99(8):084014, 2019.

\bibitem{Ripley:2020vpk}
Justin~L. Ripley and Frans Pretorius.
\newblock {Dynamics of a $\mathbb Z_2$ symmetric EdGB gravity in spherical
  symmetry}.
\newblock {\em Class. Quant. Grav.}, 37(15):155003, 2020.

\bibitem{Lara:2021piy}
Guillermo Lara, Miguel Bezares, and Enrico Barausse.
\newblock {UV completions, fixing the equations, and nonlinearities in
  k-essence}.
\newblock {\em Phys. Rev. D}, 105(6):064058, 2022.

\bibitem{Yunes:2013dva}
Nicol\'as Yunes and Xavier Siemens.
\newblock {Gravitational-Wave Tests of General Relativity with Ground-Based
  Detectors and Pulsar Timing-Arrays}.
\newblock {\em Living Rev. Rel.}, 16:9, 2013.

\bibitem{Berti:2015itd}
Emanuele Berti et~al.
\newblock {Testing General Relativity with Present and Future Astrophysical
  Observations}.
\newblock {\em Class. Quant. Grav.}, 32:243001, 2015.

\bibitem{Clough:2016ymm}
Katy Clough, Eugene~A. Lim, Brandon~S. DiNunno, Willy Fischler, Raphael
  Flauger, and Sonia Paban.
\newblock {Robustness of Inflation to Inhomogeneous Initial Conditions}.
\newblock {\em JCAP}, 09:025, 2017.

\bibitem{Clough:2017efm}
Katy Clough, Raphael Flauger, and Eugene~A. Lim.
\newblock {Robustness of Inflation to Large Tensor Perturbations}.
\newblock {\em JCAP}, 05:065, 2018.

\bibitem{Figueras:2022zkg}
Pau Figueras, Tiago Fran\c{c}a, Chenxia Gu, and Tom\'as Andrade.
\newblock {Endpoint of the Gregory-Laflamme instability of black strings
  revisited}.
\newblock {\em Phys. Rev. D}, 107(4):044028, 2023.

\bibitem{Bantilan:2012vu}
Hans Bantilan, Frans Pretorius, and Steven~S. Gubser.
\newblock {Simulation of Asymptotically AdS5 Spacetimes with a Generalized
  Harmonic Evolution Scheme}.
\newblock {\em Phys. Rev. D}, 85:084038, 2012.

\bibitem{Bantilan:2017kok}
Hans Bantilan, Pau Figueras, Markus Kunesch, and Paul Romatschke.
\newblock {Nonspherically Symmetric Collapse in Asymptotically AdS Spacetimes}.
\newblock {\em Phys. Rev. Lett.}, 119(19):191103, 2017.

\bibitem{Bantilan:2020xas}
Hans Bantilan, Pau Figueras, and Lorenzo Rossi.
\newblock {Cauchy Evolution of Asymptotically Global AdS Spacetimes with No
  Symmetries}.
\newblock {\em Phys. Rev. D}, 103(8):086006, 2021.

\bibitem{Figueras:2023ihz}
Pau Figueras and Lorenzo Rossi.
\newblock {Non-linear instability of slowly rotating Kerr-AdS black holes}.
\newblock 11 2023.

\end{thebibliography}
\bibliographystyle{unsrt}

\end{document}